\documentclass[12pt,preprint]{aastex}

\newcommand{\pinv}{{\small PI}}
\newcommand{\asca}{{\small \it ASCA}}
\newcommand{\sis}{{\small SIS}}
\newcommand{\gis}{{\small GIS}}
\newcommand{\rosat}{{\small \it ROSAT}}

\newcommand{\sax}{{\small \it Beppo-SAX}}
\newcommand{\wfi}{{\small WFI}}
\newcommand{\nfi}{{\small NFI}}
\newcommand{\lecs}{{\small LECS}}
\newcommand{\mecs}{{\small MECS}}
\newcommand{\wfc}{{\small WFC}}
\newcommand{\grbm}{{\small GRBM}}
\newcommand{\asm}{{\small ASM}}
\newcommand{\batse}{{\small BATSE}}
\newcommand{\rxte}{{\small \it RXTE}}
\newcommand{\pca}{{\small PCA}}
\newcommand{\hete}{{\small \it HETE-2}}

\newcommand{\agn}{{\small AGN}}

\newcommand{\fwhm}{{\small FWHM}}

\newcommand{\xmm}{{\it XMM-Newton}}
\newcommand{\rgs}{{\small RGS}}

\newcommand{\epic}{{\small EPIC}}
\newcommand{\ccd}{{\small CCD}}
\newcommand{\mos}{{\small MOS}}
\newcommand{\pn}{{\small PN}}
\newcommand{\om}{{\small OM}}
\newcommand{\sas}{{\small SAS}}
\newcommand{\chandra}{{\it Chandra}}

\newcommand{\hetg}{{\small HETG}}
\newcommand{\heg}{{\small HEG}}
\newcommand{\meg}{{\small MEG}}
\newcommand{\letg}{{\small LETG}}
\newcommand{\leg}{{\small LEG}}

\newcommand{\aciss}{{\small ACIS-S}}

\newcommand{\ciao}{{\small CIAO}}
\newcommand{\nasa}{{\small NASA}}
\newcommand{\xspec}{{\small XSPEC}}
\newcommand{\astroe}{{\small \it ASTRO-E2}}

\newcommand{\swift}{{\it Swift}}
\newcommand{\xrt}{{\small XRT}}
\newcommand{\grb}{{\small GRB}}
\newcommand{\xrf}{{\small XRF}}
\newcommand{\ipn}{{\small IPN}}
\newcommand{\integral}{{\it Integral}}
\newcommand{\ibis}{{\small IBIS}}
\newcommand{\mc}{{\small MC}}

\newcommand{\kev}{ke~\hspace{-0.18cm}V}
\newcommand{\ut}{{\small UT}}
\newcommand{\snr}{{\small SNR}}

\newcommand{\ergflux}{erg~cm$^{-2}$~s$^{-1}$}
\newcommand{\phflux}{photons~cm$^{-2}$~s$^{-1}$}
\newcommand{\ud}[2]{\mbox{$^{+ #1}_{- #2}$}}
\newcommand{\fekal}{\mbox{Fe K$\alpha$}}
\newcommand{\ssig}{\mbox{$\sigma$}}
\newcommand{\sn}{{\small S/N}}

\begin{document}

\title{A Search for Discrete X-ray Spectral Features in a Sample of Bright
       $\gamma$-ray Burst Afterglows}

\author{Masao Sako\altaffilmark{1,2,3,4}, Fiona A. Harrison\altaffilmark{2},
        \& Robert E. Rutledge\altaffilmark{1,5}}

\email{masao@slac.stanford.edu, fiona@srl.caltech.edu,
       rutledge@crust.physics.mcgill.ca}

\altaffiltext{1}{Theoretical Astrophysics,
                 California Institute of Technology,
                 MC 130-33, Pasadena, CA 91125}
\altaffiltext{2}{Space Radiation Laboratory,
                 California Institute of Technology,
                 MC 220-47, Pasadena, CA 91125}
\altaffiltext{3}{Present address:
                 KIPAC/SLAC,
                 2575 Sand Hill Road M/S 29,
                 Menlo Park, CA 94025}
\altaffiltext{4}{\chandra\ Postdoctoral Fellow}
\altaffiltext{5}{Present address:
                 Department of Physics,
                 Rutherford Physics Building,
                 McGill University,
                 3600 University Street,
                 Montreal, Quebec, H3A 2T8, Canada}

\received{}
\revised{}
\accepted{}

\slugcomment{Submitted to The Astrophysical Journal}
\shorttitle{X-ray Lines in GRB afterglows}
\shortauthors{Sako, Harrison, \& Rutledge}

\begin{abstract}

  We present uniform, detailed spectral analyses of $\gamma$-ray burst (\grb)
  X-ray afterglows observed with \asca, \sax, \chandra, and \xmm, and
  critically evaluate the statistical significances of X-ray emission and
  absorption features in these spectra.  The sample consists of 21 X-ray
  afterglow observations up to and including that of \grb040106 with spectra
  of sufficient statistical quality to allow meaningful line searches, chosen
  here somewhat arbitrarily to be detections with more than 100 total (source
  plus background) counts.  This sample includes all nine X-ray afterglows
  with published claims of line detections.  Moderate resolution spectra are
  available for 16 of the 21 sources, and for the remaining five the
  \chandra\ transmission grating spectrometers obtained high-resolution data.
  All of the data are available from the public archive.  We test a simple
  hypothesis in which the observed spectra are produced by a power-law
  continuum model modified by photoelectric absorption by neutral material
  both in our Galaxy and possibly also local to the burst.  As a sample, these
  afterglow spectra are consistent with this relatively simple model.
  However, since the $\chi^2$ statistic is not sensitive to weak and/or
  localized fluctuations, we have performed Monte Carlo simulations to search
  for discrete features and to estimate their significances.  Our analysis
  shows that there are four afterglows (\grb011211, \grb030227, \grb021004,
  and \grb040106) with line-like features that are significant at the $3
  \sigma$ level.  We cautiously note that, in two cases, the features are
  associated with an unusual background feature; in the other two, the
  fractional magnitudes of the lines are small, and comparable to the expected
  level of systematic uncertainty in the spectral response.  In addition, none
  of the statistically significant features are seen in more than one detector
  or spectral order where available.  We conclude that, to date, no credible
  X-ray line feature has been detected in a \grb\ afterglow.  Finally, in a
  majority of cases, we find no evidence for significant absorbing columns
  local to the \grb\ host galaxy, implying there is little evidence from X-ray
  observations that \grb\ preferentially explode in high-density environments.

\end{abstract}


\section{Introduction}
\label{section_intro}

  Detections of emission and absorption features in \grb\ X-ray afterglows
  have been claimed in a number of observations.  Iron emission lines from
  neutral as well as ionized material have been reported in the spectra of
  \grb 970508 \citep{piro99}, \grb 970828 \citep{yoshida99, yoshida01}, \grb
  991216 \citep{piro00}, and \grb 000214 \citep{antonelli00}.
  \citet{reeves02} claim detections of several emission lines from mid-$Z$
  elements in the afterglow spectrum of \grb011211, observed with \xmm.
  Similar features have been reported in a selected time interval of the \xmm\
  spectrum of \grb030227 \citep{watson03}, and in the \chandra\
  high-resolution spectrum of \grb020813 \citep{butler03b}.  \citet{watson02a}
  also argue that the \xmm\ spectra of both \grb001025 and \grb010220 are
  better fit by a thermal emission line model in collisional ionization
  equilibrium compared to a smooth power-law continuum model.  If these
  interpretations are correct, line identifications, measurements of their
  equivalent widths, velocity shifts, and their temporal behavior provide
  extremely useful diagnostics for inferring the physical conditions,
  including the geometry and dynamics of the burst environment, as well as the
  nature of the progenitor star (see, .e.g, \citealt{lazzati99, rees00,
  weth00, paerels00, meszaros01, lazzati02a, lazzati02b, ballantyne01,
  ghisellini02, kallman03, kumar03}; see, also, \citealt{boettcher03} for a
  recent review).

  The statistical significances and, hence, the reality of these detections
  are subject to confirmation.  All of the features observed to date were
  found by the original authors to be significant at the $\sim 3\sigma$ level,
  assuming that the line identifications and their energies are known a priori
  (i.e., single trial), with the exception of an emission line in the
  \grb991216 spectrum, reported to be detected at 4.7$\sigma$ \citep{piro00}.
  In the case of \grb011211, where the detection of multiple emission lines is
  claimed, the significance is estimated to be as high as $99.97 - 99.98$\%
  based on Monte Carlo simulations and $\Delta \chi^2$ tests \citep{reeves03,
    reeves02}.  \citet{borozdin02}, however, criticize the data reduction and
  background subtraction procedures performed by \citet{reeves02}, arguing
  that the single-trial significance of even the strongest feature is only
  3.1$\sigma$.  \citet{rutledge02} criticize the statistical analysis method
  adopted by both \citet{reeves03} and \citet{reeves02}, concluding that the
  lines would not be discovered in a blind search.  Interestingly,
  \citet{watson02b} report that the afterglow spectrum of \grb020322, one of
  the brightest afterglows observed with \xmm, does not show any emission
  lines.  The \chandra\ Low-Energy Transmission Grating (\letg) observation of
  \grb020405 \citep{mirabal03}, as well as the well-exposed High-Energy
  Transmission Grating (\hetg) observation of \grb021004
  \citep{grb021004_chandra, butler03b} also appear not to show evidence for
  any strong discrete spectral features.

  Given the low significance of the individual detections, and the varied
  techniques used to analyze the spectra, it is difficult to assess the
  reality of these features from the literature alone.  The primary problem is
  that the line energies are a priori unknown, and this is often not properly
  accounted for in deriving the true statistical significance.  There is no
  consistent theoretical model which unambiguously predicts the line energies.
  In some cases, features are interpreted as near-neutral Fe K$\alpha$
  fluorescent emission, sometimes as highly-ionized Fe recombination lines,
  and others report thermal line emission from mid-$Z$ elements with
  abundances often adjusted to super-solar values (in the case of thermal
  models) to match observations.  In most cases, arbitrary blueshifts are
  invoked to adjust the closest atomic transition to match the observed energy
  of detected excesses \citep{reeves03,reeves02}.  In cases where the \grb\
  redshift is not known, it is chosen to match line energies to probable
  atomic transitions.  This means that the line energies are, in fact, free
  parameters, and this must be corrected for in deriving the detection
  significance.

  A secondary issue is the statistical method applied to argue for the
  detection of lines.  A large fraction of the claims in the literature are
  based on using the $F$-test.  This test compares the $\chi^2$ values derived
  from two different models (e.g. with and without lines), and uses an
  analytic statistical model to determine if the difference represents an
  unusually good improvement.  A probabilistic confidence determines if the
  improvement implies one model is more likely than the other to be correct.
  However, \citet{protassov02} point out that the analytic $F$-test is not
  applicable to emission line detection, and will give both false positives
  and false negatives.  Claims of line detections based on the $F$-test are
  therefore subject to scrutiny.

  The goal of this paper is to present a comprehensive, uniform data analysis
  of high-quality X-ray afterglow spectra.  Numerous theoretical papers have
  been written on the interpretation and modeling of claimed emission line
  detections, and we do not attempt to validate or dispute specific
  interpretations.  Rather, we perform our analysis independent of any
  particular model.  We present detailed spectral analyses of 14 previously
  published datasets, in addition to 7 observations for which no detailed
  analysis has been published in a refereed journal.

  The paper is organized as follows.  In \S\ref{section_obs}, we present our
  sample and briefly summarize some of the relevant properties and previous
  observational analyses of each of the bursts.  In \S\ref{section_data}, we
  describe the data reduction procedures we adopt for the various instrument
  configurations used in the observations.  Results from the spectral fits are
  presented in \S\ref{section_ana}.  We describe our Monte Carlo simulations
  in \S\ref{section_mc_sims}.  The results are discussed in
  \S\ref{section_mc_results} along with a brief comment on each of the sources
  with reported emission line detections.  Finally, we discuss some of the
  implications of our results in \S\ref{section_summary}.

\section{Observations}
\label{section_obs}

  The sample we have analyzed consists of 21 \grb\ X-ray afterglows up to and
  including \grb 040106, and includes the brightest sources that allow
  meaningful constraints on discrete spectral features, and all sources for
  which the detection of emission lines has been reported.  The source list
  and some of the relevant information are summarized in
  Table~\ref{tbl:sample}.  Below we provide a brief description of the X-ray
  observations and previous analyses of each event, along with other
  outstanding features.

  {\bf \grb970508}.  Initially detected by the \sax\ \wfi\ \citep{cfp+97},
  \grb970508 is the second burst with an optical afterglow detection, and was
  the first event with a measured redshift ($z = 0.835$; \citealt{bloom98}).
  The \sax\ \nfi\ observed the error box starting \ut\ May 9.1375 (0.43 days
  after the burst), detecting a relatively bright ($F_{2 - 10~\rm{\kev}} \sim
  10^{-12}$ erg~cm$^{-2}$~s$^{-1}$) X-ray afterglow.  The \sax\ \nfi\
  continued to monitor the X-ray afterglow, performing a total of four
  observations, the last taking place six days after the burst \citep{piro99}.

  \citet{piro99} reported a narrow line feature at 3.5~\kev, corresponding to
  \ion{Fe}{26} Ly$\alpha$ line emission at the measured host redshift, during
  part of the afterglow.  The afterglow did not decay smoothly, but the X-ray
  lightcurve as measured by the \nfi\ shows a bump, beginning a day after the
  event, lasting about two days.  By breaking the first observation into two
  parts (denoted ``1a'' and ``1b''), prior to and after the ``bump'',
  \citet{piro99} find evidence for line emission in the first segment at
  99.3\% ($2.7\sigma$; using an $F$-test) confidence, assuming the known
  redshift.  No line emission is found in the second part of the pointing, or
  in any of the subsequent observations.

  {\bf \grb970828}.  This event is notable for being the prototype dark
  \grb. A fading X-ray afterglow was discovered, but no associated optical
  transient was found despite extensive followup.  Initially positioned by the
  \rxte\ All Sky Monitor (\asm; \citealt{rws+97}), {\em ASCA} followed up the
  location beginning 1.2 days after the event, detecting a moderately bright
  ($F_{2 - 10~\rm{\kev}} = 4 \times 10^{-13}$ erg~cm$^{-2}$~s$^{-1}$) X-ray
  afterglow \citep{yoshida99}.  Like \grb970508, \grb970828 exhibited
  variability in the X-ray superimposed on the power-law decay.

  Dividing the observation into three segments, denoted A, B, and C,
  \citet{yoshida99} found an excess above a power-law spectrum centered at
  $5.04^{+0.23}_{-0.31}$~\kev\ with a width of $0.31^{+0.38}_{-0.31}$~\kev\ in
  segment B with 98.3\% confidence ($2.4 \sigma$; using an $F$-test), which
  coincided with flaring activity in the afterglow.  Both A and C have spectra
  consistent with an absorbed power law.  Initially \citet{yoshida99}
  identified the line with Fe K$\alpha$ fluorescence emission, however the
  subsequent measurement of the redshift of 0.9578 for the host
  \citep{djorgovski01} places the rest-frame energy at 9.87~\kev, forcing
  \citet{yoshida01} to re-identify the feature with an \ion{Fe}{26}
  recombination continuum with no associated line emission.

  {\bf \grb991216}.  This event was a very bright \batse\ trigger
  \citep{grb991216_batse}.  The \pca\ on \rxte\ scanned the error circle, and
  discovered a bright (5 mCrab) X-ray source identified as the \grb\ afterglow
  \citep{tmm+99}.  The discovery of an associated optical transient
  \citep{umh+99} lead to a redshift measurement ($z = 1.02$;
  \citealt{vrh+99}).  \chandra\ followed up the optical afterglow position
  with the \hetg\ in a 9.65~ksec observation 37 hours after the burst,
  detecting a bright ($F_{2 - 10~\rm{\kev}} = 2 \times 10^{-12}$
  erg~cm$^{-2}$~s$^{-1}$) X-ray afterglow \citep{pgg+99}.

  \citet{piro00} analyzed the \hetg\ spectrum, and found excess emission at
  two energies, which they interpret as due to highly ionized iron.  The most
  prominent feature, which they claim to be significant at the $4.7 \sigma$
  level, lies at $3.49 \pm 0.06$ \kev\ and is much broader than the instrument
  resolution, with a gaussian width of $\sigma = 0.23 \pm 0.07$ \kev.  This
  is, by far, the highest significance claimed for any emission line of any
  \grb.  The second, marginally significant feature lies at $4.4 \pm 0.5$
  \kev\ (99.5\% confidence, or $2.8 \sigma$; $F$-test).  If associated with
  the iron recombination continuum (threshold rest energy at 9.28 \kev), the
  implied redshift is $z = 1.11 \pm 0.11$, and if the lower energy feature is
  associated with H-like Ly$\alpha$ of Fe, the implied $z = 1.0 \pm 0.02$.
  These are both consistent with the measured $z$ of 1.02.

  {\bf \grb000210}.  This is another event categorized as a dark \grb.  The
  \sax\ \wfc\ positioned this burst, which was among the brightest detected by
  that mission \citep{grb000210_grbm}.  Deep optical searches performed
  $\sim16$ hours after the event failed to locate an optical afterglow
  \citep{gjo+00}.  Both the \sax\ \nfi\ as well as \chandra\ \aciss\ performed
  followup observations, and both detected the X-ray afterglow; \sax\
  observations started $\sim$8 hours after the event \citep{grb000210_nfi},
  and a \chandra\ pointing began 21 hours after \citep{grb000210_chandra}.
  Deep optical observations of the \chandra\ position revealed a host galaxy,
  with a measured redshift of 0.8463 \citep{pfg+02}.  A radio afterglow was
  also detected \citep{pfg+02}, and a $\sim2.5\sigma$ sub-mm excess toward the
  \grb000210 has been reported \citep{bck+03}.

  The X-ray afterglow appears unremarkable, following a smooth $t^{-1.38}$
  decay.  Fits to the joint \nfi/\chandra\ data reveal a spectrum consistent
  with an absorbed power law \citep{pfg+02}.

  {\bf \grb000214}.  This burst was detected by the {\em BeppoSAX} \grbm\ and
  positioned by the \wfc\ \citep{grb000214_wfc1, grb000214_wfc2}.  Followup
  with the \nfi\ beginning 12 hours after the burst revealed an X-ray
  afterglow with a flux of $F_{2 - 10~\rm{\kev}} \sim 8 \times 10^{-13}$
  erg~cm$^{-2}$~s$^{-1}$ \citep{antonelli00}.  The X-ray lightcurve is
  unremarkable, following a power-law decay.  No optical counterpart was
  found, however the followup was complicated by the moon, and the limits are
  not particularly constraining.  No redshift has been derived due to the lack
  of an accurate localization, required for host identification.

  \citet{antonelli00} fit the X-ray spectrum of the entire \nfi\ observation.
  They find an absorbed power law to be a poor fit (chance probability $P =
  0.02$), with an excess above the power law at an energy between $4 -
  5$~\kev.  By adding a narrow gaussian line, they obtain an acceptable fit,
  with the line centroid being at $4.7 \pm 0.2$~\kev, with an equivalent width
  of $\sim 2.1$~\kev, with 99.73\% confidence ($3 \sigma$, using an $F$-test).
  Although no independent redshift was measured \citet{antonelli00}
  interpreted this line as a redshifted Fe K$\alpha$ line.

  {\bf \grb000926}.  The Interplanetary Network (\ipn) discovered \grb000926
  \citep{grb000926_ipn}.  The optical afterglow of this 25~s long event was
  identified less than a day later \citep{gcc+00, dfp+00}.  The redshift,
  measured from optical absorption features, is $2.0379 \pm 0.0008$
  \citep{cgh+03}.  The afterglow was well-monitored in the optical
  \citep{hys+01}, and was detected in the {\small IR} \citep{dip+00,fyn+01}
  and radio \citep{hys+01}.

  The \sax\ Medium Energy Concentrating Spectrometer (\mecs) instrument
  (sensitive from $1.3 - 10$~\kev) discovered the X-ray afterglow
  \citep{pir+01}.  The X-ray source was weak, and so we do not analyze the
  \sax\ data.  \citet{ggp+00} observed the source for 10~ksec with \chandra\
  using the \aciss 3 backside-illuminated chip on Sept $29.674 - 29.851$.
  \chandra\ again observed \grb000926 in a 33~ksec long ToO taken Oct $10.176
  - 10.76$ \ut, also with \aciss 3.  The afterglow was clearly detected in
  each of these observations. \citet{pir+01} fit the combined \chandra\ and
  \nfi\ spectrum, finding an absorbed power law to be an acceptable fit, with
  no evidence for any spectral features.

  {\bf \grb001025}.  This burst (also called \grb001025{\small A}, to
  differentiate from a second \grb\ observed during the same \ut\ day), which
  lasted about 15~s, was localized by the \rxte\ \asm\ \citep{grb001025_asm}.
  No optical counterpart was identified to a limit of $R = 24.5$
  \citep{fmm+00}.  \xmm\ observed the location of \grb001005 starting about
  1.9 days after the burst with a total \epic\ exposure of about 25~ksec, and
  found two X-ray sources in the \asm\ error circle \citep{ass+00}.  The
  brighter of the two, detected at $F_{0.2 - 10~\rm{\kev}} = 4.4 \times
  10^{-14}$~erg~cm$^{-2}$~s$^{-1}$, is considered the most likely afterglow
  candidate, however some uncertainty remains over this identification.
  \citet{watson02a} find the source to have decreased in flux during the
  observation at 99.8\% confidence, while \citet{borozdin02} find the source
  to be consistent with constant flux.  The lack of an optical counterpart to
  the X-ray source to $R = 24.5$ would tend to rule out an \agn, strengthening
  the association with the \grb.

  \citet{borozdin02} analyzed the spectrum of the afterglow candidate, and
  found an acceptable fit using a power law with Galactic absorption.
  \citet{watson02a}, however, find a collisionally ionized plasma model to be
  a significantly better fit to the spectrum.  In the thermal plasma model,
  the redshift is allowed to vary to obtain the best fit.  According to
  \citet{watson02a}, the plasma model is a better fit due to discrete
  fluctuations observed between $\sim 0.5$ and 2~\kev, which these authors
  interpret as due to the blend of lines from Mg, Si, S, Ar, and Ni~L.  The
  fit temperature of 3.4~\kev\ is determined by the continuum shape, and the
  redshift corresponding to the best fit is $z$ =
  0.70\ud{0.3}{0.1}\footnote{\citet{watson02a} give this redshift in the text;
    but in their Table~1, they state the best-fit is $z=0.53$ (90\% confidence
    range 0.5-0.55).  Our own fit to the given line centroids with
    uncertainties using the line identifications of \citet{watson02a}, which
    does not include continuum uncertainties, gives $z=0.65 \pm 0.04$.}, from
  reported observed line energies (with detection significances of the
  individual lines) of 0.80\ud{0.04}{0.05} ($4 \times 10^{-4}$), $1.16 \pm
  0.05$ ($8 \times 10^{-4}$), $1.64 \pm 0.07$ ($8 \times 10^{-4}$), $2.2 \pm
  0.1$ (0.02), and 4.7\ud{0.8}{0.4} \kev\ (0.12), corresponding to
  \ion{Mg}{12} (1.46 \kev), \ion{Si}{14} (1.99~\kev), \ion{S}{16} (2.60~\kev),
  \ion{Ar}{17} (3.30~\kev), and \ion{Ni}{28} (8.10~\kev), respectively.  While
  the probabilities for detecting the individual lines were given, the method
  for deriving these probabilities was not.

  {\bf \grb010220}. This burst, positioned by \sax\ \citep{grb010220_sax1,
  grb010220_sax2, grb010220_sax3}, has no identified optical counterpart.
  Searches reached to 23.5 in R, but were complicated by the position of the
  burst in the Galactic plane \citep{bfp+01}.  \xmm\ observed the error circle
  starting 14.8 hours after the event.  Like \grb001025, the X-ray counterpart
  is uncertain, with four sources discovered in the error circle.  None of
  these is clearly variable, and the brightest has been associated with the
  afterglow \citep{watson02a}.

  \citet{watson02a} have analyzed the spectrum of the probable counterpart,
  and find an absorbed power law to be an acceptable fit.  They also claim a
  significant deviation from a power-law model at around 3.9 \kev, and that
  adding an unresolved gaussian line with equivalent width $1.8^{+0.8}_{-1.2}$
  \kev\ improves the fit at $>99$\% (2.6$\sigma$) confidence (as with \grb
  0001025, this reference gives the confidence level for detection of this
  line, but does not describe the method by which the confidence level was
  determined).  According to their fits, a thermal plasma model is
  significantly preferred over the absorbed power law, and allowing the Ni
  abundance to vary from solar, the 3.9~\kev\ can be accounted for with this
  element.  It is unclear how \citet{watson02a} identify Ni (rest energy
  7.47~\kev) with the feature, since no optical redshift has been measured.
  The required redshift assuming the 3.9 \kev\ line to be due to Ni~K$\alpha$
  fluorescence is $z=0.92.$

  {\bf \grb011030}.  This X-ray rich \grb\ (or X-ray flash -- \xrf) was
  originally detected by \sax\ \citep{xrf011030_sax1, xrf011030_sax2}.  While
  it is not clear whether \xrf s and \grb s are different manifestations of
  the same phenomenon, \xrf s exhibit afterglows with similar characteristics
  to \grb\ afterglows.  A radio afterglow was detected $\sim 9$ days after the
  burst \citep{taylor01}.  No optical afterglow was detected for this event.
  \chandra\ performed two followup observations, 10.5 \citep{hyf+01}, and 20
  days after the event.  With \aciss, \chandra\ detected an X-ray source with
  a flux of $F_{2 - 10~\rm{\kev}} 2.4 \times 10^{-13}$ erg~cm$^{-2}$~s$^{-1}$
  in the error box, which faded by more than an order of magnitude between the
  two pointings.  This source is certainly the afterglow associated with the
  initial flash.  The spectrum, on preliminary analysis, appears consistent
  with a power law with Galactic absorption \citep{hyf+01}.

  {\bf \grb011211}.  This X-ray rich \grb\ was localized by \sax, and is
  notable, at 270~s duration, for being the longest event detected by the
  \wfc\ \citep{gh+01}.  The identification of an optical afterglow lead to the
  measurement of the absorption-line redshift of $z = 2.140$ \citep{hsg+02}.
  \xmm\ began observing the error circle starting $\sim$11~hours after the
  event, and discovered an X-ray afterglow \citep{grb011211_xmm1} with a
  time-averaged flux of $F_{0.22 - 10~\rm{\kev}} = 1.7 \times 10^{-13}$
  erg~cm$^{-2}$~s$^{-1}$ \citep{reeves03}.

  Considerable controversy surrounds the analysis of the X-ray spectrum of
  this event.  \citet{reeves02} consider only the \pn\ data, divide the
  observations into time sections, and claim that in the first 5~ksec
  interval, an absorbed power law marginally fits the data.  They find that
  adding lines from partially ionized Mg, Si, S, Ar and Ca, with a mean
  redshift of 1.9 (implying a blueshift, possibly due to an outflow, of $\sim
  0.1 c$) improves the fit at 99.5\% confidence.  The corresponding observed
  (rest-frame) line centroids are 0.44, 0.71, 0.88, 1.22 and 1.46~\kev.  They
  find the most significant of the lines (0.71, 0.88, and 1.22 \kev) are
  detected with 98\% (2.3\ssig), 97\% (2.2\ssig), and 93\% (1.8\ssig)
  confidence for detection at a single energy, respectively, employing the
  $F$-test.  \citet{borozdin02} analyze the same 5~ksec interval both for the
  \pn\ data alone, and for the combined \pn, \mos 1 and \mos 2 data.  For the
  combined set they find a good fit to an absorbed power law with Galactic
  absorption, with no improvement from adding lines.  For the \pn\ data alone,
  they find marginal improvement for adding lines at the energies found by
  \citet{reeves02}, and they suggest the difference is due to the selection of
  background regions.  \citet{rutledge02} independently analyzed the
  statistical significance of the features, finding that significance is
  marginal.

  {\bf \grb020321}.  This weak \grb\ was discovered in the \sax\ \wfc\ using
  ground trigger logic \citep{gmf+02}.  A very faint variable object was
  detected in a followup observation with the \nfi\ $\sim$6 hours after the
  event \citep{gan02}.  Deep optical imaging failed to find any variable
  source either at the \nfi\ position, or the \wfc\ error box to a limiting
  magnitude of $R \sim 24$ \citep{pdm02}.  A refined analysis of the \nfi\
  data showed the detection to be marginal, and an \xmm\ observation taking
  place $10.3 - 24.2$ hours after the \grb\ failed to detect the \nfi\ source,
  but discovered another variable X-ray source in the \wfc\ error circle.
  This source was not detected in a \chandra\ observation taking place 9.9
  days after the \grb, and is considered the likely (although not certain)
  counterpart of the \grb\ \citep{ikh+02}.  We analyze the data from this
  source in this paper.

  {\bf \grb020322}. This weak \sax\ burst was followed up with the \nfi\ 6.5
  hours after the event, and a new, relatively bright X-ray source was found
  in the error circle \citep{gan02b}.  This was confirmed, in
  \xmm\ observations starting about 15.5 hours after the \grb\ to be the
  fading afterglow. The X-ray flux, at $F_{0.2 - 10~\rm{\kev}} = 3.5 \times
  10^{-13}$ \ergflux\ made this a high signal-to-noise \xmm\ detection. A very
  weak optical counterpart was detected \citep{bmh+02}, and confirmed to be
  the afterglow.  The host, at 27th magnitude \citep{bfr+02} is very faint,
  and no redshift has been measured.  \citet{watson02b} analyzed the spectrum,
  and find it to be consistent with an absorbed power law, with column in
  excess of the Galactic value.

  {\bf \grb020405}. This bright \ipn\ burst \citep{grb020405_ipn} was very
  well studied, with bright optical \citep{pkb+03} and radio \citep{bsf+03}
  transients, evidence for a late-time red bump in the optical decay possibly
  due to a supernova component \citet{pkb+03}, optical polarization detections
  \citep{cmg+03}, and a measured redshift of $z = 0.690$ \citep{masetti02a,
    price03}.  Detailed optical/{\small NIR} spectrophotometric and
  polarimetric analyses were published by \citet{masetti03}.
  \citet{mirabal03} triggered a long 50~ksec {\em Chandra} \letg\ observation,
  which started 1.68 days after the \grb.  They detected an X-ray afterglow
  with flux $F_{0.2 - 10~\rm{\kev}} = 1.36 \times 10^{-12}$ \ergflux\ 1.71
  days after the event.  They analyze the X-ray spectrum, finding it to be
  consistent with a featureless power law with the suggestion of an absorbing
  column in excess of Galactic.

  {\bf \grb020813}.  This typical long-duration \grb\ was localized by \hete\
  \citep{grb020813_hete}.  The rapid dissemination of the position enabled
  prompt identification of an optical counterpart \citep{fbp+02}, and the
  redshift $z=1.255$ has been measured both in absorption and in emission
  \citep{bsc+03}.  \chandra\ observed the region with the \hetg\ starting 17.7
  hours after the burst, with a total integration of 77~ksec \citep{vmf+02}.
  The mean flux of the X-ray afterglow during this observation was $F_{0.6 -
  6~\rm{\kev}} = 2.2 \times 10^{-12}$ \ergflux.  A preliminary reduction
  showed no obvious spectral features \citep{vmf+02}.  Subsequently,
  \citet{butler03b} reported evidence at $3.3 \sigma$ significance for a line at
  1.3~\kev, which they interpret as due to H-like sulfur, blueshifted (with
  respect to the host redshift) by $0.1c$.

  {\bf \grb021004}. The position of this {\em HETE-2} \grb\ was disseminated
  quickly \citep{grb021004_hete}, and a bright optical afterglow discovered 9
  minutes after the event \citep{fps+03}.  The redshift was soon determined to
  be $z = 2.323$ from the identification of the Ly$\alpha$ line
  \citep{chornok02}.  The \chandra\ \hetg\ observed the afterglow starting
  20.5 hours after, with a total exposure of 87~ksec.  Preliminary reduction
  of the data showed a fading X-ray source with mean flux of $F_{2 -
  10~\rm{\kev}} = 4.3 \times 10^{-13}$ \ergflux\ \citep{sh02}.  This analysis
  also showed the spectrum to be consistent with a power law, with no evidence
  for absorption in excess of the Galactic value.

  {\bf \grb030226}.  An optical afterglow of this average brightness, 100~s
  long \hete\ burst \citep{ssg+03} was discovered 2.6 hours after the trigger
  \citep{fcp03}.  Absorption spectroscopy identified the redshift as $z =
  1.99$ \citep{aow+03}.

  \chandra\ observed this event with \aciss 3 starting 37.1 hours after the
  trigger, with a total exposure of 34~ksec.  A relatively weak; $F_{2 -
  10~\rm{\kev}} = 3.2 \times 10^{-14}$ \ergflux\ averaged over the
  observation, X-ray afterglow was clearly detected at the position of the
  optical transient.  A preliminary analysis of the spectrum \citep{pfh+03}
  found it to be consistent with an absorbed power law, with absorption in
  excess of Galactic.  A subsequent analysis disputed the N$_H$ value, finding
  it consistent with the Galactic value \citep{sf03}.

  {\bf \grb030227}. This weak, long-duration \grb\ was positioned by
  \integral\ \ibis\ \citep{grb030227_integral1, grb030227_integral2}, and
  followed up in the X-ray by \xmm\ with a 13 hour pointing beginning 8 hours
  after the event \citep{mgt+03}.  After the detection of a fading X-ray
  source with average flux of $F_{0.2 - 10~\rm{\kev}} = 8.7 \times 10^{-13}$
  \ergflux, a relatively faint optical counterpart was discovered in the \xmm\
  error circle \citep{sfk+03}.  No redshift has yet been measured.

  \citet{mgt+03} analyzed the spectrum from the entire observation, and find
  it to be consistent with an absorbed power law, with a column in excess of
  the Galactic value.  They also find weak evidence for an emission feature at
  1.67~\kev, which they postulate could be due to iron, without asserting a
  detection.  In a subsequent analysis, \citet{watson03} divided the
  observation up into four segments.  In the first three, the spectrum in
  consistent with an absorbed power law, but in the last 11~ksec, they claim
  the presence of 5 emission lines at centroid energies of 0.62, 0.86, 1.11,
  1.35, and 1.66 \kev, with significance ranging from $1.7 - 3.8 \sigma$,
  using an $F$-test, for detection of lines at single energies depending on
  how they estimate the uncertainty.  Assuming a redshift of 1.35, the line
  energies correspond to H- and He-like lines of Mg, Si, S, Ar, and Ca.
  \citet{watson03} also claim that the continuum fit for zero emission-line
  flux is ruled out at the $4.4 - 4.7 \sigma$ significance.  For comparative
  purposes, \citet{mgt+03} estimate the significance of the feature at $E =
  1.67\ud{0.01}{0.03}$~\kev\ to be 3.2\ssig\ using an $F$-test.

  {\bf \grb030328}.  The position of this typical long-duration \hete\ \grb\
  was rapidly disseminated \citep{vcv+03}, enabling both the early
  identification of an optical transient, and a \chandra\ grating observation
  beginning only 15.3 hours after the initial trigger
  \citep{pp03}.  Optical spectroscopy
  identified the source redshift as $z = 1.52$ \citep{martini03}.

  The \letg\ observed the event for 94~ksec.  A preliminary analysis
  \citep{butler03a} found that during the observation the mean source flux was
  $F_{0.5 - 3~\rm{\kev}} = 1.9 \times 10^{-13}$ \ergflux, and that the
  spectrum is consistent with a power law with Galactic absorption.

  {\bf \grb030329}. The rapid dissemination of the position of this
  \hete\ event \citep{grb030329_hete} enabled rapid identification of a bright
  optical counterpart, and the measurement of the redshift, $z = 0.169$
  \citep{greiner03}.  The proximity and brightness of this famous burst
  enabled the detection of a Type 1c supernova, {\small SN}2003dh,
  superimposed on the afterglow, securing the association of some
  long-duration events with the deaths of massive stars \citep{garnavich03,
    stanek03, hjorth03}.  The X-ray afterglow was bright enough to be detected
  by \rxte\ approximately 5 hours after the burst \citep{grb030329_rxte}.
  \xmm\ observed the position twice, in pointings separated by nine days.
  Temporal and spectral analyses of the first observation and an analysis of
  the temporal behavior through the second observation were published by
  \cite{tiengo03} and \citet{tiengo04}, respectively.  The first observation
  had a 30~ksec exposure, while the second was 39~ksec.  No detailed line
  search analysis has yet been published of these data.

  {\bf \grb031203}. This long burst was localized by \integral\ \ibis\
  \citep{grb031203_integral}, and a followup observation by \xmm\ started only
  $\sim 6$ hours after the burst.  Of the two bright sources that were
  detected inside the \ibis\ error circle \citep{grb031203_xmm1}, one of them
  appears to be not consistent with the \rosat\ upper limit, suggesting that
  this is the afterglow \citep{grb031203_rosat}.  The redshift of the possible
  host galaxy was determined to be $z=0.105$ \citep{grb031203_red}, coincident
  with a variable radio source \citep{grb031203_radio1, grb031203_radio2}.  A
  spectacular X-ray dust echo was detected for the first time in a \grb\
  \citep{grb031203_dust1, grb031203_dust2}.

  A second \xmm\ observation was performed starting approximately $\sim 2.9$
  days after the burst.  \citet{grb031203_xray} claim that the X-ray light
  curve steepens from $-0.4 \pm 0.1$ to $-0.9 \pm 0.1$ approximately one day
  after the burst, although we cannot confirm this in our independent
  analysis.  Both observations appear to be consistent with a single power-law
  decay time slope of $-0.4 \pm 0.1$, which is one of the most gradual decay
  ever detected.

  {\bf \grb040106}. \xmm\ observed this long \grb\ localized by
  \integral\ \ibis\ \citep{grb040106_integral} starting $\sim 6$ hours after
  the burst.  A fading X-ray source was soon discovered \citep{grb040106_xmm}.
  This burst was the brightest ever observed by \xmm, with $\sim 23000$ total
  photons detected in the $0.2 - 10$~\kev\ range.  A likely optical
  counterpart has been identified \citep{masetti04}, but no spectroscopic
  redshift has been measured to date.

  A summary of reported emission line detections, including their
  identifications and the quoted significances, is listed in
  Table~\ref{tbl:reported}.  This table lists two additional sources
  (\grb001025A and \grb010220) that are not listed in Table~1 of
  \citet{boettcher03}.

\section{Data Reduction}
\label{section_data}

  In this section, we describe the reduction procedures that we adopted for
  the various observatory/instrument configurations.  Table~\ref{tbl:fits} lists,
  for each dataset, the total number of counts detected inside the extraction
  region (source plus background) and the estimated number of background
  counts.  In a few cases, we split the observation into two time intervals to
  match those adopted in previous publications.

\subsection{\chandra}

  Three sources (\grb991216, \grb020813, and \grb021004) were observed with
  the \chandra\ \citep{weisskopf_chandra} \hetg\ \citep{canizares_hetg} with
  the \aciss\ \citep{garmire_acis} placed at the focal plane.  The \hetg\
  consists of two separate grating arrays; the Medium Energy Gratings (\meg)
  and the High Energy Gratings (\heg), which are optimized in the approximate
  wavelength ranges $\lambda = 5 - 20$ \AA\ and $\lambda = 1.5 - 15$ \AA,
  respectively.  The spectral resolving power of the \heg\ is $R \equiv
  \lambda/\Delta\lambda_{F\!W\!H\!M} \approx 1000 \times (\lambda/10$~\AA) and
  is approximately twice that of the \meg, for a total of $\approx 1350$ and
  $\approx 700$ useful resolution elements for the HEG and MEG, respectively.
  Two other sources (\grb020405 and \grb030328) were observed with the \letg\
  \citep{brinkman_letg} with the \aciss\ at the focal plane.  This
  configuration optimizes the sensitivity in the $\lambda = 5 - 30$ \AA\
  region with a resolving power that is approximately less than half that of
  the \meg, with $\approx 400$ resolution elements.


  The data were retrieved from the \chandra\ Data
  Archive\footnote{http://asc.harvard.edu/cda/} and were processed using
  \ciao\ v2.3\footnote{http://asc.harvard.edu/ciao/}.  We use sky coordinates
  of the aspect-corrected Level 1 events to determine the location of the peak
  X-ray flux in the zeroth order image, which we use to measure the dispersion
  angles and assign wavelengths to each of the dispersed events.  Only events
  with \verb^grade=0,2-6^ and \verb^status=0^ were used in the analyses.
  Background events produced during several serial readout frames on \aciss4
  were removed using \verb2destreak2, which significantly reduces the noise
  level on the positive dispersed orders.  Source events were then spatially
  extracted using a 4.8$\arcsec$ filter in the cross-dispersion direction.  We
  then use a standard pulse-height filter to separate the first order events.
  Ancillary response files were generated based on these extraction regions.
  The relative effective area may be uncertain by $\sim 10$\% above $\sim
  2$~\AA\ and $\sim 20$\% below $\sim
  2$~\AA\footnote{http://asc.harvard.edu/proposer/POG/html/}.

  We use also the zeroth order spectrum, which has an intrinsic spectral
  resolution that is lower than that of the \hetg\ by an order of magnitude or
  more, but contains roughly an equal number of photons compared to the
  combined first order dispersed events.  Source photons are extracted from a
  circular region with a radius of $4$\arcsec.  Background events are
  extracted from several large off-axis regions on the \aciss\ chips.  The
  positive and negative first order events are combined, which are then fit
  simultaneously with the zeroth order spectrum.  The relative normalizations
  of the zeroth and first order spectra are allowed to vary within $20$\%.

  Four sources (\grb000210, \grb000926, \grb011030, and \grb030226) were
  observed with the \aciss\ at the focal plane with no gratings.  The spectral
  reduction procedure is identical to that of the zeroth order image in the
  \hetg/\aciss\ configuration described above.  The effective area is $\sim
  15$ and $\sim 7$ times higher at $E = 1.5 ~\rm{\kev}$ and $6 ~\rm{\kev}$,
  respectively, than that with the \hetg\ placed in front.  The spectral
  resolving power of \aciss\ is $R \approx E_{\rm \kev}/(0.110 + 0.012~E_{\rm
  \kev})$, where $E_{\rm{\kev}}$ is the energy in \kev, for $\approx 190$
  resolution elements in the $0.5 - 8$~\kev\ range.  None of these four
  datasets has been published to date.

\subsection{\xmm}

  Nine afterglows (\grb001025, \grb010220, \grb011211, \grb020321, \grb020322,
  \grb030227, \grb030329, \grb031203, and \grb040106) were observed with \xmm\
  Observatory \citep{xmm}.  \xmm\ consists of three X-ray telescopes with the
  European Photon Imaging Camera (\epic) \pn\ \citep{epic_pn} at one of the
  focal planes and the \mos\ \citep{epic_mos} \ccd s at the remaining two.
  Behind the two mirror assemblies that focus light onto the \mos\ are two
  identical sets of Reflection Grating Spectrometers (\rgs; \citealt{rgs}).
  The \epic\ \pn\ and \mos\ consist of \ccd\ arrays with spectral resolving
  powers of $\sim 15~E_{\rm{\kev}}^{1/2}$ for both the \pn\ and \mos\ for
  $\approx$ 220 resolution elements in the $0.2 - 8$~\kev\ range.  The \pn\
  chips are sensitive in the $\sim 0.2 - 10$~\kev\ range with a peak effective
  area of $\sim 1200 ~\rm{cm}^{2}$ at $\sim 2$~\kev.  The \mos\ chips are
  sensitive in the $\sim 0.5 - 10$~\kev\ with roughly half the effective area
  per array of the \pn.  Data obtained with the Optical Monitor (\om) are not
  used for any of our analyses.

  The data were retrieved from the \xmm\ Science
  Archive\footnote{http://xmm.vilspa.esa.es/external/xmm\_data\_acc/xsa/index.shtml}
  and were processed with the \xmm\ Science Analysis
  Software\footnote{http://xmm.vilspa.esa.es/external/xmm\_sw\_cal/sas\_frame.shtml}
  (\sas).  We use only data collected on the \pn\ and \mos\ detectors.  The
  \rgs\ data are not used, since the signal-to-noise ratios (\sn) of each of
  the bursts are too low to provide useful constraints.  Events on bad/hot
  pixels and pixels on the edge of the \ccd\ chips were excluded from the
  analysis by selecting events with only \verb^FLAG=0^.  Event grade
  selections were made using \verb^PATTERN<=4^ and \verb^PATTERN<=12^ for the
  \pn\ and \mos, respectively.  Source events from both the \pn\ and
  \mos\ were extracted from circular region centered on the source with a
  radius of $30 - 45 \arcsec$ depending on the level of the background flux
  during the observation.  Background events were taken from large source-free
  regions (typically $\sim 2 - 3 \arcmin$ in radius) that lie on the same
  \ccd\ chip.  Redistribution matrices and effective area curves were computed
  for each observation and each instrument using the \sas\ tools \verb^rmfgen^
  and \verb^arfgen^, respectively.  The \pn\ and \mos\ spectra are fit
  simultaneously, allowing the relative normalizations to vary within
  5\%\footnote{see, e.g.,
    http://xmm.vilspa.esa.es/docs/documents/CAL-TN023.ps.gz}.

  For \grb 010220, \citet{watson02a} found 20~ksec (of 43~ksec exposure) in
  which the background flaring was sufficiently low for analysis, while find
  find only $\sim 12$~ksec.  The data of \grb011211 are split into two time
  intervals; (1) the first 5~ksec during which detections of soft X-ray lines
  were reported by \citet{reeves02} and (2) the remaining 24~ksec where no
  lines were seen.  The data of \grb030227 were analyzed using data collected
  over the entire observation, and using data only collected during the last
  11~ksec, during the period reported by \citet{watson03} to contain emission
  lines.  \grb030329 was observed at two epochs separated by $\sim 9$ days,
  and the spectra are analyzed separately.  For the remaining six sources
  (\grb001025, \grb010220, \grb020321, \grb020322, \grb0301203, and
  \grb040106), we use data collected over the entire observation.

\subsection{\sax}

  Two sources (\grb970508 and \grb000214) were observed with the \sax\ \nfi,
  which consists of the Low Energy Concentrator Spectrometer (\lecs;
  \citealt{sax_lecs}) and two working units of the Medium Energy Concentrator
  Spectrometer (\mecs; \citealt{sax_mecs}).  The \lecs\ is sensitive to X-rays
  in the range $0.1 - 4.0 ~\rm{\kev}$ with an effective area of $40
  ~\rm{cm}^2$ at $E \sim 2 ~\rm{\kev}$.  The \mecs\ units are sensitive a
  slightly higher energy range ($1.3 - 10 ~\rm{\kev}$) with a combined peak
  effective area of $90 ~\rm{cm}^2$ at $E \sim 5 ~\rm{\kev}$.  The spectral
  resolving power of both the \lecs\ and the \mecs\ can be represented
  approximately by $R = 10~(E/6.4~\rm{\kev})^{0.42}$, for $\approx 15$ energy
  independent resolution elements ($0.1 - 3.0$~\kev, \lecs;
  \citealt{sax_lecs}) and $\approx 60$ resolution elements ($1.3 - 8$~\kev,
  \mecs; \citealt{sax_mecs}).

  The data reduction was performed with {\small HEASOFT}
  v5.2\footnote{http://heasarc.gsfc.nasa.gov/docs/software/lheasoft/}.  The
  \mecs\ and \lecs\ events were extracted from circular regions with radii of
  $3 \arcmin$ and $8 \arcmin$, respectively, centered on the source.  Since
  the background depends on the detector location, background spectra were
  generated from deep exposures of blank fields using an identical extraction
  region.  The background flux levels were normalized by comparing the count
  rates in the outer source-free regions.  For both observations, the
  \mecs\ background fluxes were found to be within 5\%.  Pre-made response
  matrices and effective area
  files\footnote{ftp://ftp.asdc.asi.it/pub/sax/cal/responses/} were used for
  the analyses.

  \grb970508 was observed with \sax\ during multiple epochs within one week of
  the burst \citep{piro98}.  The presence of an intense iron line was reported
  during its early afterglow ($<~1~\rm{day}$; time interval denoted as ``1a''
  in \citet{piro99}), and disappeared after $\sim~1~\rm{day}$ (interval
  denoted as ``1b'').  We analyze these spectra separately.

\subsection{\asca}

  The afterglow of \grb970828 was observed by the \asca\ Observatory
  \citep{asca} and the data were originally published by \citet{yoshida99}.
  \asca\ carries four focusing mirrors, with two \ccd\ imaging spectrometers
  (\sis0 and \sis1) and two gas scintillation imaging spectrometers (\gis2 and
  \gis3) at the focal plane.  The data were retrieved from the {\small
    HEASARC}
  archive\footnote{http://heasarc.gsfc.nasa.gov/db-perl/W3Browse/w3browse.pl}.

  \citet{yoshida99} identify a feature during a $\sim 20 ~\rm{ksec}$ interval
  of flaring activity, which is denoted as time segment ``B'' by
  \citet{yoshida99}.  The line was claimed to be absent during the other time
  intervals.  We, therefore, sum the remaining data (time segments ``A'' and
  ``C'') and analyze them collectively.

  The data reduction was again performed using {\small HEASOFT} v5.2.  The
  data were screened through standard criteria: rejection of hot flickering
  pixels and data contaminated by bright earth, grade selections, etc.  Source
  events were extracted from circular regions with a radii of $3\arcmin$ and
  $5\arcmin$ for the \sis\ and \gis, respectively.  The background flux level
  was steady during the entire observation so we use the total exposure to
  create background spectra for each instrument using off-axis source-free
  regions.  The data from all four units were fit simultaneously in the energy
  ranges $0.5 - 10 ~\rm{\kev}$ and $0.7 - 10 ~\rm{\kev}$ for the \sis\ and
  \gis, respectively.

\section{Spectral Analysis}
\label{section_ana}

  Our primary method for searching for line features is through Monte Carlo
  simulations, however for general spectral characterization, and for
  comparison with other work, we have performed standard spectral analysis,
  fitting the raw spectrum to an absorbed power law model.

  We performed all of the spectral fitting with the
  \xspec\
  v.11\footnote{http://heasarc.gsfc.nasa.gov/docs/software/lheasoft/xanadu/xspec/index.html}
  spectral fitting package.  We binned the spectra for all sources so that
  each spectral bin contains at least 15 - 20 counts (the number depending on
  the total counts detected) except for the grating spectra, which were binned
  uniformly by a certain factor so that each bin typically contains at least a
  few counts on average\footnote{We choose not to bin the grating spectra to
    contain at least $\sim 15$ counts per bin, since this typically results in
    an undersampling of the instrument resolution by an order of magnitude or
    more in most parts of the spectra.  We note, however, that the continuum
    parameters derived from spectra of both binning schemes are nearly
    identical and any subsequent analyses and their results are not affected
    by our choice of binning.}.  We then fit each of the background subtracted
  spectra using a power-law model with both Galactic and intrinsic absorption
  by cold material.  For the \grb\ with known redshifts, we fix the intrinsic
  absorber to be at the measured redshift.  Otherwise, we set $z = 1$, which
  roughly represents an average value of all \grb\ with known redshifts.  The
  model contains three free parameters; the intrinsic column density
  $N_{\rm{H}}^{\rm{host}}$, the power-law photon index $\Gamma$, and its
  normalization $F_{\rm{1 \kev}}$ defined as the flux in units of
  photons~cm$^{-2}$~s$^{-1}$~\kev$^{-1}$ at 1 \kev.  The functional form of
  the continuum model can be written as,
\begin{equation}
  F(E) = F_{\rm{1 \kev}}~E_{\rm{\kev}}^{-\Gamma}~
         e^{-(N_{\rm{H}}^{\rm{host}}~\sigma_{\rm{abs}}(E,z)+
         N_{\rm{H}}^{\rm{Gal}}~\sigma_{\rm{abs}}(E))},
\end{equation}
  where $E_{\rm{\kev}}$ is the photon energy in units of \kev\ and
  $\sigma_{\rm{abs}}(E)$ is the energy-dependent absorption cross section of
  interstellar material.  The Galactic hydrogen column densities are derived
  from \ion{H}{1} maps by \citet{dickey90}.  We assume that the metal
  abundances relative to hydrogen are those given by \citet{anders89}.

  We manually incorporate a fix to the low-energy quantum efficiency
  degradation of the \chandra\ \aciss\ using the \verb2acisabs2
  model\footnote{http://www.astro.psu.edu/users/chartas/xcontdir/xcont.html}.
  This effect, due to accumulation of molecular contamination onto either the
  optical filters and/or the \ccd\ chips, is time-dependent and reduced the $E
  \sim 0.7 ~\rm{\kev}$ effective area by $\sim 20 - 30$\% during the
  \grb020813 and \grb021004 observations and by $\sim 40$\% during the
  \grb030328 observation compared to pre-launch values.  The \grb991216
  observation was performed only $\sim 5$ months after launch and the data are
  therefore not expected to be affected significantly.

\begin{figure}
  \includegraphics[scale=0.6,angle=-90]{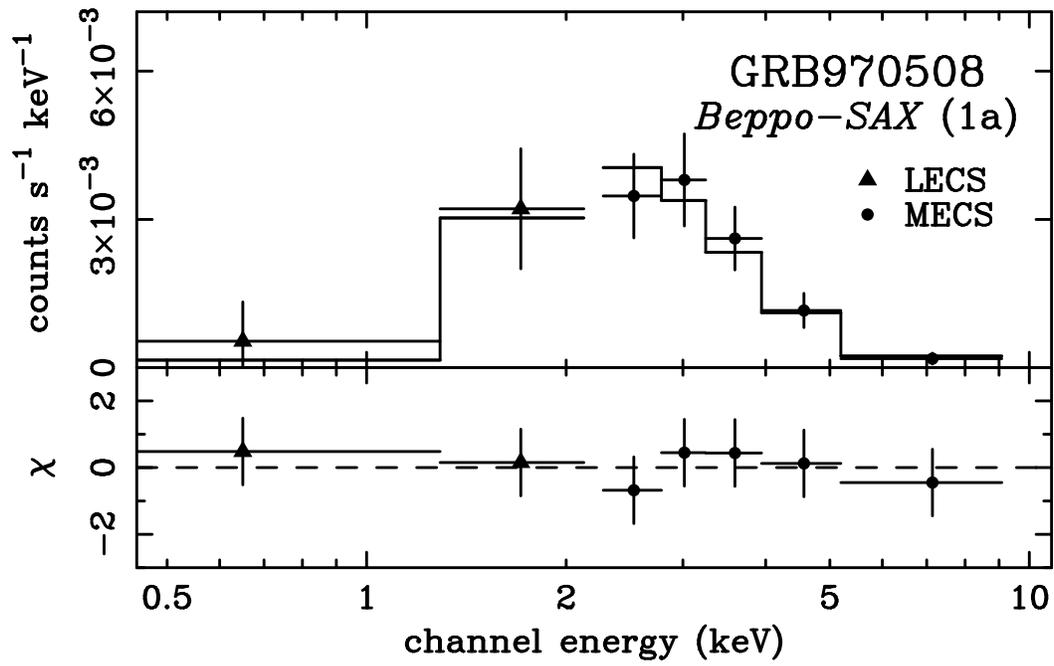}
  \caption{The observed X-ray spectrum of \grb970508 time segment 1a.  In this
           figure and the similar ones that follow, the top panel shows the
           data with the best-fit model superimposed.  The lower panel shows
           the resulting $\chi$ for each of the spectral channels, except in
           the high-resolution \chandra\ grating data where we plot the flux
           residuals.}
  \label{fig970508_1}
\end{figure}

\begin{figure}
  \includegraphics[scale=0.6,angle=-90]{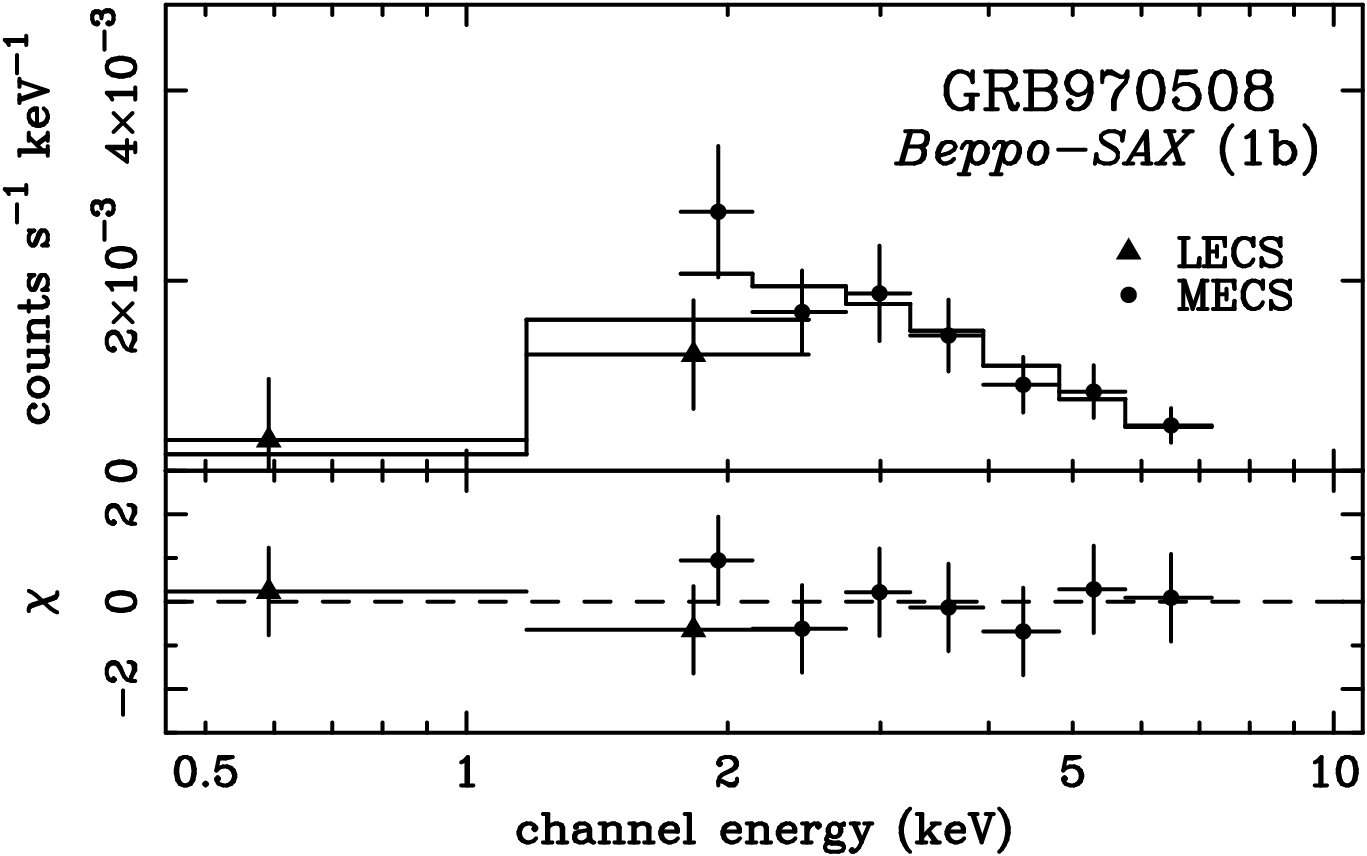}
  \caption{Same as in Figure~\ref{fig970508_1} for time segment 1b.}
  \label{fig970508_2}
\end{figure}

\begin{figure}
  \includegraphics[scale=0.6,angle=-90]{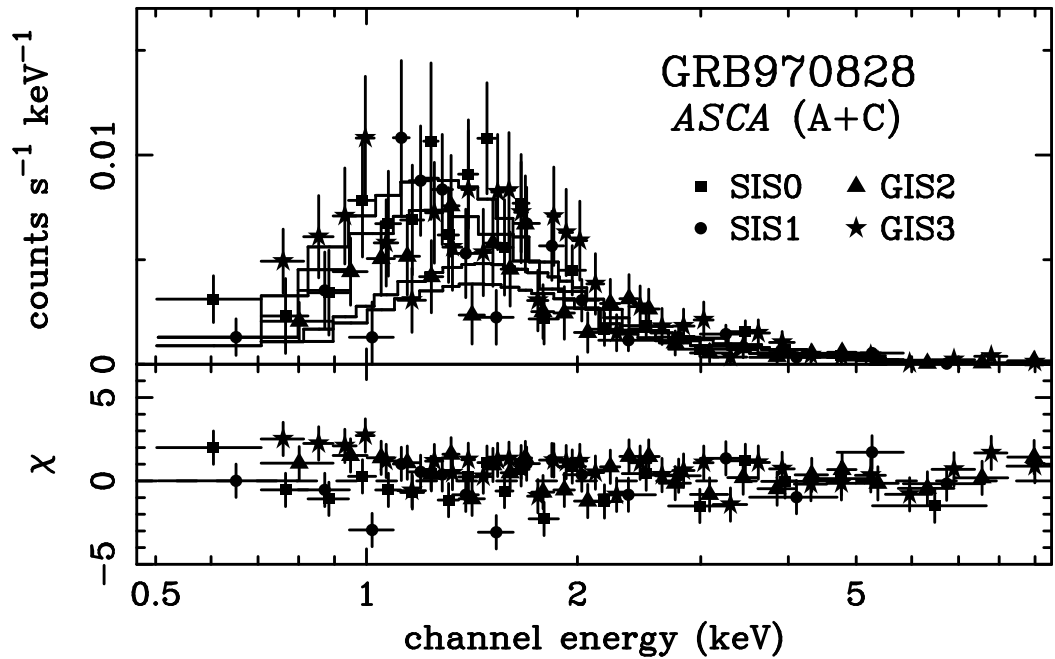}
  \caption{Same as in Figure~\ref{fig970508_1} for \grb970828 time segments A
           and C.}
  \label{fig970828_ac}
\end{figure}

\begin{figure}
  \includegraphics[scale=0.6,angle=-90]{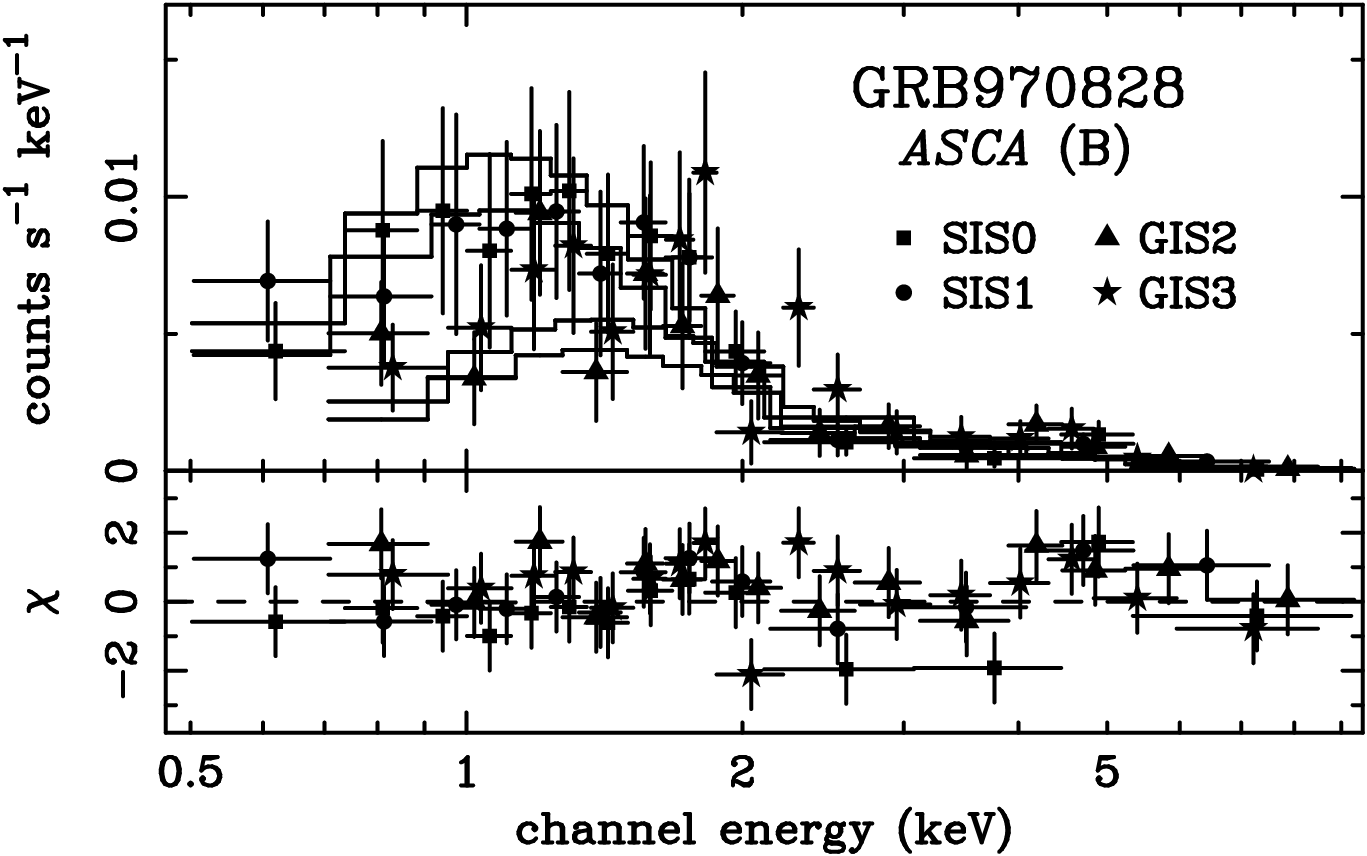}
  \caption{Same as in Figure~\ref{fig970508_1} for \grb970508 time segment B.}
  \label{fig970828_b}
\end{figure}

\begin{figure}
  \includegraphics[scale=0.6,angle=-90]{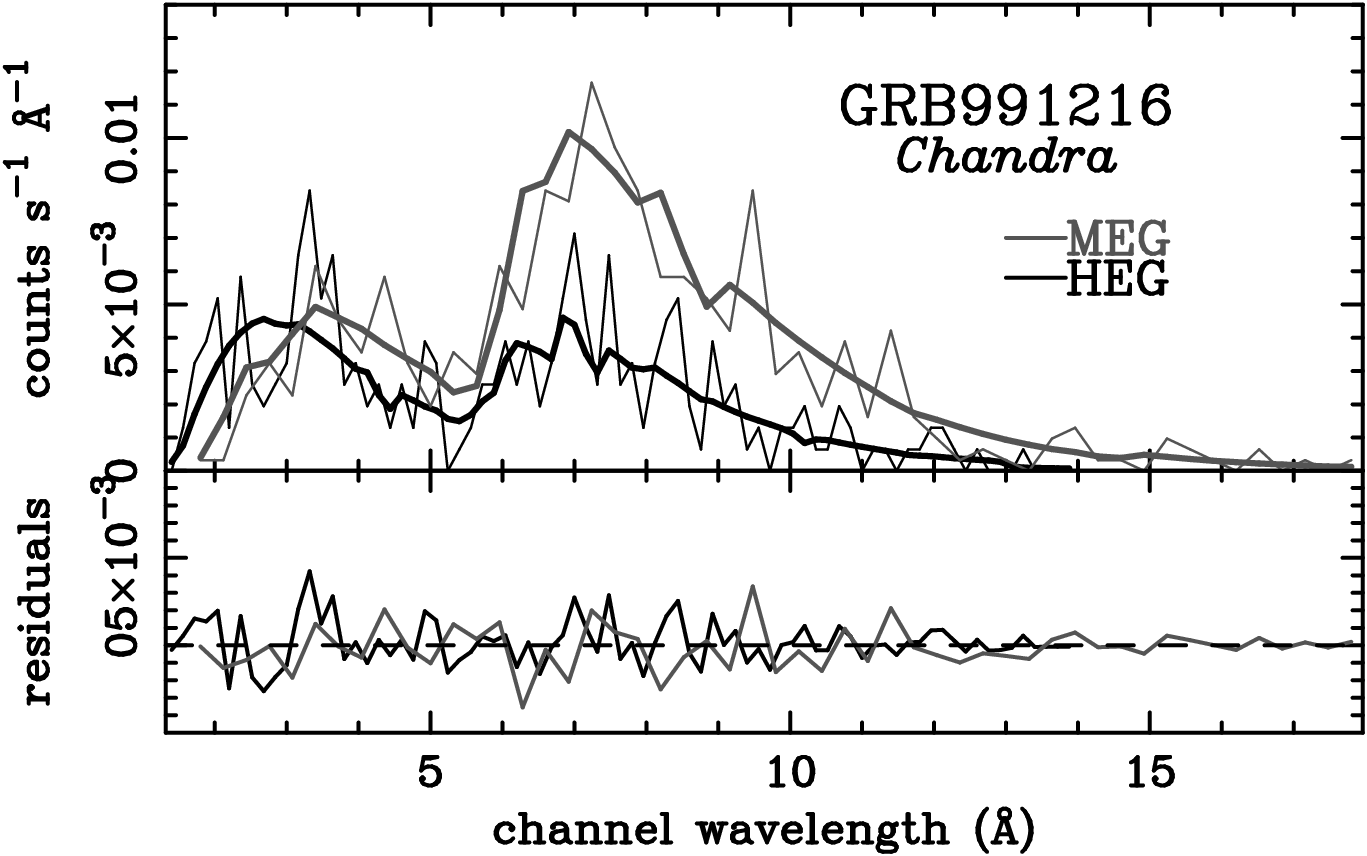}
  \caption{Same as in Figure~\ref{fig970508_1} for the dispersed spectrum of
           \grb991216.}
  \label{fig991216_m1}
\end{figure}

\begin{figure}
  \includegraphics[scale=0.6,angle=-90]{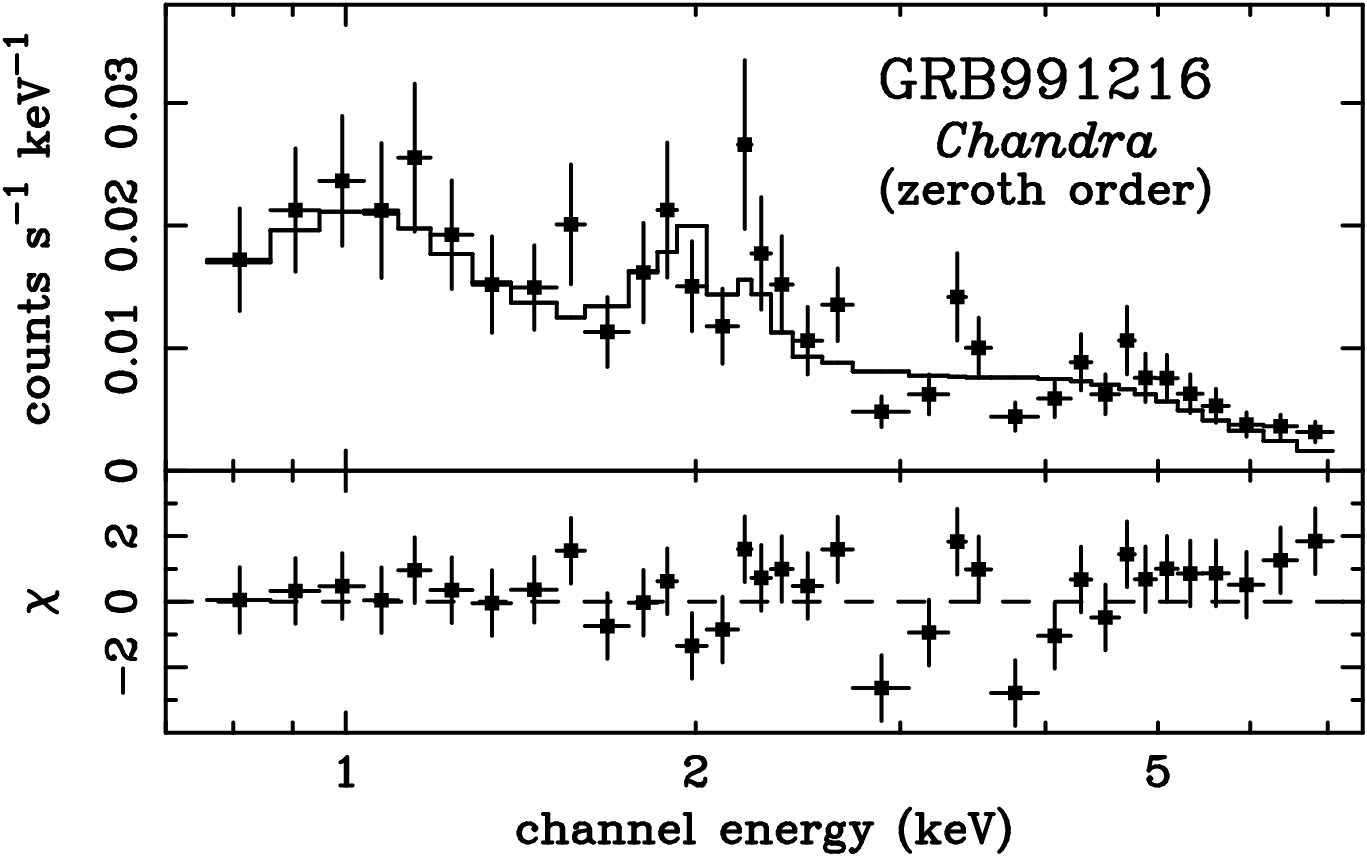}
  \caption{Same as in Figure~\ref{fig970508_1} for the zeroth order spectrum of
           \grb991216.}
  \label{fig991216_m0}
\end{figure}

\begin{figure}
  \includegraphics[scale=0.6,angle=-90]{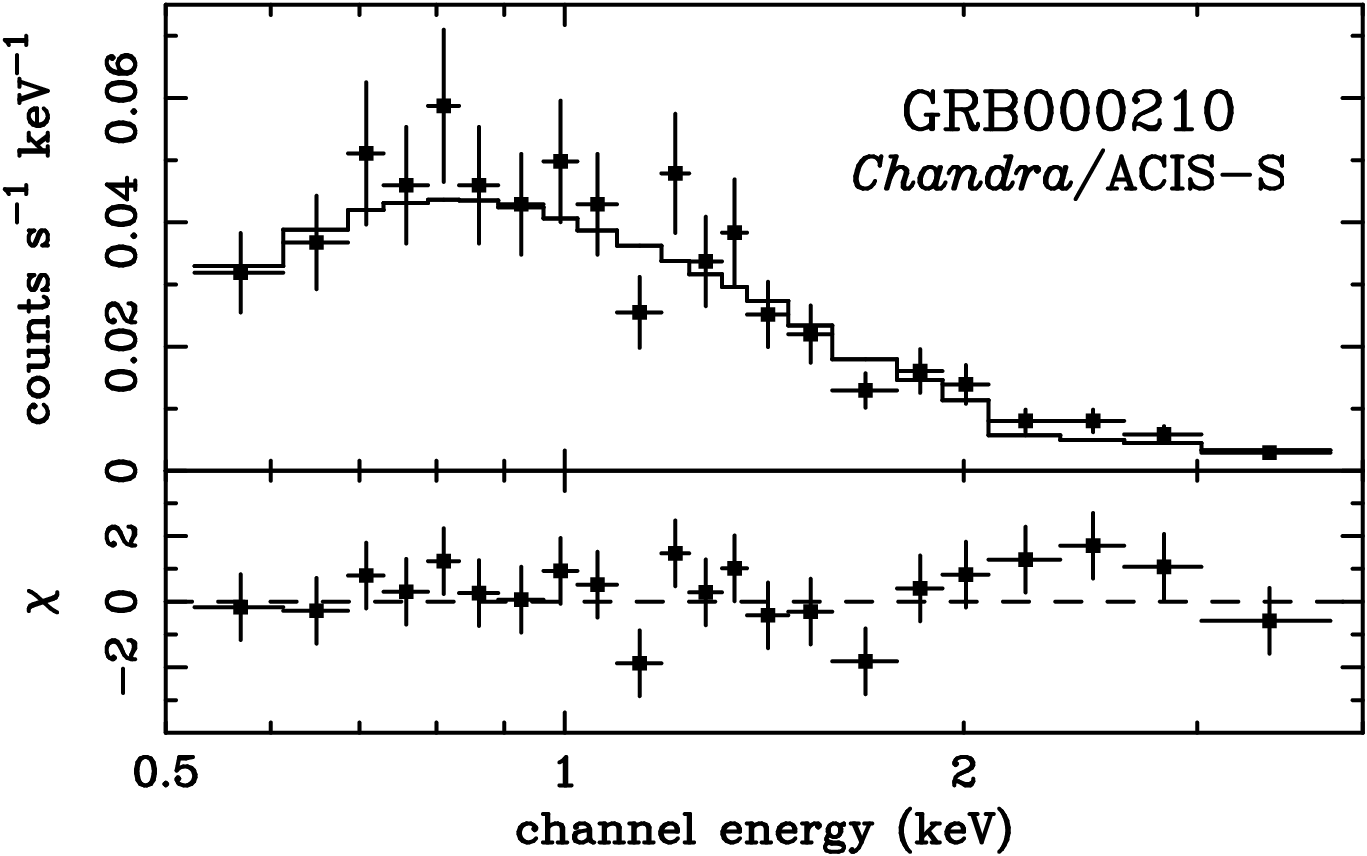}
  \caption{Same as in Figure~\ref{fig970508_1} for \grb000210.}
  \label{fig000210}
\end{figure}

\begin{figure}
  \includegraphics[scale=0.6,angle=-90]{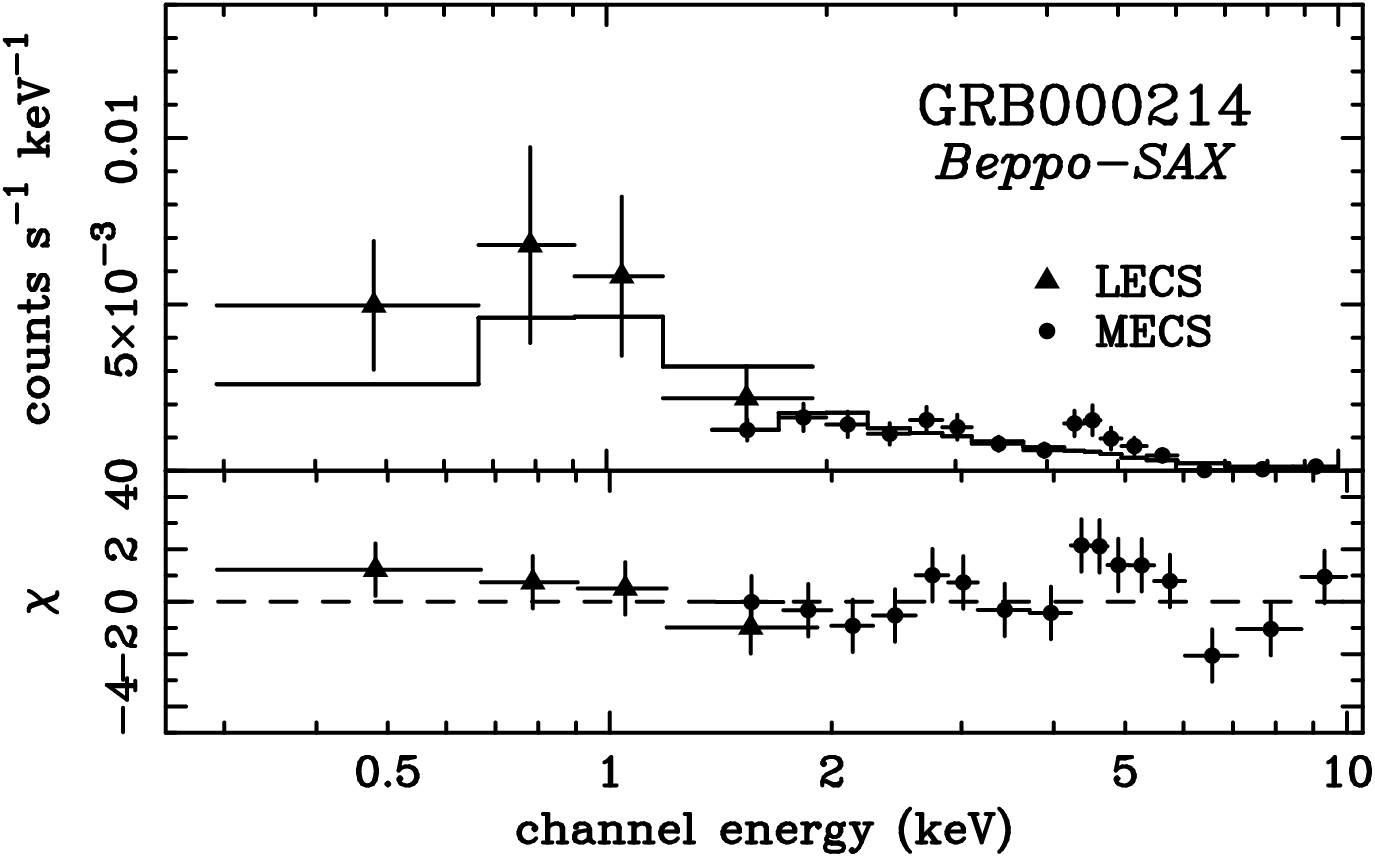}
  \caption{Same as in Figure~\ref{fig970508_1} for \grb000214.}
  \label{fig000214}
\end{figure}

\begin{figure}
  \includegraphics[scale=0.6,angle=-90]{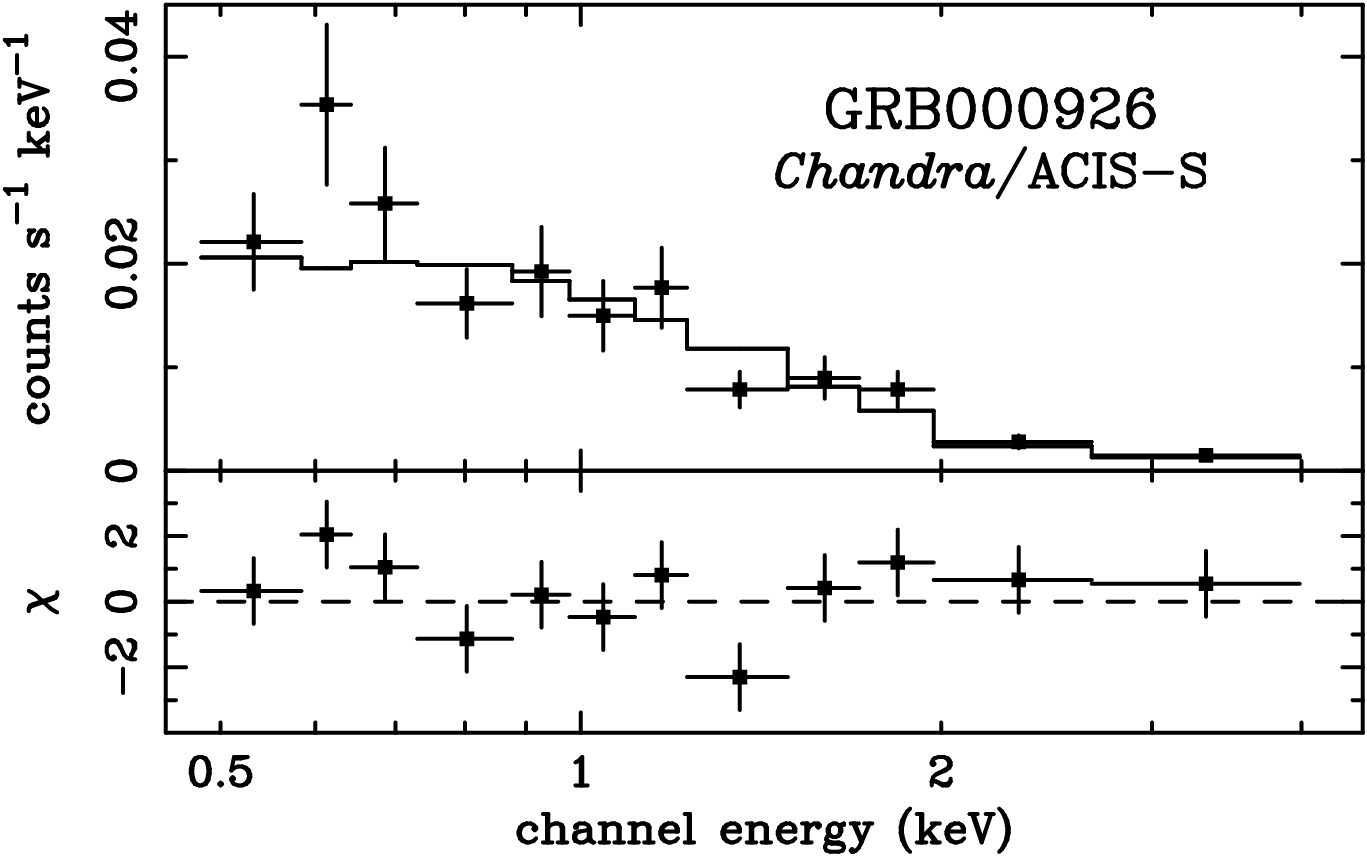}
  \caption{Same as in Figure~\ref{fig970508_1} for \grb000926.}
  \label{fig000926}
\end{figure}

\begin{figure}
  \includegraphics[scale=0.6,angle=-90]{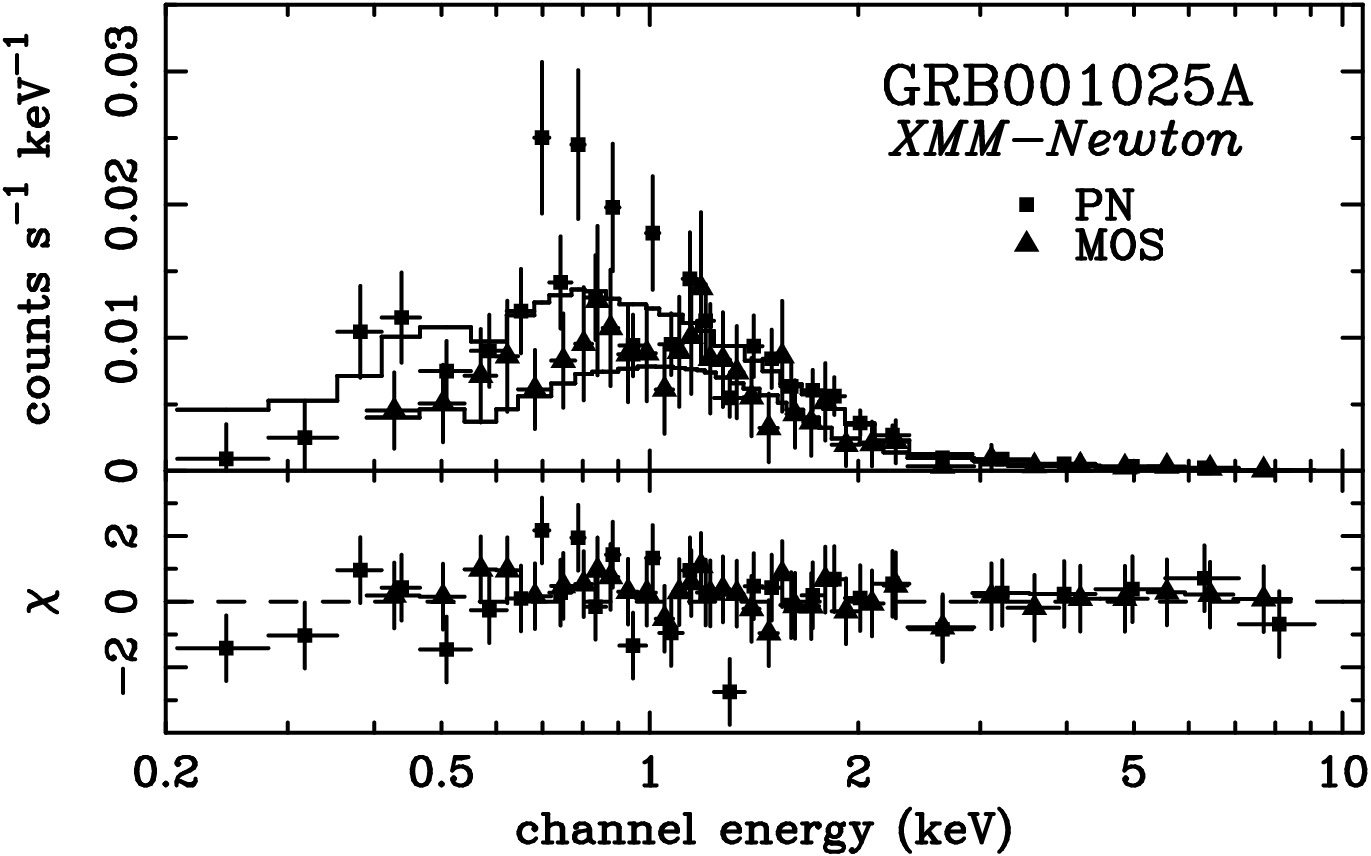}
  \caption{Same as in Figure~\ref{fig970508_1} for \grb001025.}
  \label{fig001025}
\end{figure}

\begin{figure}
  \includegraphics[scale=0.6,angle=-90]{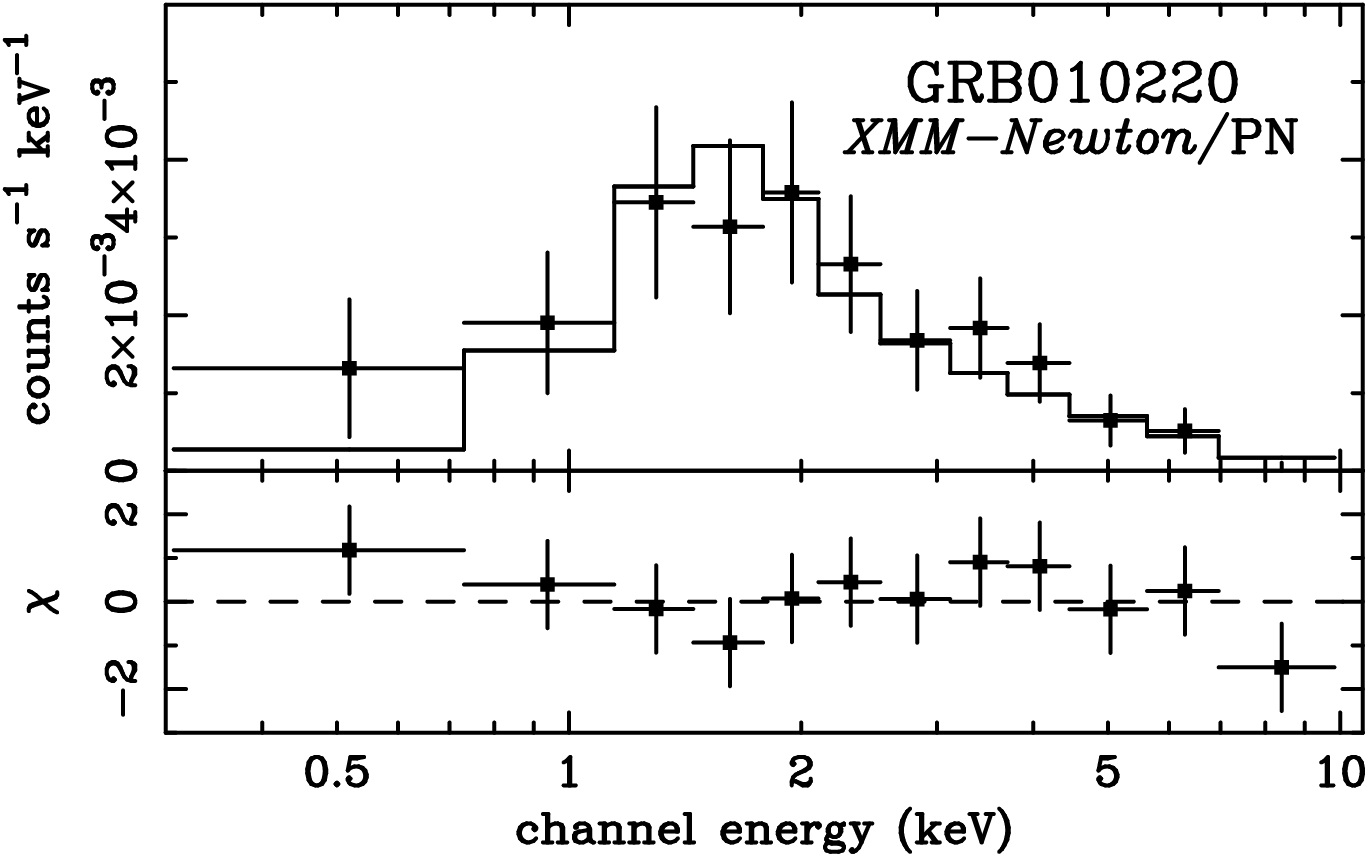}
  \caption{Same as in Figure~\ref{fig970508_1} for \grb010220.}
  \label{fig010220}
\end{figure}

\begin{figure}
  \includegraphics[scale=0.6,angle=-90]{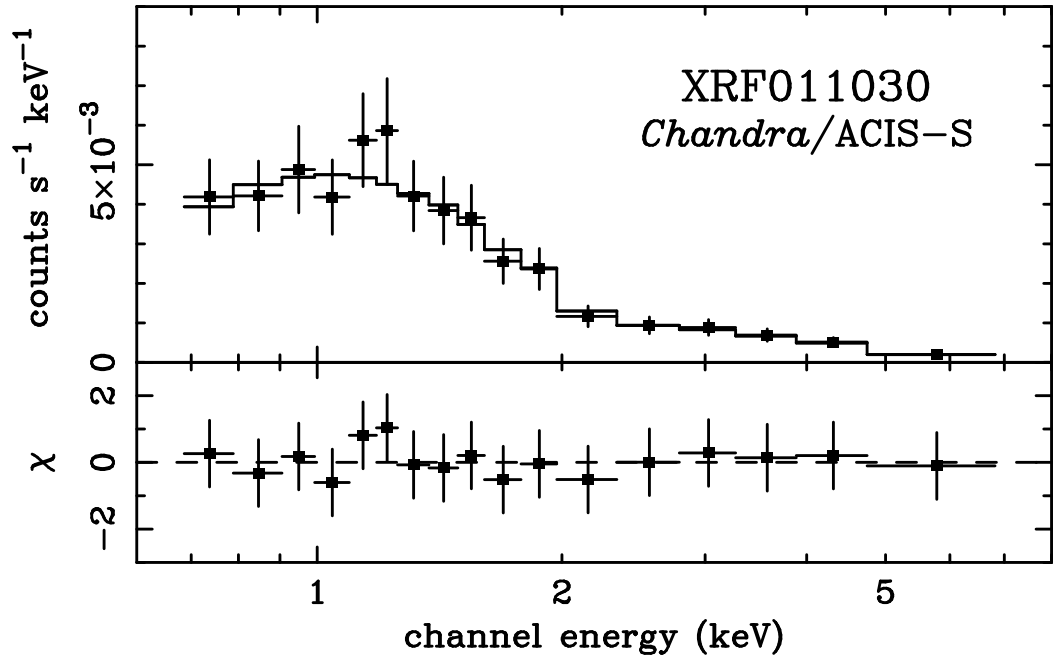}
  \caption{Same as in Figure~\ref{fig970508_1} for {\small XRF}011030.}
  \label{fig011030}
\end{figure}

\begin{figure}
  \includegraphics[scale=0.6,angle=-90]{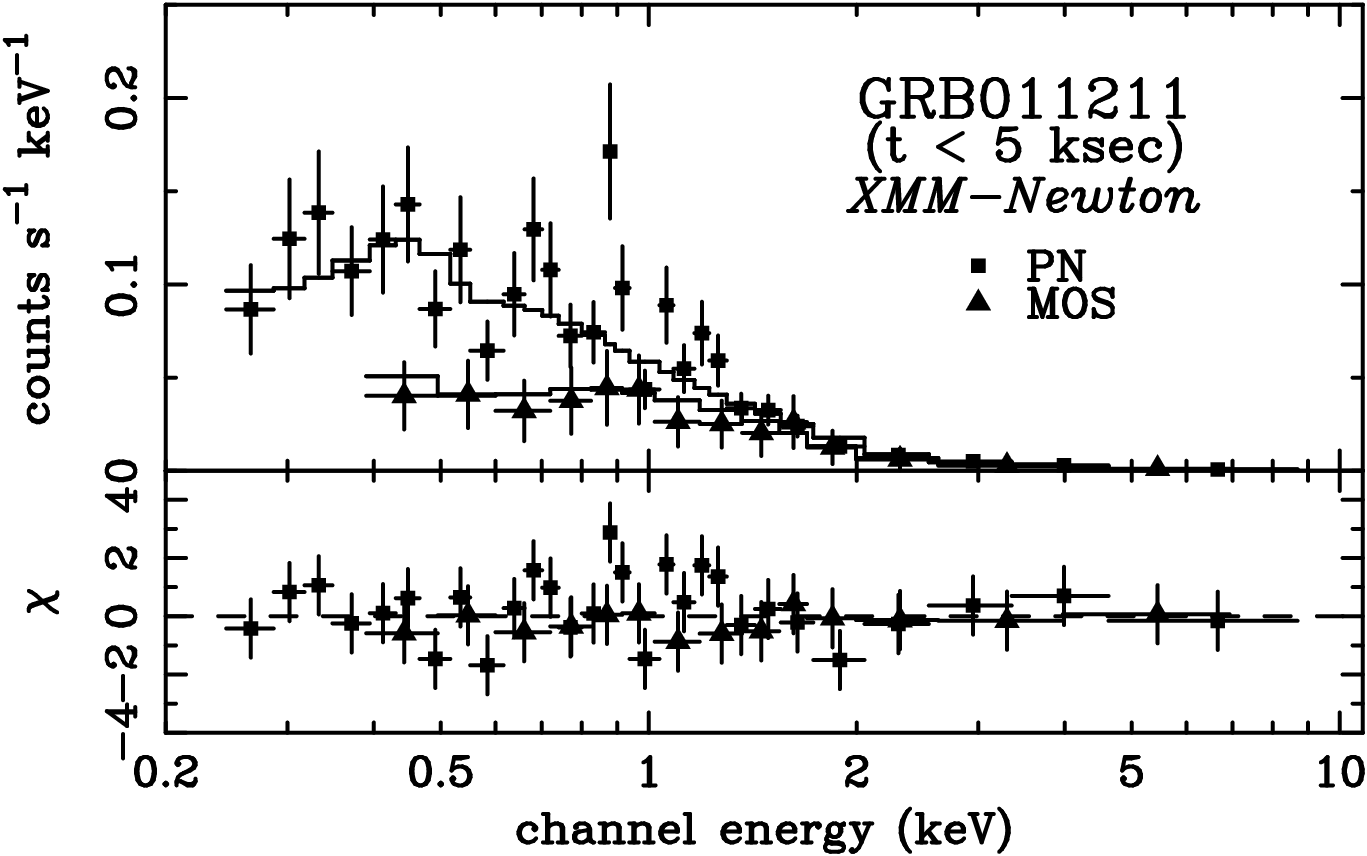}
  \caption{Same as in Figure~\ref{fig970508_1} for \grb011211 time segment 1.}
  \label{fig011211_1}
\end{figure}

\begin{figure}
  \includegraphics[scale=0.6,angle=-90]{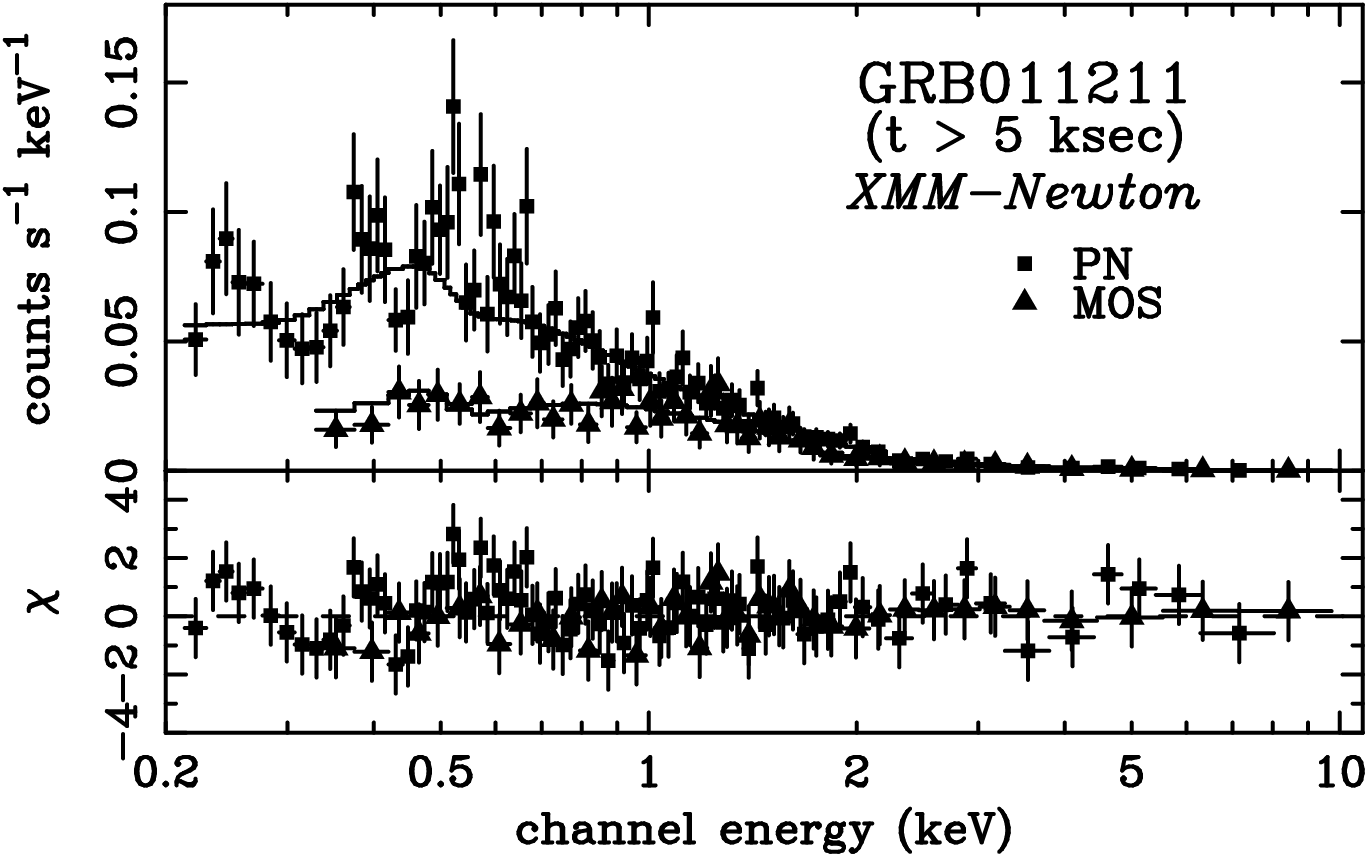}
  \caption{Same as in Figure~\ref{fig970508_1} for \grb011211 time segment 2.}
  \label{fig011211_2}
\end{figure}

\begin{figure}
  \includegraphics[scale=0.6,angle=-90]{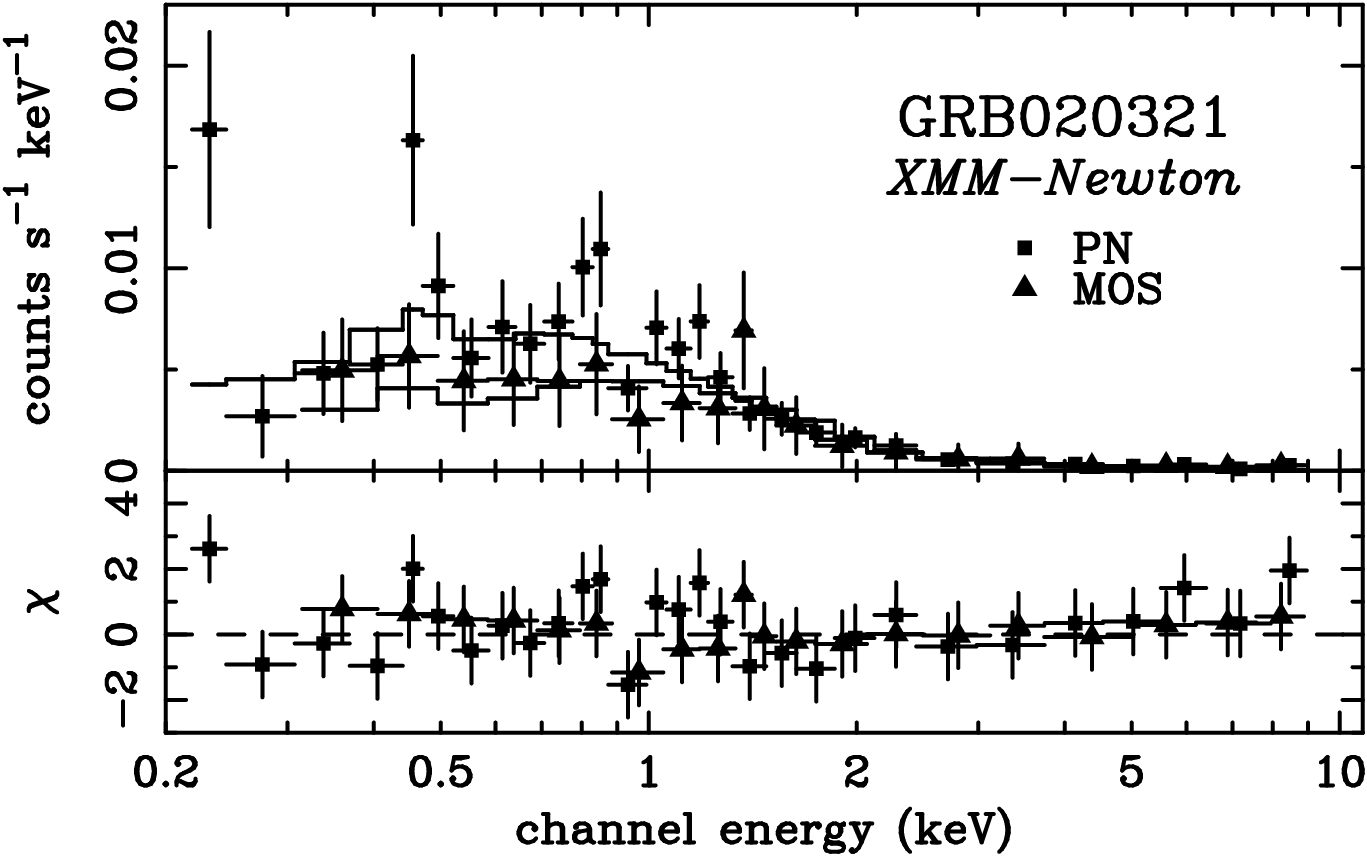}
  \caption{Same as in Figure~\ref{fig970508_1} for \grb020321.}
  \label{fig020321}
\end{figure}

\begin{figure}
  \includegraphics[scale=0.6,angle=-90]{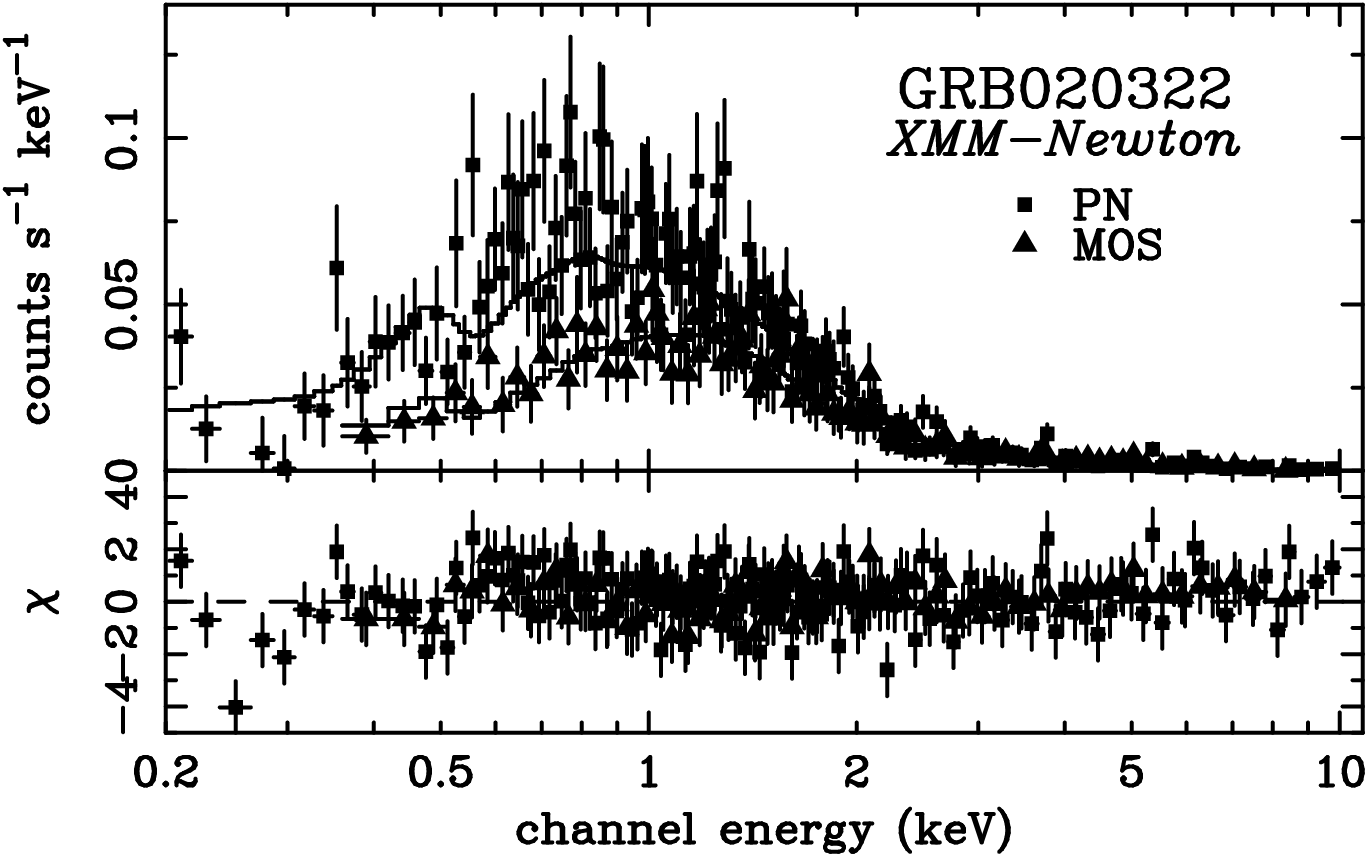}
  \caption{Same as in Figure~\ref{fig970508_1} for \grb020322.}
  \label{fig020322}
\end{figure}

\begin{figure}
  \includegraphics[scale=0.6,angle=-90]{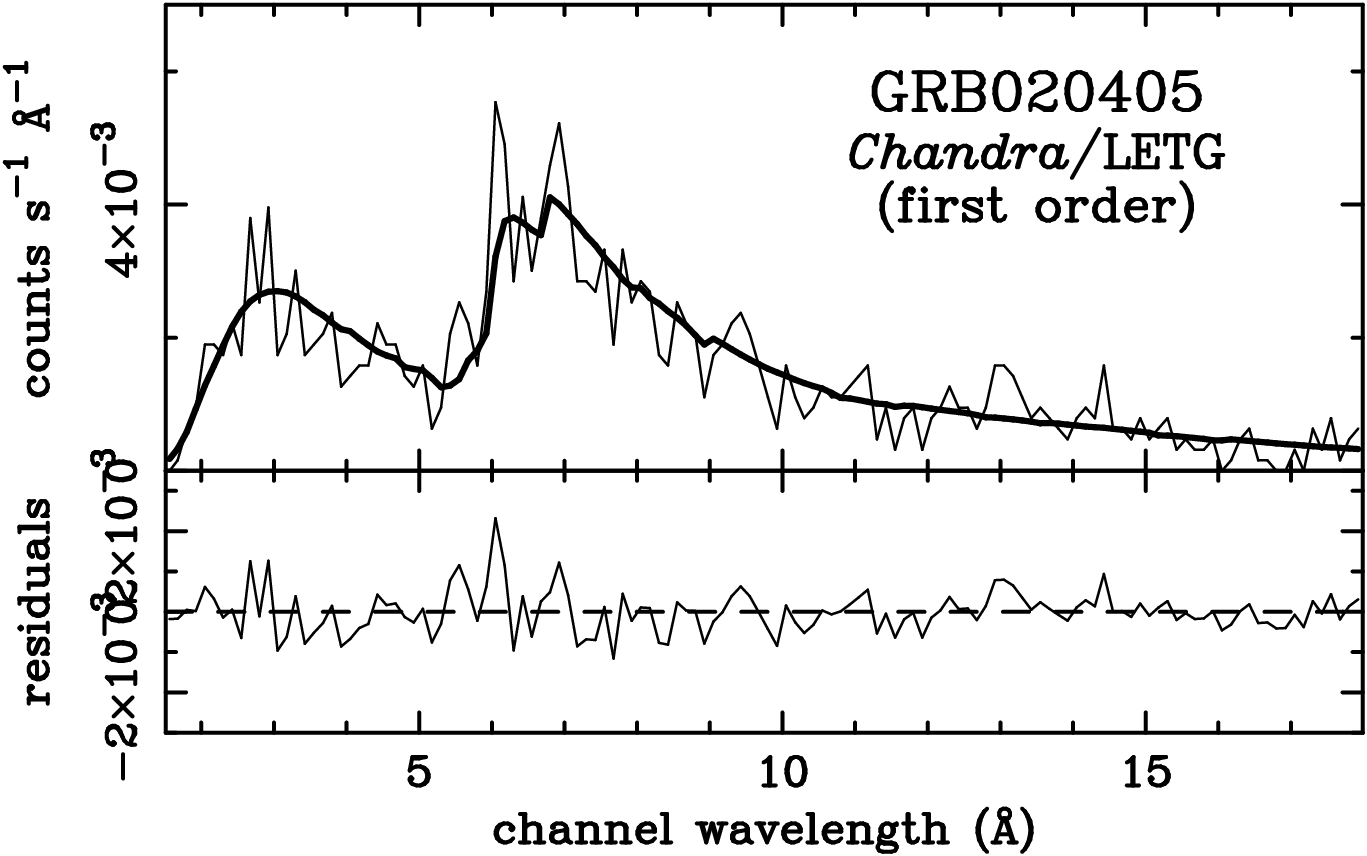}
  \caption{Same as in Figure~\ref{fig970508_1} for the dispersed spectrum of
           \grb020405.}
  \label{fig020405_m1}
\end{figure}

\begin{figure}
  \includegraphics[scale=0.6,angle=-90]{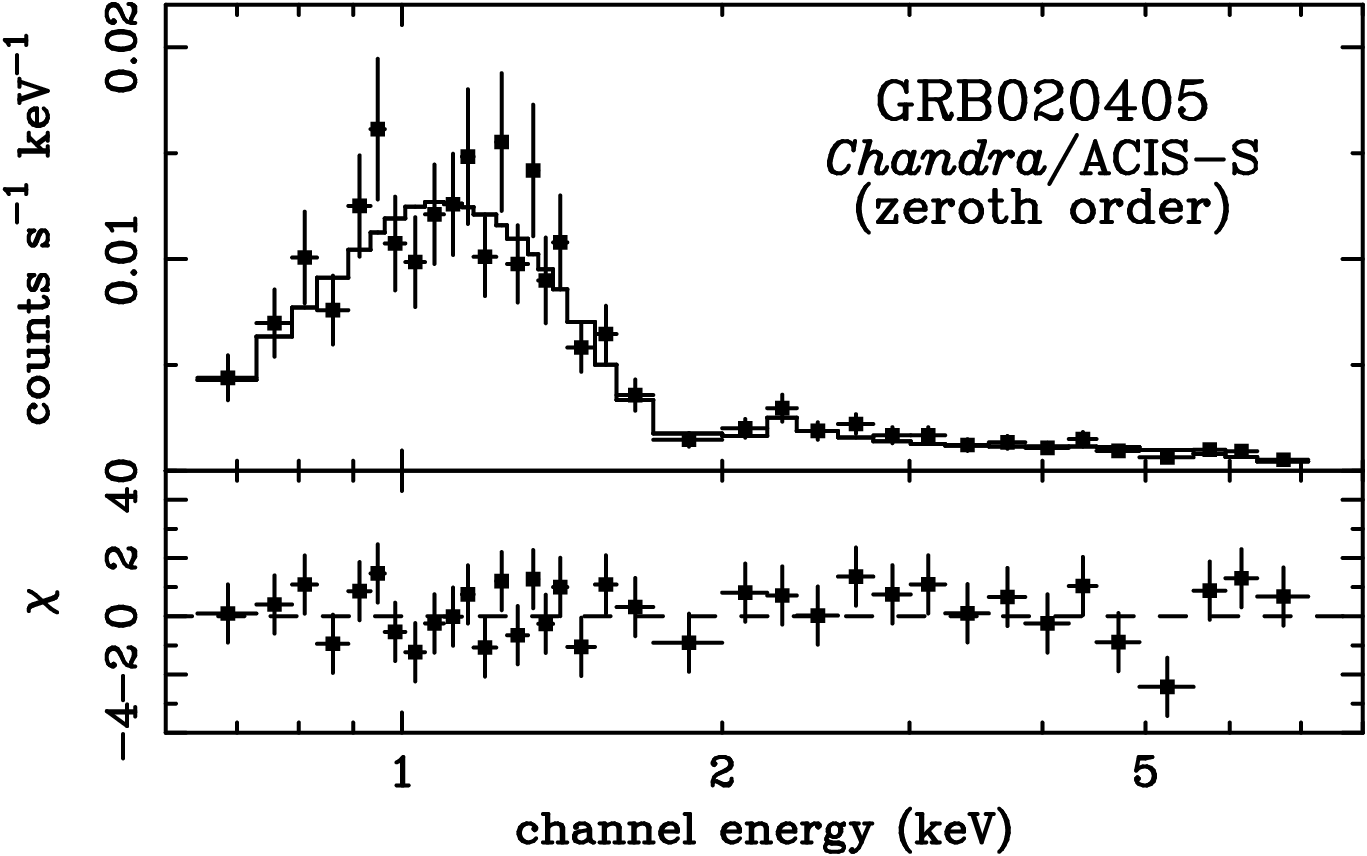}
  \caption{Same as in Figure~\ref{fig970508_1} for the zeroth order spectrum
           of \grb020405.}
  \label{fig020405_m0}
\end{figure}

\begin{figure}
  \includegraphics[scale=0.6,angle=-90]{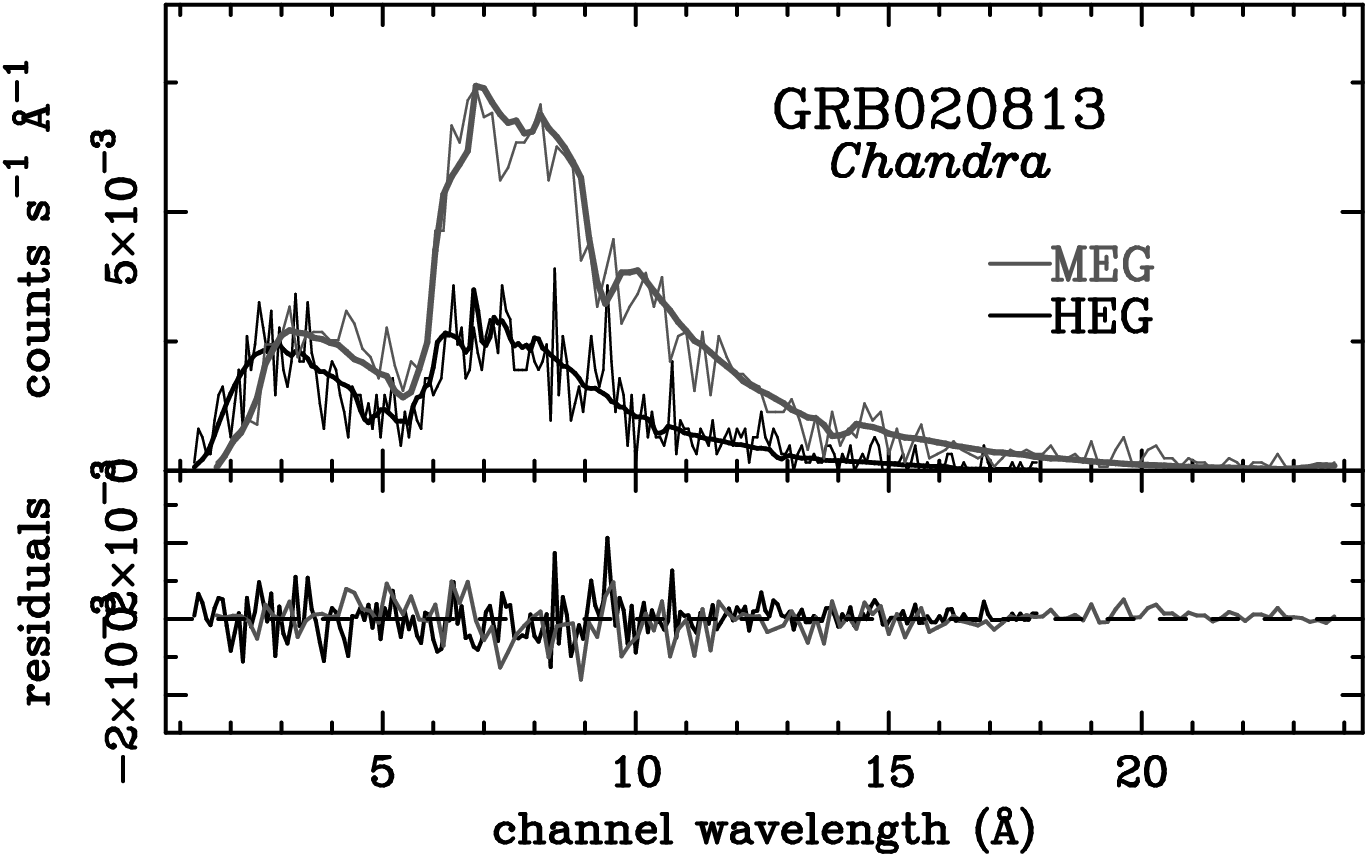}
  \caption{Same as in Figure~\ref{fig970508_1} for the dispersed spectrum of
           \grb020813.}
  \label{fig020813_m1}
  \end{figure}

\begin{figure}
  \includegraphics[scale=0.6,angle=-90]{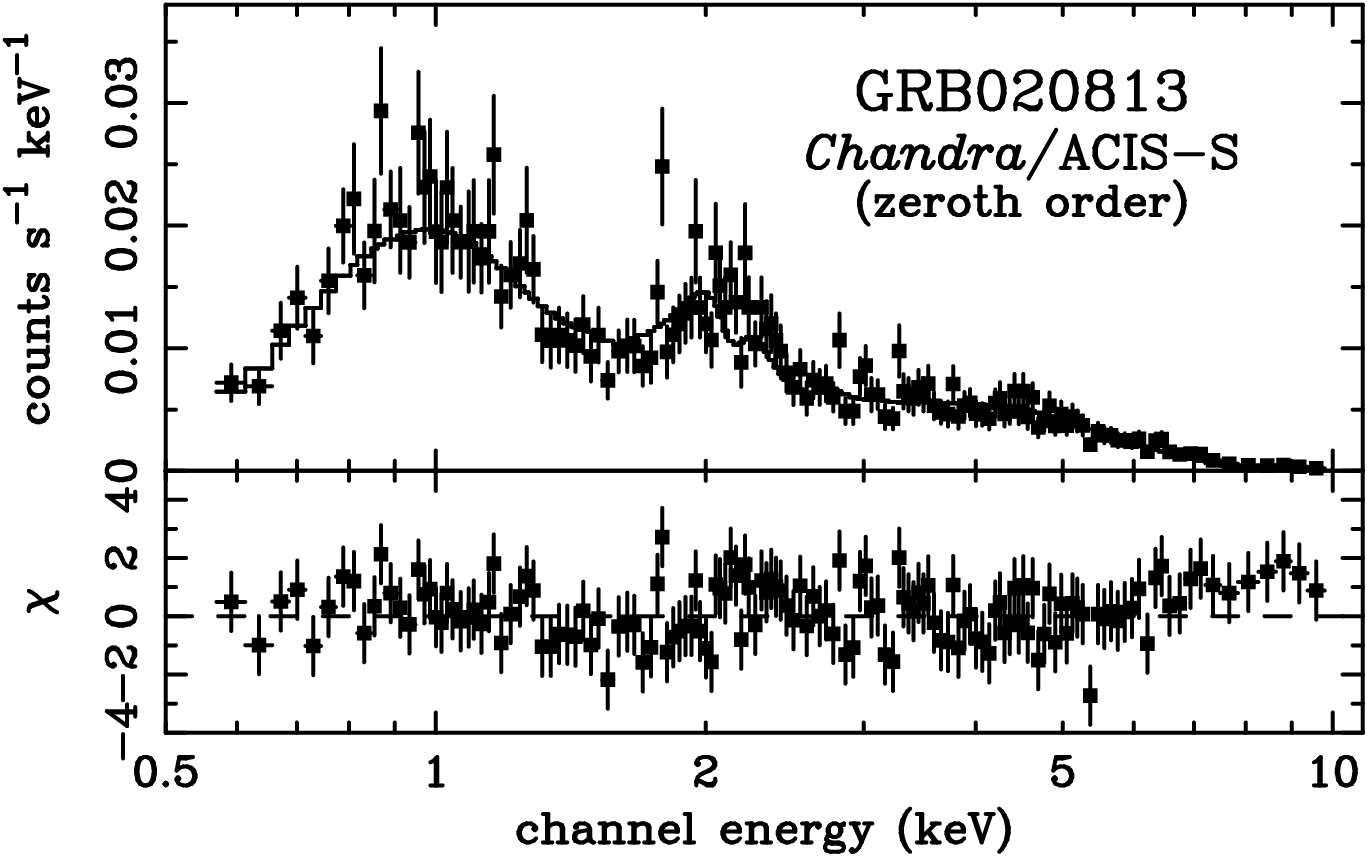}
  \caption{Same as in Figure~\ref{fig970508_1} for the zeroth order spectrum
           of \grb020813.}
  \label{fig020813_m0}
\end{figure}

\begin{figure}
  \includegraphics[scale=0.6,angle=-90]{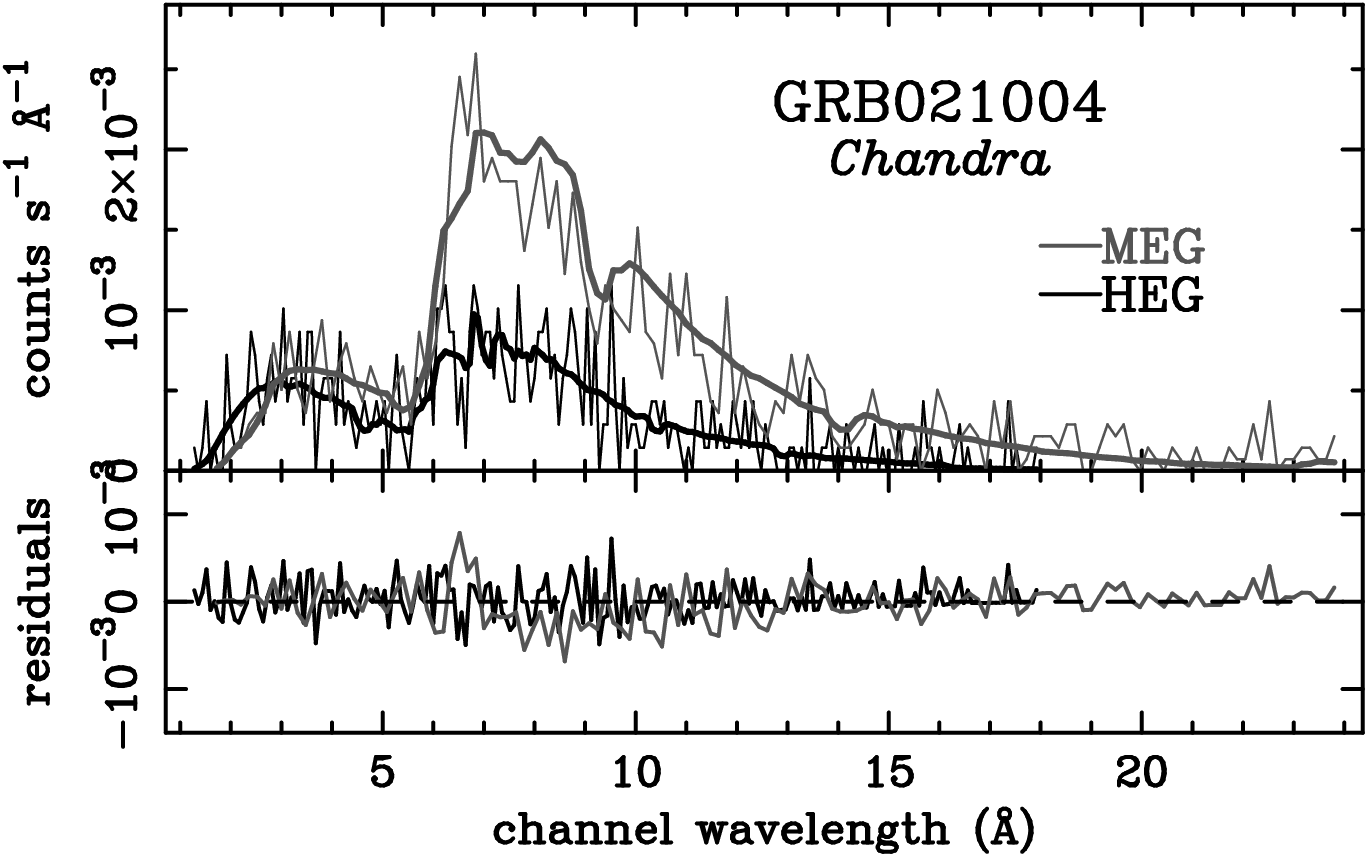}
  \caption{Same as in Figure~\ref{fig970508_1} for the dispersed spectrum of
           \grb021004.}
  \label{fig021004_m1}
\end{figure}

\begin{figure}
  \includegraphics[scale=0.6,angle=-90]{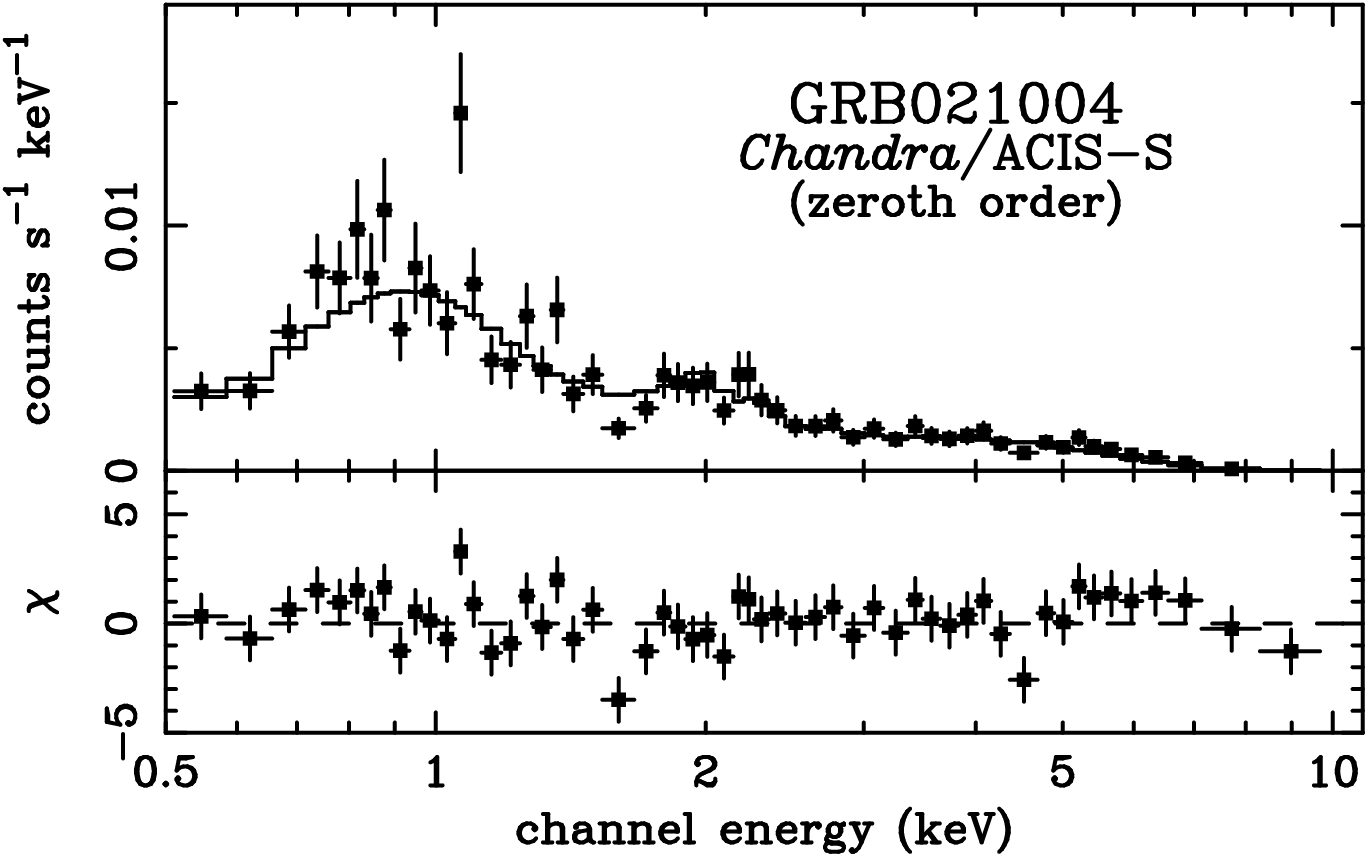}
  \caption{Same as in Figure~\ref{fig970508_1} for the zeroth order spectrum
           of \grb021004.}
  \label{fig021004_m0}
\end{figure}

\begin{figure}
  \includegraphics[scale=0.6,angle=-90]{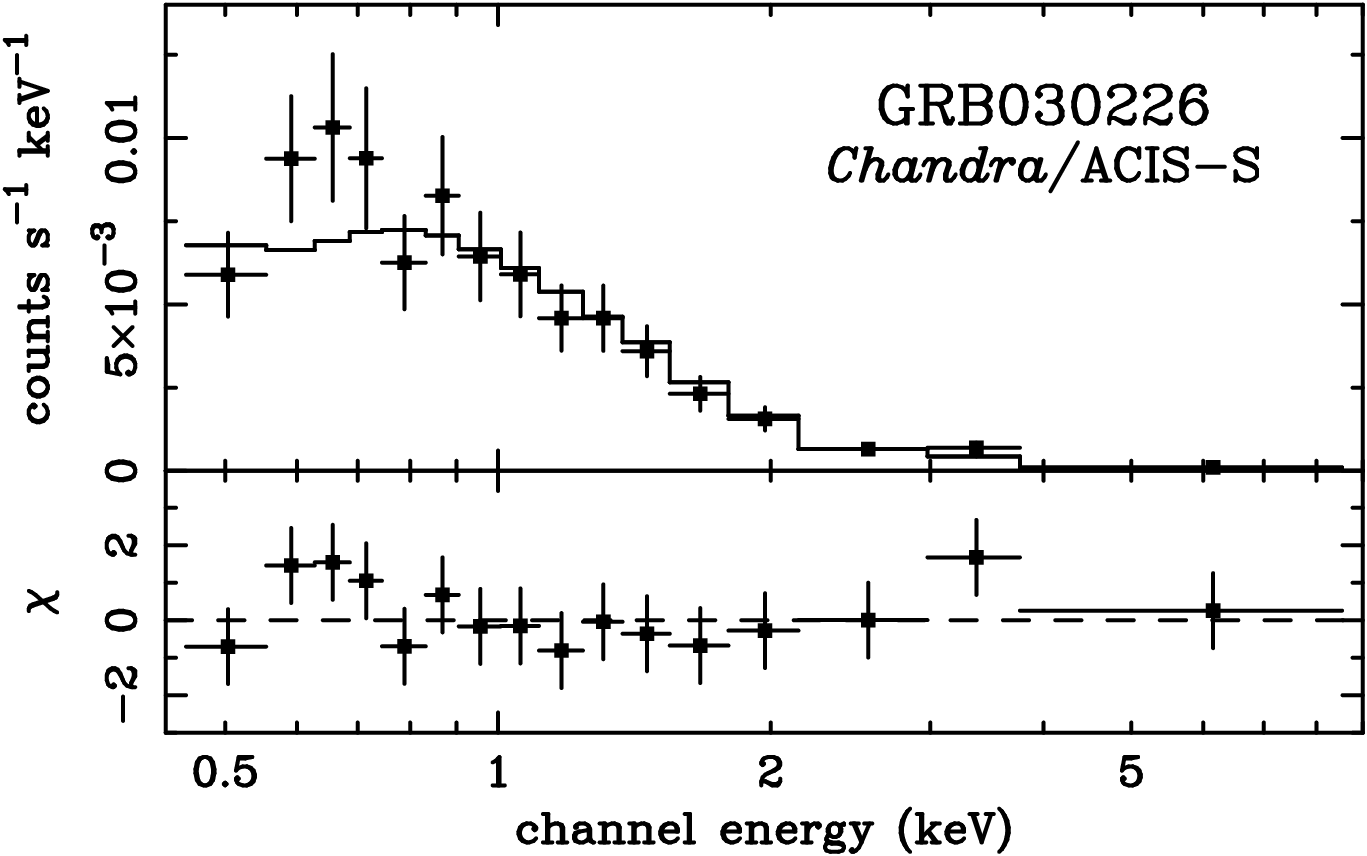}
  \caption{Same as in Figure~\ref{fig970508_1} for \grb030226.}
  \label{fig030226}
\end{figure}

\begin{figure}
  \includegraphics[scale=0.6,angle=-90]{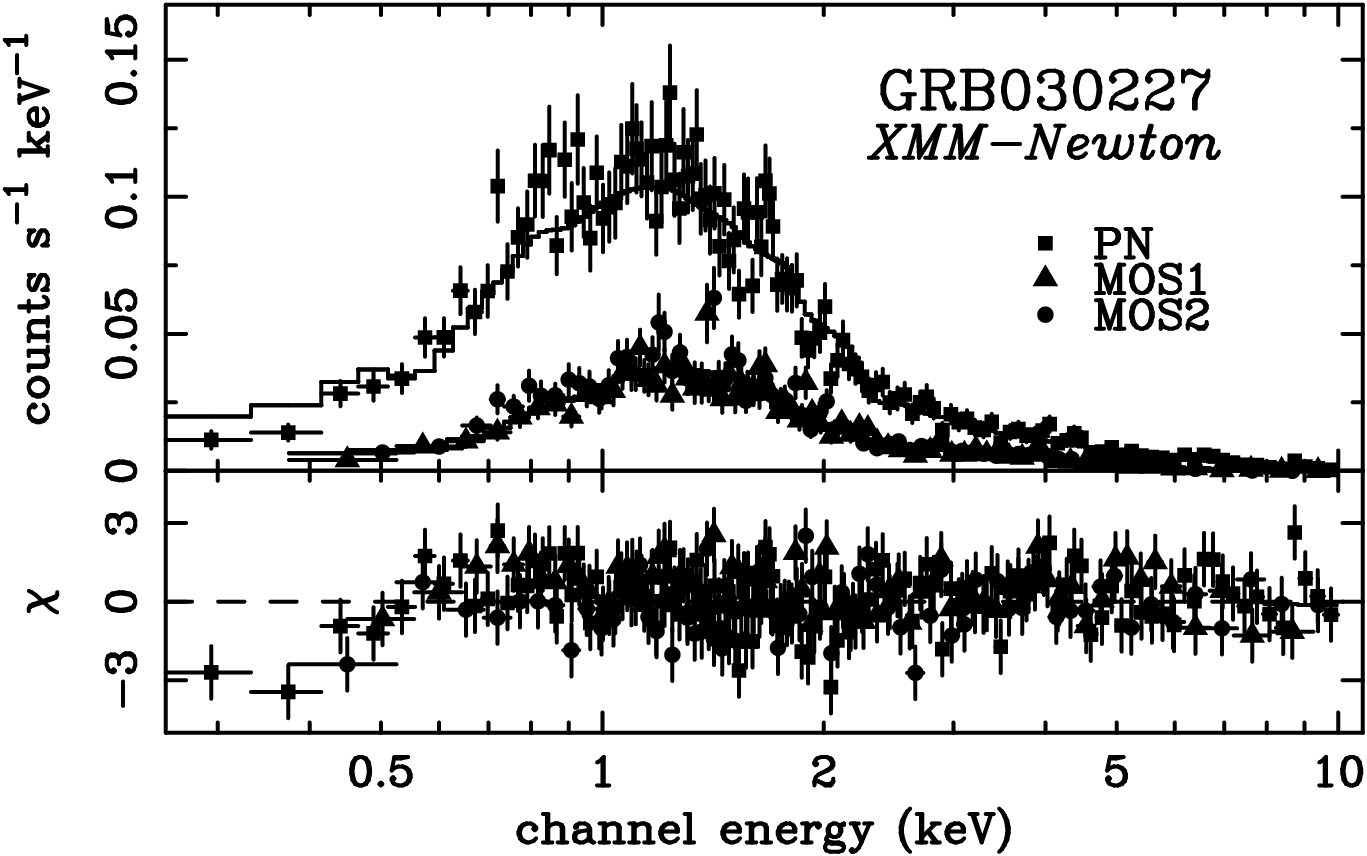}
  \caption{Same as in Figure~\ref{fig970508_1} for the dispersed spectrum of
           \grb030227.}
  \label{fig030227_m1}
\end{figure}

\begin{figure}
  \includegraphics[scale=0.6,angle=-90]{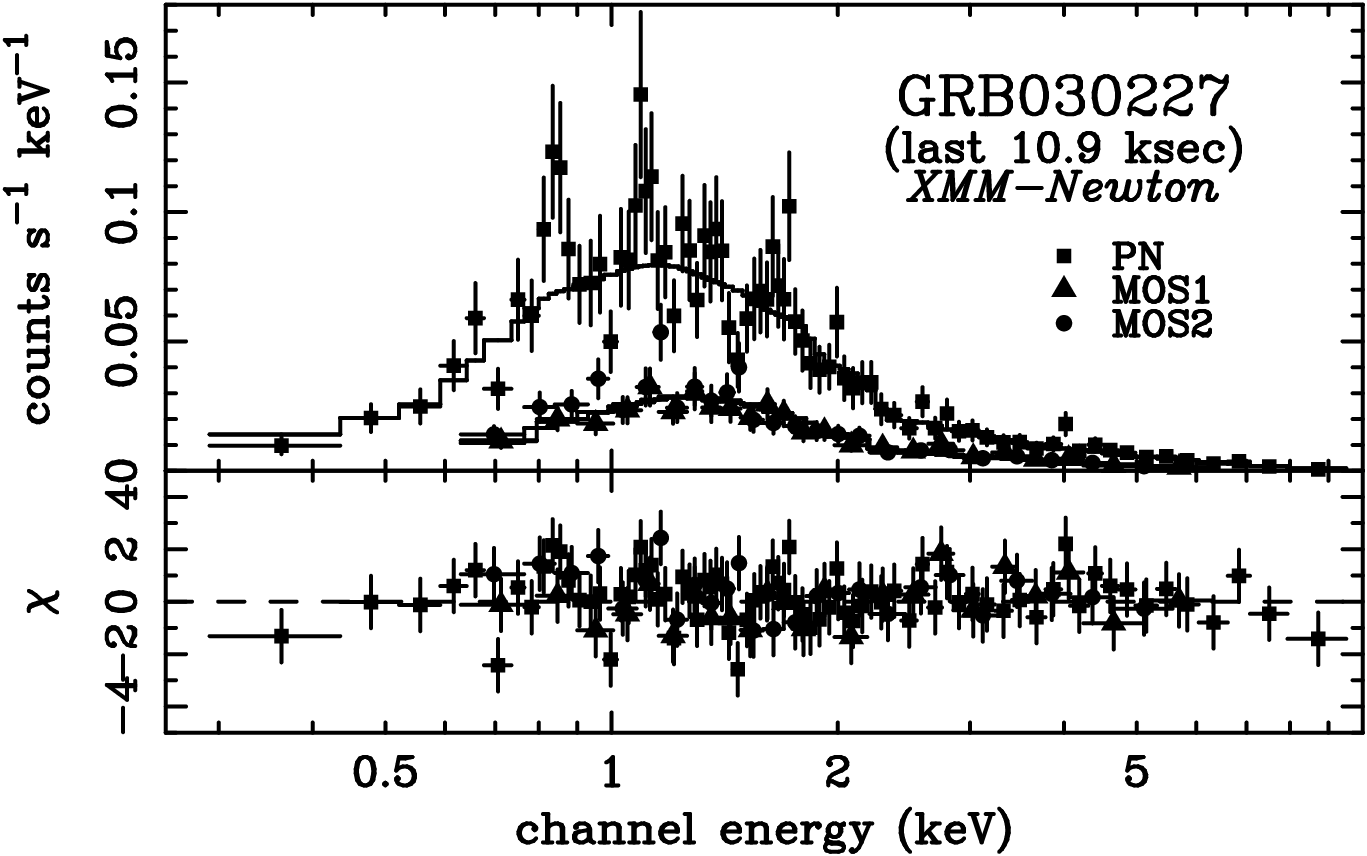}
  \caption{Same as in Figure~\ref{fig970508_1} for the zeroth order spectrum
           of \grb030227.}
  \label{fig030227_m0}
\end{figure}

\begin{figure}
  \includegraphics[scale=0.6,angle=-90]{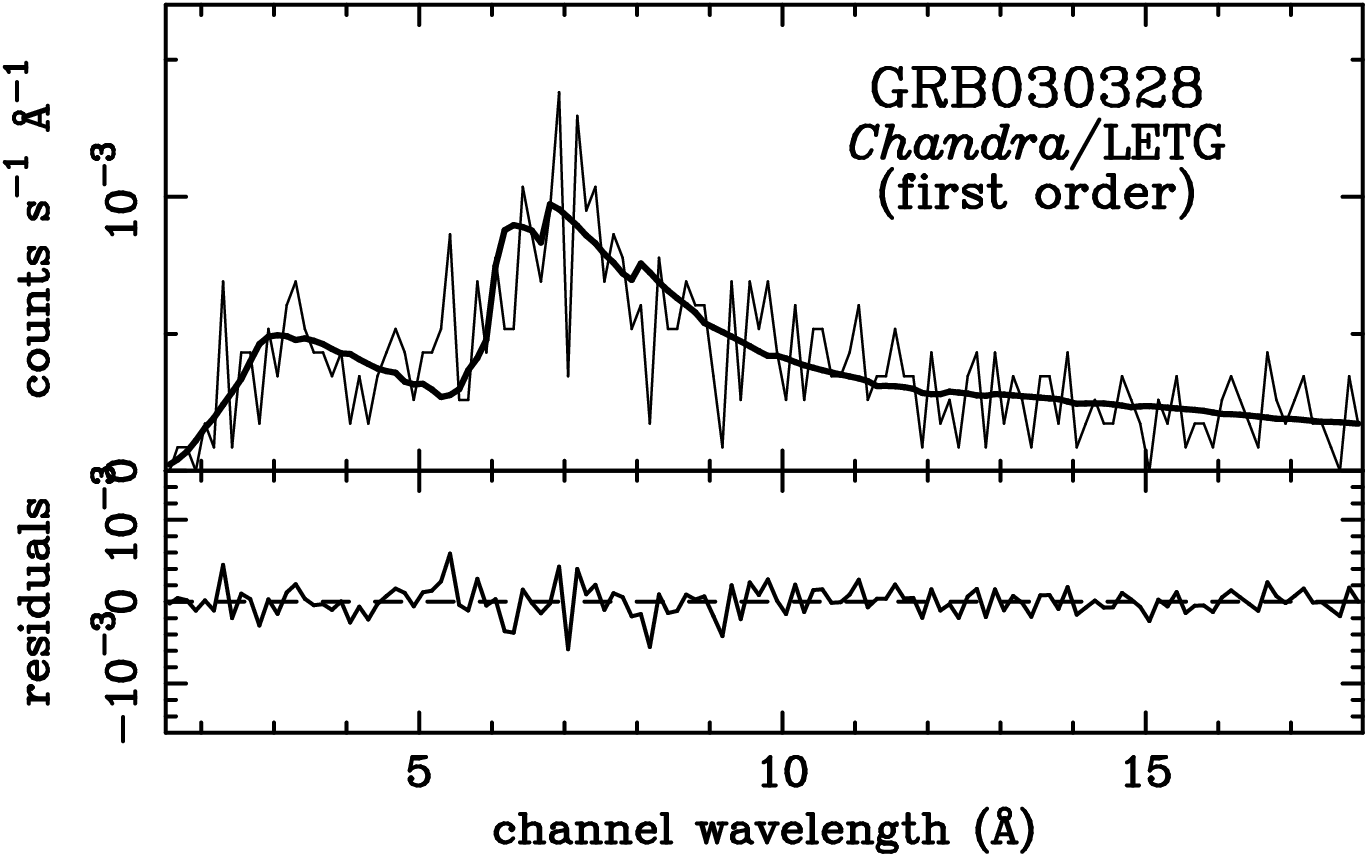}
  \caption{Same as in Figure~\ref{fig970508_1} for the dispersed spectrum of
           \grb030328.}
  \label{fig030328_m1}
\end{figure}

\begin{figure}
  \includegraphics[scale=0.6,angle=-90]{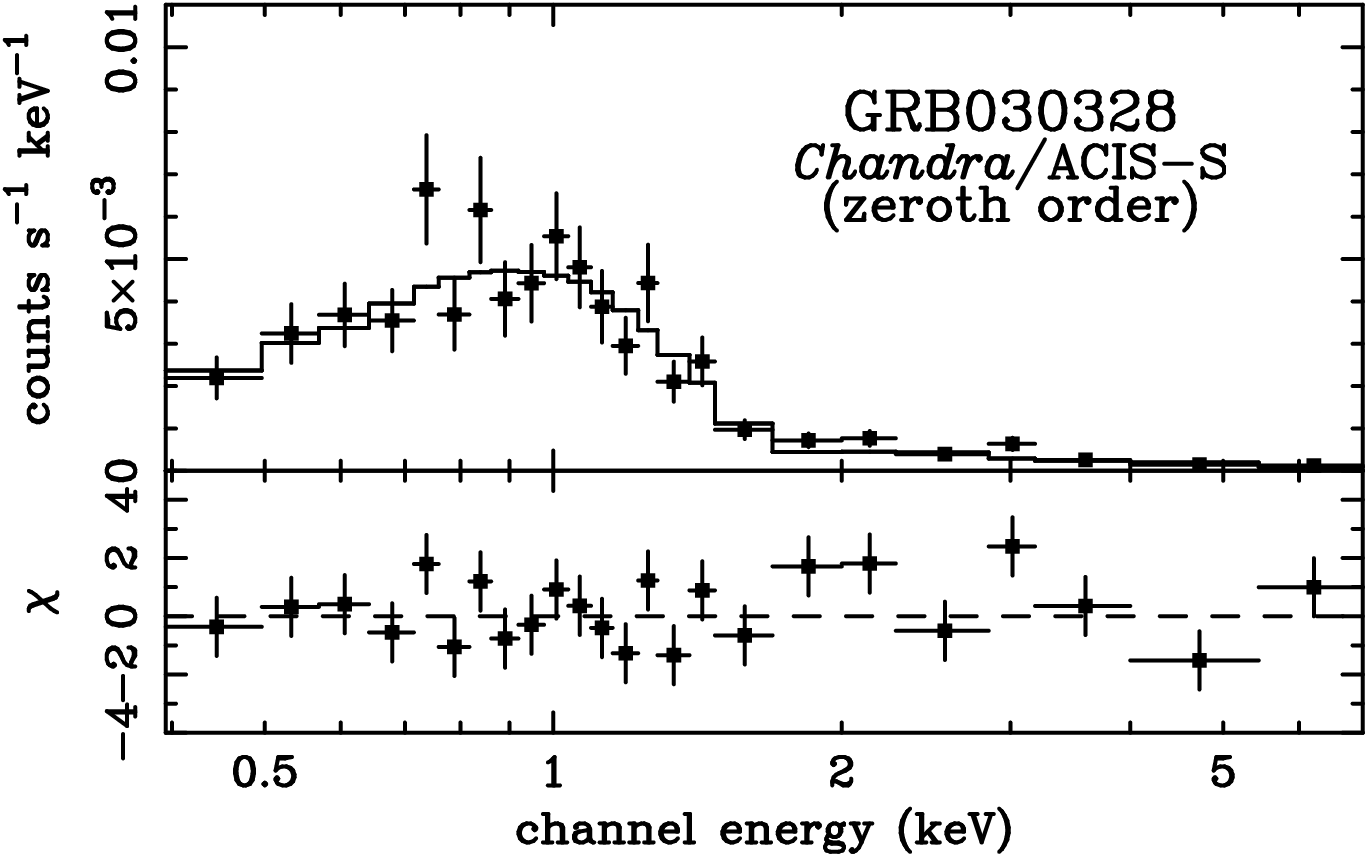}
  \caption{Same as in Figure~\ref{fig970508_1} for the zeroth order spectrum of
           \grb030328.}
  \label{fig030328_m0}
\end{figure}

\begin{figure}
  \includegraphics[scale=0.6,angle=-90]{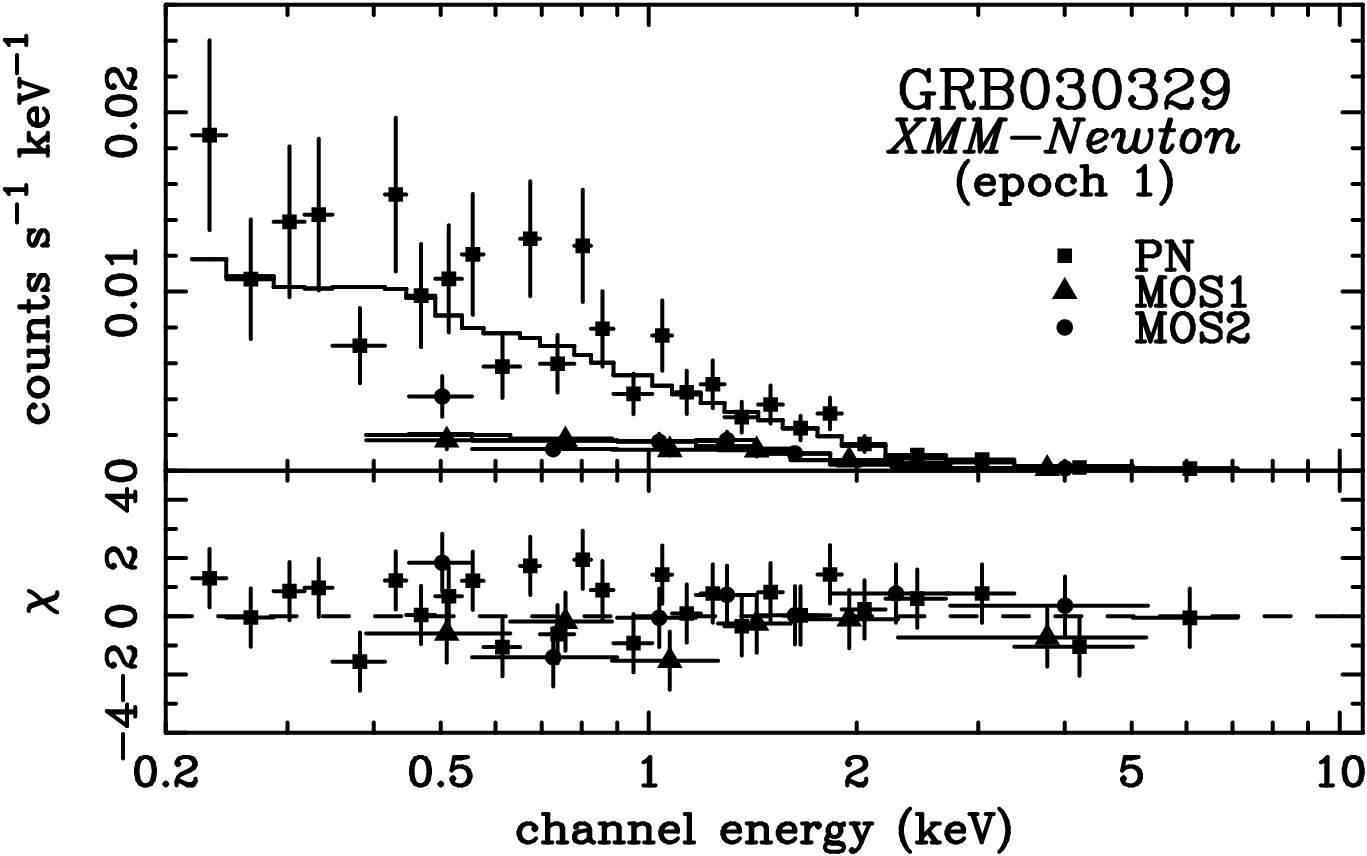}
  \caption{Same as in Figure~\ref{fig970508_1} for \grb030329 epoch 1.}
  \label{fig030329_1}
\end{figure}

\begin{figure}
  \includegraphics[scale=0.6,angle=-90]{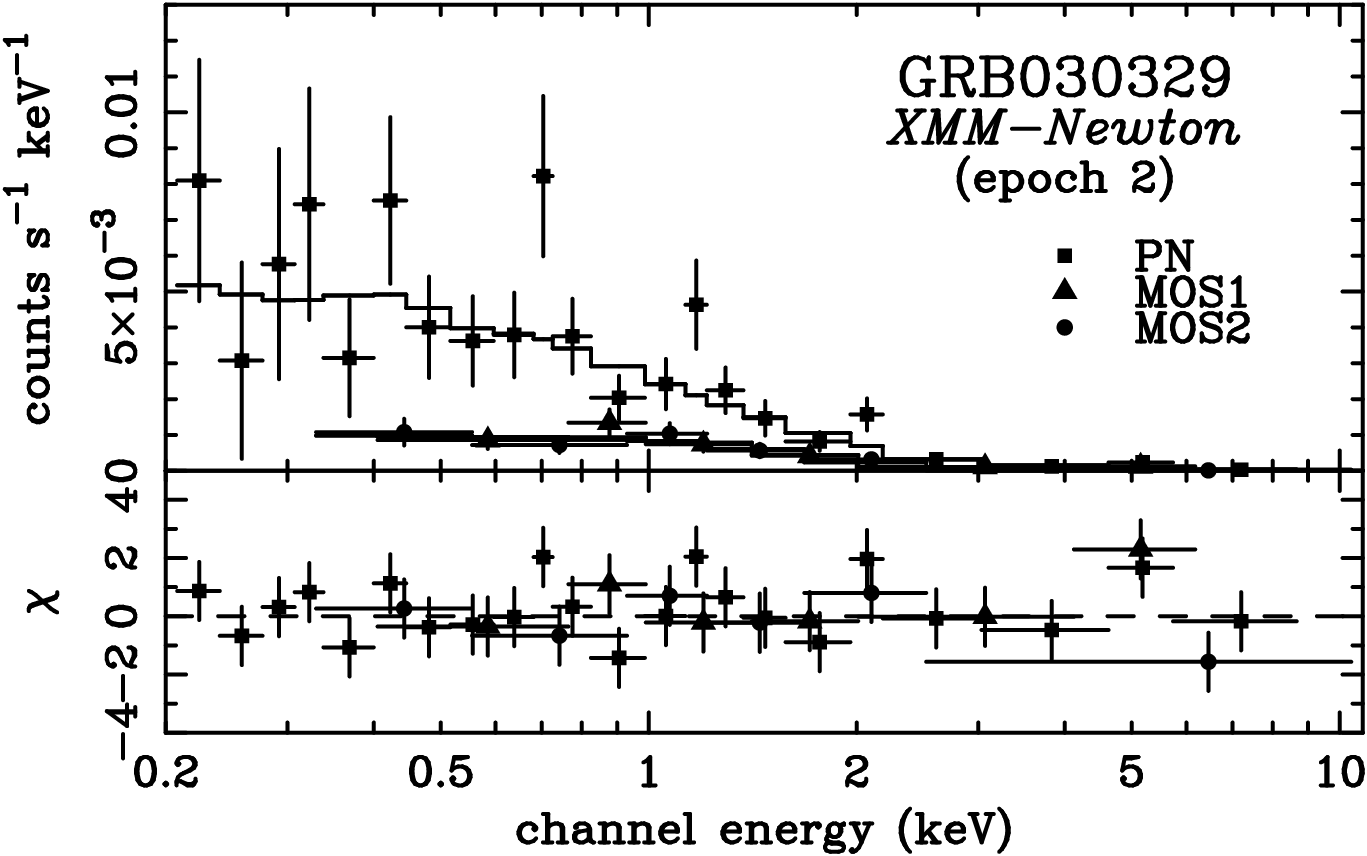}
  \caption{Same as in Figure~\ref{fig970508_1} for \grb030329 epoch 2.}
  \label{fig030329_2}
\end{figure}

\begin{figure}
  \includegraphics[scale=0.6,angle=-90]{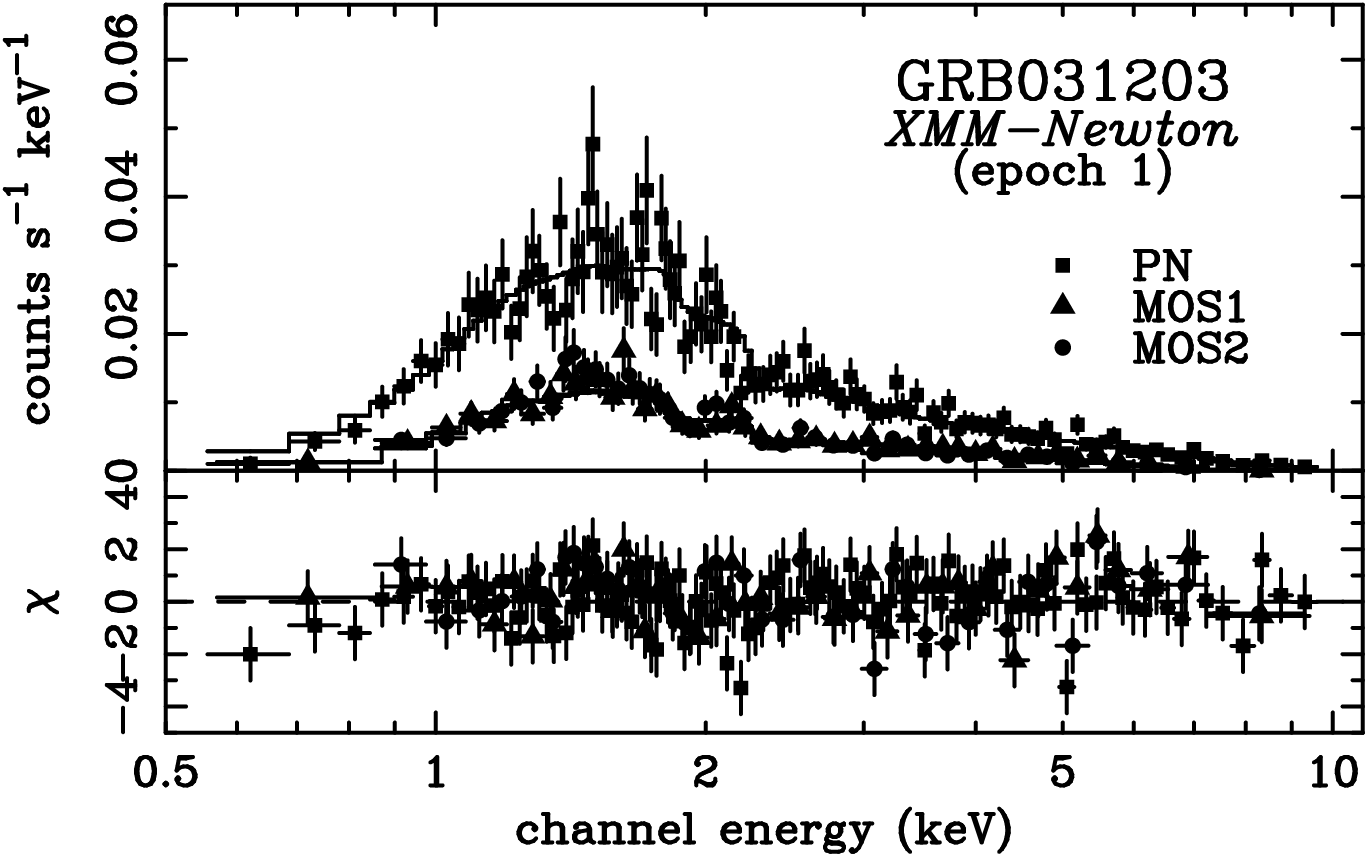}
  \caption{Same as in Figure~\ref{fig970508_1} for \grb031203 epoch 1.}
  \label{fig031203_1}
\end{figure}

\begin{figure}
  \includegraphics[scale=0.6,angle=-90]{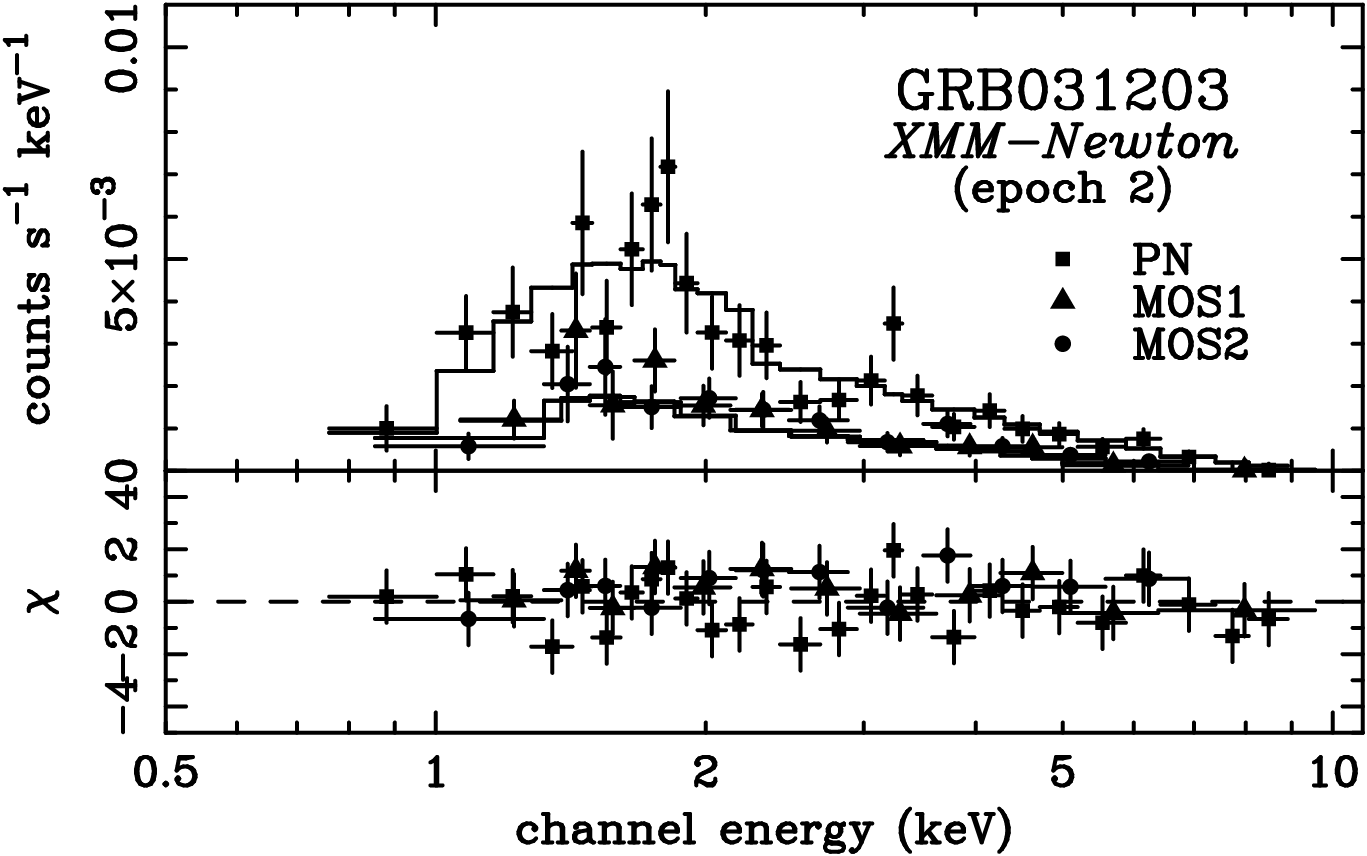}
  \caption{Same as in Figure~\ref{fig970508_1} for \grb031203 epoch 2.}
  \label{fig031203_2}
\end{figure}

\begin{figure}
  \includegraphics[scale=0.6,angle=-90]{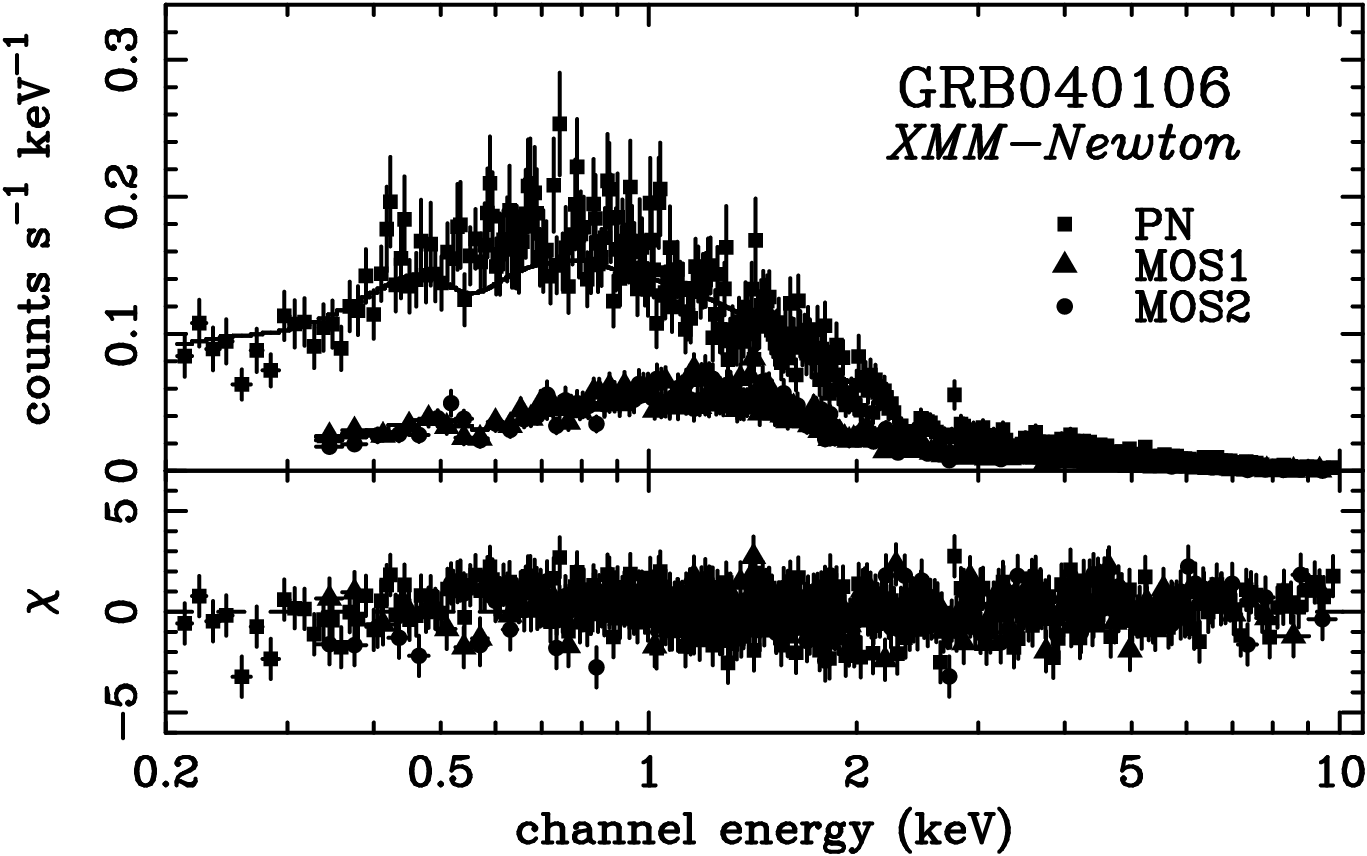}
  \caption{Same as in Figure~\ref{fig970508_1} for \grb040106.}
  \label{fig040106}
\end{figure}

  Figures~\ref{fig970508_1} -- \ref{fig040106} show the binned spectra, and
  $\chi$ value for each spectral bin for the respective best-fit model.
  Table~\ref{tbl:fits} lists the best-fit parameters and the resulting
  $\chi^2$ and null-hypothesis probability, $P(\chi,\nu)$.  We do not list
  $P(\chi,\nu)$ for the dispersed grating spectra because they were fit with a
  $C$-statistic due to the low number counts detected per spectral bin.  The
  sample as a whole is consistent with the hypothesis of absorbed power law
  X-ray emission; out of 27 separate spectra that are fit using
  $\chi^2$-minimization, four have $P(\chi,\nu) < 0.1$, consistent with what
  one would expect statistically from a uniform sample.  If we consider the
  four spectra with $P(\chi,\nu) < 0.1$; the zeroth order spectrum of
  \grb991216, \grb000214, the zeroth order spectrum of \grb021004, and the
  zeroth order spectrum of \grb030328 (Figures~\ref{fig991216_m0},
  \ref{fig000214}, \ref{fig021004_m0}, and \ref{fig030328_m0}, respectively),
  only \grb991216 and \grb000214 appear to show discrete residuals suggestive
  of statistically significant line features.  Examination of the other
  spectra shows one other interesting case, \grb970828 segment B
  (Figure~\ref{fig970828_b}), where the residuals appear to be concentrated in
  a broad feature (although $P(\chi,\nu)$ is acceptable).  As we discuss in
  the following section, however, the $\chi^2$ statistic is not sensitive to
  weak and/or localized fluctuations, so the $\chi^2$ values are not in
  general particularly useful for line searches.  Similarly, $F$-tests yield
  unreliable values for the significances of line-like features as described
  in detail by \citet{protassov02}.

\section{Monte Carlo Simulations}
\label{section_mc_sims}

  As discussed in the previous section, the spectra of most sources are
  consistent with an absorbed power-law model.  However, since the $\chi^2$
  statistic is not sensitive to small but systematic deviations that would
  result, for example, from the presence of a weak narrow line, the true null
  hypothesis probability may be lower than that inferred from the $\chi^2$
  statistic.  In addition, the presence of a weak line may be overlooked when
  the spectra are binned to contain at least 20 counts per bin so that the
  $\chi^2$ statistic can be used for parameter fitting.

  In this section, we describe the Monte Carlo (\mc) simulations that are
  performed to test whether any of the discrete, unresolved line-like features
  are statistically significant.  The method is identical to that adopted by
  \citet{rutledge02}, which was used to estimate the multi-trial statistical
  significance of the reported soft X-ray features observed in the spectrum of
  \grb 011211 \citep{reeves02}.  We briefly summarize the procedures below.

  For each source spectrum and time interval, we perform $10^5$ simulations
  based on the respective best-fit models.  We denote the detected number of
  photons (source plus background events that lie within the extraction region
  and time selection) by $C_{\rm{tot}}$ and the estimated number of background
  photons by $C_{\rm{bkg}}$, where $C_{\rm{tot}} > C_{\rm{bkg}}$.  Since we do
  not know the true number of source photons and hence the source flux, we
  simulate a fixed number of source plus background photons.  The number of
  background photons is estimated using a Poisson deviate with an average of
  $C_{\rm{bkg}}$ counts\footnote{There are several alternative approaches one
  might adopt in choosing the number of counts to generate in each \mc\
  simulation.  One possibility is to use a Poisson deviate with an average of
  $C_{\rm{tot}}$ and $C_{\rm{bkg}}$ counts for the total and background
  portion of the spectra, respectively.  This approach, however, assumes that
  the observed flux is in fact the true source flux, which is, in general, not
  the case.  We, therefore, choose to fix the total number of simulated counts
  to the observed value, since this is the only true observable parameter.
  Adopting a Poisson deviate of the observed number of counts would, in any
  case, result in a lower significance of any discrete feature, so our derived
  significances assuming a fixed number of counts can be regarded as upper
  limits.}.  Background events are randomly selected from analytic background
  spectral models, which are derived from fits to the spectra obtained from
  off-axis regions of the same data sets (\chandra\ and \xmm) or from long
  blank field observations (\asca\ and \sax).

  We generate counts spectra $I_j$ for each simulation and convolve it
  through a matched filter to compute the $C(E_i)$ values defined as,
\begin{equation}
  C(E_i) = \sum_{j[E_i + 3\sigma(E_i)]}^{j[E_i - 3\sigma(E_i)]} I_j
           \frac{1}{\sqrt{2\pi}\sigma(E_i)}~{\rm exp}\left[-\frac{1}{2}
           \left(\frac{E_i-E_j}{\sigma(E_i)}\right)^2\right] \delta E_j,
\end{equation}
  where $E_i$ is the energy in \kev\ of \pinv\ channel $i$ and $\sigma(E_i)$
  is the energy-dependent gaussian width of the instrument response kernel
  (see, \citealt{rutledge02}).  For each instrument, the shape of the response
  kernel is extracted at five energies -- 0.5, 1, 2, 5, and 10 \kev\ -- from
  their respective response matrices and fit to a gaussian model.  The
  best-fit $\sigma(E)$ at these five points are then fit to a quadratic
  function of energy of the form,
\begin{equation}
  \label{eq:sigmaE}
  \sigma(E) = c_0 + c_1 E_{\rm{\kev}} + c_2 E_{\rm{\kev}}^2,
\end{equation}
  where $E_{\rm{\kev}}$ is the energy in \kev.  The values for $c_i$ derived
  in this manner for each of the instrument are listed in Table~\ref{tbl:kernel},
  and they reproduce the energy dependence of $\sigma$ to within $\sim 20$\%
  across the entire bandpass.  The response kernels, in general, are
  asymmetric and non-gaussian.  However, a gaussian reproduces the core of the
  kernel fairly well, and the results are not sensitive to their exact shapes
  outside of the core.  For the \heg\ and \meg, we adopt a cubic polynomial of
  the form,
\begin{equation}
  \label{eq:sigmaL}
  \sigma(\lambda) = c_0 + c_1 \lambda_{\mbox{\small \AA}} + c_2
    \lambda_{\mbox{\small \AA}}^2 + c_3 \lambda_{\mbox{\small \AA}}^3,
\end{equation}
  where the $\lambda_{\mbox{\small \AA}}$ is the wavelength in \AA.  We adopt
  the parameters $c_i$ as listed on the {\small MIT HETG} web
  site\footnote{http://space.mit.edu/HETG/technotes/ede\_0102/ede\_0102.html}.
  We list them in Table~\ref{tbl:kernel} for completeness.  Finally, for the
  \leg, we assume a constant wavelength dispersion of $\sigma = 0.0195$~\AA,
  which is a good approximation in the $2 - 20$~\AA\ region.

  As emphasized by \citet{rutledge02}, this procedure maximizes the
  signal-to-noise of line-like features with intrinsic widths that are smaller
  than or comparable to the instrument response.  For spectra acquired with
  the \ccd, the search is sensitive to lines with widths narrower than $v \sim
  10,000~\rm{km~s}^{-1}$.  For the handful of high-resolution spectra in our
  sample, however, the velocity range spanned by adopting the instrument width
  is rather limited ($v \sim 1000~\rm{km~s}^{-1}$).  In these cases, we use
  widths that are up to 16 times the instrument width to search for
  potentially broader lines as well.

\begin{figure}
  \includegraphics[scale=0.6,angle=-90]{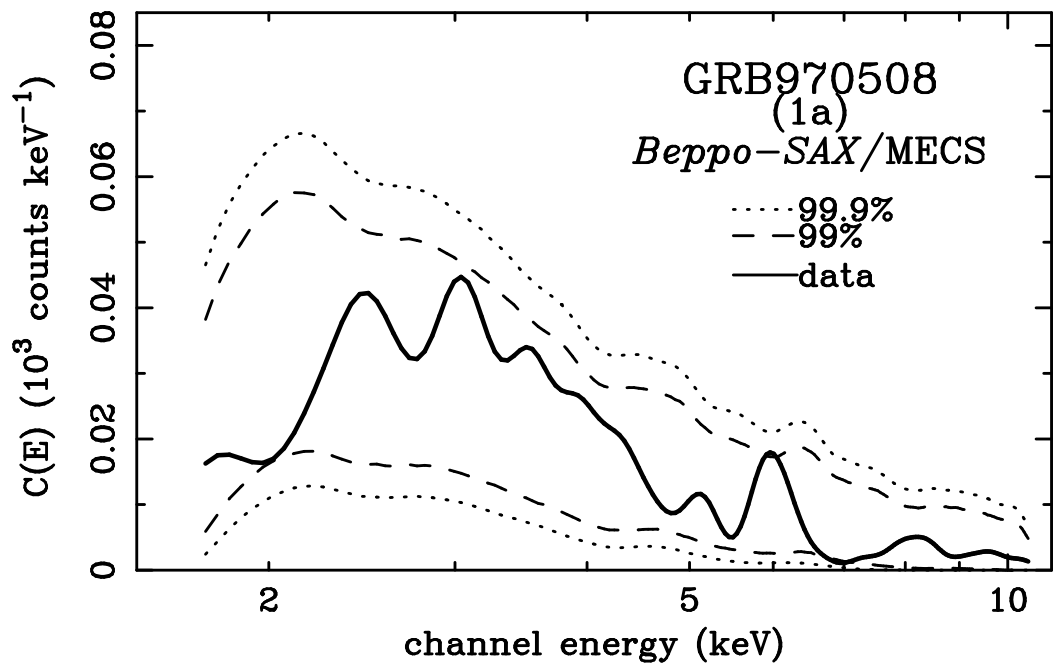}
  \caption{Upper and lower limits for $C(E)$ derived from Monte Carlo
           simulations based on the best-fit continuum model parameters for
           \grb970508 time section 1a.  The following figures show the same
           for the bursts listed in Table~\ref{tbl:fits}.  The data are
           represented as thick solid curves, while single-trial 99\%, 99.9\%,
           and 99.99\% limits as functions of the energy are shown as dashed,
           dotted, and thin gray solid curves, respectively.  Features in
           excess of the 99.9\% limits in our analyses are marked with arrows.
           Previously claimed detections by other authors are labeled by
           vertical dashed lines.  Finally, the vertical dotted lines
           represent the energies where a feature is detected at $>99.9$\%
           single-trial confidence in other available instruments or spectral
           orders.}
  \label{fig970508mc_1}
\end{figure}

\begin{figure}
  \includegraphics[scale=0.6,angle=-90]{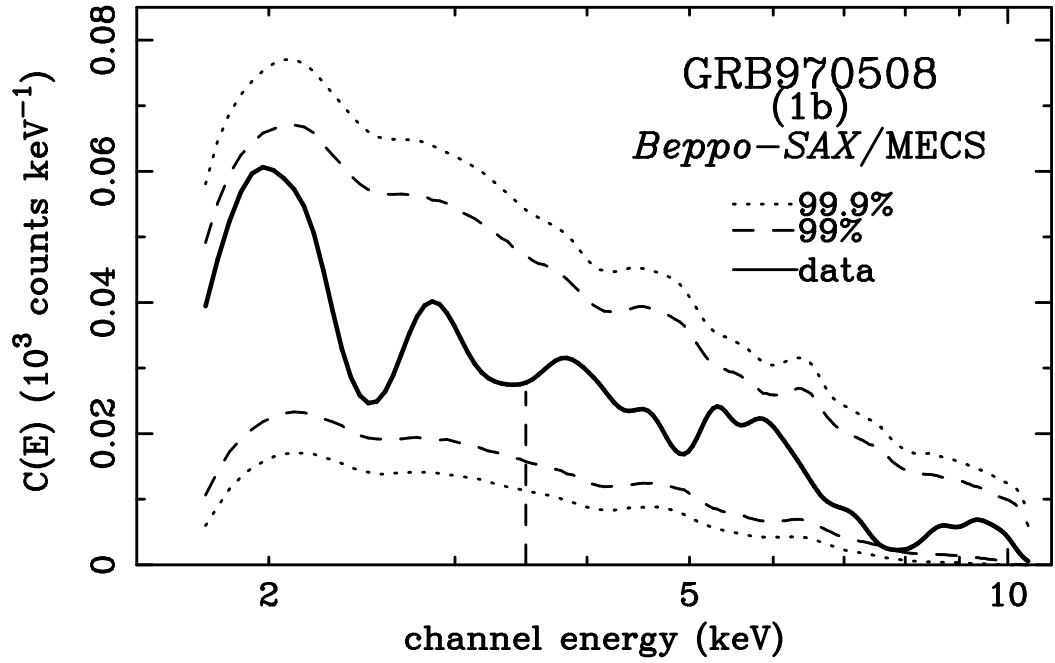}
  \caption{Same as in Figure~\ref{fig970508mc_1} for \grb970508 section 1b.}
  \label{fig970508mc_2}
\end{figure}

\begin{figure}
  \includegraphics[scale=0.6,angle=-90]{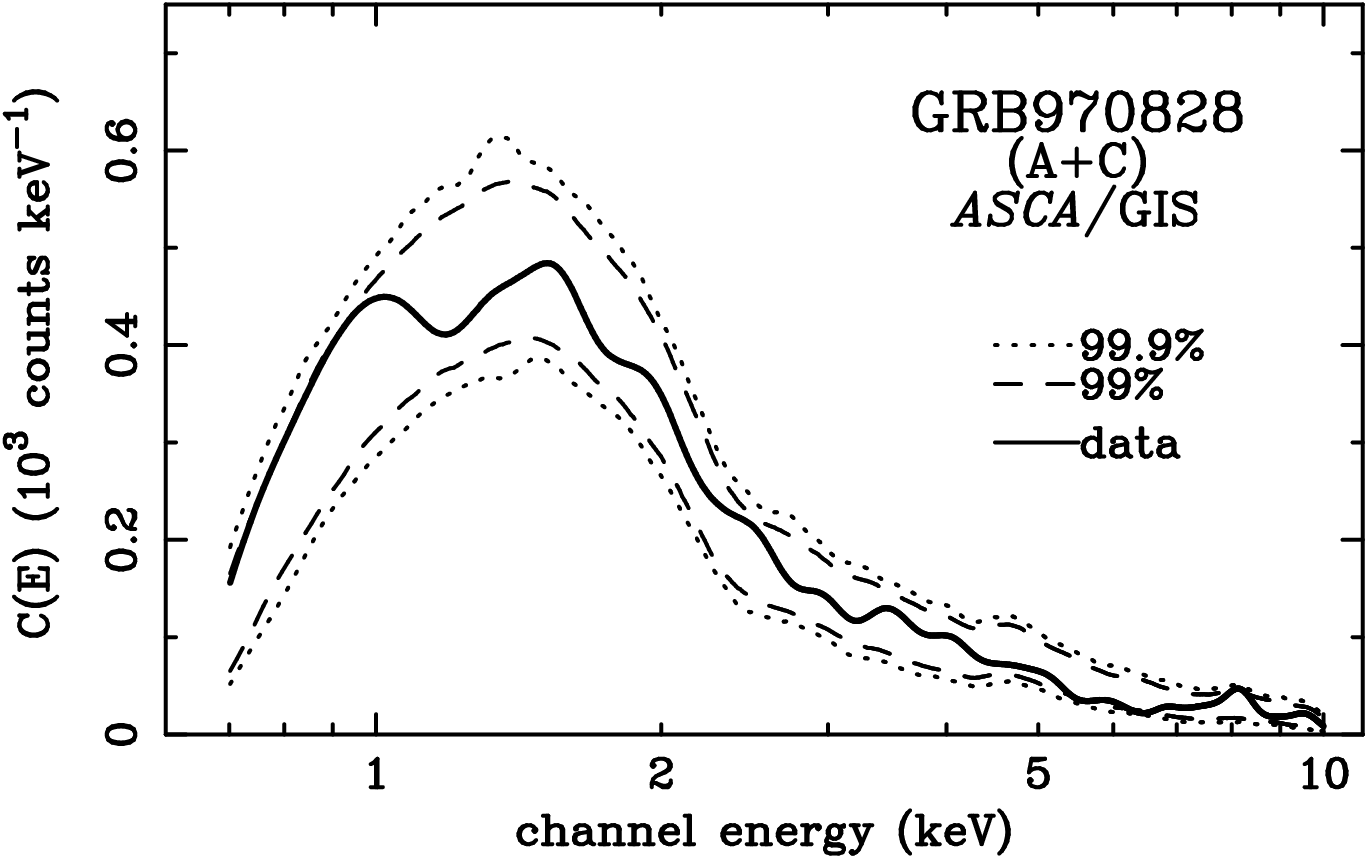}
  \caption{Same as in Figure~\ref{fig970508mc_1} for \grb970828 GIS time section A and C.}
  \label{fig970828mc_gis_ac}
\end{figure}

\begin{figure}
  \includegraphics[scale=0.6,angle=-90]{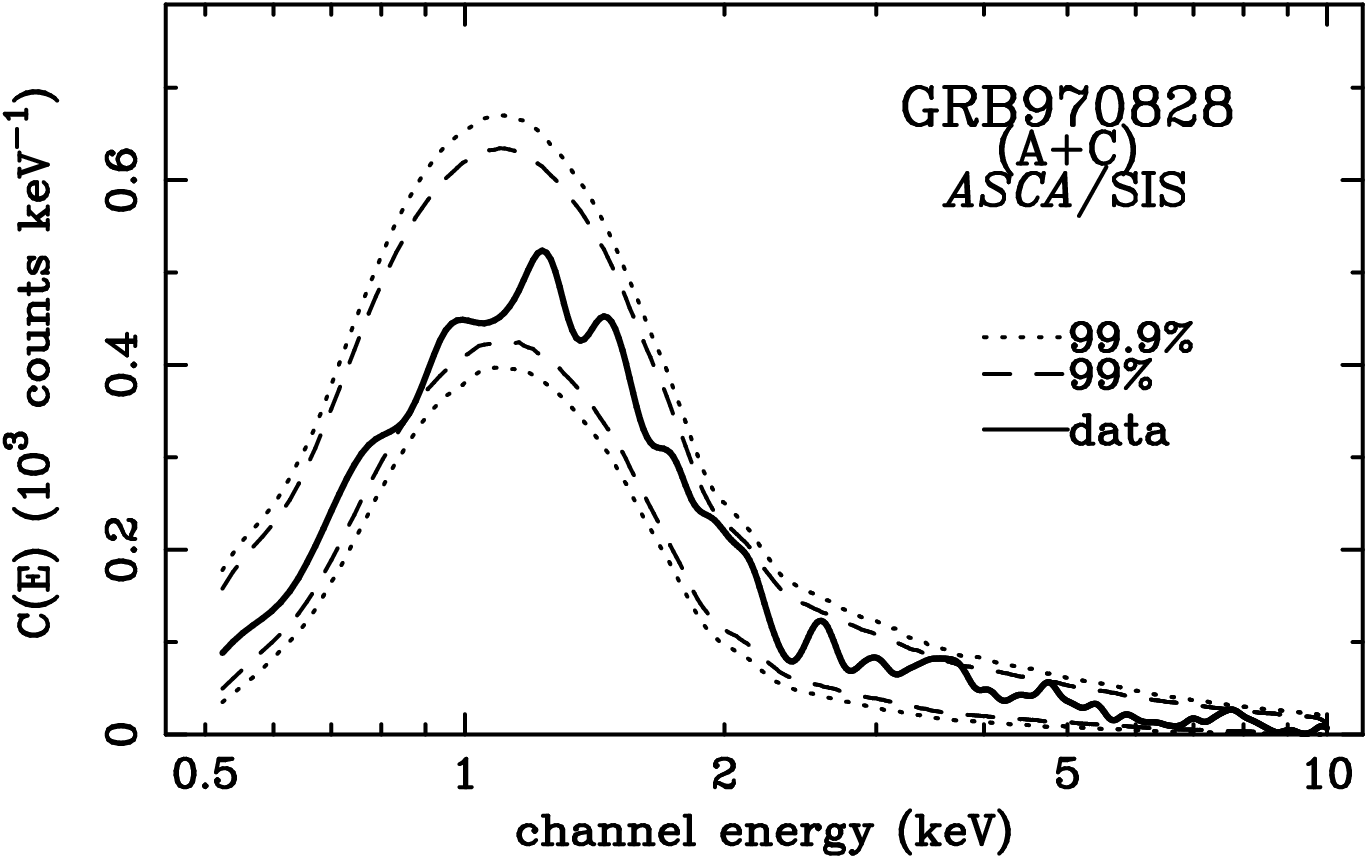}
  \caption{Same as in Figure~\ref{fig970508mc_1} for \grb970828 SIS time section A and C.}
  \label{fig970828mc_sis_ac}
\end{figure}

\begin{figure}
  \includegraphics[scale=0.6,angle=-90]{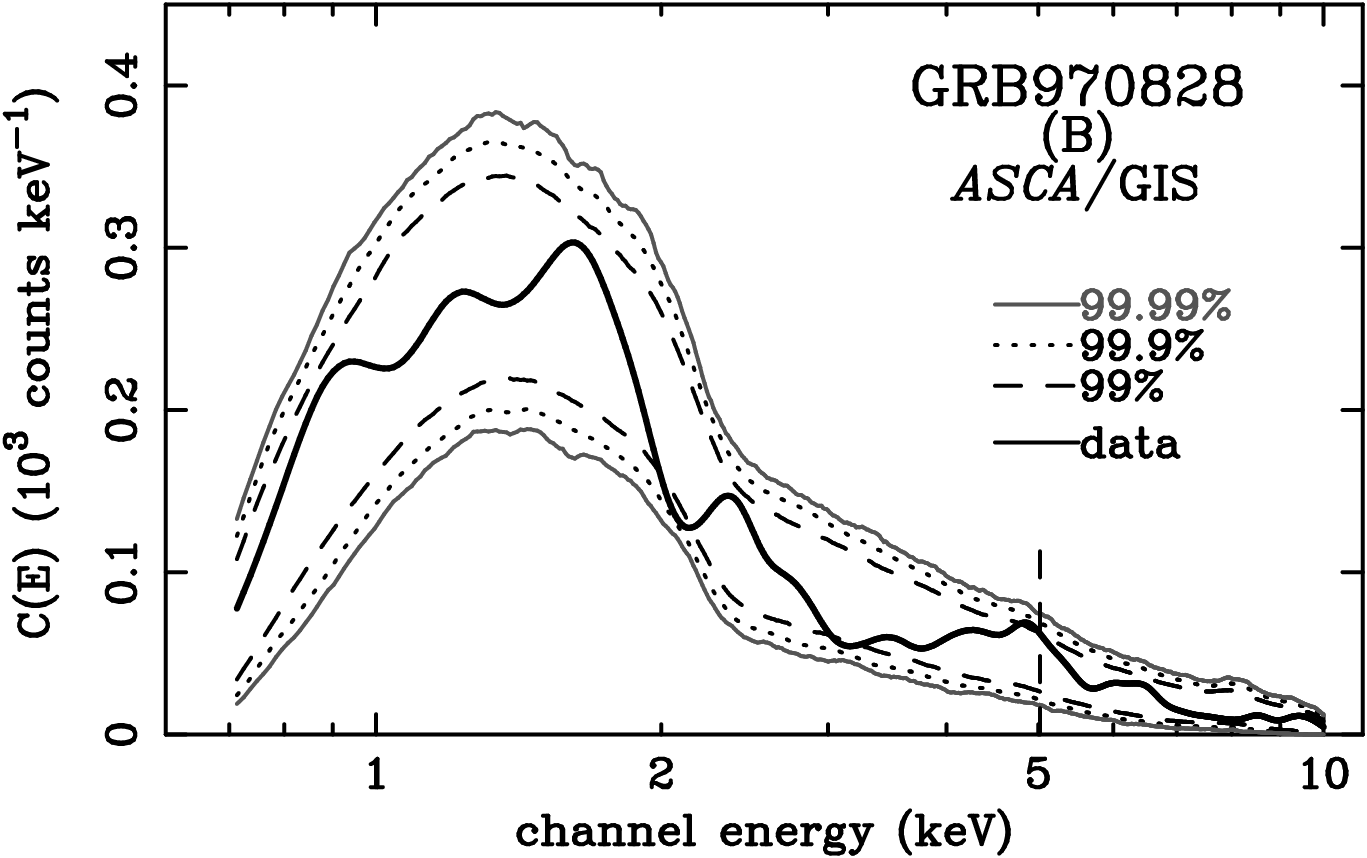}
  \caption{Same as in Figure~\ref{fig970508mc_1} for \grb970828 GIS time section B.}
  \label{fig970828mc_gis_b}
\end{figure}

\begin{figure}
  \includegraphics[scale=0.6,angle=-90]{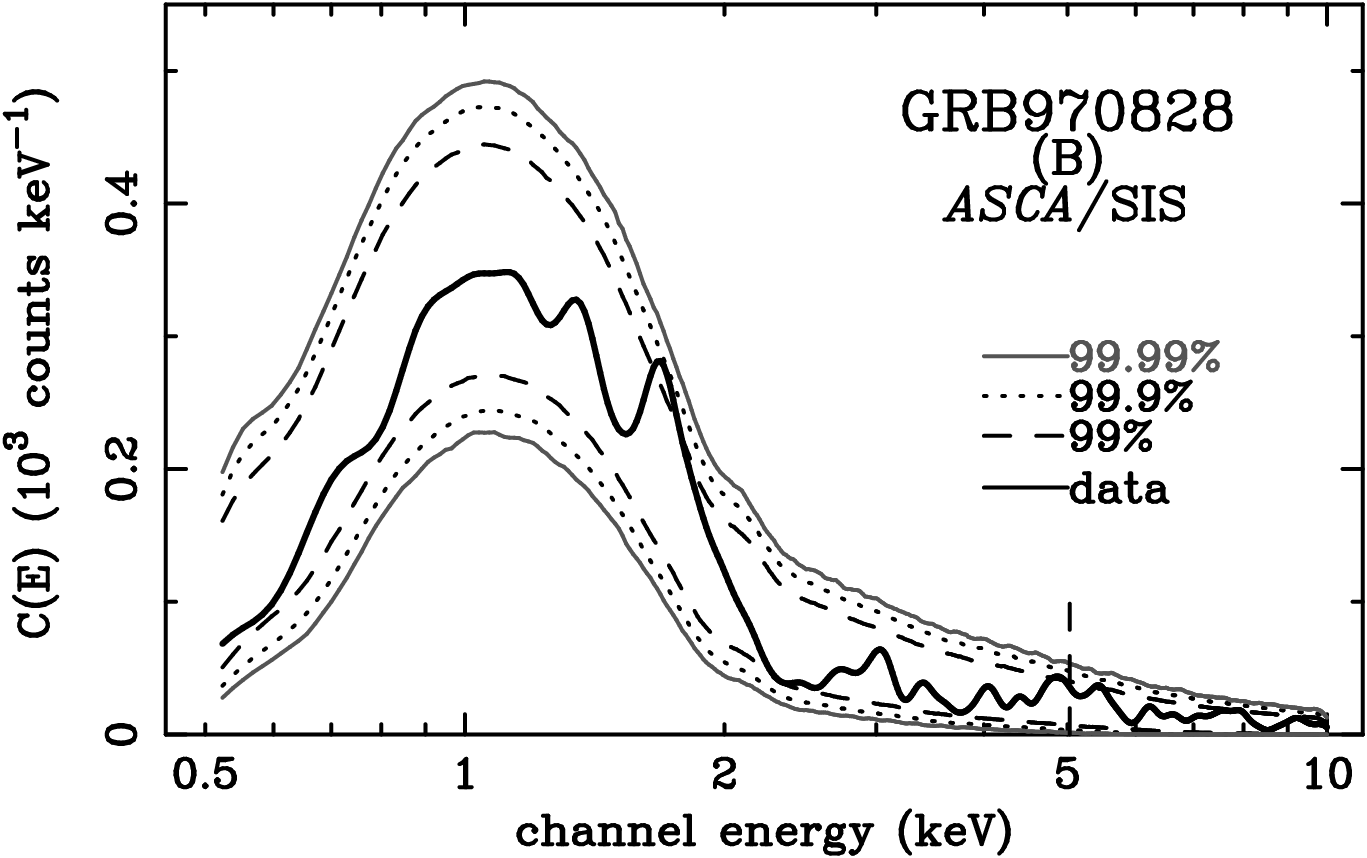}
  \caption{Same as in Figure~\ref{fig970508mc_1} for \grb970828 SIS time section B.}
  \label{fig970828mc_sis_b}
\end{figure}

\begin{figure}
  \includegraphics[scale=0.6,angle=-90]{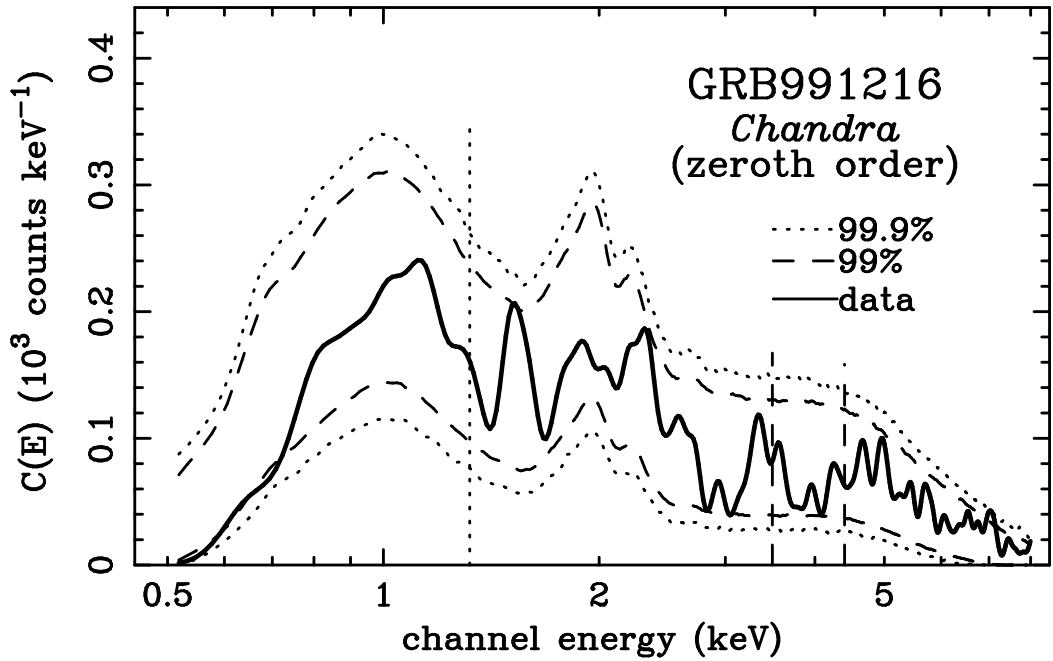}
  \caption{Same as in Figure~\ref{fig970508mc_1} for \grb991216 zero order.}
  \label{fig991216mc_zero}
\end{figure}

\begin{figure}
  \includegraphics[scale=0.6,angle=-90]{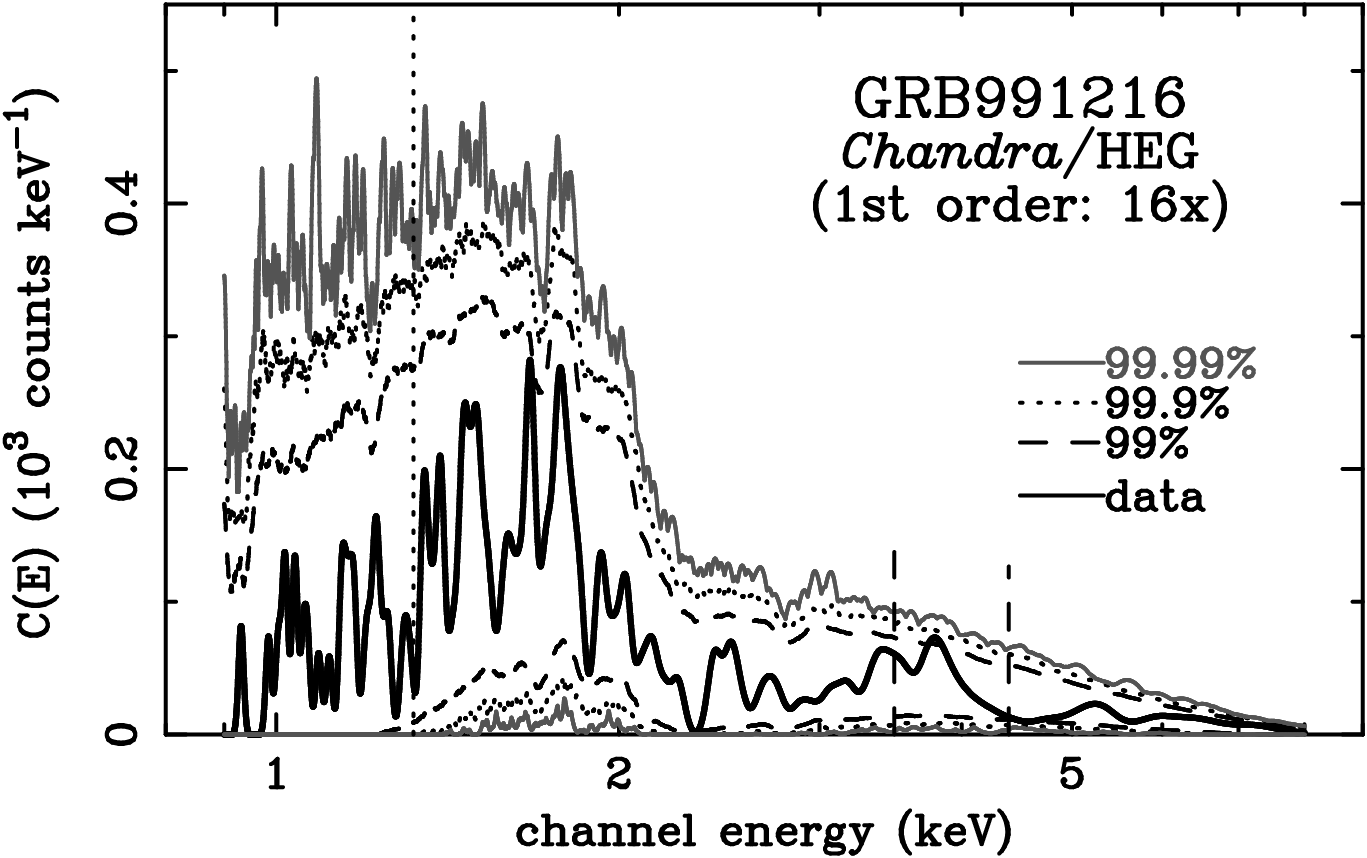}
  \caption{Same as in Figure~\ref{fig970508mc_1} for \grb991216 first order HEG.}
  \label{fig991216mc_heg}
\end{figure}

\begin{figure}
  \includegraphics[scale=0.6,angle=-90]{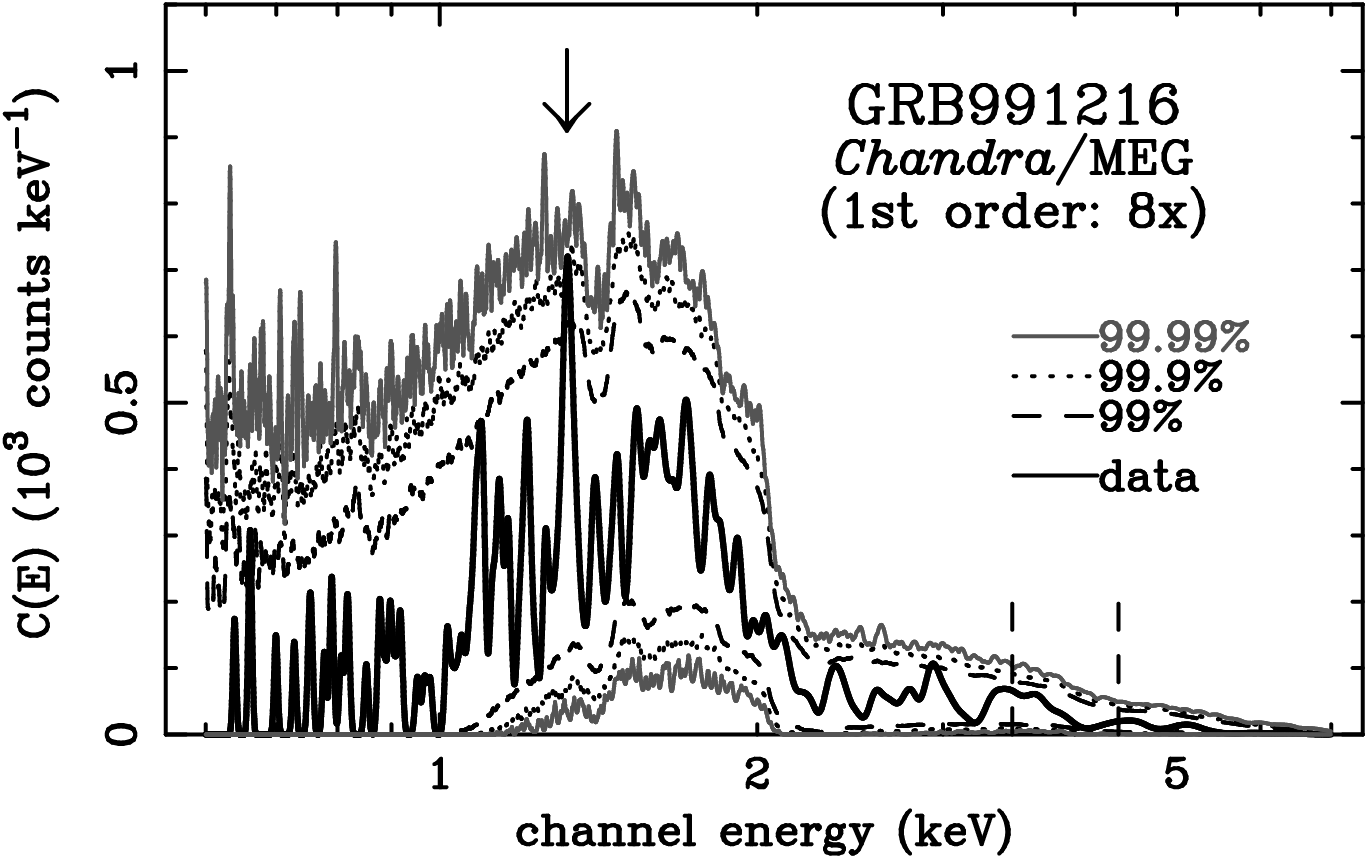}
  \caption{Same as in Figure~\ref{fig970508mc_1} for \grb991216 first order MEG.}
  \label{fig991216mc_meg}
\end{figure}

\begin{figure}
  \includegraphics[scale=0.6,angle=-90]{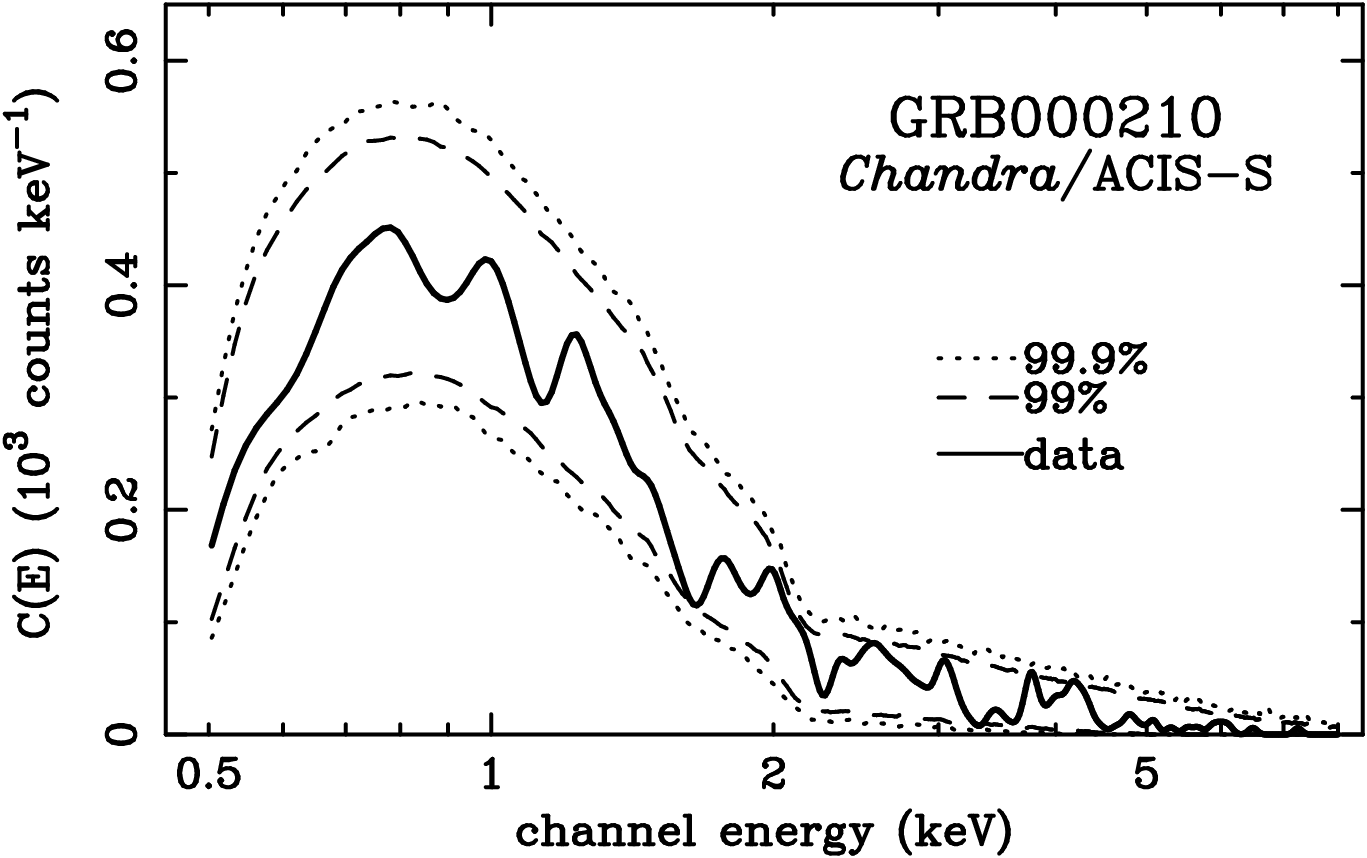}
  \caption{Same as in Figure~\ref{fig970508mc_1} for \grb000210 ACIS.}
  \label{fig000210mc}
\end{figure}

\begin{figure}
  \includegraphics[scale=0.6,angle=-90]{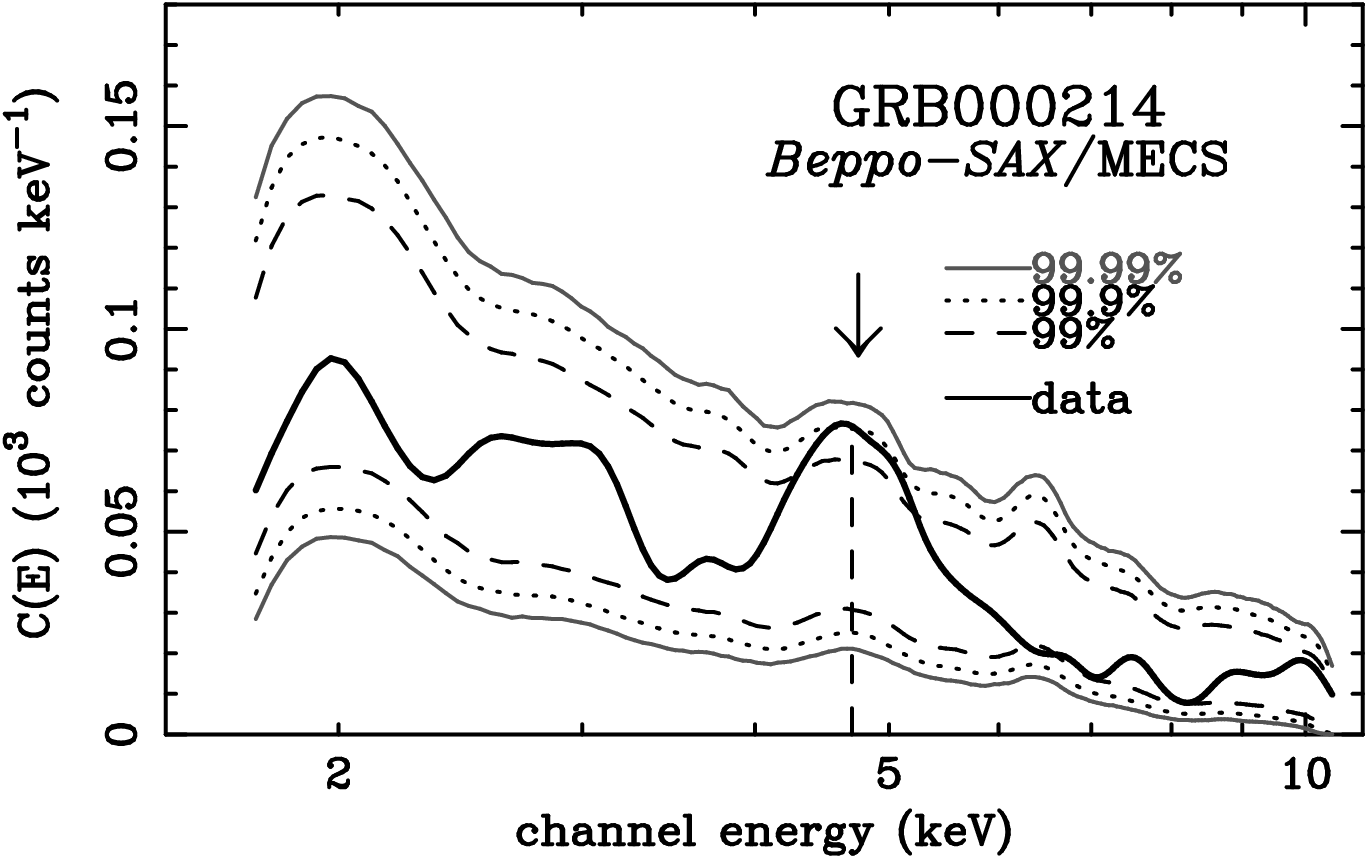}
  \caption{Same as in Figure~\ref{fig970508mc_1} for \grb000214 MECS.}
  \label{fig000214mc}
\end{figure}

\begin{figure}
  \includegraphics[scale=0.6,angle=-90]{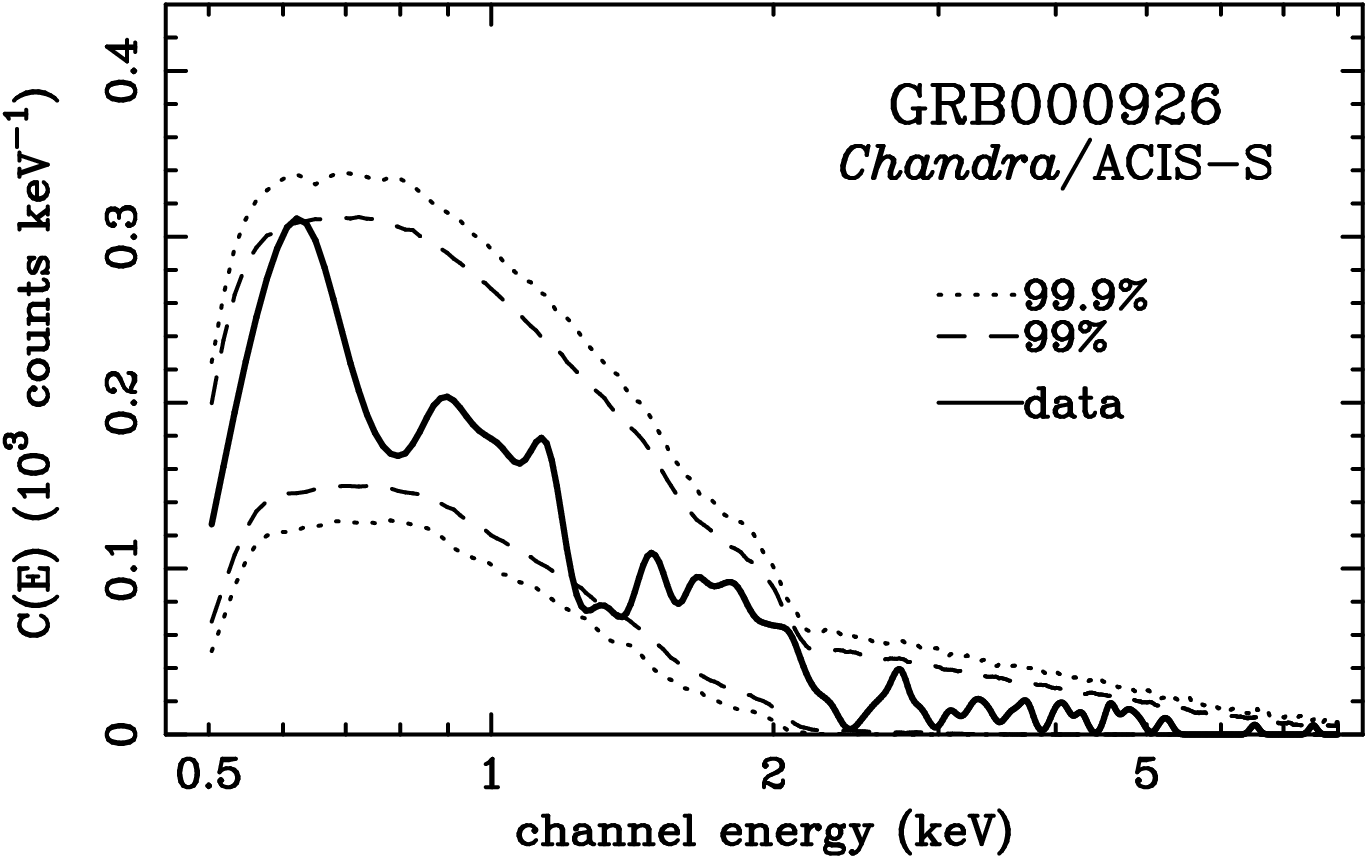}
  \caption{Same as in Figure~\ref{fig970508mc_1} for \grb000926 ACIS.}
  \label{fig000926mc}
\end{figure}

\begin{figure}
  \includegraphics[scale=0.6,angle=-90]{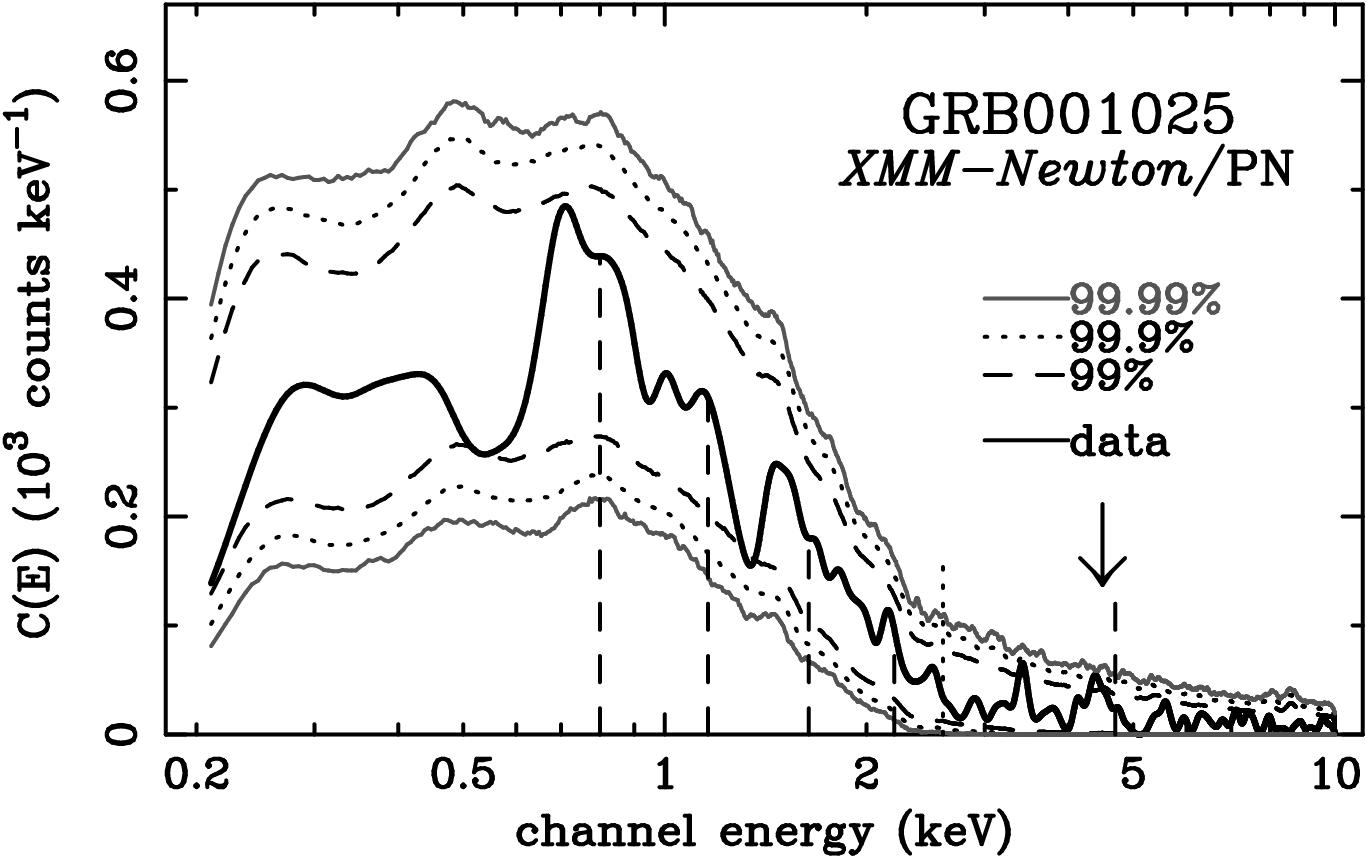}
  \caption{Same as in Figure~\ref{fig970508mc_1} for \grb001025 PN.}
  \label{fig001025mc_pn}
\end{figure}

\begin{figure}
  \includegraphics[scale=0.6,angle=-90]{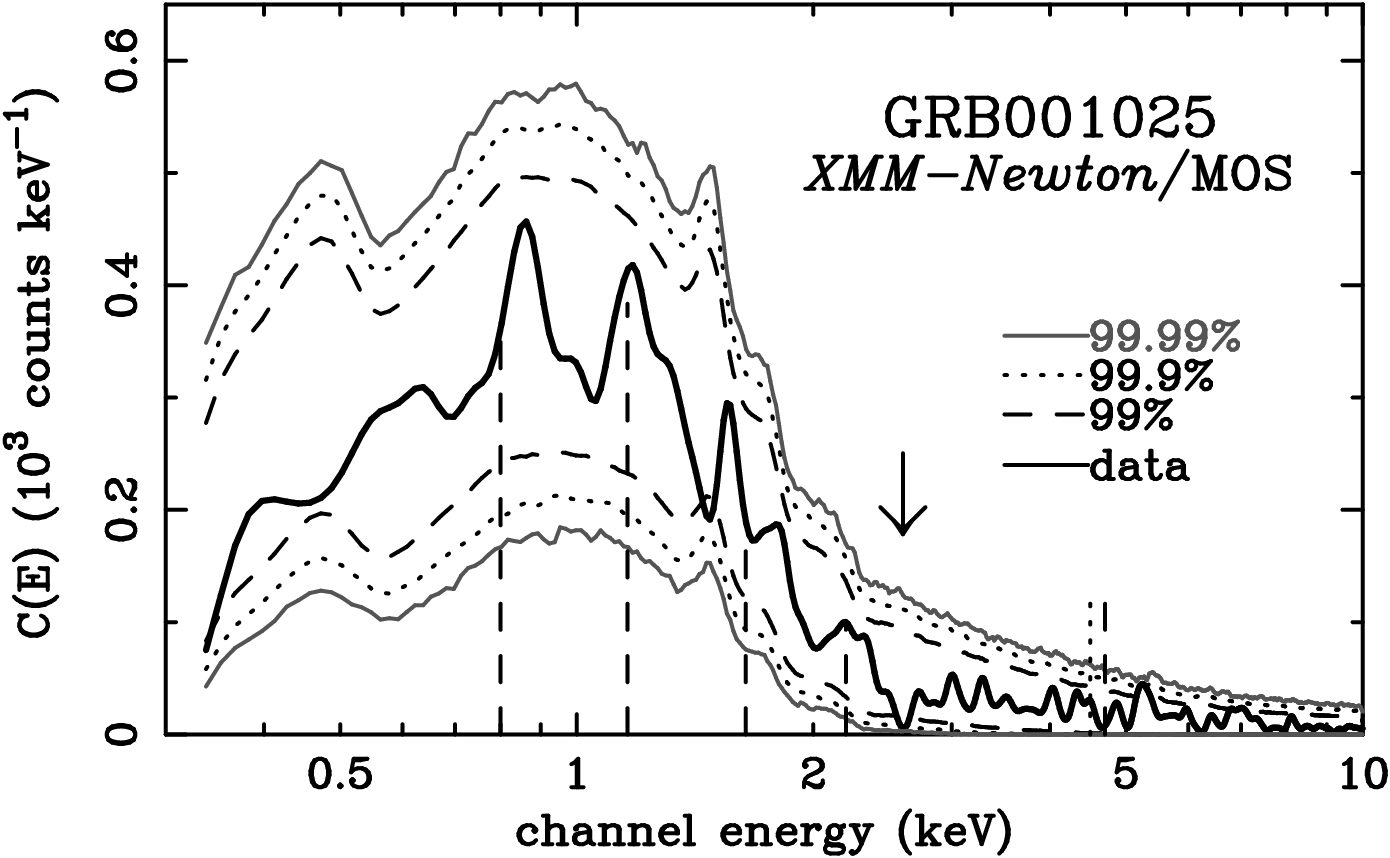}
  \caption{Same as in Figure~\ref{fig970508mc_1} for \grb001025 MOS.}
  \label{fig001025mc_mos}
\end{figure}

\begin{figure}
  \includegraphics[scale=0.6,angle=-90]{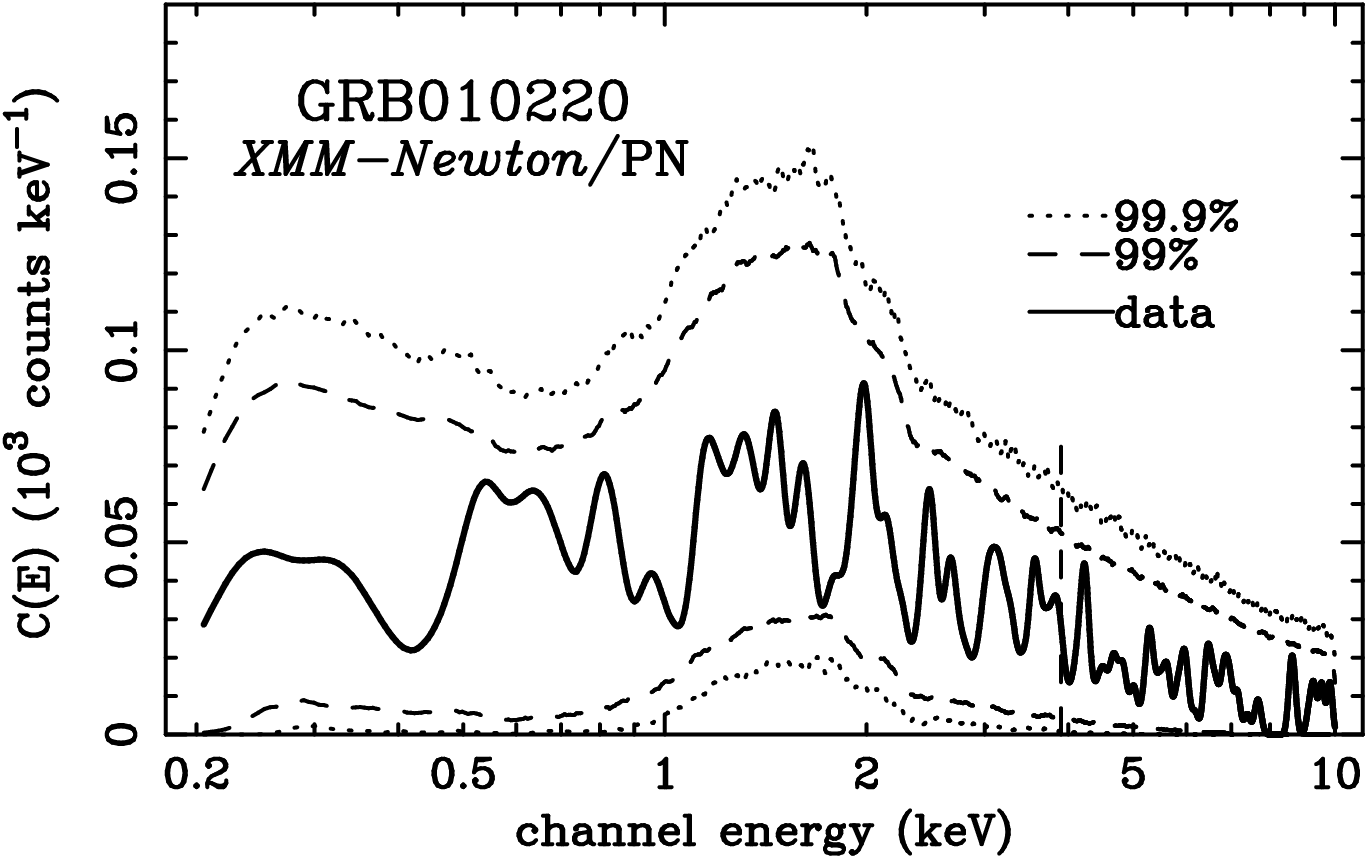}
  \caption{Same as in Figure~\ref{fig970508mc_1} for \grb010220 PN.}
  \label{fig010220mc}
\end{figure}

\begin{figure}
  \includegraphics[scale=0.6,angle=-90]{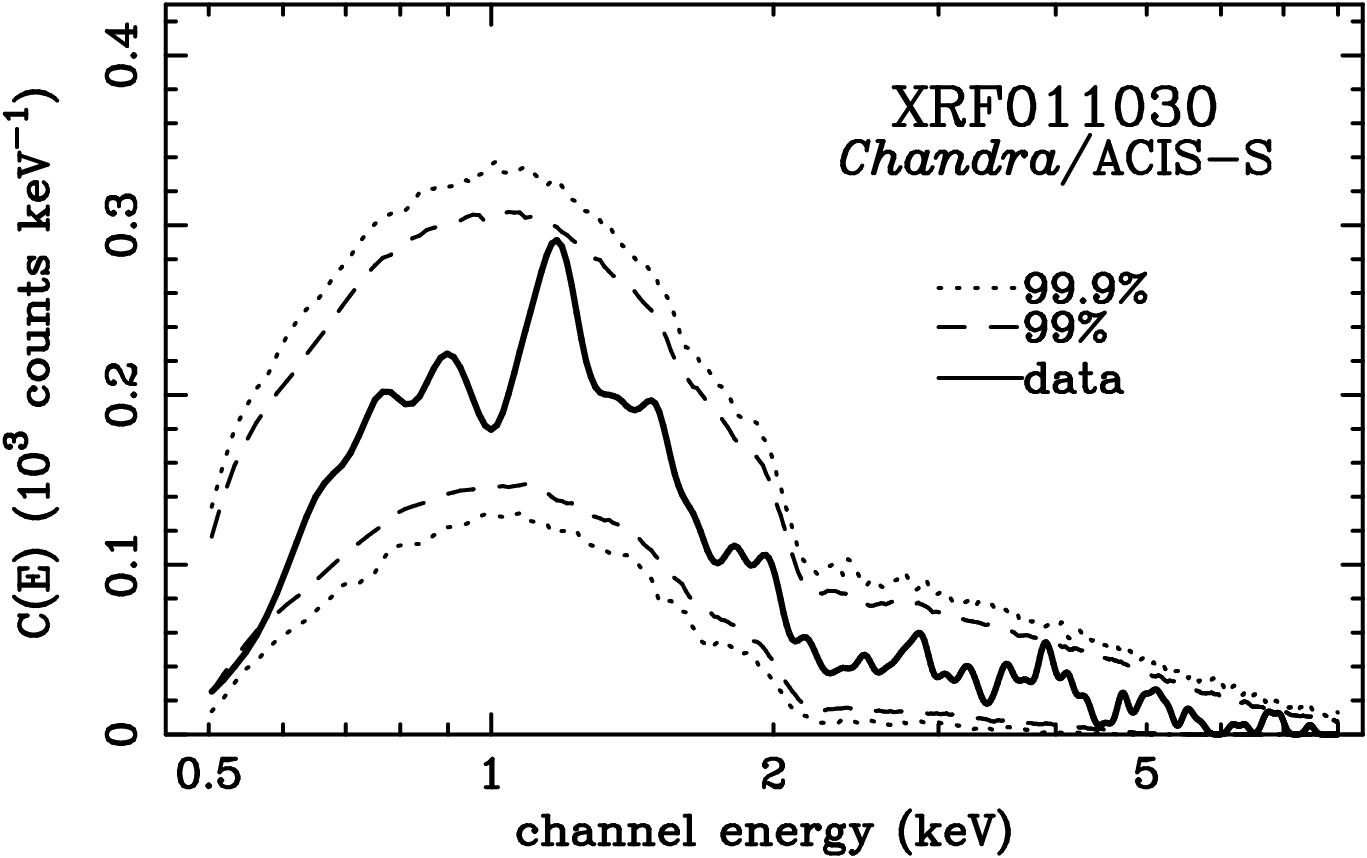}
  \caption{Same as in Figure~\ref{fig970508mc_1} for \grb011030 ACIS.}
  \label{fig011030mc}
\end{figure}

\begin{figure}
  \includegraphics[scale=0.6,angle=-90]{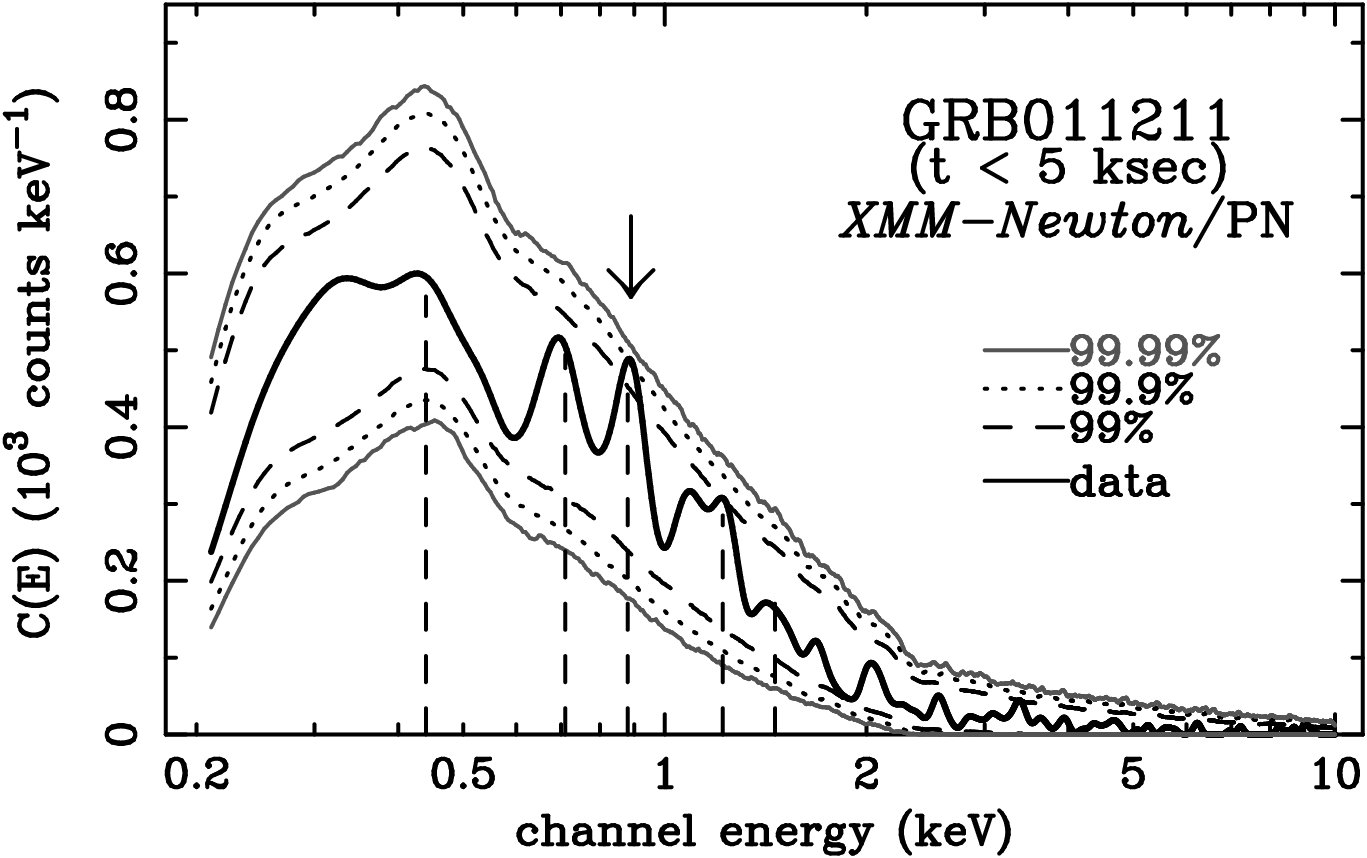}
  \caption{Same as in Figure~\ref{fig970508mc_1} for \grb011211 PN first 5 ksec.}
  \label{fig011211mc_pn_1}
\end{figure}

\begin{figure}
  \includegraphics[scale=0.6,angle=-90]{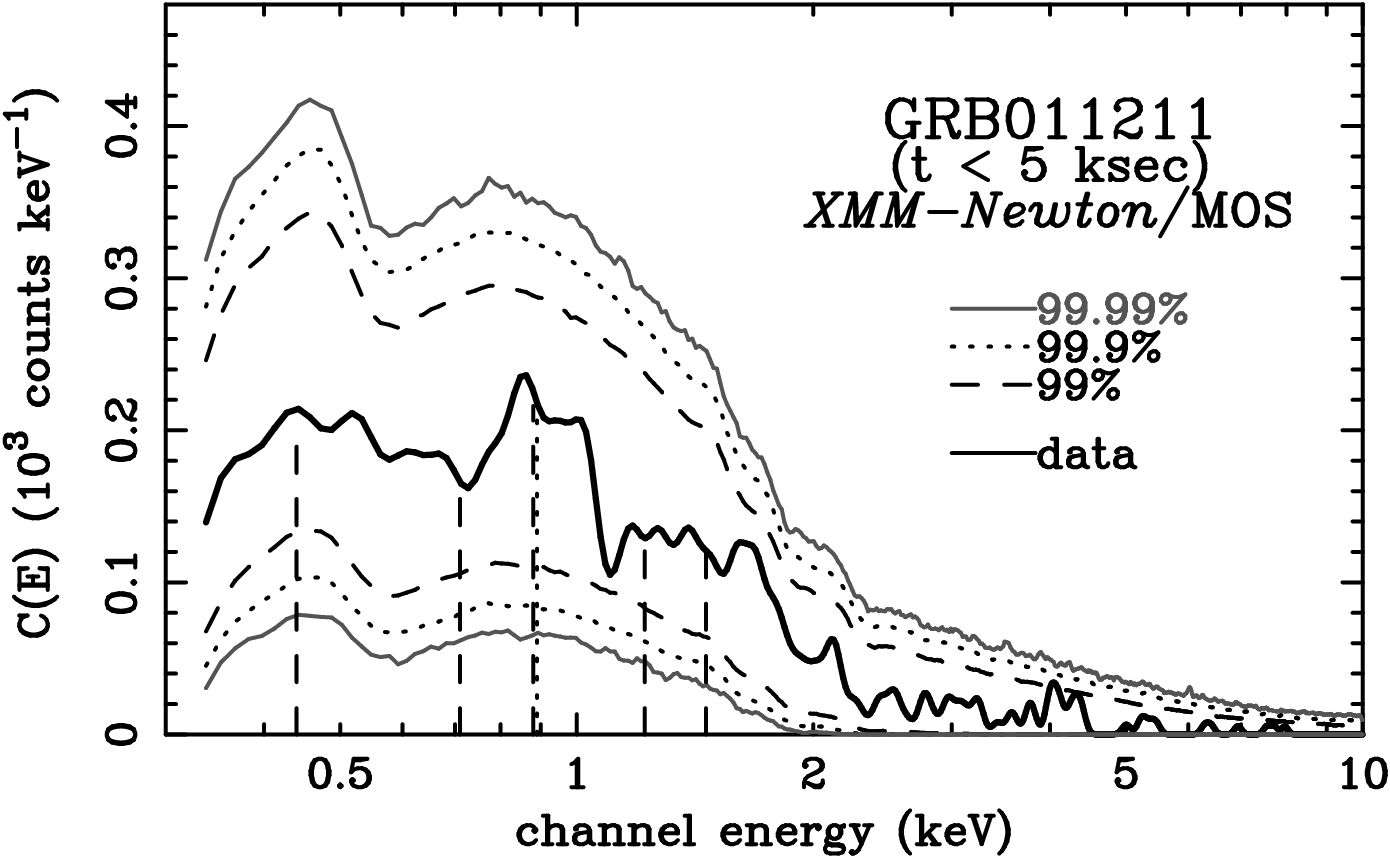}
  \caption{Same as in Figure~\ref{fig970508mc_1} for \grb011211 MOS first 5 ksec.}
  \label{fig011211mc_mos_1}
\end{figure}

\begin{figure}
  \includegraphics[scale=0.6,angle=-90]{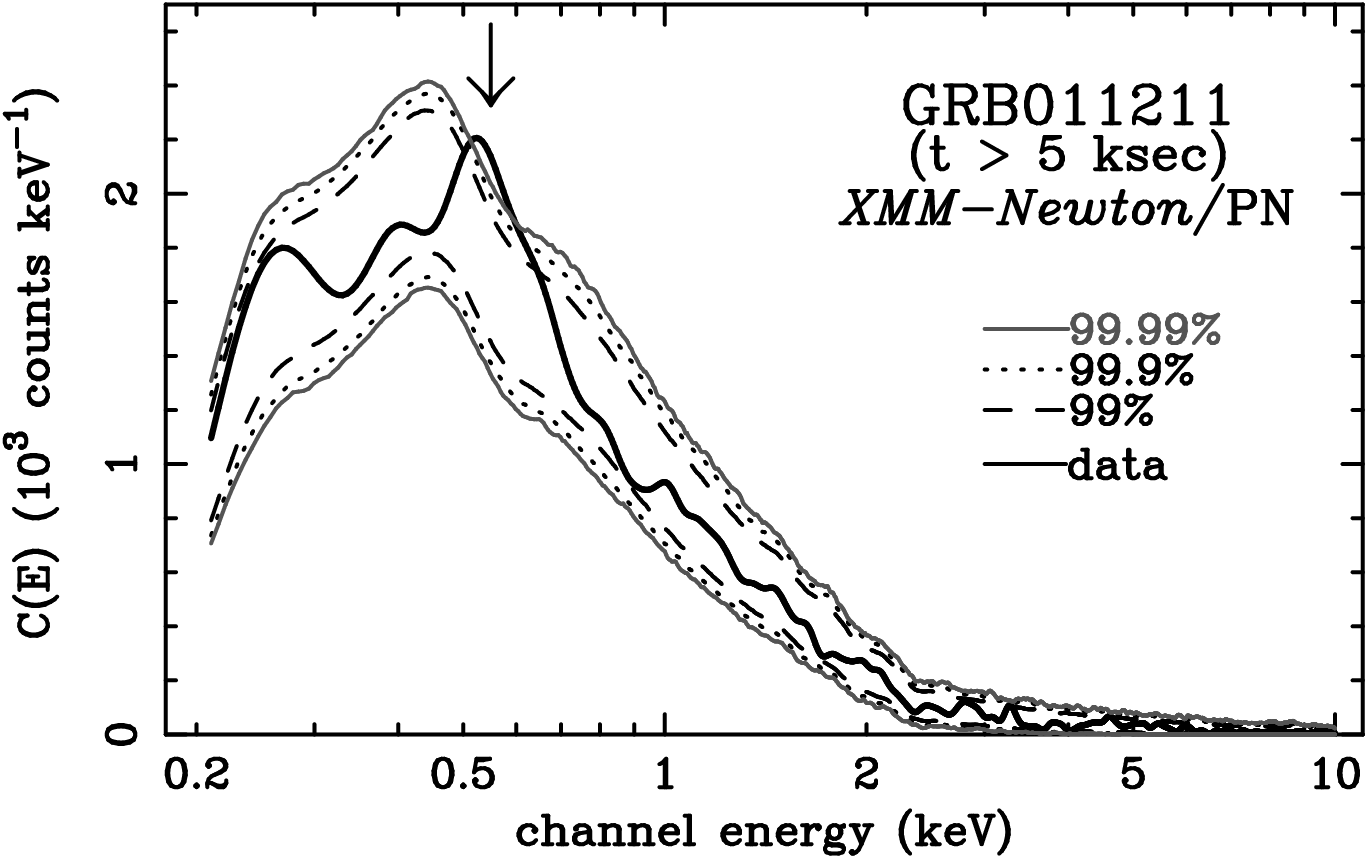}
  \caption{Same as in Figure~\ref{fig970508mc_1} for \grb011211 PN $t > 5$~ksec.}
  \label{fig011211mc_pn_2}
\end{figure}

\begin{figure}
  \includegraphics[scale=0.6,angle=-90]{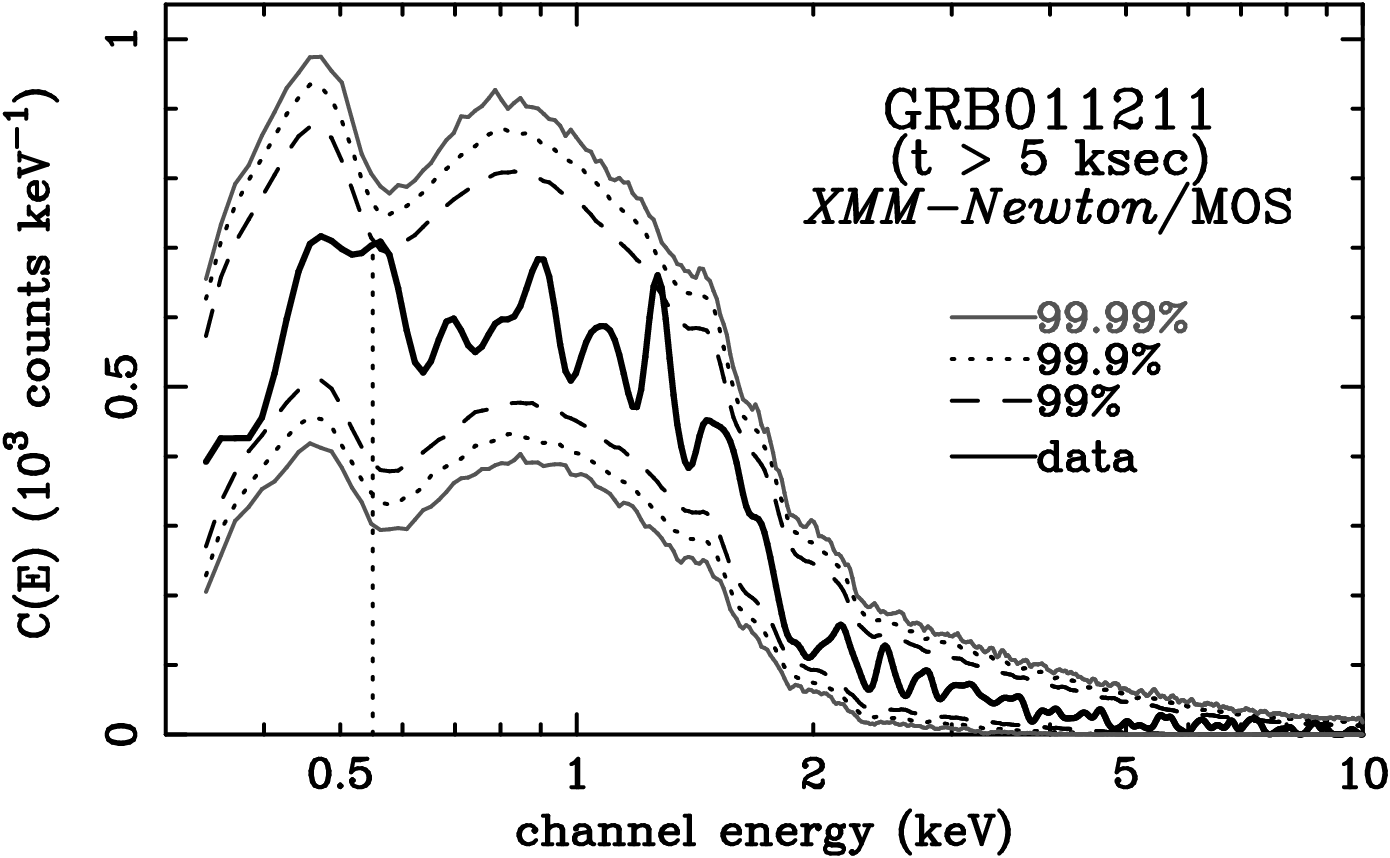}
  \caption{Same as in Figure~\ref{fig970508mc_1} for \grb011211 MOS $t > 5$~ksec.}
  \label{fig011211mc_mos_2}
\end{figure}

\begin{figure}
  \includegraphics[scale=0.6,angle=-90]{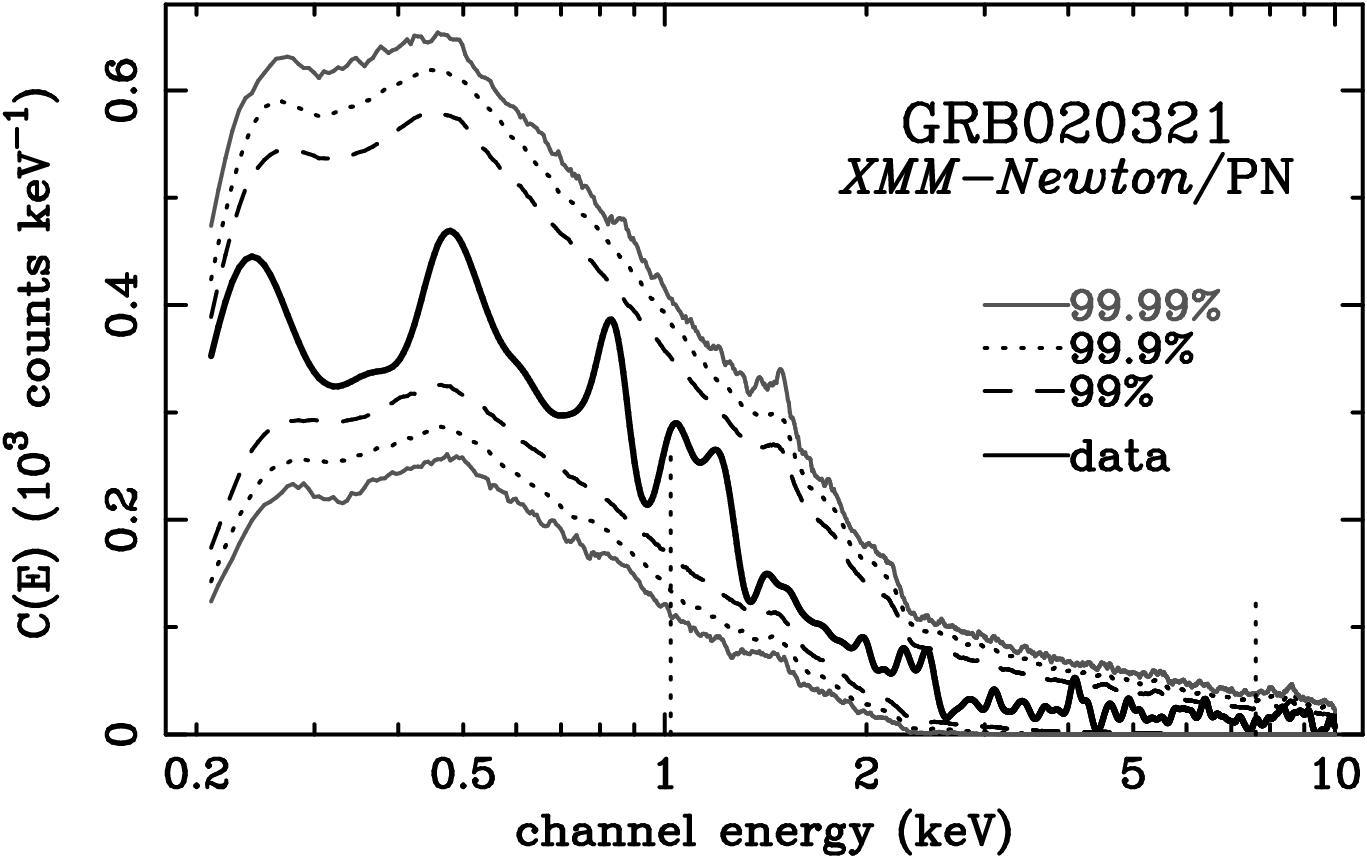}
  \caption{Same as in Figure~\ref{fig970508mc_1} for \grb020321 PN.}
  \label{fig020321mc_pn}
\end{figure}

\begin{figure}
  \includegraphics[scale=0.6,angle=-90]{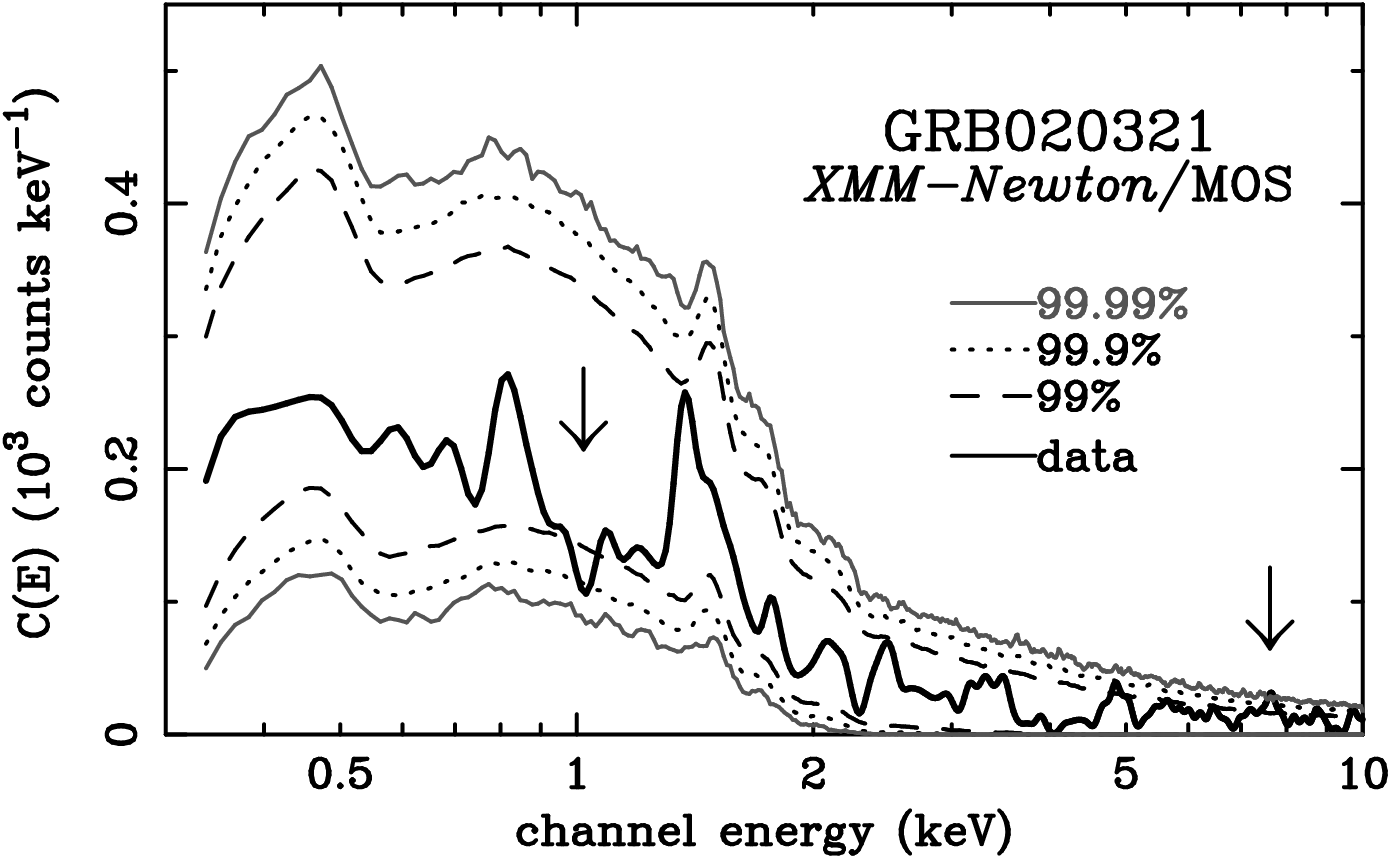}
  \caption{Same as in Figure~\ref{fig970508mc_1} for \grb020321 MOS.}
  \label{fig020321mc_mos}
\end{figure}

\begin{figure}
  \includegraphics[scale=0.6,angle=-90]{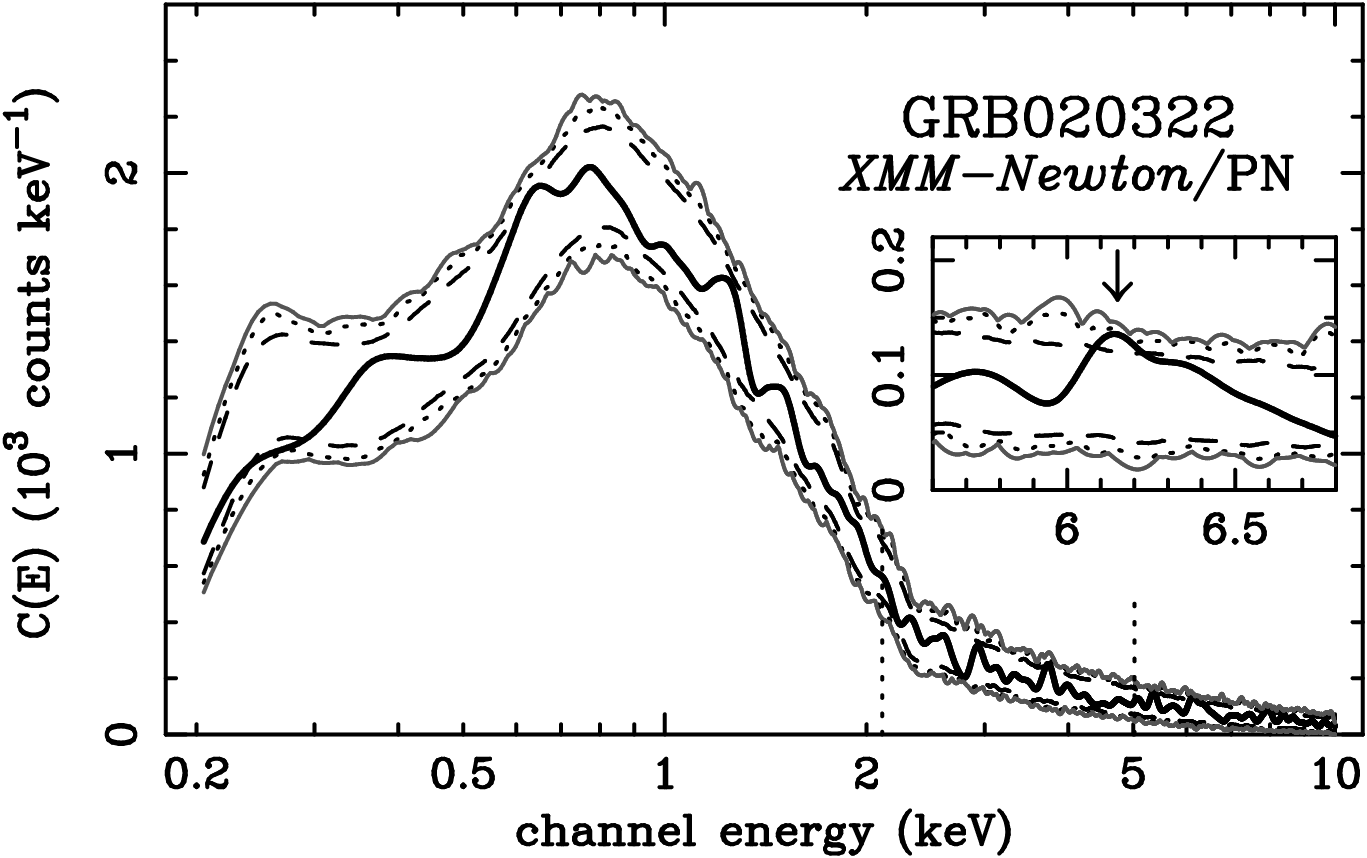}
  \caption{Same as in Figure~\ref{fig970508mc_1} for \grb020322 PN.}
  \label{fig020322mc_pn}
\end{figure}

\begin{figure}
  \includegraphics[scale=0.6,angle=-90]{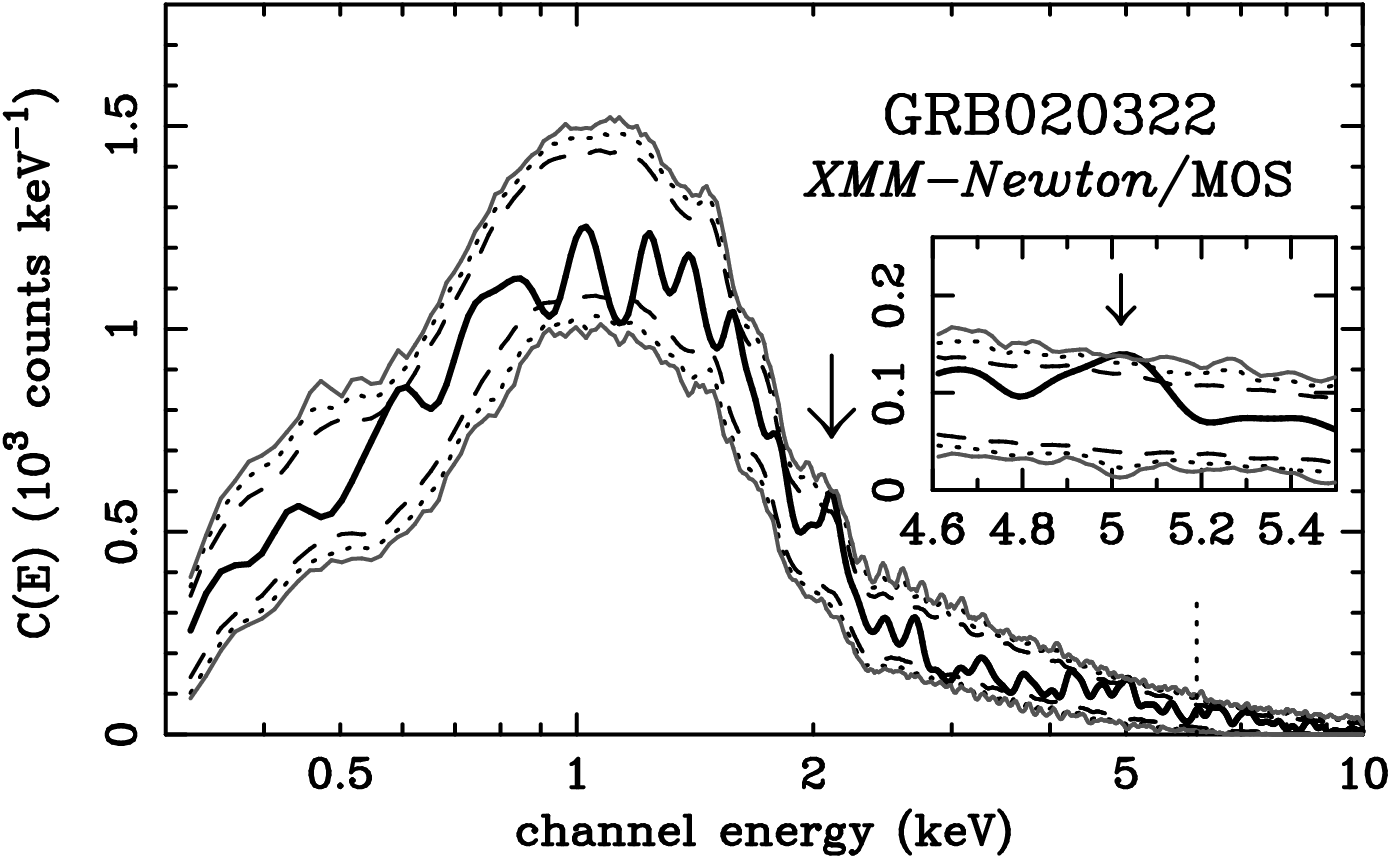}
  \caption{Same as in Figure~\ref{fig970508mc_1} for \grb020322 MOS.}
  \label{fig020322mc_mos}
\end{figure}

\begin{figure}
  \includegraphics[scale=0.6,angle=-90]{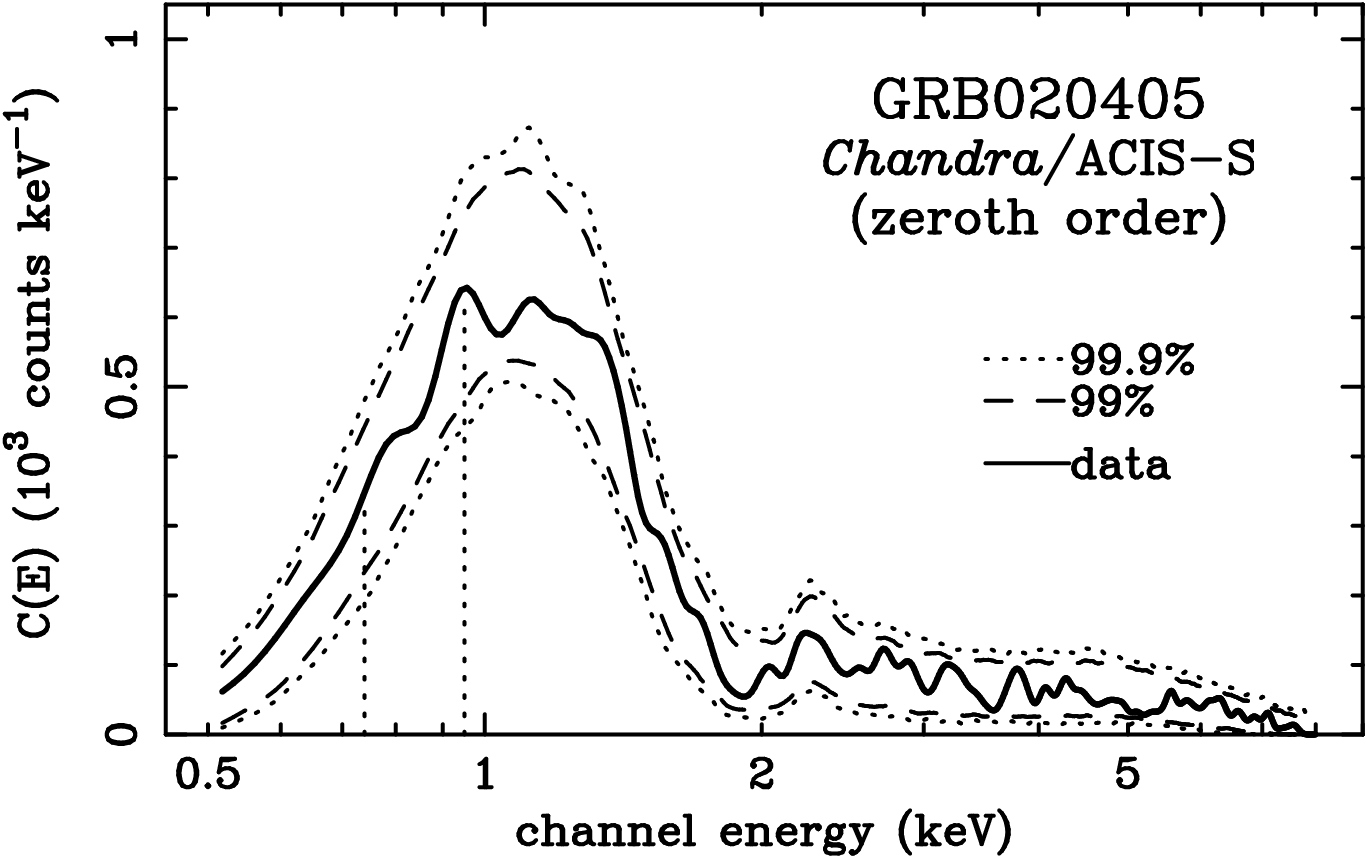}
  \caption{Same as in Figure~\ref{fig970508mc_1} for \grb020405 zero order.}
  \label{fig020405mc_zero}
\end{figure}

\begin{figure}
  \includegraphics[scale=0.6,angle=-90]{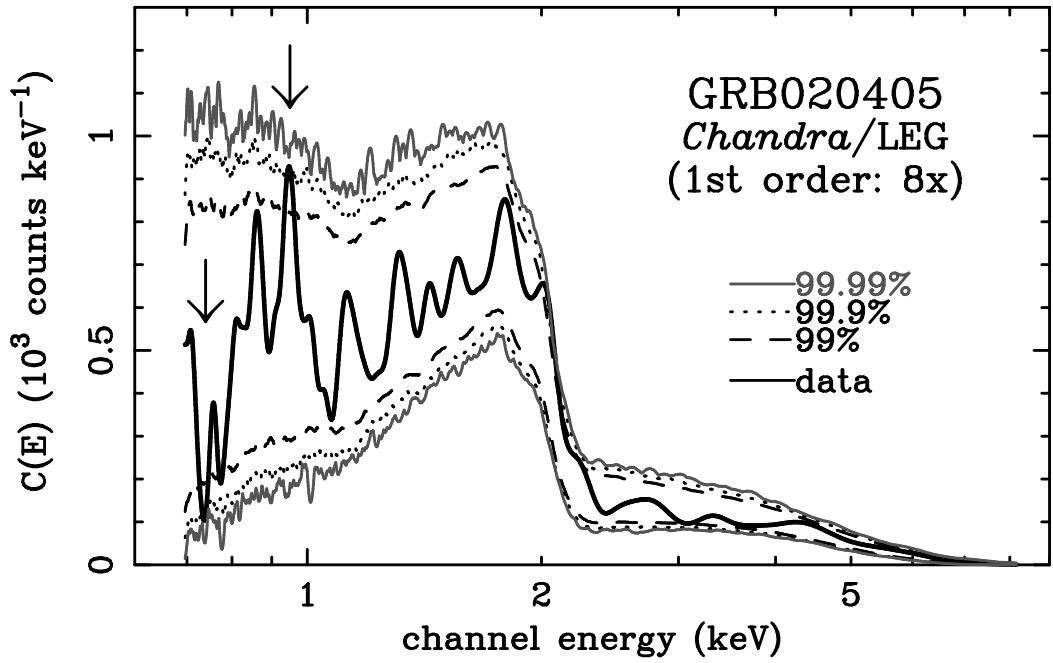}
  \caption{Same as in Figure~\ref{fig970508mc_1} for \grb020405 dispersed.}
  \label{fig020405mc_leg}
\end{figure}

\begin{figure}
  \includegraphics[scale=0.6,angle=-90]{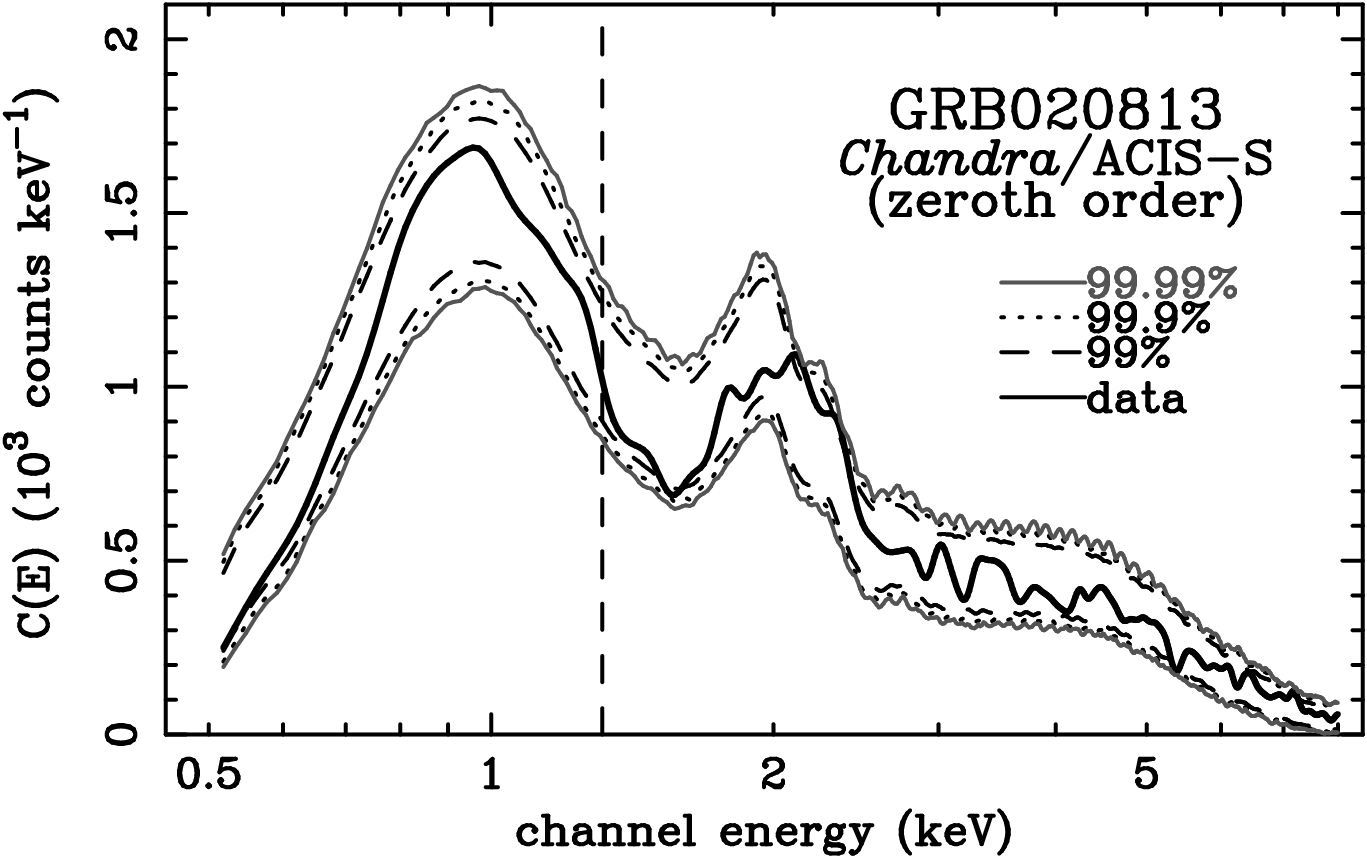}
  \caption{Same as in Figure~\ref{fig970508mc_1} for \grb020813 zero order.}
  \label{fig020813mc_zero}
\end{figure}

\begin{figure}
  \includegraphics[scale=0.6,angle=-90]{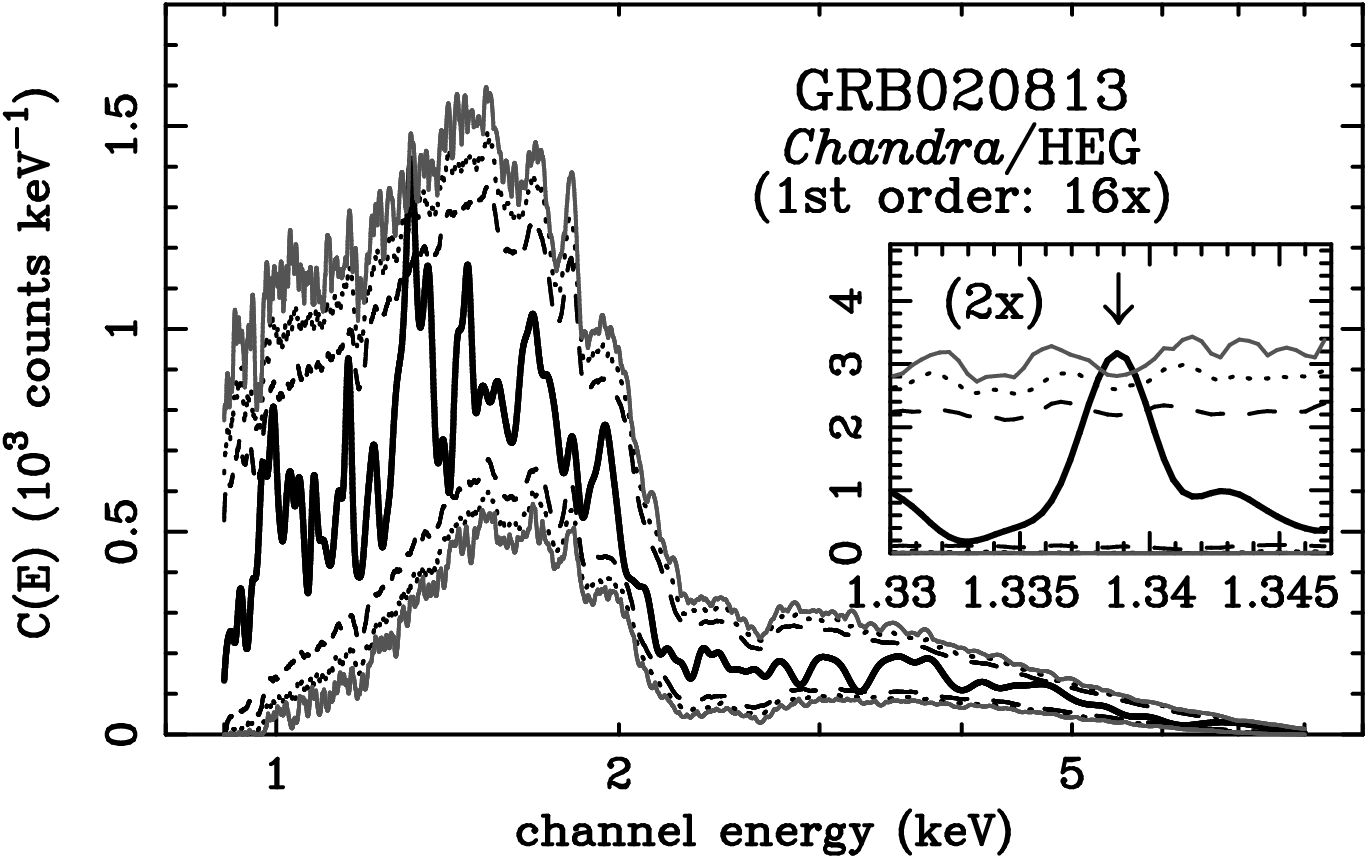}
  \caption{Same as in Figure~\ref{fig970508mc_1} for \grb020813 HEG dispersed spectrum.}
  \label{fig020813mc_heg}
\end{figure}

\begin{figure}
  \includegraphics[scale=0.6,angle=-90]{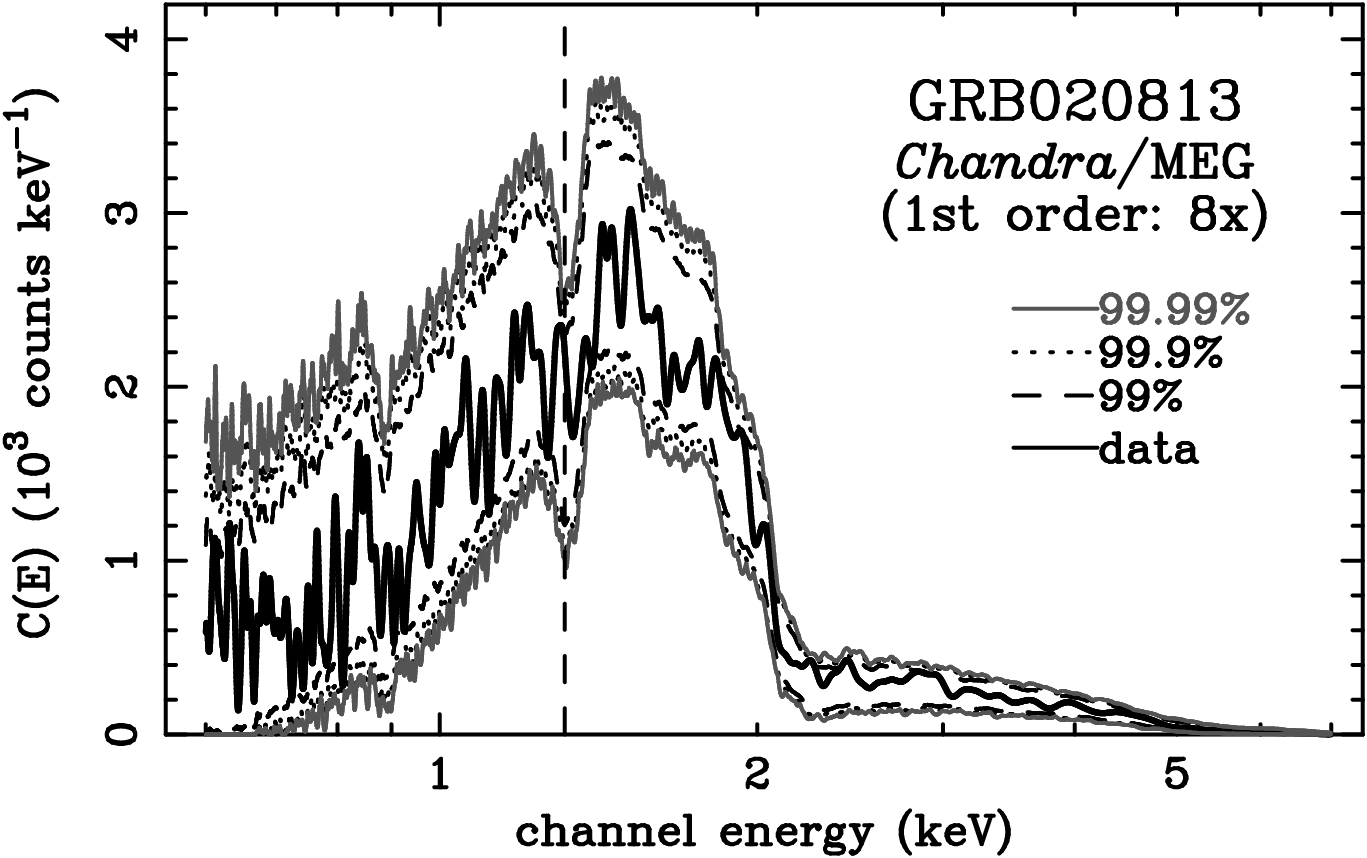}
  \caption{Same as in Figure~\ref{fig970508mc_1} for \grb020813 MEG dispersed spectrum.}
  \label{fig020813mc_meg}
\end{figure}

\begin{figure}
  \includegraphics[scale=0.6,angle=-90]{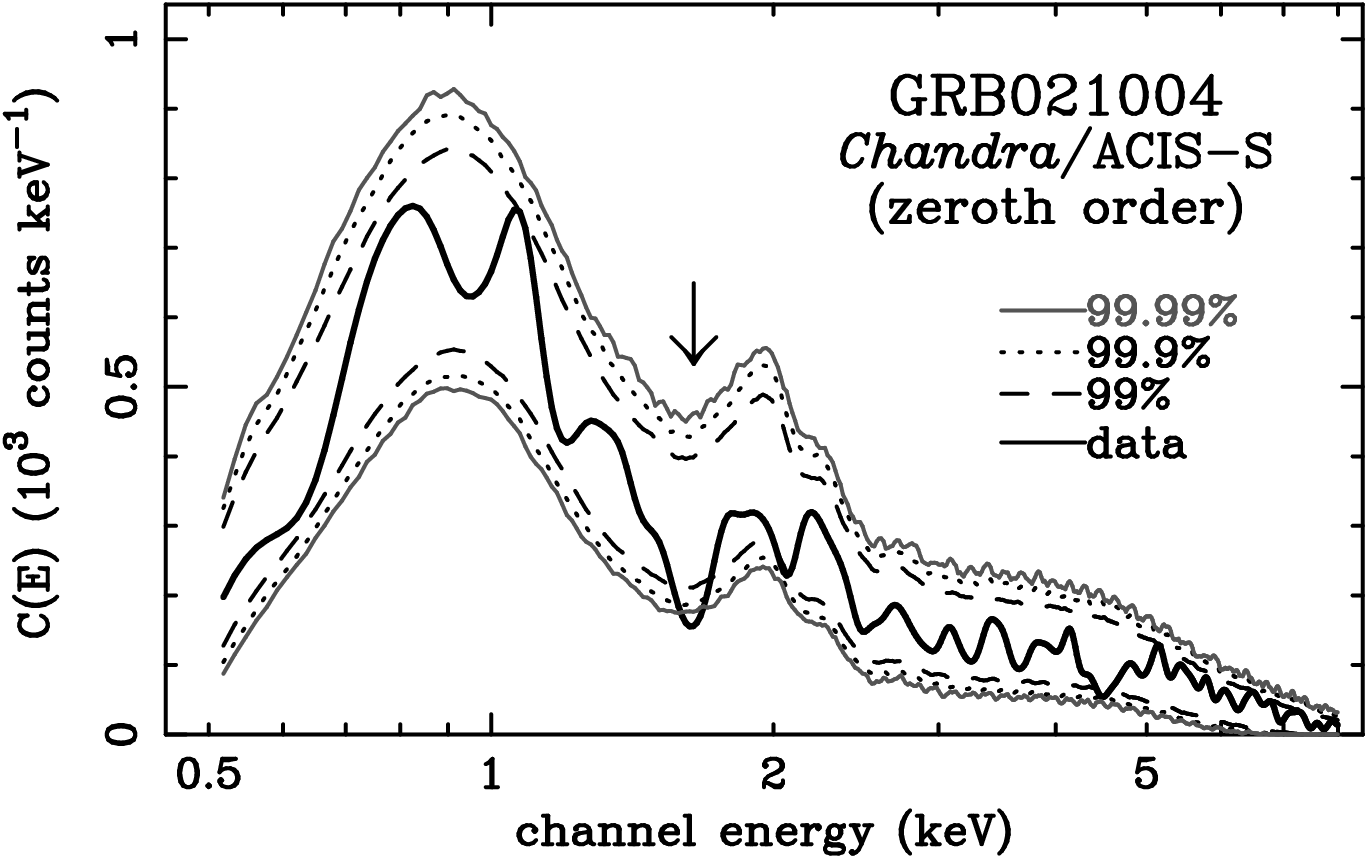}
  \caption{Same as in Figure~\ref{fig970508mc_1} for \grb021004 zero order.}
  \label{fig021004mc_zero}
\end{figure}

\begin{figure}
  \includegraphics[scale=0.6,angle=-90]{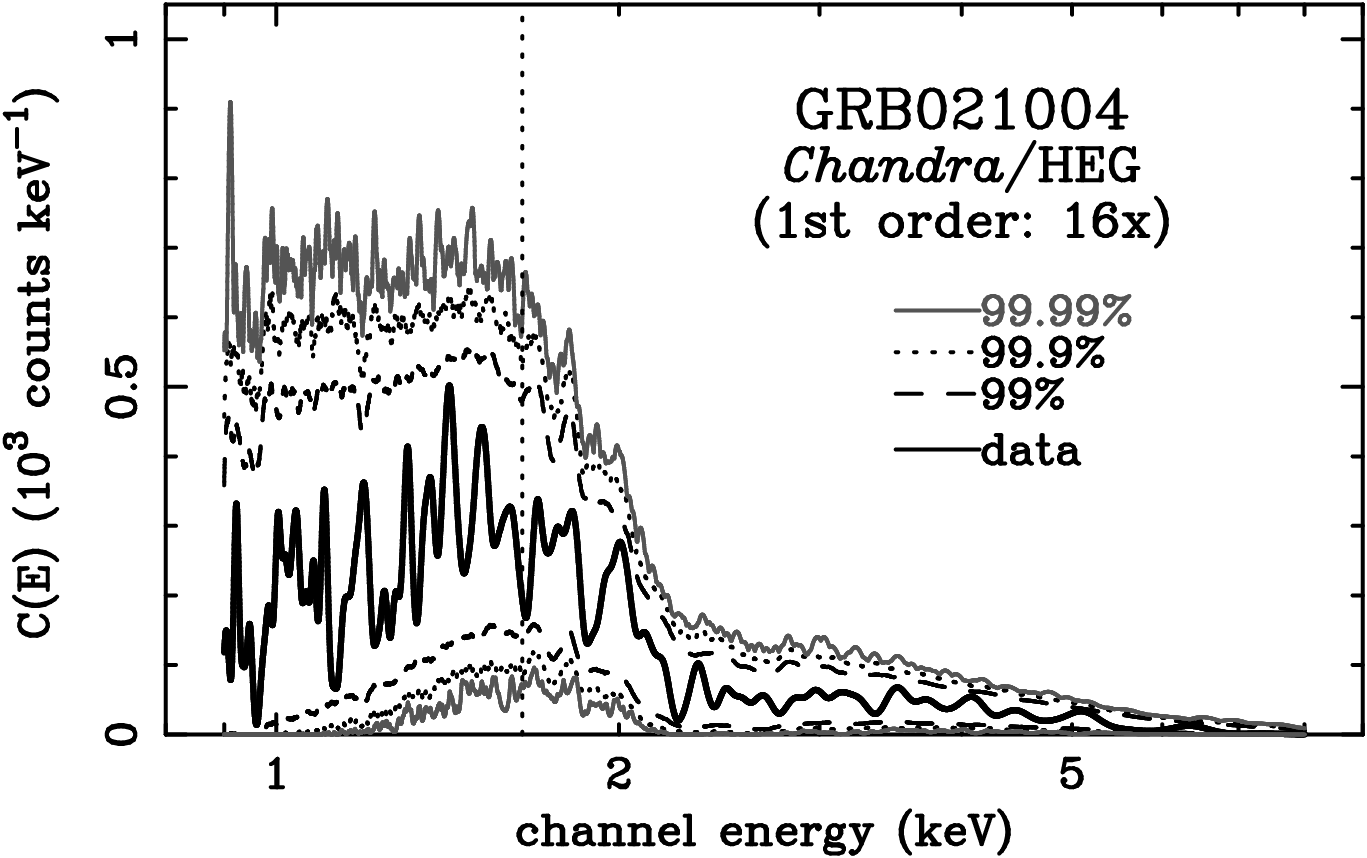}
  \caption{Same as in Figure~\ref{fig970508mc_1} for \grb021004 HEG dispersed spectrum.}
  \label{fig021004mc_heg}
\end{figure}

\begin{figure}
  \includegraphics[scale=0.6,angle=-90]{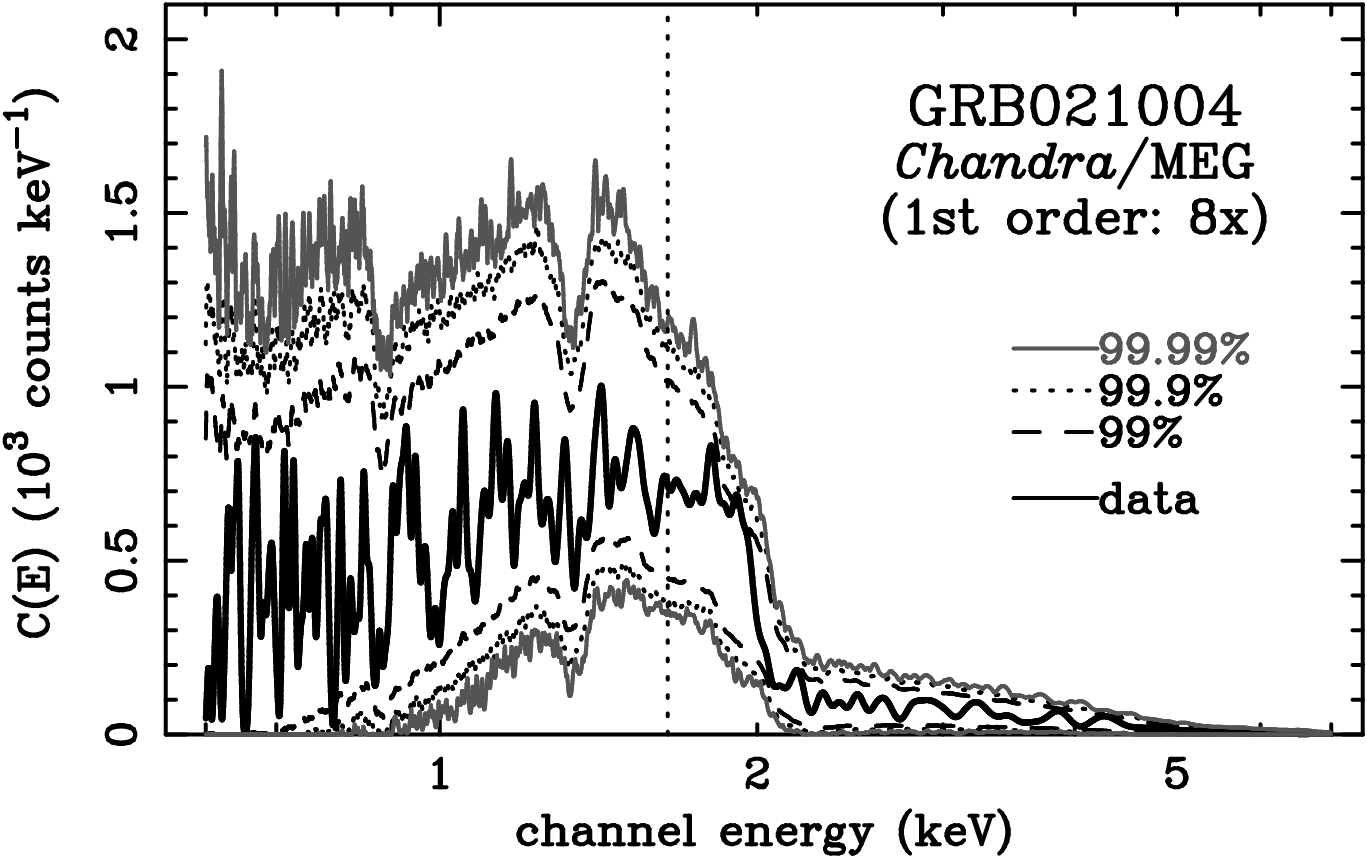}
  \caption{Same as in Figure~\ref{fig970508mc_1} for \grb021004 MEG dispersed spectrum.}
  \label{fig021004mc_meg}
\end{figure}

\begin{figure}
  \includegraphics[scale=0.6,angle=-90]{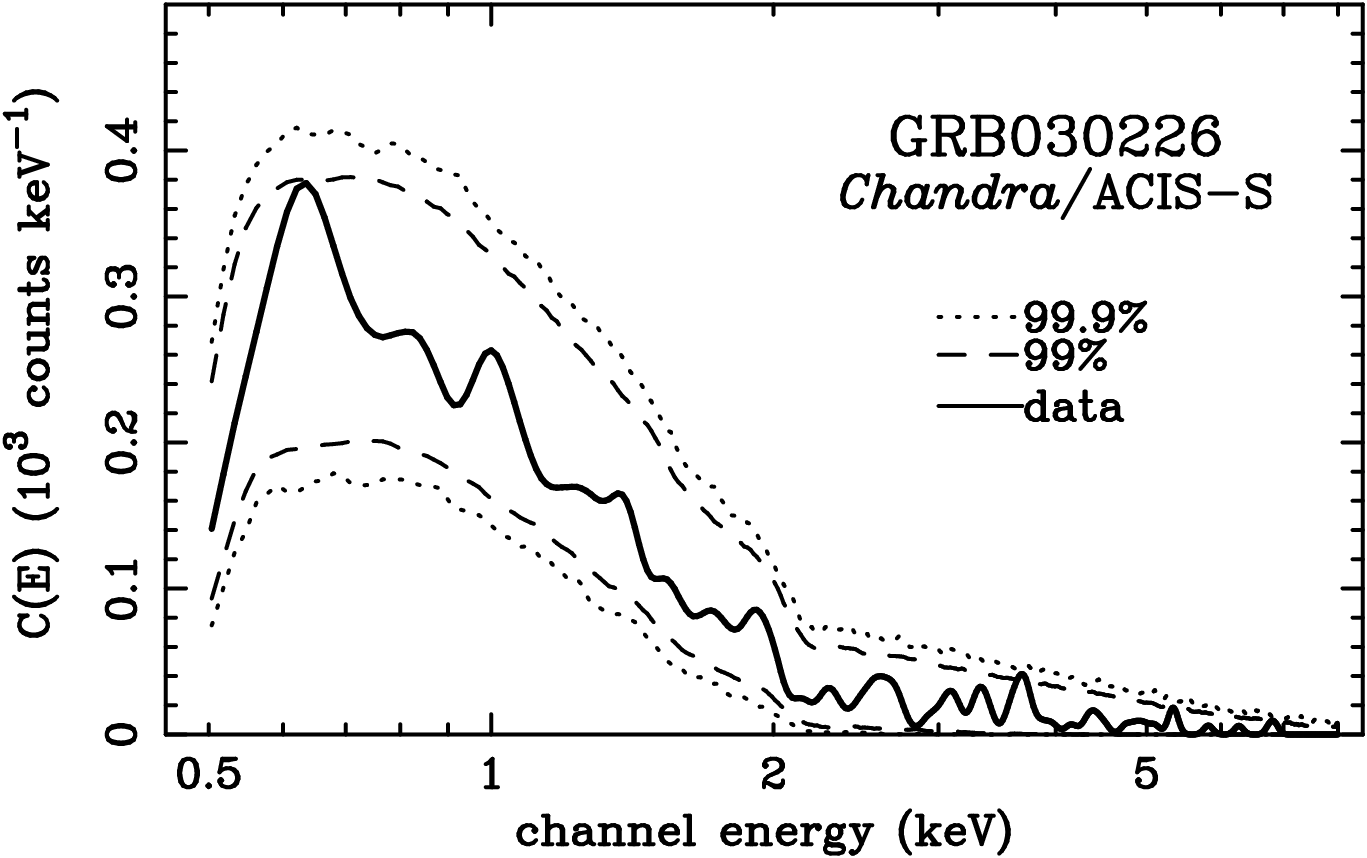}
  \caption{Same as in Figure~\ref{fig970508mc_1} for \grb030226 ACIS.}
  \label{fig030226mc}
\end{figure}

  For each spectral bin, we then sort the $10^5$ $C(E_i)$ values in increasing
  order and draw the 1000th, 100th, and 10th highest values as the 99\%,
  99.9\%, and 99.99\% single-trial significance levels, respectively, as a
  function of energy.  Similarly, we take the 1000th, 100th, and 10th lowest
  points, which represent 99\%, 99.9\%, and 99.99\% lower limits, to identify
  possible absorption features in the spectra\footnote{Note that, particularly
  for the 99.99\% confidence limits, the \mc\ function $C(E_i)$ can be Poisson
  limited (that is, the few number of simulations produce strong
  point-to-point variations).  When the data approach the confidence limit
  where the these variations can be seen in the simulations, we increased the
  number of simulations.  In selected parts of the spectrum where the the
  number of counts per \pinv\ bin is extremely low, the $C(E_i)$ can only
  possess a small set of discrete values, which results in wiggles in the
  confidence curves (see, e.g., the $99.99$\% curves in the \pn\ spectra above
  $\sim 2$~\kev, as well as those of the \chandra\ grating spectra) that
  cannot be smoothed out by increasing the number of simulations.  The data,
  however, are convolved with the matched-filter in the exact same manner as
  described above, so the derived significances are robust.}.  The data are
  convolved in the exact same manner as the simulated events, and we compare
  the values of each energy bin to the upper and lower bounds.
  Figures~\ref{fig970508mc_1} -- \ref{fig040106mc_mos} show the results for
  each source.

  We search through the matched-filtered spectrum for line-like features that
  exceed or fall below a certain confidence level.  If a significant feature
  is found, the single-trial significance at the peak of the feature is
  recorded.  We then return to the \mc\ simulations and count the ones that
  contain one or more features that meet or exceed that same single-trial
  limit {\it at any arbitrary energy}.  The number of such simulations are
  recorded and divided by the total number of simulations performed to yield
  the true multi-trial significance.  Qualitatively, the multi-trial
  probability of detecting a line is higher by a factor of approximately the
  total number of resolution elements in the detector bandpass.  For example,
  an emission feature with a single-trial significance of $99.99$\% (i.e.,
  occurs once in $10^4$ random trials at the specified energy) in a detector
  with $100$ resolution elements is seen approximately once in $\sim 100$
  random trials at an arbitrary energy.  The energy dependence of the
  resolving power and the effective area, however, complicates this and the
  simplest way to address this is through \mc\ simulations.

  This method provides a robust estimate of the statistical significances of
  discrete, unresolved features observed in the spectra, assuming the energy
  of the feature is not known {\em a priori}.  \citet{reeves03},
  \citet{watson02a} and \citet{watson03} adopt a different method, where each
  of the simulated spectra are fit by models with and without emission lines.
  The changes in $\chi^2$ are, then, recorded and compared to the observed
  $\Delta \chi^2$ derived from the data.  This method should, in principle,
  yield the probability that the observed features are due to chance
  coincidences.  In practice, however, the significances derived using this
  method will be overestimated since, as described by \citet{rutledge02}, an
  automated $\chi^2$-minimization is almost never capable of finding a true
  \emph{global} minimum when the $\chi^2$ surface is complicated and contain
  multiple local minima, for example, when the model involves multiple
  emission lines.  Therefore, the inferred $\Delta \chi^2$ between the models
  with and without emission lines for each simulation is underestimated.  This
  is why the statistical significance derived by \citet{reeves03} for the
  features in the \grb011211\ spectrum is significantly higher compared to
  that of \citet{rutledge02}.

  Finally, we make explicit our conventions for quoting an "equivalent
  gaussian sigma" for a given significance.  When the \mc\ simulation of $M$
  trials produces $N$ values in excess of that observed at any given energy
  (single-trial), we give the probability of having observed such a high value
  produced from noise alone as $p_{\rm excess}=N/M$.  This probability
  includes all phase-space for the observable, both above and below the
  median.  Note that if our observed value is exactly at the median, then
  $p_{\rm excess} = 0.5$.

  We then quote a "equivalent gaussian sigma" for this probability.  This
  provides a short-hand and intuitive denotation of the significance of the
  observation.  We adopt the convention that, for a probability $p_{\rm
  excess}$ of producing a value at or above the observed value from noise
  alone, the equivalent gaussian sigma $n$ is found from
\begin{equation}
  \label{eq:ges1}
  p_{\rm excess} = \frac{1}{\sqrt{2\pi}} \int_{n}^\infty e^{-x^{2}/2} dx,
\end{equation}
  which means that, for a measured value at the median of the simulated
  distribution ($p_{\rm excess}$=0.5), this would be $0\sigma$ in excess of
  the median value.  We would call this feature to be signficant at the
  $0\sigma$ level -- and it would not be regarded as a detection of anything.
  In the case that $p_{\rm excess}=2 \times 10^{-5}$, then $n=-4\sigma$ -- or,
  a 4$\sigma$ deficit in counts beneath the median, which would be a
  significant detection of an absorption line-like feature.

  When multi-trial signficances as estimated by counting the $N'$ simulations
  that show emission features with single-trial significances above a certain
  value, the equivalent gaussian sigma $n'$ is derived from
\begin{equation}
  \label{eq:ges2}
  p_{\rm excess} = \sqrt{\frac{2}{\pi}}\int_{n'}^\infty e^{-x^{2}/2} dx.
\end{equation}
  which is a a one-sided distribution -- i.e., if all of the \mc\ simulations
  show an excess, then the significance is $0 \sigma$.  In comparison with
  Equ.~\ref{eq:ges1}, Equ.~\ref{eq:ges2} gives a slightly higher value for the
  equilvalent gaussian sigma for the same $p_{\rm excess}$.
  \citet{antonelli00}, \citet{watson02a}, \citet{butler03b}, and
  \citet{watson03} have adopted Equ.~\ref{eq:ges2} in deriving the equivalent
  gaussian sigma.  The differences are relatively small especially at higher
  significances.

\section{Results of the Line Searches}
\label{section_mc_results}

  As shown in Figures~\ref{fig970508mc_1} -- \ref{fig040106mc_mos}, there are
  a number of discrete line-like features that exceed the $99.9$\% limit,
  which approximately corresponds to $3.1\sigma$ \emph{at that particular
  energy} (single trial) assuming a gaussian distribution of counts.  In 26
  cases listed in Table~\ref{tbl:search_results}, the single-trial confidence
  exceeds $99.9$\% ($3.1 \sigma$)\footnote{Note our use of nomenclature.  We
  use the term ``XX\% confidence'', when XX\% of the signals produced by
  background alone are at or below that detected.}.

  As we have noted earlier, we do not know {\em a priori} the expected
  energies of the lines, even in cases where the source redshift is known,
  since in some models the line emitting material can have an arbitrary
  velocity.  In such cases, the relevant quantity is the chance probability of
  detecting \emph{any} feature that exceeds the $99.9$\% limit ($3.3 \sigma$,
  multiple trials).  As shown in the final column of
  Table~\ref{tbl:search_results}, we find signals which meet this confidence
  limit in four datasets: in \grb011211 (at time $t>$5~ksec), \grb030227 (the
  complete dataset), \grb021004, and in \grb040106.

  Below, we first discuss our results in comparison with previous reports of
  line detection, followed by discussion of the four datasets with features
  with multi-trial probabilities with $>3.3\sigma$ gaussian equivalent
  significance.

\subsection{Comparison with Previous Reports of Line Detection}

  Below we compare our results to cases where line detections have been
  claimed in the published literature.
  
  {\bf \grb970508:} \citet{piro99} reported an unresolved line feature at
  3.5~\kev\ with a significance of $99.3$\% ($2.7$\ssig) during time interval
  1a during part of the afterglow.  We find no evidence for a significant
  excess at 3.5~\kev, or any energy (see Figure~\ref{fig970508mc_1}) during
  the same time interval.  The single-trial confidence at 3.5~\kev\ is $\sim
  60$\% ($\sim 0.25\sigma$).  The corresponding multi-trial confidence limit
  is only $\sim 15$\% ($\sim 0.2$\ssig).

  {\bf \grb970828:} We find no evidence of any single-energy feature similar
  to the one claimed by \citet{yoshida99} during time interval B.  However,
  the reported feature was claimed to be resolved with a centroid of $E =
  5.04^{+0.23}_{-0.31} ~\rm{\kev}$ and a width of $\sigma =
  0.31^{+0.38}_{-0.31} ~\rm{\kev}$ \citep{yoshida99,yoshida01}.  We see an
  excess at a similar energy in the both the \sis\ and \gis\ data
  (Figure~\ref{fig970828mc_sis_b} and \ref{fig970828mc_gis_b}), with
  single-trial significances of 99.2\% (2.4\ssig) and 99.4\% (2.5\ssig),
  respectively.  Our estimate of the multi-trial significance is only $\sim
  70$\% ($\sim 1.0$\ssig) in each instrument.  Searches for lines with larger
  widths resulted in those with lower significance.  It is interesting,
  however, that the residuals of the power-law model fit {\it visually} show
  excesses in all four detectors (see Figure~\ref{fig970828_b}).  We also note
  that the \sn\ in this spectral range is only of order unity, which implies
  that the estimated significance is highly-dependent on the exact shape and
  magnitude of the local background spectrum.  The reality of this excess
  feature should, therefore, be taken with caution.

  {\bf \grb991216:} In the case of the \grb991216 \chandra\ spectrum, where a
  4.7$\sigma$ detection of an emission line at $E \sim 3.5~\rm{\kev}$ has been
  claimed \citep{piro00}, we instead find only a $99.0$\% ($\sim 2.3\sigma$)
  confidence single-trial fluctuation in the dispersed \heg\ spectrum at a
  width of 16 times it's instrument resolution or \fwhm\ $\sim 0.7$~\kev\ (see
  Figure~\ref{fig991216mc_heg}), which is close to value quoted in
  \citet{piro00}.  This fluctuation is at a slightly higher energy ($\sim
  3.8$~\kev).  Our multi-trial confidence limit of this feature is 20\%
  ($0.3$\ssig).  We do not find any feature near $4.4 \pm 0.5$~\kev, as
  reported as a marginal detection by \citet{piro00}.  The zeroth order
  spectrum appears to show an excess feature centered at $E \sim 3.4$~\kev,
  which is below that seen in the dispersed spectrum, (see
  \ref{fig991216mc_zero}), but is not significant.  Searches for significant
  features that are narrower than 16 times the \heg\ resolution resulted in a
  non-detection.

  At least part of the source for the discrepancy between the results of our
  analysis and that of \citet{piro00} is the adopted continuum level.
  \citet{piro00} use a power-law slope and flux determined from the zeroth
  order spectrum.  However, as can be seen in their Figure 1, this continuum
  systematically underestimates the flux across the entire \hetg\ bandpass.
  Any fluctuation is, therefore, in excess of the adopted continuum flux.  If
  we allow for a $\sim 20$\% calibration uncertainty between the zeroth and
  first orders, the spectra of both orders are well-represented by a single
  continuum and, as a consequence, the single-trial significance of the line
  decreases to $2.3\sigma$ at a single energy.  Inspection of the \grb020813
  and \grb021004 spectra, both of which were collected in the same instrument
  configuration, also show a consistent systematic $\sim 20$\% discrepancy in
  the normalizations between the zeroth and first orders.

  Incidentally, we note that a feature seen in the \meg\ at $E \sim
  1.32~\rm{\kev}$ is of relatively higher significance than the $E \sim
  3.5~\rm{\kev}$ feature reported by \citet{piro00}, although it is still
  extremely marginal.  The formal significance is 99.91\% single trial ($\sim
  3.1$\ssig) and 89.2\% multi-trial confidence ($\sim 1.6$\ssig).

  {\bf \grb000214:} \citet{antonelli00} find a feature at $E \sim
  4.7~\rm{\kev}$ feature which they claim, using an $F$-test to be significant
  at $99.73$\% at a single energy.  In our simulations, we find a broad excess
  at similar energy (Figure~\ref{fig000214mc}), significant at $\sim 99.92$\%
  ($\sim 3.2\sigma$) at a single energy.  This is slightly higher than found
  by \citet{antonelli00}.  However, since the redshift is not known, the
  appropriate multi-trial significance is $99.03$\% ($\sim 2.6\sigma$).

  {\bf \grb001025:} \citet{watson02a} find a plasma model to be a better fit
  to the spectrum of this \grb\ than an absorbed power law, due to a soft
  excess observed between $\sim 0.5$ and 2~\kev.  We find no significant
  excess at these energies (Figures~\ref{fig001025}, \ref{fig001025mc_pn}, and
  \ref{fig001025mc_mos}).  The single-trial probabilities are all $< 99.9$\%
  confidence, which is below that found with identical instrumentation for
  \grb011211 (see below).  The multi-trial significance of the most
  significant feature ($\sim 0.7$~\kev\ excess seen in the \pn) is $\sim
  65$\%.

  {\bf \grb010220:} \citet{watson02a} claim the existence of a feature in the
  \pn\ spectrum of this afterglow at 3.9~\kev.  As shown in
  Figures~\ref{fig010220} and \ref{fig010220mc}, we find no evidence for
  features at any energy being detected with single-energy significance
  $>$99\% confidence (2.3$\sigma$). The multi-trial significance limit would
  be, consequently, less than that found for \grb001025 (see above), which is
  $< 65$\% confidence ($< 0.9$\ssig).

  {\bf \grb011211:} In \grb011211 the presence of several soft X-ray emission
  lines from mid-$Z$ elements during the first $5~\rm{ksec}$ of the
  observation has been claimed \citep{reeves03, reeves02}.  We find one
  feature in the \pn\ spectrum at $E = 0.85 ~\rm{\kev}$ with $99.9$\%
  single-trial significance (see Figure~\ref{fig011211mc_pn_1}), coincident
  with a line \citet{reeves02} identified with 97\% confidence.  This is
  consistent with the results of \citet{rutledge02} and \citet{borozdin02}.
  The chance probability of finding such a feature at an arbitrary energy is
  $\sim 15$\% ($1.4 \sigma$).  \label{sec:pre}

  In the spectra acquired during the latter ($t > 5$~ksec) time interval,
  there is an excess feature in the \pn\ spectrum
  (Figure~\ref{fig011211mc_pn_2}) near the instrumental oxygen edge just above
  $E \sim 0.5~\rm{\kev}$.  This feature is statistically significant at the
  $>99.99$\% level at an arbitrary energy, so the formal confidence is
  $>3.89\sigma$.  The feature is not seen in the \mos\ spectrum, which as an
  exposure (in cm$^2$~s) a factor $\sim 3$ smaller at this energy.  We cannot
  exclude the possibility that the excess is due to calibration uncertainty
  near the oxygen edge, although no such features have been reported in other
  high \snr\ data, nor did we find an excess at this energy in our analysis of
  the continuum source 3C273 (see below).  What makes this particular case
  somewhat suspicious is that the background spectrum shows an excess similar
  to the one seen in the \grb\ spectrum.

  {\bf \grb020813:} \citet{butler03b} have reported evidence for a line at
  1.3~\kev.  We do find a narrow fluctuation at approximately this energy in
  the \heg\ spectrum at $\sim 99.99\%$ confidence
  (Figure~\ref{fig020813mc_heg}), single trial.  The multi-trial confidence
  is, however, only 97.86\% (2.3 $\sigma$).  Note that the decrease in
  confidences between single-trial and multi-trial is greater for grating
  observations, due to the larger number of independent resolution elements in
  grating spectra.

  Inspection of the raw counts in each of the dispersion orders shows an
  excess in only the $m = +1$ order of the \heg.  Of the four spectral orders
  that are considered ($m = \pm 1$ in \heg\ and \meg), this is the one with
  the lowest effective area in this spectral range.  The feature is not
  detected in any of the other spectral orders including the zeroth order.  In
  particular, the effective area of the \meg\ $m = -1$ order is $\sim 4$ times
  larger than that of the \heg\ $m = +1$, which makes the reality of this
  detection somewhat suspicious.

\begin{figure}
  \includegraphics[scale=0.6,angle=-90]{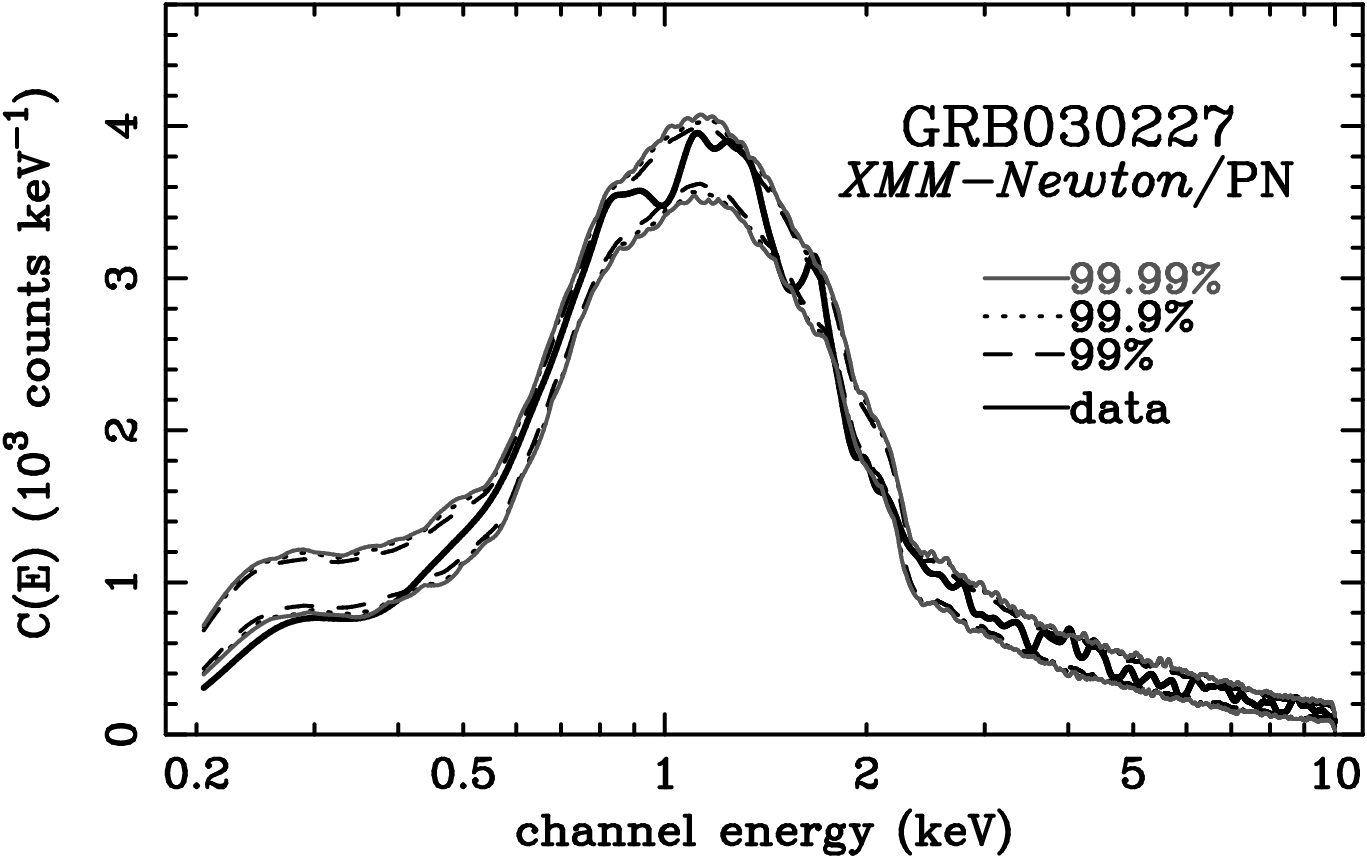}
  \caption{Same as in Figure~\ref{fig970508mc_1} for \grb030227 PN cumulative spectrum.}
  \label{fig030227mc_pncum}
\end{figure}

\begin{figure}
  \includegraphics[scale=0.6,angle=-90]{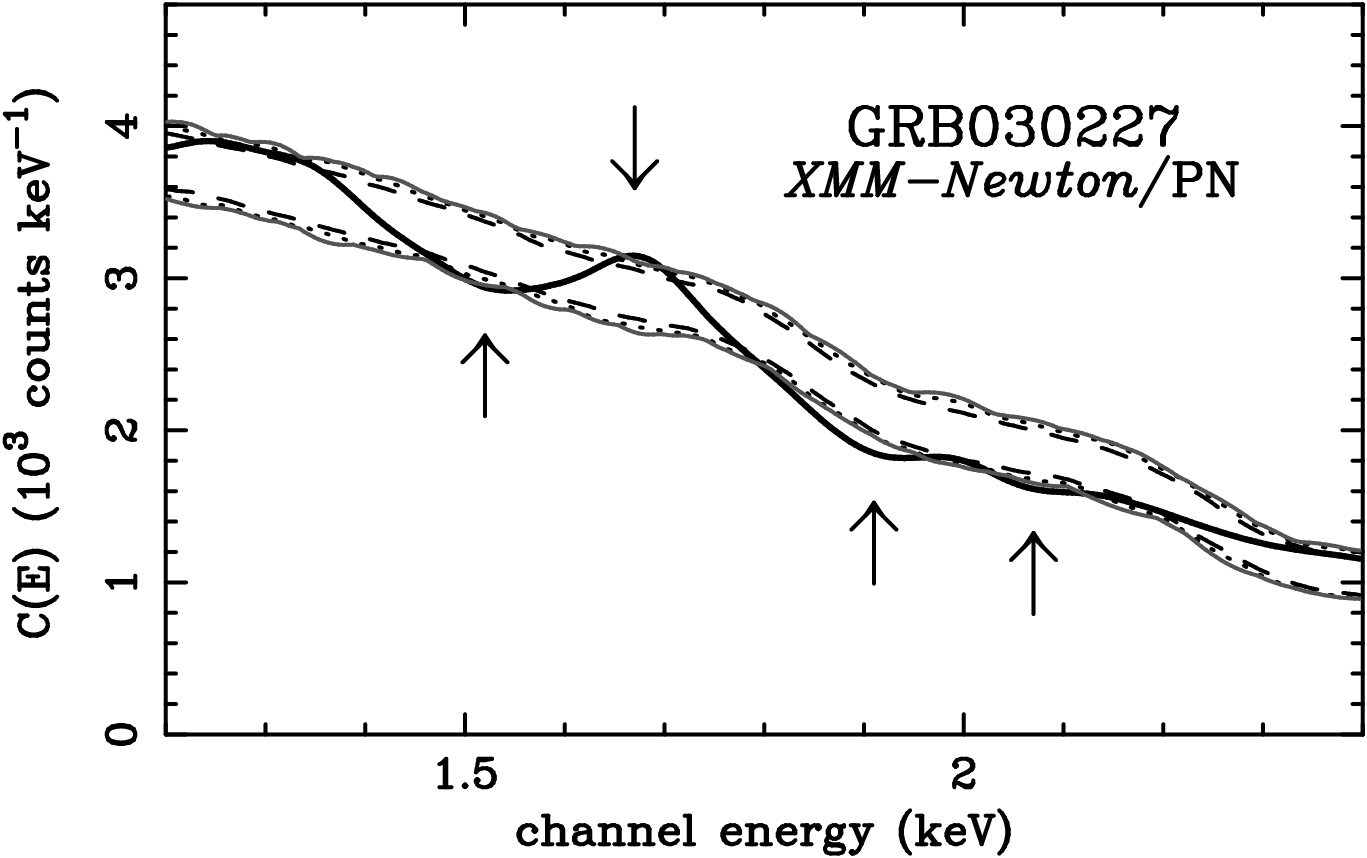}
  \caption{A blow-up of Figure~\ref{fig030227mc_pncum} in the 1.2 -- 2.4~\kev\ region.}
  \label{fig030227mc_pncum_zoom1}
\end{figure}

\begin{figure}
  \includegraphics[scale=0.6,angle=-90]{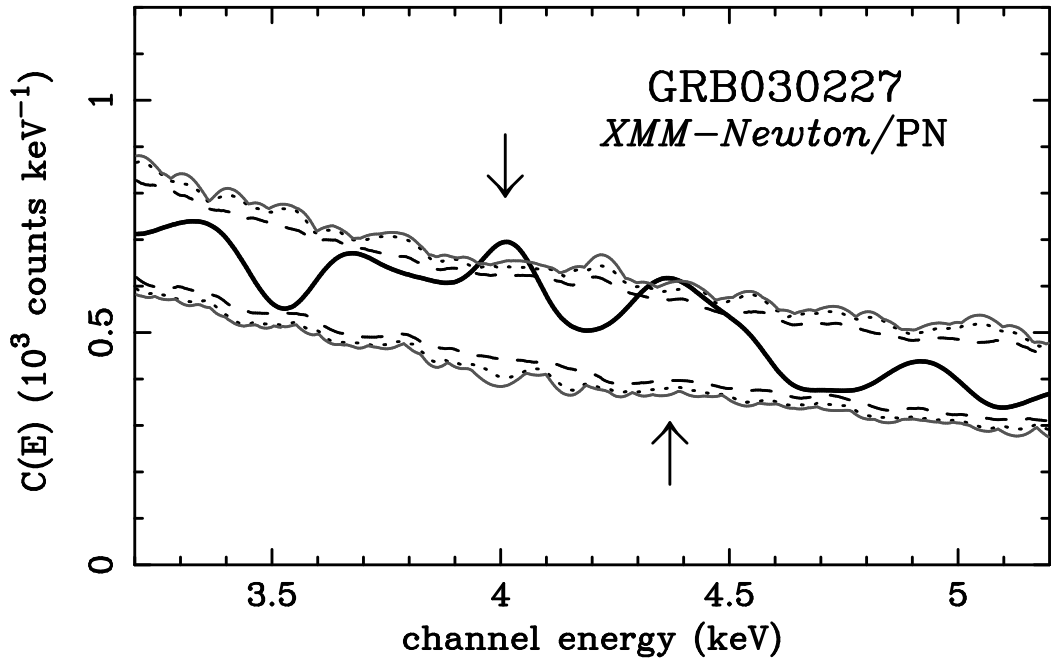}
  \caption{A blow-up of Figure~\ref{fig030227mc_pncum} in the 3.2 -- 5.2~\kev\ region.}
  \label{fig030327mc_pncum_zoom2}
\end{figure}

\begin{figure}
  \includegraphics[scale=0.6,angle=-90]{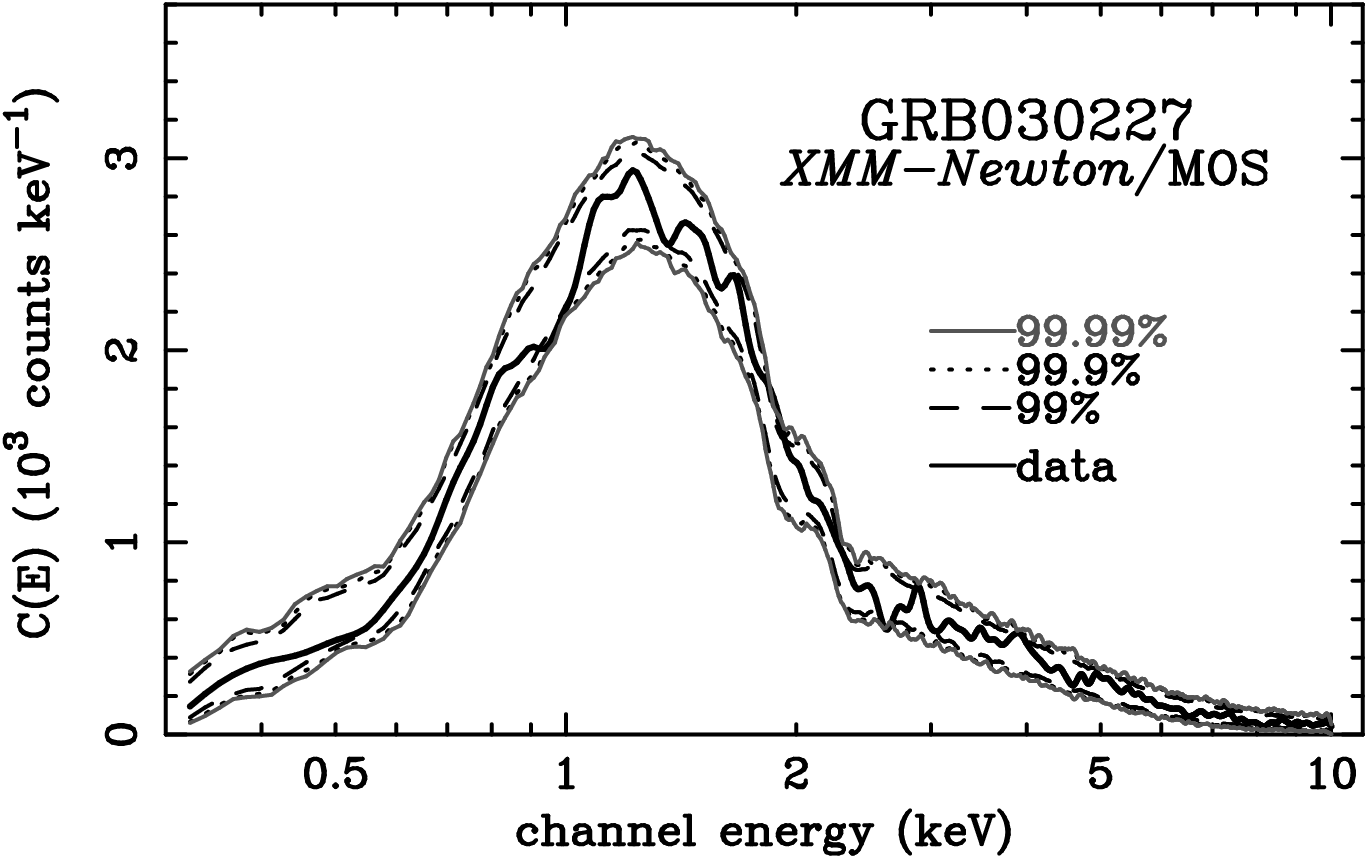}
  \caption{Same as in Figure~\ref{fig970508mc_1} for \grb030227 MOS cumulative spectrum.}
  \label{fig030227mc_moscum}
\end{figure}

\begin{figure}
  \includegraphics[scale=0.6,angle=-90]{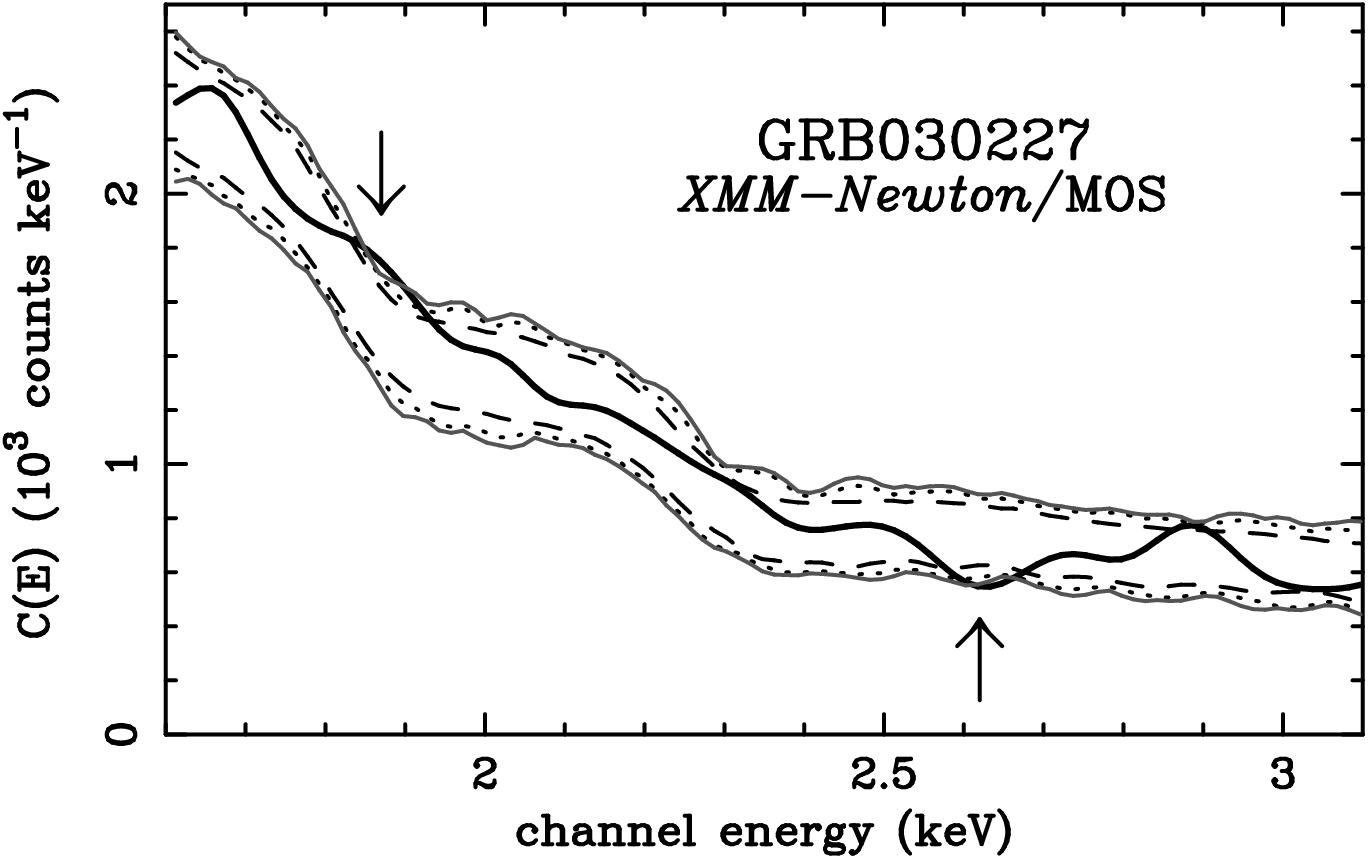}
  \caption{A blow-up of Figure~\ref{fig030227mc_moscum} in the 2.1 -- 3.1~\kev\ region.}
  \label{fig030227mc_moscum_zoom1}
\end{figure}

\begin{figure}
  \includegraphics[scale=0.6,angle=-90]{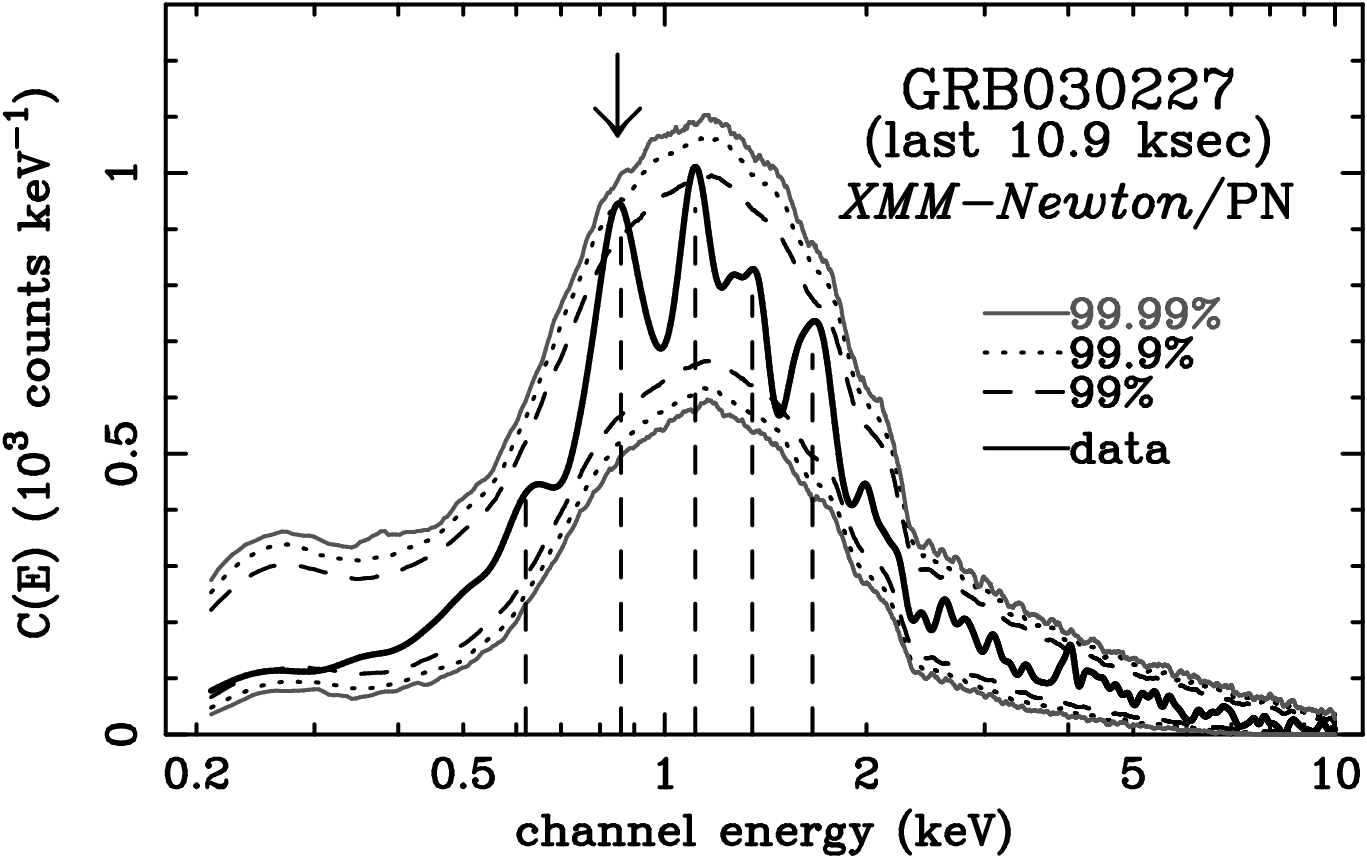}
  \caption{Same as in Figure~\ref{fig970508mc_1} for \grb030227 PN last 10.9 ksec.}
  \label{fig030327mc_pnlast}
\end{figure}

\begin{figure}
  \includegraphics[scale=0.6,angle=-90]{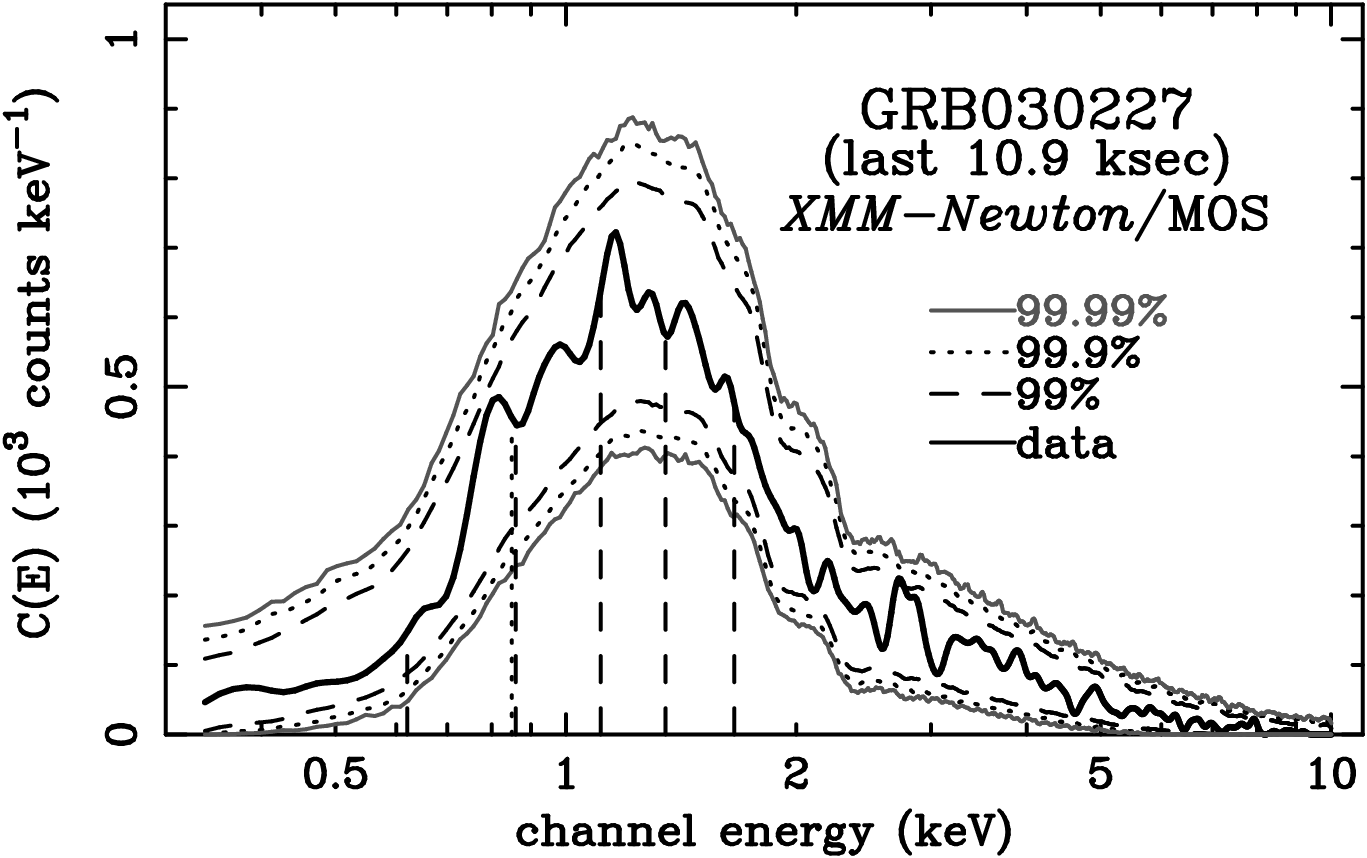}
  \caption{Same as in Figure~\ref{fig970508mc_1} for \grb30227 MOS last 10.9 ksec.}
  \label{fig030327mc_moslast}
\end{figure}

\begin{figure}
  \includegraphics[scale=0.6,angle=-90]{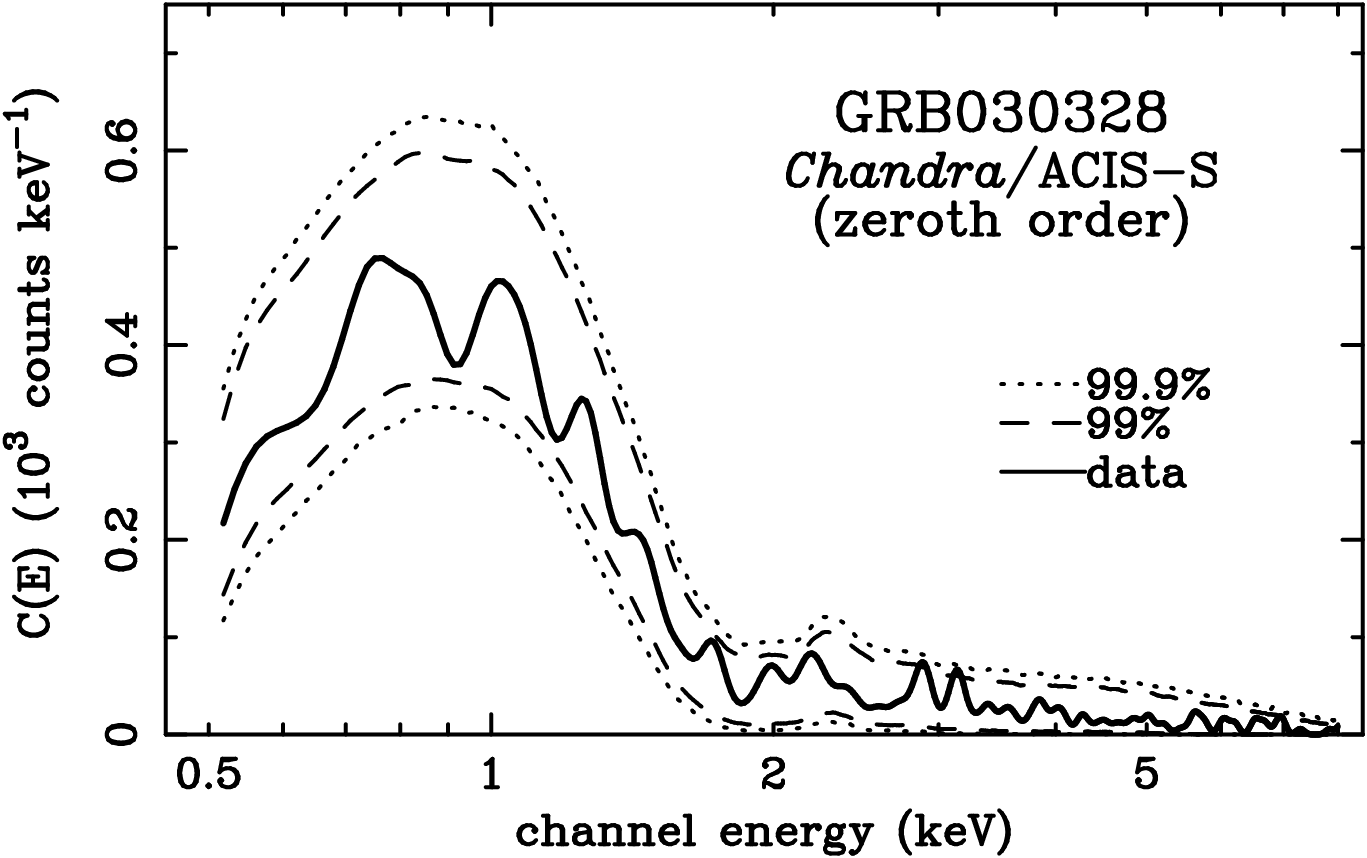}
  \caption{Same as in Figure~\ref{fig970508mc_1} for \grb030328 zero order.}
  \label{fig030328mc_zero}
\end{figure}

\begin{figure}
  \includegraphics[scale=0.6,angle=-90]{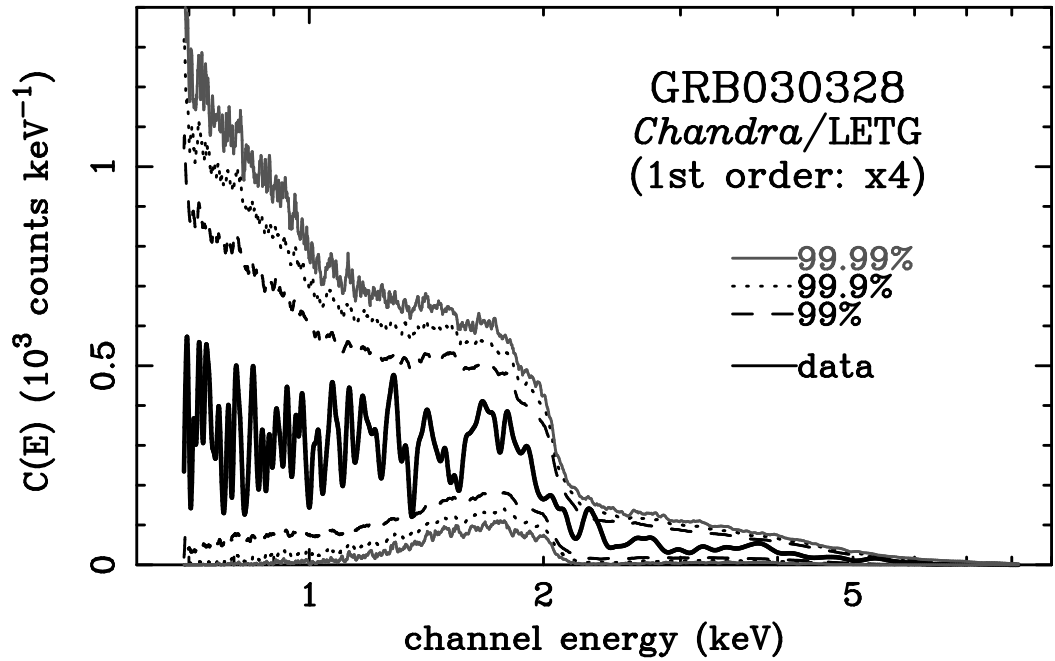}
  \caption{Same as in Figure~\ref{fig970508mc_1} for \grb030328 LEG dispersed.}
  \label{fig030328mc_leg}
\end{figure}

\begin{figure}
  \includegraphics[scale=0.6,angle=-90]{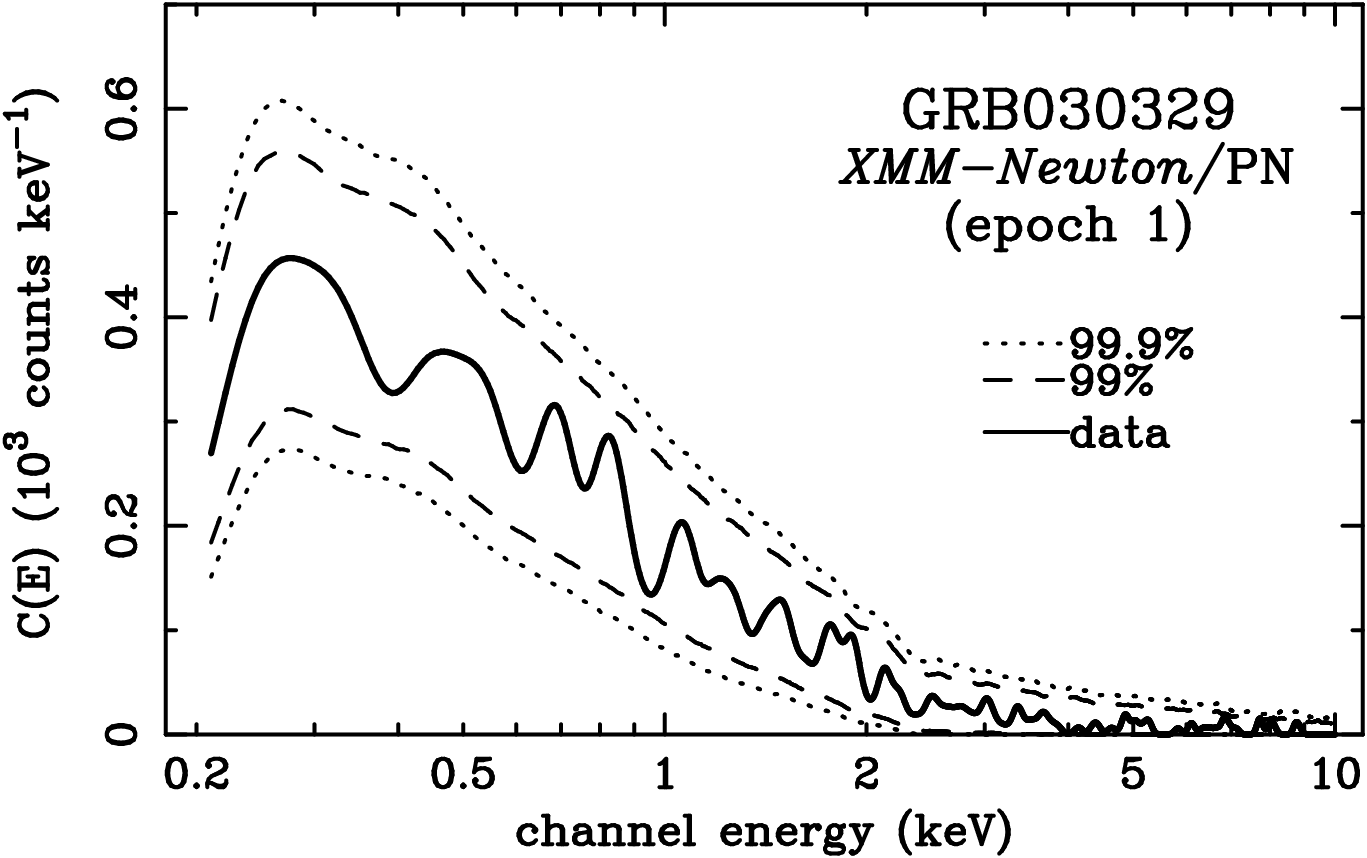}
  \caption{Same as in Figure~\ref{fig970508mc_1} for \grb030329 PN time segment 1.}
  \label{fig030329mc_pn_1}
\end{figure}

\begin{figure}
  \includegraphics[scale=0.6,angle=-90]{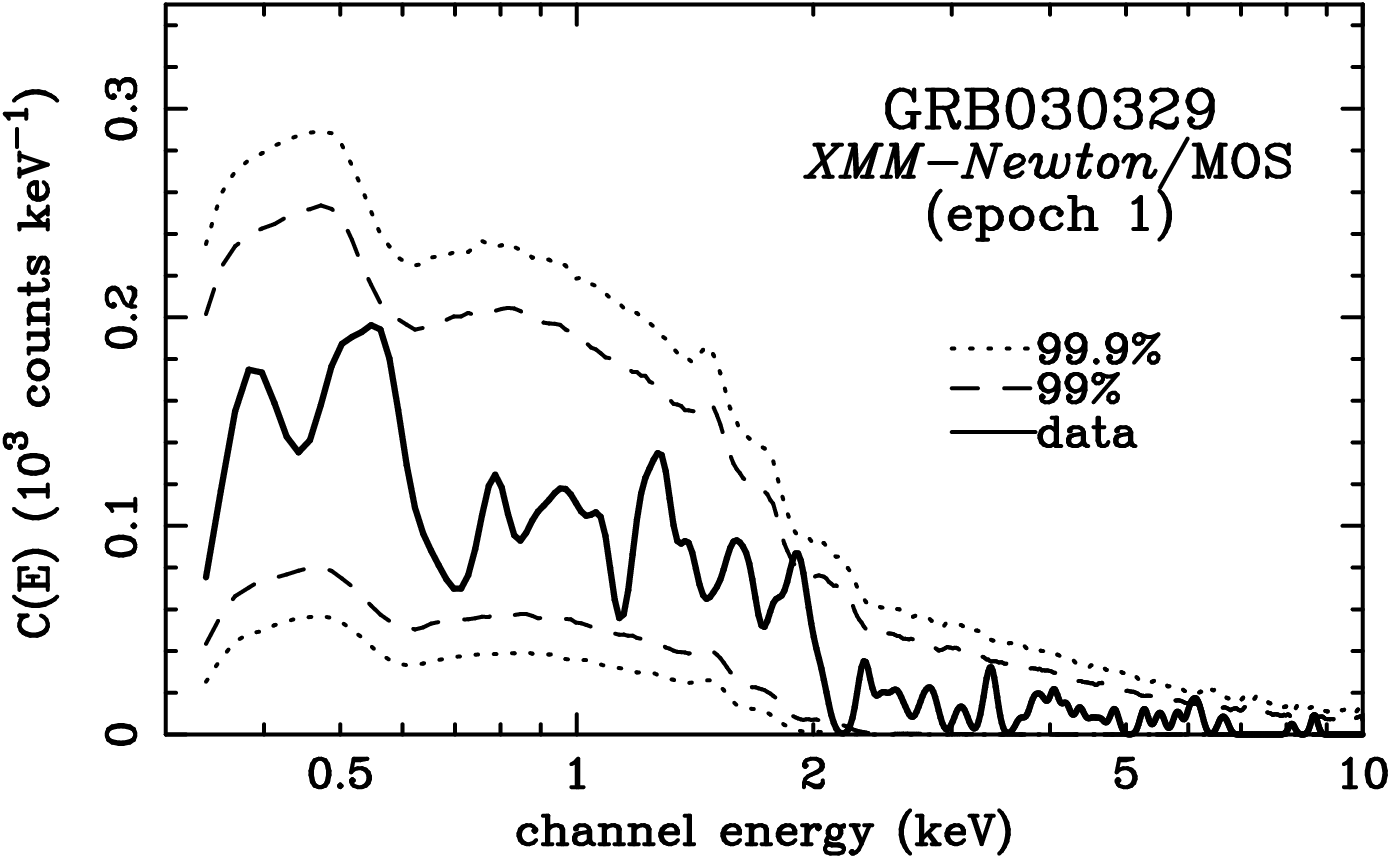}
  \caption{Same as in Figure~\ref{fig970508mc_1} for \grb030329 MOS time segment 1.}
  \label{fig030329mc_mos_1}
\end{figure}

\begin{figure}
  \includegraphics[scale=0.6,angle=-90]{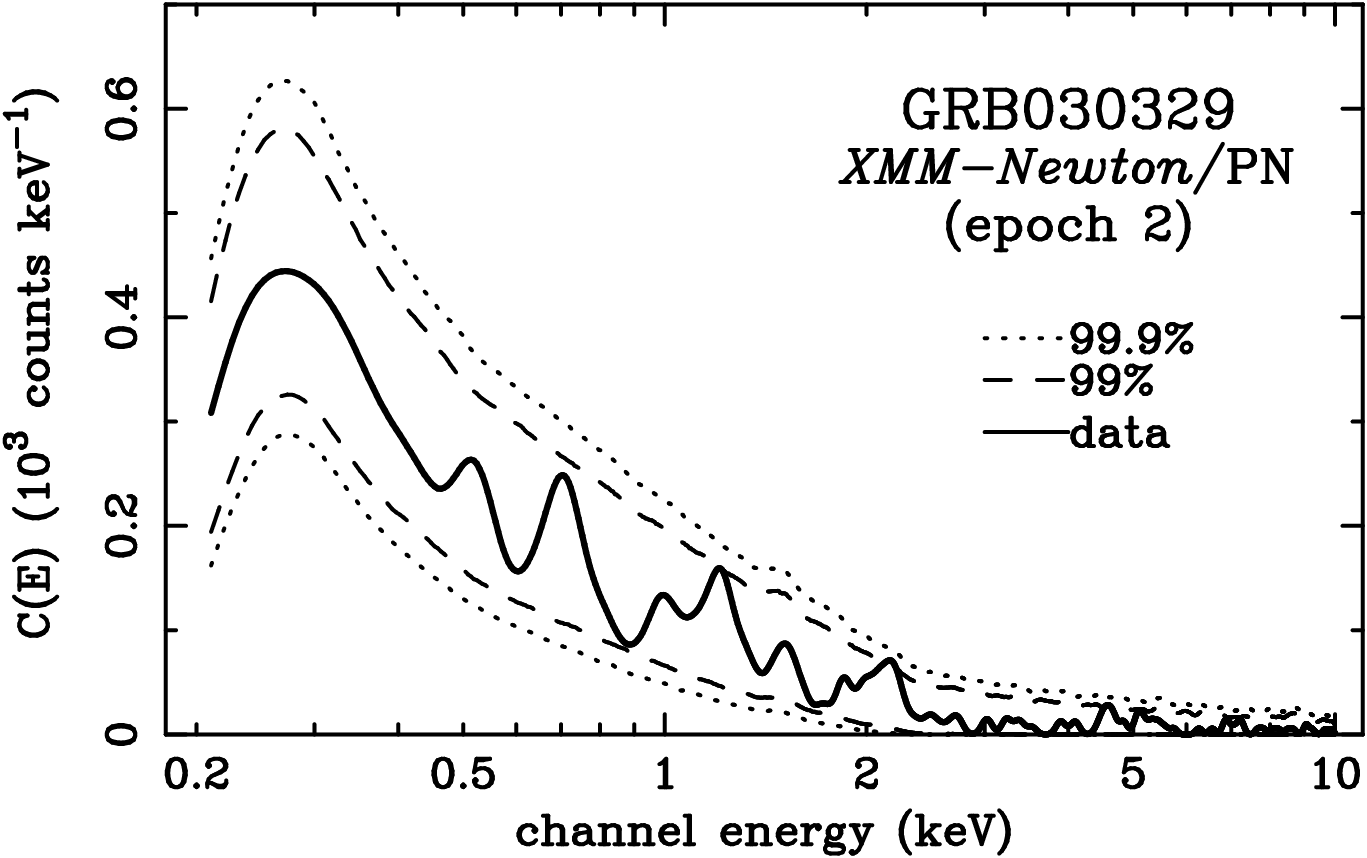}
  \caption{Same as in Figure~\ref{fig970508mc_1} for \grb030329 PN time segment 2.}
  \label{fig030329mc_pn_2}
\end{figure}

\begin{figure}
  \includegraphics[scale=0.6,angle=-90]{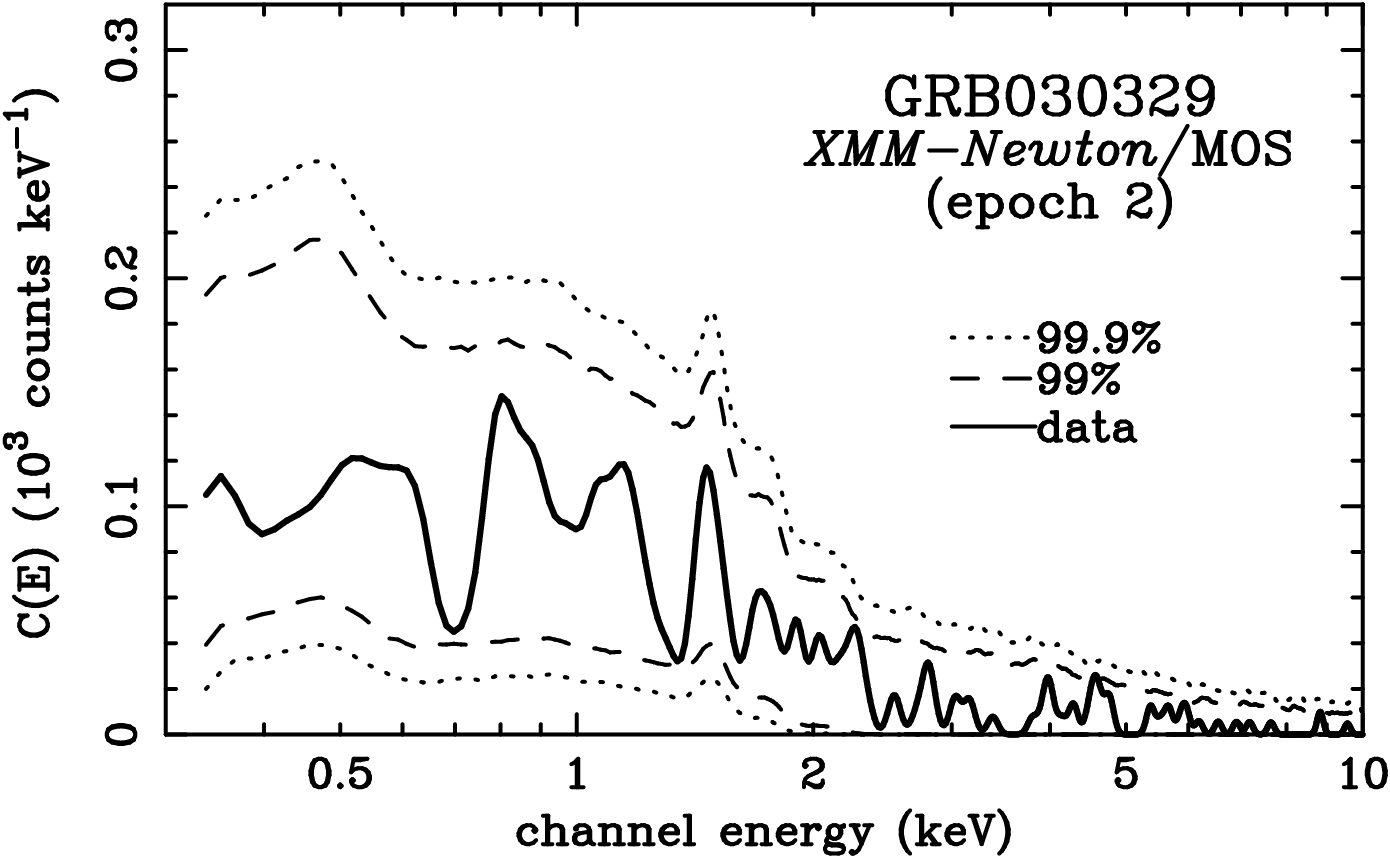}
  \caption{Same as in Figure~\ref{fig970508mc_1} for \grb030329 MOS time segment 2.}
  \label{fig030329mc_mos_2}
\end{figure}

\begin{figure}
  \includegraphics[scale=0.6,angle=-90]{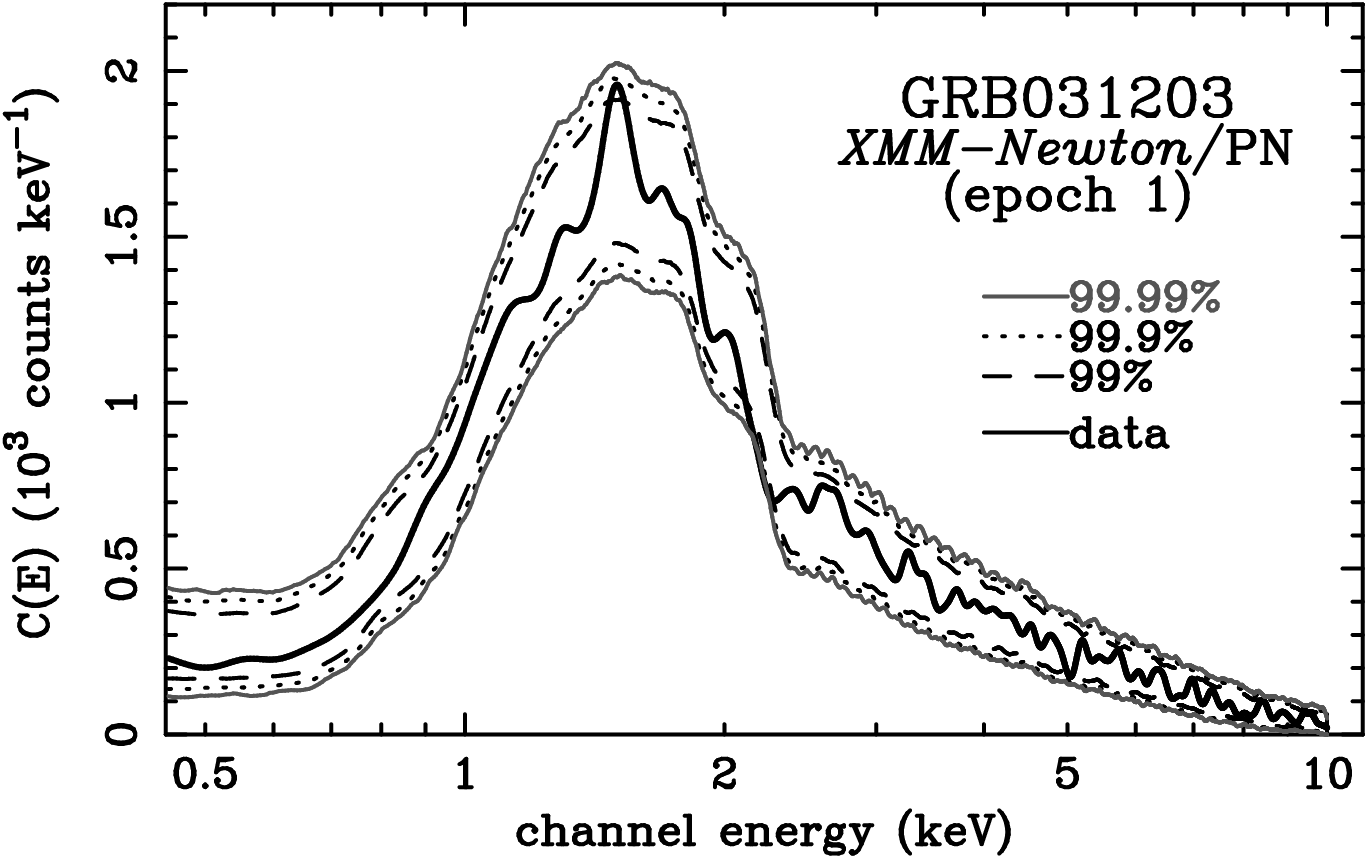}
  \caption{Same as in Figure~\ref{fig970508mc_1} for \grb031203 PN time segment 1.}
  \label{fig031203mc_pn_1}
\end{figure}

\begin{figure}
  \includegraphics[scale=0.6,angle=-90]{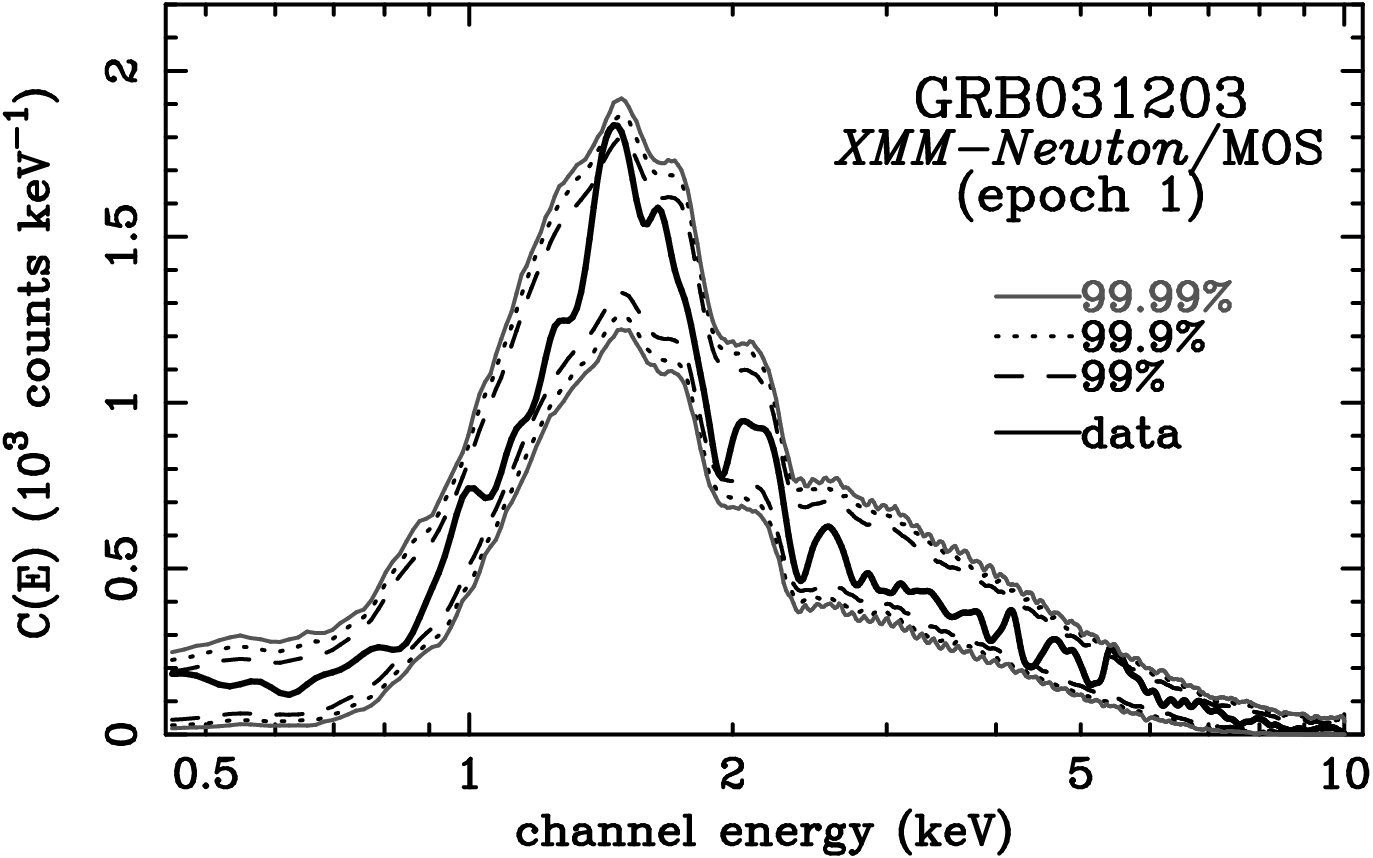}
  \caption{Same as in Figure~\ref{fig970508mc_1} for \grb031203 MOS time segment 1.}
  \label{fig031203mc_mos_1}
\end{figure}

\begin{figure}
  \includegraphics[scale=0.6,angle=-90]{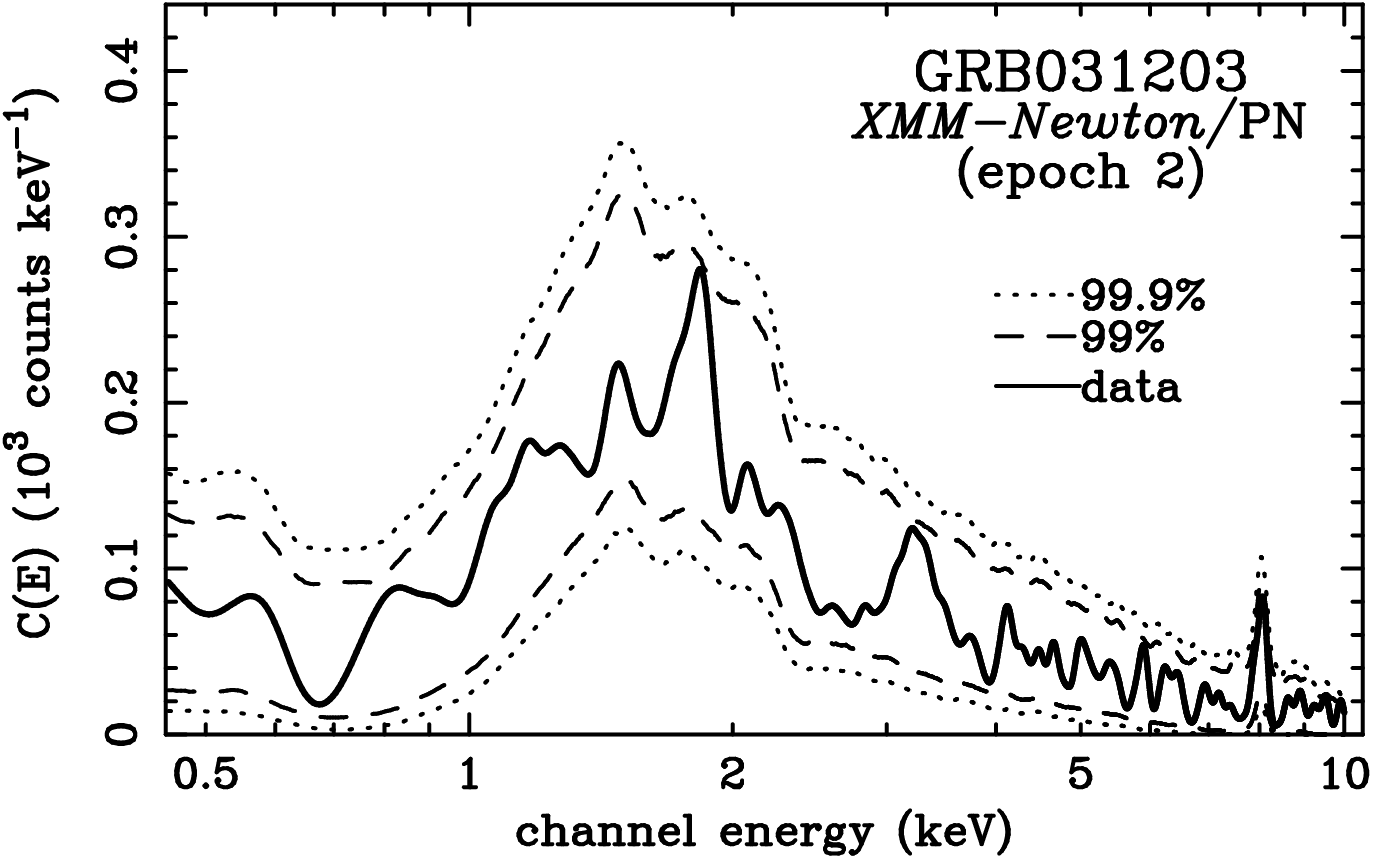}
  \caption{Same as in Figure~\ref{fig970508mc_1} for \grb031203 PN time segment 2.}
  \label{fig031203mc_pn_2}
\end{figure}

\begin{figure}
  \includegraphics[scale=0.6,angle=-90]{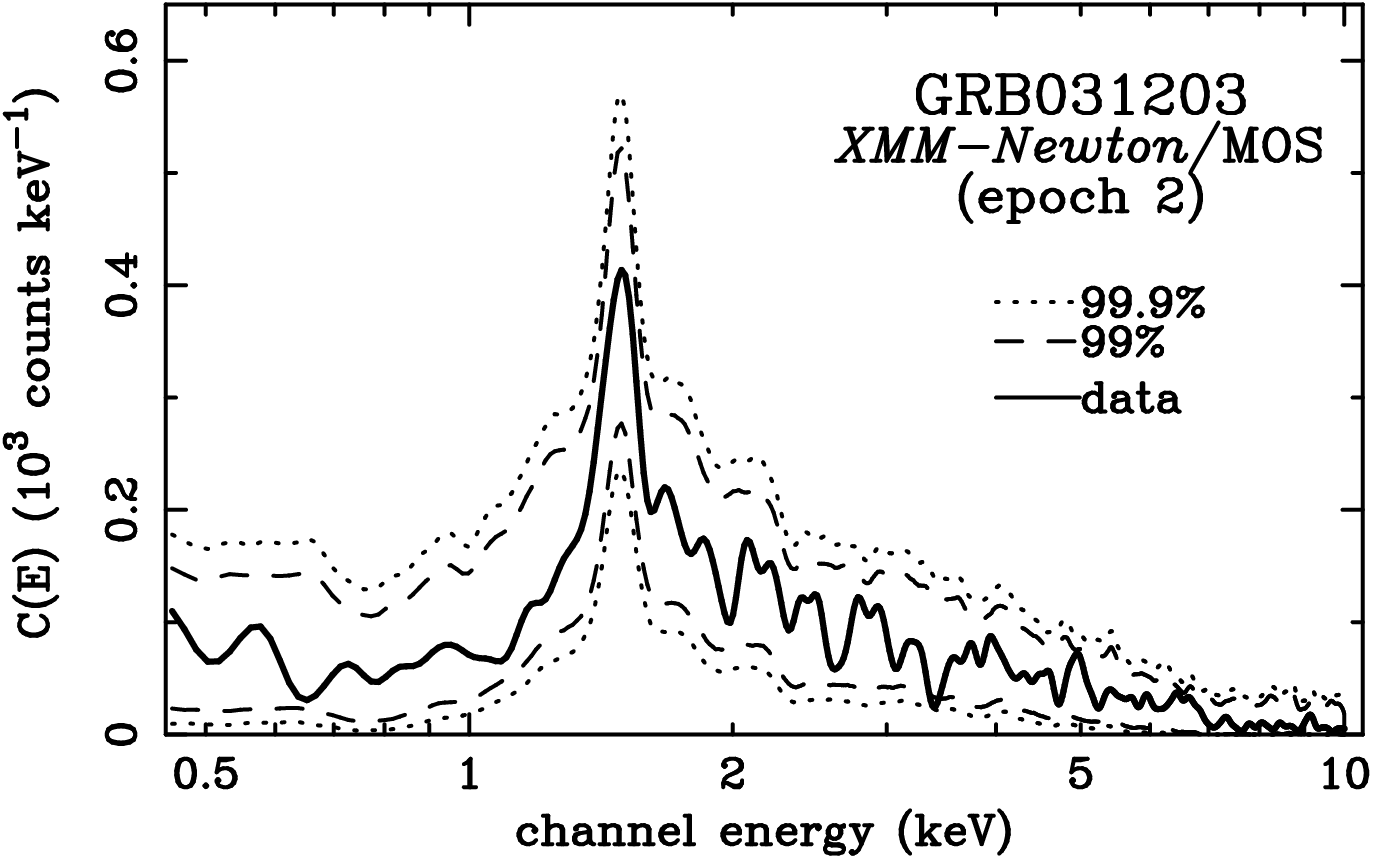}
  \caption{Same as in Figure~\ref{fig970508mc_1} for \grb031203 MOS time segment 2.}
  \label{fig031203mc_mos_2}
\end{figure}

\begin{figure}
  \includegraphics[scale=0.6,angle=-90]{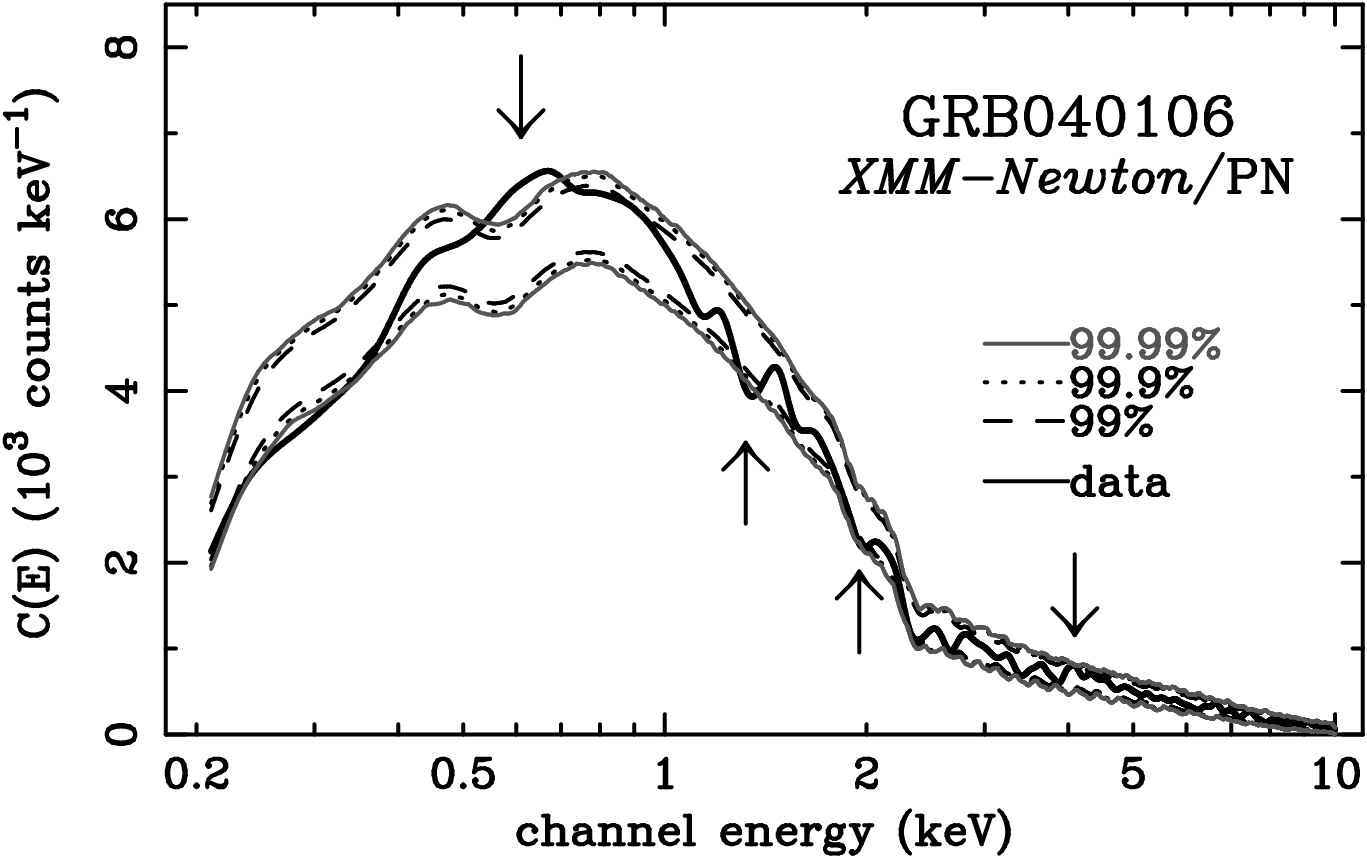}
  \caption{Same as in Figure~\ref{fig970508mc_1} for \grb040106 PN.}
  \label{fig040106mc_pn}
\end{figure}

\begin{figure}
  \includegraphics[scale=0.6,angle=-90]{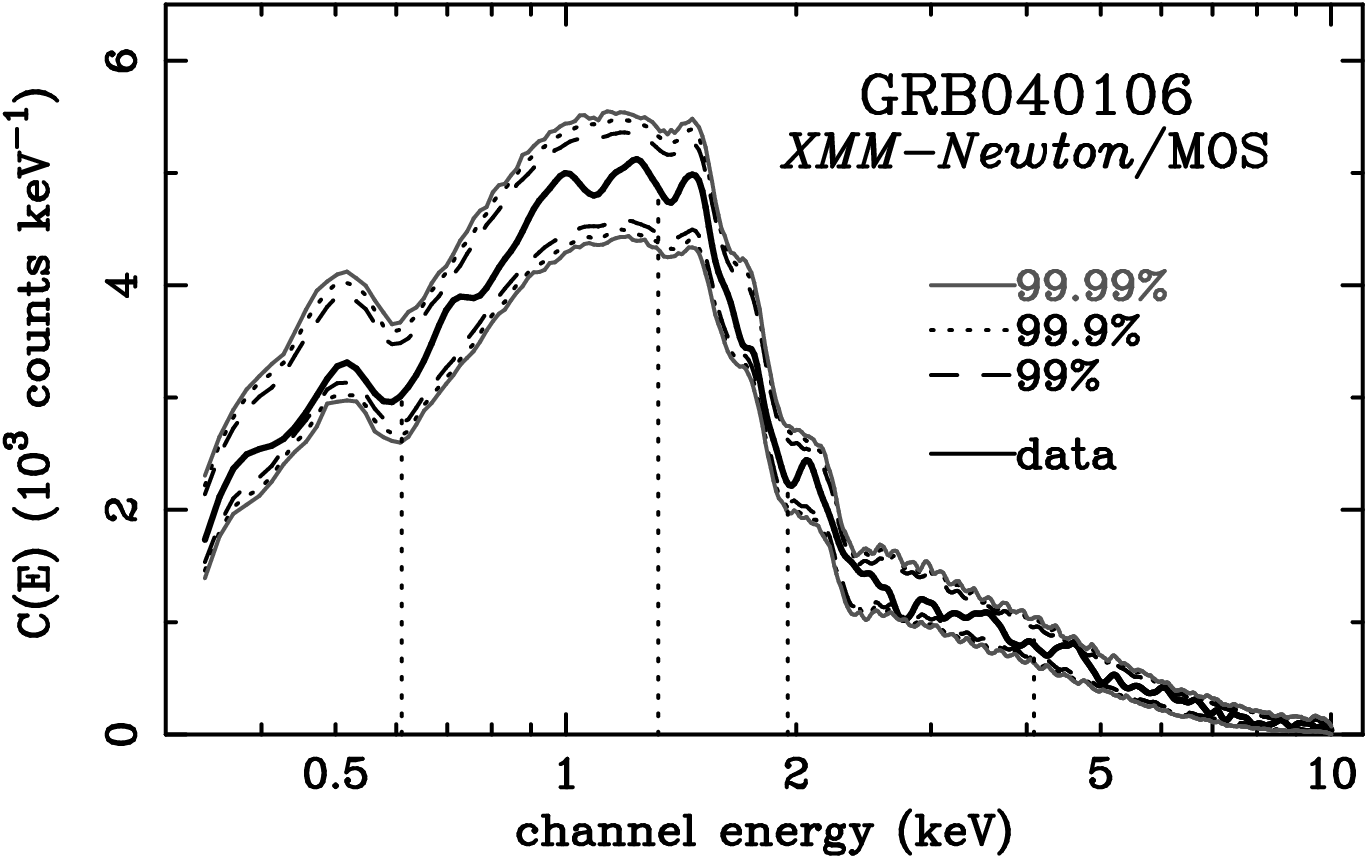}
  \caption{Same as in Figure~\ref{fig970508mc_1} for \grb040106 MOS.}
  \label{fig040106mc_mos}
\end{figure}

  {\bf \grb030227:} This is the second highest signal-to-noise observation in
  our sample only after \grb040106.  \citet{watson03} find four emission lines
  between 0.62 and 1.67~\kev\ in the last 11~ksec of the \pn\ spectrum of this
  event.  In our analysis (Figures~\ref{fig030227mc_pncum} --
  \ref{fig030327mc_moslast}), we find no significant (at $>99.9$\%) features
  in the spectrum during this time interval (a feature near 0.84~\kev\ is
  close to 99.9\% confidence, single trial).  The most significant feature,
  with single-trial detection confidence of $\la 99.9$\%, has multi-trial
  detection confidence of $< 85$\% ($< 1.4$\ssig).

  We do, however, find several fluctuations that are in excess of the 99.99\%
  limits in the cumulative \pn\ and \mos\ spectra
  (Figures~\ref{fig030227mc_pncum} and \ref{fig030227mc_moscum}).  One of
  these is in the $1.60 - 1.69$~\kev\ range, coincident with a feature
  identified previously by \citet{mgt+03}.  However, the multi-trial
  significance is marginal (98.93\% confidence detection; 2.6\ssig).  Two
  features at $3.95 - 4.06$~\kev\ and $4.33 - 4.39$~\kev\ have multi-trial
  significance greater than 3.3\ssig\ (3.54\ssig\ and 3.62\ssig,
  respectively).  The former is particularly interesting, since there is a
  feature at approximately the same energy ($E \sim 3.9~\rm{\kev}$) in the
  \mos\ as well, just above the 99.9\% limit.

  The background spectra of the \pn\ or the \mos\ both do not show discrete
  features near this energy.  To check for other possible calibration
  uncertainties, we have analyzed the \xmm\ data of a featureless continuum
  source, 3C273, to see whether excesses of similar significance do indeed
  appear in the spectra of other sources.  We have extracted only a small
  fraction of the events so that the spectrum contains the same number of
  counts in the 1 -- 5 \kev\ range as in the \grb030227 spectrum.  We fit the
  data to an absorbed power-law model and performed \mc\ simulations to derive
  formal significances of the local fluctuations.  Interestingly, no features
  in excess of 99.99\% single-trial confidence were detected in either the
  \pn\ or the \mos\ data of 3C273.

  In summary, in most cases where lines have been claimed, we either find the
  fluctuations to be of lower significance, or to be absent.  A 3\ssig\
  feature at a given energy in a spectrum with few counts visually appears
  more significant (i.e., higher equivalent width) than in a spectrum of high
  statistical quality.  These fluctuations are also transient and would appear
  only during a segment of the total exposure.  This trend is similar to the
  ones seen in many observations in the sense that (1) the ``lines'' are
  transient, (2) almost always $\sim 3\sigma$ irrespective of the statistical
  quality of the data, and (3) high equivalent width.

\subsection{Other Sources that Exhibit Line-like Fluctuations}

  As summarized in Table~\ref{tbl:search_results}, several sources other than
  those reported in the literature show features in excess/deficit of the
  99.9\% single-trial upper/lower limits.  Six of these features in four
  spectra meet a 3$\sigma$ multi-trial significance cut-off:

  \begin{enumerate} 

  \item {\bf \grb011211:} As discussed in \S\ref{sec:pre}, this source shows a
        highly significant excess in the $0.51 - 0.64$~\kev\ energy range,
        with a multi-trial significance in excess of 3.89\ssig.  If modeled as
        a gaussian line, the feature is unresolved with a centroid of $E =
        0.58$~\kev.  The observed line flux and equivalent width ($EW$) in the
        \pn\ data is $F = 1.1 \times 10^{-5}$~\phflux\ and $EW = 70$~eV,
        respectively.  This is one of the two most statistically significant
        excess in the present analysis.  The 1\ssig\ upper limit in the \mos\
        data is $EW = 27$~eV.  The rest frame energy for this feature is $E =
        1.6 - 2.0$~\kev, which corresponds to a K-shell transition array of
        neutral to hydrogen-like Si.  The excess at this same energy in the
        background spectrum (discussed above), however, makes this case
        somewhat suspicious.  We leave further analysis with a refined
        calibration to future work.

  \item {\bf \grb021004:} One feature in the zeroth order spectrum at $E =
        1.60 - 1.69$~\kev\ is significant at $> 99.99$\% ($> 3.72$\ssig)
        single-trial and at $99.92$\% ($3.35$\ssig) multi-trial.  This feature
        lies {\it below} the lower 99.99\% limit, which implies an absorption
        feature.  The line is again unresolved, with a centroid of
        $E = 1.64$~\kev.  The absorption line flux and equivalent widths are
        $F = -1.0 \times 10^{-6}$~\phflux\ and $EW = -91$~eV, respectively.
        The rest frame energy of $E = 5.3 - 5.6$~\kev\ does not
        correspond to any strong atomic transition.  An obvious feature at the
        same energy is not observed in the dispersed high resolution spectrum,
        with a 1\ssig\ lower limit of $-26$~eV, but is still consistent with
        the zeroth order spectrum.

  \item {\bf \grb030227:} We find six features in the \pn\ spectrum and two in
        the \mos\ spectrum which have single trial confidence levels
        $>$99.99\%; two of these have multi-trial confidences levels
        $>3\sigma$ observed in the \pn\ detector.  The centroids are at $E =
        4.00$~\kev\ and $E = 4.36$~\kev, with emission line fluxes of $F = 1.0
        \times 10^{-6}$~\phflux\ and $F = 1.1 \times 10^{-6}$~\phflux,
        respectively.  The corresponding equivalent widths are $EW = 66$~eV
        and $EW = 83$~eV.  The upper limits derived from the \mos\ spectra are
        $EW = 93$~eV and $EW = 32$~eV for the $4.00$~\kev\ and $4.36$~\kev\
        features, respectively.  The host redshift is not known and so the
        features cannot be identified with known atomic transitions.

  \item {\bf \grb040106:} Three features in the \pn\ and one in the \mos\ have
        single-trial significances in excess of 99.9\%.  The feature at $\sim
        0.6$~\kev\ is formally significant at $>3.89$~\ssig, and its structure
        is similar to the one detected in the $t > 5$~ksec of \grb011211.  The
        corresponding flux observed in the \pn\ detector is $F = 1.4 \times
        10^{-5}$~\phflux\ ($EW = 39$~eV) with a line centroid of $E =
        0.66$~\kev.  The upper limit derived from the \mos\ data is $F = 9.1
        \times 10^{-6}$~\phflux\ ($EW = 26$~eV).  The $\sim 1.33$~\kev\
        feature is an absorption line-like feature with a multi-trial
        confidence of $99.89$\% or 3.26\ssig.  The absorption flux observed in
        the \pn\ spectrum is $F = -2.2 \times 10^{-6}$~\phflux\ ($EW =
        -18$~eV), and the lower limit in the \mos\ data is $F = 9 \times
        10^{-7}$~\phflux\ ($EW = 7$~eV).  Once gain, since no redshift
        measurement of the host is available to date, we cannot identify any
        of the features with strong atomic transitions.

\end{enumerate}

\subsection{Measurements of Equivalent Hydrogen Column Density in the Host}

  \citet{gw01} analyzed eight \grb\ X-ray afterglows, of which only one of
  them (\grb970508) is part of our sample, and found that they show evidence
  for absorption in excess of the Galactic line of sight values.  If the
  absorbing material is at the redshift of the \grb, the columns are large, so
  that combined with low $A_V$ as measured from multi-color optical data,
  \citet{gw01} take this to be evidence for dense ionized material surrounding
  the progenitor.  The low $A_V$ values can be accounted for by assuming the
  material is ionized by radiation from the explosion.

  If we consider all of our measurements, 14 out of the 21 total have a host
  column with either a best-fit value of zero, or within 2\ssig\ of zero
  (Table~\ref{tbl:fits}).  In four cases, \grb970508, \grb020322, \grb020405,
  and \grb030227, there is strong evidence for non-zero host $N_H$ at $\geq 4
  \sigma$, where the error includes only statistical uncertainty, but does not
  include any estimate of the error in the assumed Galactic $N_H$.  Of the
  three most significant (\grb020322, \grb030227, and \grb031203) the host
  redshift of two (\grb020322 and \grb030227) is, unfortunately, unknown (we
  assume $z = 1$), so the measurement is problematic.  \grb031203 does appear
  to exhibit a column density that is slightly higher, though statistically
  significant, than that of \citet{dickey90}, but it is consistent with the
  dust echo measurements as described in \citet{grb031203_dust1}.  Therefore,
  in a majority of cases, we find no evidence for significant absorbing
  columns local to the burst.  Of course, columns less than a few times
  10$^{21}$~cm$^{-2}$ are difficult to measure given the significant redshifts
  for the events.

  \citet{stratta04} examined 13 afterglows observed with \sax, which includes
  the two sources presented in this paper, and found that 2 sources show
  highly-significant absorption above the Galactic value.  Both \grb 970508
  and \grb 000214 do not appear to show any obvious signatures of substantial
  host galaxy absorption, which is consistent with what we find.  The
  upper-limits they derive for the remaining 11 sources are comparable to or
  above the magnitude of the detections, as are the upper-limits we derive,
  implying that the non-detections are due to instrumental sensitivity.

\subsection{Consistency with Prior Detection Claims}

  Our analysis at first glance appears to be inconsistent with previous work,
  in that we do not confirm any prior claimed detection to be significant.  In
  addition, the most significant deviations we find have not been previously
  reported, even in spectra with published analyses.  In some cases we find
  deviations at the reported energy, but at statistical significance
  substantially lower than claimed.  In a few cases we do not see positive
  fluctuations at the claimed energy, and in three cases we find fluctuations
  at $> 3$\ssig\ that have not been reported in spectra analyzed by others.
  An important point to note is that our analyses, by no means, rule out the
  possibility of discrete features in any of the spectra.  The presence of
  weak features is certainly possible, but, in most cases, they are not
  required by the data in a model-independent way.

  In most cases we can understand the origin of the discrepancy.  In
  \grb991216 we disagree with the continuum level adopted by \citet{piro00}.
  If the continuum model lies below most of the data points (see Figure 1 of
  \citealt{piro00}), the inferred significance of the line would consequently
  be overestimated, although we find that renormalizing the continuum to the
  one adopted by \citet{piro00} yields a single-trial significance of $\sim
  3.2$\ssig, which is still lower than claimed by those authors ($\sim
  4.7$\ssig).  In most other cases, we have probably adopted different
  background regions for subtraction, so that if reported features are
  background fluctuations, we would expect to find them at different energy,
  or at different level depending on the exact background region used.  In any
  case, a robust detection should not be highly-sensitive to the exact nature
  of the assumed background spectrum.

  Our analysis results are in rough agreement with most of those published
  recently by \citet{butler04}, who have also re-analyzed a small subset of
  the data presented here using a similar statistical approach.  There are,
  however, notable differences in the quoted multi-trial significances of the
  features in the high-resolution grating spectra of \grb991216 and
  \grb020813.  They find that the claimed feature in the first order spectrum
  of \grb991216 \citep{piro00} is significant at $> 3.7$\ssig\ multi-trial,
  while our analysis yields a significance of at most $\sim 1.3$\ssig\ (see,
  also, Figure~\ref{fig991216mc_heg} and~\ref{fig991216mc_meg}).  Similarly,
  they find that the significance of the line claimed in \grb020813 by
  \citet{butler03b} is $3.5$\ssig\ multi-trial, while we find it to be
  significant at only $2.7$\ssig.  These discrepancies are most likely due to
  the fact that \citet{butler04} choose to simulate counts based on their
  spectral fit to unbinned spectra, which predicts a count rate that is {\it
    lower} than the observed value.  This means that fluctucation present in
  the spectrum appear preferentially as positive emission line-like
  fluctuations.  We believe that our approach of fixing the number of
  simulated counts to the obseved value is more robust, unless there are good
  reasons to believe {\it a priori} that the true continuum level is lower
  than what one infers from a blind fit.

  We point out that a true multi-trial significance should also account for
  the total number of spectra inspected, and depends on whether the
  observation was divided into several time segments or if data from more than
  one detector with comparable \sn\ and spectral coverage were available.  It
  appears that none of the authors have properly derived the significances of
  the features that are seen only in selected time segments of a longer
  observation and in a single detector.  For example, if a $3 \sigma$ line is
  detected only in the \pn\ spectrum aqcuired in the first 5~ksec of a 40~ksec
  exposure, then the true multi-trial probability that this is not due to a
  statistical fluctuation is $\sim 2 \times 40/8 = 10$ times lower or $\sim
  2.2 \sigma$ (the factor of $\sim 2$ comes from the fact that the two
  \mos\ cameras combined provide data of comparable sensitivity).  If the
  \sn\ for each time segment and/or detector are not identical, one must
  carefully weigh each trial by estimating its sensitivity to the detection of
  the line seen in the other dataset(s).

  In all cases, the authors have not explicitly stated all of the time
  segments used in their line searches.  As stated earlier, none of the
  features are observed in more than one detector and this has not been
  factored into their derived significances.  We have not factored these
  criteria into the quoted significance for the features that we detect
  either, since we make no claims as to the reality, interpretation, or
  ubiquity of features such as the ones that appear in \grb011211, \grb021004,
  \grb030227, or \grb040106.  Rather, we find these to be motivation for
  significantly higher signal-to-noise observations, and we call into question
  prior statements regarding the detection of features with a specific
  interpretation.
 
\section{Summary and Conclusions}
\label{section_summary}

  We have performed detailed, uniform analyses of 21 of the brightest \grb\
  X-ray afterglows, and have derived the statistical significances of
  line-like fluctuations that appear in the spectra.  We find that in most
  cases, the data are well-fit by an absorbed power-law spectral model, with
  the significance of any deviations being marginal.  This includes several
  cases where either lines, or deviations from a power-law continuum have
  previously been reported at reasonable significance.  In particular, this
  includes \grb970508, \grb970828, \grb991216, and \grb000214 where features
  from Fe emission have been reported, \grb011211, \grb010220, and \grb030227
  where a thermal emission lines from low-$Z$ elements have been claimed, and
  \grb001025, where one prior analysis found the spectrum to be better
  described by a thermal emission model than by a power law.

  There are four interesting exceptions that we find are worth attention; the
  cumulative spectrum of \grb030227, the $t > 5~\rm{ksec}$ spectrum of
  \grb011211, the zeroth order spectrum of \grb021004, and \grb040106.  In
  addition, there is one interesting case (\grb970828), where a broad excess
  is visible in all four detectors on \asca, although the formal significance
  is marginal and depends highly on the assumed background spectrum.

  The absorption line-like feature seen in the zeroth order spectrum of
  \grb021004 is formally significant at $3.35$\ssig\ multi-trial.  This
  feature corresponds to an energy of $\sim 5.5$~\kev\ in the rest frame of
  the \grb, and does not have an obvious identification.  It is not observed
  in the dispersed spectrum.  We also note that there is a local minimum in
  the effective area at this energy, which makes the feature appear more
  significant visually, as in the case for \grb991216, where the claimed
  emission line lies at a local maximum of the efficiency curve.

  The $t > 5~\rm{ksec}$ interval of the \grb011211 spectrum and the cumulative
  spectrum of \grb040106 both exhibit features at $E \sim 0.5~\rm{\kev}$.
  These are by far the most statistically significant in our sample with
  multi-trial significances exceeding $3.89$\ssig.  The host's known redshift
  of \grb011211 places the feature at a rest frame of $E \sim 1.7$~\kev, which
  may be identified as Si K fluorescence, but could easily be due to other
  elements moving at arbitrary velocities.  As noted above, the observed
  energies correspond also to a local fluctuation in the background spectrum as
  well as an instrumental absorption edge of atomic or molecular oxygen.

  In \grb030227, \citet{watson03} report the detection of multiple emission
  lines during the last $\sim 11~\rm{ksec}$ of the observation.  We do not
  confirm this result and we find that the strongest feature at $E \sim
  0.85$~\kev\ is significant at $\la 99.9$\% for a single trial.  However, the
  cumulative spectrum of the entire observation does appear to show
  fluctuations with substantially higher significance than those reported by
  \citet{watson03}.  One of these, at $E \sim 4$~\kev, is significant at
  3.54\ssig.  Unfortunately there is no measured redshift for this event, so
  the rest-frame energy is unknown.

  Although our analysis rules out all of the reported features, we do find
  some evidence, in four cases, for marginally significant spectral features
  at energies that do not suggest a clear interpretation.  We cannot rule out
  the possibility that relatively weak line features are present in the X-ray
  spectra of some, if no all, \grb s.  Given the diagnostic power of such
  detections, longer exposures of bright events are certainly warranted.  Our
  work does, however, call into question the reality of all of the features
  detected to-date, as well as the interpretation of these features as either
  iron, or specific low-$Z$ elements.

  \swift, which will observe a large number of \grb s early on, when they are
  bright, has the potential and opportunity to confirm the existence of X-ray
  line emission.  Although the effective area of the \swift\ X-ray Telescope
  (\xrt\footnote{http://www.swift.psu.edu/xrt/details/}) is a factor of $\sim
  10$ lower than that of the \xmm\ \pn\ camera and a factor of $\sim 20$ lower
  than the combined \epic\ effective area, it is capable of collecting data
  starting only minutes after the burst compared to $\sim 0.5$~days for both
  \xmm\ and \chandra.  For a flux decay time slope of $\alpha = 1 - 2$, where
  the X-ray flux $F_X \propto t^{-\alpha}$, a typical burst will be $2 - 3$
  orders of magnitude brighter during the beginning of a \swift\ observation
  and the \xrt\ data on average will be of higher statistical quality by
  roughly an order of magnitude.  The late afterglow $\ga 0.5$~days after the
  burst will still be better studied with the \ccd s on \xmm, \chandra, as
  well as \astroe.  If, however, the lines are narrow ($v \la
  1000~\rm{km~s}^{-1}$), the \chandra\ gratings, the \xmm\ \rgs, and the X-ray
  Spectrometer on \astroe\ will provide more sensitive measurements on most
  bursts.  Clearly, in any case, this is not a straightforward measurement,
  and a carefully planned observation strategy is required to maximize the
  detection probability.  Monitoring of the early X-ray flux with \swift\ and
  follow-up observations of the brightest afterglows with \xmm, \chandra, and
  \astroe\ will minimize the risk of obtaining an underexposed spectrum,
  although this does not by any means guarantee the detection of lines.  Of
  course, the ultimate goal is to take one step further and actually use X-ray
  spectroscopic diagnostics to infer important physical information about the
  burst.  Note, however, that the physical conditions of the X-ray emission
  regions during the early afterglow (accessible only with \swift) can be
  vastly different from those approximately a day after the burst, and it is
  not clear whether a direct comparison of measurements will be meaningful.

  We finally point out that spectral identifications is also not a
  straightforward task.  This is obvious when only a single significant
  feature is detected, but is also true even when multiple lines are detected
  in the spectrum.  The energies of the hydrogen-like Ly$\alpha$ lines of
  neighboring abundant metals, for example, are spaced roughly equally and
  they can be misinterpreted or their identifications can at best be
  ambiguous.  If a significant amount of iron is present, the spectrum may
  exhibit broad bumps that may dominate over narrow isolated lines from other
  elements and may not appear as prominent line emission.  Although arbitrary
  bulk velocities can be invoked to match transitions of specific elements,
  any constraint on the burst redshift (e.g., from optical absorption/emission
  line measurements) is extremely valuable.

  If the combination of {\em Swift}, \xmm, \chandra, and \astroe\ is unable to
  find similar features in spectra with large numbers of counts, then the
  $\sim 3$\ssig\ detections found here are likely spurious.  If this is the
  case, tight upper limits on the line fluxes and equivalent widths will
  provide invaluable constraints on the geometry and chemical abundances of
  the burst environment, which can then be used as quantitative tests of
  \grb\ afterglow models.

  \acknowledgements The authors thank Derek B. Fox for useful discussions and
  Nicola Masetti for carefully reading the manuscript.  The authors also thank
  the anonymous referee for helping improve the presentation of the paper.
  {\small} {\small MS} was supported by \nasa\ through \chandra\ Postdoctoral
  Fellowship Award Number {\small PF}1-20016 issued by the \chandra\ X-ray
  Observatory Center, which is operated by the Smithsonian Astrophysical
  Observatory for and behalf of \nasa\ under contract {\small NAS}8-39073.
  {\small FAH} acknowledges support from a Presidential Early Career Award.
  This research has made use of data obtained from the High Energy
  Astrophysics Science Archive Research Center ({\small HEASARC}), provided by
  \nasa's Goddard Space Flight Center.

\clearpage

\begin{deluxetable}{cccrcl}
  \tabletypesize{\scriptsize}
  \tablewidth{0pt}
  \tablecaption{Observations of Bright GRB X-ray Afterglows\label{tbl:sample}}
  \tablehead{
    \colhead{Burst} &
    \colhead{Mission} &
    \colhead{Instrument(s)} & 
    \colhead{$T_{\rm{start}}$\tablenotemark{a}} &
    \colhead{Exposure\tablenotemark{b}} &
    \colhead{$z$\tablenotemark{c}}
  }
  \startdata
    970508  & {\it Beppo-SAX}  & LECS \& MECS & 0.4340  & 15.4 & 0.835 \\
    970828  & {\it ASCA}    & SIS \& GIS   & 1.1906  & 33.8 & 0.9578 \\
    991216  & \chandra & HETG/ACIS-S  & 1.5215  & 9.65 & 1.02 \\
    000210  & \chandra & ACIS-S       & 0.7778  & 8.93 & 0.8463 \\
    000214  & {\it Beppo-SAX}  & LECS \& MECS & 0.5237  & 43.5 & \nodata \\
    000926  & \chandra & ACIS-S       & 2.6861  & 10.2 & 2.066 \\
    001025  & \xmm     & EPIC-PN      & 1.9092  & 23.7 & \nodata \\
    \nodata & \nodata  & EPIC-MOS     & 1.8814  & 37.1 & \nodata \\
    010220  & \xmm     & EPIC-PN      & 0.6247  & 12.1 & \nodata \\
    011030  & \chandra & ACIS-S       & 10.4592 & 46.6 & \nodata \\
    011211  & \xmm     & EPIC-PN      & 0.5029  & 26.8 & 2.140 \\
    \nodata & \nodata  & EPIC-MOS     & 0.4851  & 31.2 & \nodata \\
    020321  & \xmm     & EPIC-PN      & 0.4289  & 36.3 & \nodata \\
    \nodata & \nodata  & EPIC-MOS     & 0.4119  & 40.0 & \nodata \\
    020322  & \xmm     & EPIC-PN      & 0.6428  & 23.8 & \nodata \\
    \nodata & \nodata  & EPIC-MOS     & 0.6222  & 28.2 & \nodata \\
    020405  & \chandra & LETG/ACIS-S  & 1.6825  & 50.6 & 0.690 \\
    020813  & \chandra & HETG/ACIS-S  & 0.7367  & 76.9 & 1.254 \\
    021004  & \chandra & HETG/ACIS-S  & 0.8533  & 86.7 & 2.323 \\
    030226  & \chandra & ACIS-S       & 1.5436  & 36.4 & 1.986 \\
    030227  & \xmm     & EPIC-PN      & 0.3604  & 33.3 & \nodata \\
    \nodata & \nodata  & EPIC-MOS     & 0.3440  & 35.9 & \nodata \\
    030328  & \chandra & LETG/ACIS-S  & 0.6387  & 92.7 & 1.520\\
    030329  & \xmm     & EPIC-PN      & 37.0577 & 29.6 & 0.1685 \\
    \nodata & \nodata  & EPIC-MOS     & 37.0423 & 32.4 & \nodata \\
    \nodata & \nodata  & EPIC-PN      & 60.3962 & 38.8 & \nodata \\
    \nodata & \nodata  & EPIC-MOS     & 60.3807 & 46.5 & \nodata \\
    031203  & \xmm     & EPIC-PN      & 0.2723  & 50.7 & 0.105 \\
    \nodata & \nodata  & EPIC-MOS     & 0.2567  & 57.4 & \nodata \\
    \nodata & \nodata  & EPIC-PN      & 2.8549  & 36.5 & \nodata \\
    \nodata & \nodata  & EPIC-MOS     & 2.8414  & 34.4 & \nodata \\
    040106  & \xmm     & EPIC-PN      & 0.2552  & 37.1 & \nodata \\
    \nodata & \nodata  & EPIC-MOS     & 0.2344  & 42.6 & \nodata \\
  \enddata

  \tablenotetext{a}{start of the X-ray observation in days since the time of
                    the GRB in the observers' frame}

  \tablenotetext{b}{total exposure in ksec used in the analysis}

  \tablenotetext{c}{spectroscopic redshift of the host galaxy inferred from
                    non-X-ray data}

\end{deluxetable}

\begin{deluxetable}{cccccll}
  \tabletypesize{\scriptsize}
  \rotate
  \tablewidth{0pt}
  \tablecaption{Summary of Reported Emission Lines in the Literature\label{tbl:reported}}
  \tablehead{
    \colhead{Burst} &
    \colhead{Epoch/order} &
    \colhead{Line Energy (keV)} &
    \colhead{Line Width (keV)} &
    \colhead{Model/Identification} &
    \colhead{Significance ($\sigma$)\tablenotemark{a}} &
    \colhead{Reference} \\
  }

  \startdata
  970508   & 1a
           & 3.5
           & $\la 0.5$
           & \fekal\
           & 99.3\% (2.7$\sigma$)
           & \citet{piro99} \\
  970828   & B
           & 5.03\ud{0.23}{0.31}
           & $0.31^{+0.38}_{-0.31}$
           & Fe XXVI Ly$\alpha$ or RRC\tablenotemark{b}
           & 98.3\% (2.4$\sigma$)
           & \citet{yoshida99, yoshida01} \\
  991216   & $|m| = 1$
           & $3.49 \pm 0.06$
           & $0.23 \pm 0.07$
           & Fe XXVI Ly$\alpha$
           & 99.99974\% (4.7$\sigma$)
           & \citet{piro00} \\
  \nodata  & $m = 0$
           & $4.4 \pm 0.5$
           & \nodata
           & Fe XXVI RRC
           & 99.5\% (2.8$\sigma$)
           & \citet{piro00} \\
  000214   & \nodata
           & $4.7 \pm 0.2$
           & unresolved
           & \fekal\
           & 99.85\% (3.2$\sigma$)
           & \citet{antonelli00} \\
  001025   & \nodata
           & $0.80^{+0.04}_{-0.05}$
           & $0.10 \pm 0.03$
           & Mg XII
           & 99.94\% (3.4\ssig)
           & \citet{watson02a} \\
  \nodata  & \nodata
           & $1.16 \pm 0.05$
           & \nodata
           & Si XIV
           & 99.92\% (3.4\ssig)
           & \nodata \\
  \nodata  & \nodata
           & $1.64 \pm 0.07$
           & \nodata
           & S XVI
           & 99.92\% (3.4\ssig)
           & \nodata \\
  \nodata  & \nodata
           & $2.2 \pm 0.1$
           & \nodata
           & Ar XVIII
           & 98\% (2.3\ssig)
           & \nodata \\
  \nodata  & \nodata
           & 4.7\ud{0.8}{0.4}
           & \nodata
           & Ni XXVIII
           & 88\% (1.6\ssig)
           & \nodata \\
  010220   & \nodata
           & $\sim 3.9$
           & unresolved
           & Ni XXVIII Ly$\alpha$
           & 99.0\% (2.6$\sigma$)
           & \citet{watson02a} \\
  011211   & ($t<5$~ksec)
           & $0.44 \pm 0.04$
           & unresolved
           & Mg XI
           & 99.97\% (3.6\ssig)\tablenotemark{d}
           & \citet{reeves03} \\
  \nodata  & \nodata
           & $0.71 \pm 0.02$
           & \nodata
           & Si XIV
           & \nodata
           & \nodata \\
  \nodata  & \nodata
           & $0.88 \pm 0.02$
           & \nodata
           & S XVI
           & \nodata
           & \nodata \\
  \nodata  & \nodata
           & $1.22 \pm 0.03$
           & \nodata
           & Ar XVIII
           & \nodata
           & \nodata \\
  \nodata  & \nodata
           & $1.46 \pm 0.07$
           & \nodata
           & Ca XX
           & \nodata
           & \nodata \\
  020813   & $|m| = 1$
           & $1.31 \pm 0.01$
           & $0.008 \pm 0.004$
           & S XVI Ly$\alpha$
           & 99.90\% (3.3$\sigma$)
           & \citet{butler03b} \\
  030227   & (last 11~ksec)
           & 0.62\ud{0.03}{0.02}
           & unresolved
           & Mg XII
           & 97\% (2.2\ssig)
           & \citet{watson03} \\
  \nodata  & \nodata
           & 0.86\ud{0.02}{0.03}
           & \nodata
           & Si XIV
           & 99.98\% (3.8\ssig)
           & \nodata \\
  \nodata  & \nodata
           & $1.11 \pm 0.02$
           & \nodata
           & S XVI
           & 99.96\% (3.5\ssig)
           & \nodata \\
  \nodata  & \nodata
           & 1.35\ud{0.04}{0.03}
           & \nodata
           & Ar XVII\tablenotemark{f} or XVIII
           & 92\% (1.7\ssig)
           & \nodata \\
  \nodata  & \nodata
           & $1.66 \pm 0.04$
           & \nodata
           & Ca XIX or XX
           & 99\% (2.5\ssig)
           & \nodata \\
  \nodata  & \nodata
           & 1.67\ud{0.01}{0.03}
           & \nodata
           & \fekal\
           & 99.86\% (3.2\ssig)
           & \citet{mgt+03} \\
  \enddata

  \tablenotetext{a}{Probability of producing the observed data from noise
                    only, as reported in the reference. The $\sigma$ value is
                    the equivalent gaussian number of standard deviations
                    corresponding to the stated confidence level.}

  \tablenotetext{b}{Radiative recombination continuum}

  \tablenotetext{c}{The lines are collectively interpreted as due to thermal
                    line emission at a redshift of $z=0.7$\ud{0.3}{0.1}.}

  \tablenotetext{d}{Combined multi-trial significance of the five lines,
                    interpreted as due to thermal line emission at a redshift
                    of $z=1.91 \pm 0.06$.}

  \tablenotetext{e}{The five lines are interpreted as due to thermal line
                    emission at a redshift of $z=1.39$\ud{0.03}{0.06}.}

  \tablenotetext{f}{Correcting a typo in \citet{watson03}.}

\end{deluxetable}

\begin{deluxetable}{rcrrlccccc}
  \tabletypesize{\scriptsize}
  \tablewidth{0pt}
  \tablecaption{Continuum Spectral Parameters\label{tbl:fits}}
  \tablehead{
    \colhead{Burst} &
    \colhead{Epoch/order} &
    \colhead{$C_{\rm{tot}}$\tablenotemark{a}} &
    \colhead{$C_{\rm{bkg}}$\tablenotemark{b}} &
    \colhead{$N_{\rm{H}}^{\rm{Gal}}$\tablenotemark{c}} &
    \colhead{$N_{\rm{H}}^{\rm{host}}$\tablenotemark{d}} & 
    \colhead{$\Gamma$\tablenotemark{e}} &
    \colhead{$F_{\rm{1 keV}}$\tablenotemark{f}} &
    \colhead{$\chi^2/\nu$} &
    \colhead{$P(\chi,\nu)$\tablenotemark{g}}
  }
  \startdata
    970508 & 1a  &  137 &  34 & 0.051 & $10^{+6}_{-4}$ &
  $3.1^{+0.5}_{-0.4}$ & $2.6^{+0.5}_{-0.4} \times 10^{-3}$ & 2.6/5 & 0.76 	\\
    \nodata    & 1b  &  188 &  56 & 0.051 &  $7^{+11}_{-7}$  &
  $2.0^{+1.0}_{-0.9}$ & $3.5^{+0.4}_{-0.3} \times 10^{-4}$ & 2.4/6 & 0.88 	\\
    970828 & A+C & 1922 & 953 & 0.036 & $1.5^{+1.2}_{-1.0}$ &
  $2.4^{+0.4}_{-0.3}$ & $2.0^{+1.0}_{-0.6} \times 10^{-4}$ & 100/86 & 0.14 	\\
    \nodata    &  B  & 1215 & 515 & 0.036 & $0.0^{+0.8}_{-0.0}$ &
  $2.1^{+0.3}_{-0.3}$ & $1.7^{+0.6}_{-0.6} \times 10^{-4}$ &  58/54 & 0.33 	\\
    991216 & $|m| = 1$ &  762 & 13 & 0.20 & $0.8^{+0.4}_{-0.3}$ &
  $1.6^{+0.1}_{-0.1}$ & $8.2^{+0.5}_{-0.5} \times 10^{-4}$ & 153/130 & \nodata  \\
    \nodata & $m = 0$ &  604 & 4 & 0.20 & $0.0^{+0.3}_{-0.0}$ &
  $1.7^{+0.1}_{-0.1}$ & $6.7^{+0.5}_{-0.5} \times 10^{-4}$ & 46/32 & 0.05 	\\
    000210 & \nodata &  552 & 2 & 0.025 & $3.7^{+1.1}_{-1.1}$ &
  $1.9^{+0.1}_{-0.1}$ & $8.8^{+0.6}_{-0.7} \times 10^{-5}$ &  26/21 & 0.22 	\\
    000214 & \nodata &  378 & 134 & 0.058 & $0.0^{+1.2}_{-0.0}$ &
  $2.0^{+0.3}_{-0.3}$ & $1.4^{+0.5}_{-0.4} \times 10^{-4}$ &  28/18 & 0.07 	\\
    000926 & \nodata &  267 &   1 & 0.027 & $0.0^{+0.4}_{-0.0}$ &
  $1.9^{+0.3}_{-0.2}$ & $3.2^{+0.4}_{-0.4} \times 10^{-5}$ &  15/13 & 0.20 	\\
    001025 & \nodata & 1249 & 313 & 0.061 & $0.19^{+0.07}_{-0.06}$ &
  $2.6^{+0.4}_{-0.2}$ & $2.5^{+0.7}_{-0.5} \times 10^{-5}$ &  43/63 & 0.98 	\\
    010220 & \nodata &  247 & 124 & 0.86  & $0.1^{+1.3}_{-0.1}$ &
  $1.8^{+1.0}_{-0.4}$ & $1.9^{+0.7}_{-0.6} \times 10^{-5}$ & 6.5/9  & 0.69 	\\
    011030 & \nodata &  376 &   6 & 0.10 & $0.0^{+0.1}_{-0.0}$ &
  $1.5^{+0.3}_{-0.2}$ & $1.1^{+0.3}_{-0.1} \times 10^{-5}$ & 3.1/14 & 1.00 	\\
    011211 & $t<5$ ksec &  936 &  58 & 0.042 & $0.1^{+0.4}_{-0.1}$ &
  $2.2^{+0.2}_{-0.2}$ & $7.8^{+1.0}_{-0.7} \times 10^{-5}$ & 38/40  & 0.56 	\\
    \nodata    & $t>5$ ksec & 3046 & 294 & 0.042 & $0.2^{+0.2}_{-0.2}$ &
  $2.3^{+0.1}_{-0.2}$ & $4.6^{+0.5}_{-0.1} \times 10^{-5}$ & 99/129 & 0.98 	\\
    020321 & \nodata & 1664 &  487 & 0.082 & $0.02^{+0.06}_{-0.02}$ &
  $2.1^{+0.2}_{-0.3}$ & $9.9^{+2.3}_{-2.0} \times 10^{-6}$ & 33/45  & 0.91 	\\
    020322 & \nodata & 5791 & 1252 & 0.046 & $0.20^{+0.03}_{-0.03}$ &
  $2.3^{+0.1}_{-0.1}$ & $1.1^{+0.1}_{-0.1} \times 10^{-4}$ & 228/241 & 0.72 	\\
    020405 & $|m| = 1$ & 1242 & 84 & 0.043 & $0.4^{+0.1}_{-0.1}$ &
  $1.7^{+0.1}_{-0.1}$ & $2.6^{+0.1}_{-0.1} \times 10^{-4}$ & 1364/1318 &  \nodata \\
    \nodata & $m = 0$ & 604 & 4 & 0.043 & $0.6^{+0.3}_{-0.1}$ &
  $2.0^{+0.2}_{-0.1}$ & $3.7^{+0.8}_{-0.7} \times 10^{-4}$ & 32/33 & 0.49 	\\
    020813 & $|m| = 1$ & 4659 & 117 & 0.075 & $0.0^{+0.1}_{-0.01}$ &
  $1.8^{+0.1}_{-0.1}$ & $6.3^{+0.2}_{-0.2} \times 10^{-4}$ & 455/348 & \nodata 	\\
    \nodata & $m = 0$  & 3679 & 77 & 0.075 & $0.0^{+0.1}_{-0.0}$ &
  $1.8^{+0.1}_{-0.1}$ & $5.2^{+0.2}_{-0.2} \times 10^{-4}$ & 145/142 & 0.42 	\\
    021004 & $|m| = 1$ & 1634 & 109 & 0.011 & $0.0^{+0.1}_{-0.0}$ &
  $2.1^{+0.1}_{-0.1}$ & $2.0^{+0.1}_{-0.1} \times 10^{-4}$ & 440/348 & \nodata 	\\
    \nodata & $m = 0$ & 1252 & 5 & 0.011 & $0.0^{+0.3}_{-0.0}$ &
  $2.0^{+0.1}_{-0.1}$ & $1.6^{+0.1}_{-0.1} \times 10^{-4}$ & 77/52 & 0.02 	\\
    030226 & \nodata &  345 & 1 & 0.016 & $0.0^{+0.5}_{-0.0}$ &
  $2.1^{+0.2}_{-0.2}$ & $1.3^{+0.1}_{-0.1} \times 10^{-5}$ & 11/13 & 0.59 	\\
    030227 & total & 13676 & 2348 & 0.22 & $0.60^{+0.04}_{-0.04}$ &
  $1.97^{+0.02}_{-0.02}$ & $2.34^{+0.04}_{-0.04} \times 10^{-4}$ & 303/278 & 0.12 	\\
    \nodata & last 11 ksec & 2777 & 171 & 0.22 & $0.7^{+0.2}_{-0.2}$ &
  $1.9^{+0.1}_{-0.1}$ & $1.8^{+0.2}_{-0.2} \times 10^{-4}$ & 115/124 & 0.70 	\\
    030328 & $|m| = 1$ & 562 & 131 & 0.043 & $0.0^{+0.2}_{-0.0}$ &
  $1.9^{+0.1}_{-0.1}$ & $6.6^{+0.5}_{-0.5} \times 10^{-5}$ & 1172/1318 &  \nodata \\
    \nodata & $m = 0$ & 529 & 18 & 0.043 & $0.18^{+0.17}_{-0.15}$ &
  $2.1^{+0.2}_{-0.1}$ & $7.6^{+1.2}_{-1.2} \times 10^{-5}$ & 30/21 & 0.09 	\\
    030329 & I & 665 & 80 & 0.021 & $0.00^{+0.03}_{-0.00}$ &
  $2.0^{+0.2}_{-0.2}$ & $7.4^{+0.8}_{-0.9} \times 10^{-6}$ & 37/37 & 0.48 	\\
    \nodata & II & 543 & 143 & 0.021 & $0.02^{+0.07}_{-0.02}$ &
  $2.0^{+0.3}_{-0.2}$ & $3.9^{+0.9}_{-1.1} \times 10^{-6}$ & 34/31 & 0.32 	\\
    031203 & I & 7012 & 647 & 0.59 & $0.29^{+0.04}_{-0.03}$ &
  $1.8^{+0.1}_{-0.1}$ & $1.1^{+0.1}_{-0.1} \times 10^{-4}$ & 218/208 & 0.32 	\\
    \nodata & II & 1194 & 381 & 0.59 & $0.80^{+0.17}_{-0.16}$ &
  $1.9^{+0.1}_{-0.1}$ & $5.8^{+0.4}_{-0.4} \times 10^{-5}$ & 41/50 & 0.81 	\\
    040106 & \nodata & 23190 & 1420 & 0.086 & $0.00^{+0.02}_{-0.00}$ &
  $1.50^{+0.02}_{-0.01}$ & $1.88^{+0.05}_{-0.05} \times 10^{-4}$ & 622/603 & 0.29 \\
  \enddata

  \tablenotetext{a}{total number of source and background counts detected
  within the source extraction region}
  \tablenotetext{b}{estimated number of background counts within the
  extraction region}
  \tablenotetext{c}{Galactic column density in multiples of
  $10^{22}~\rm{cm}^{-2}$ from \citet{dickey90}}
  \tablenotetext{d}{inferred column density at the redshift of the host galaxy
  in multiples of $10^{22}~\rm{cm}^{-2}$}
  \tablenotetext{e}{photon index}
  \tablenotetext{f}{continuum flux at 1~keV in units of
  $\rm{ph~cm}^{-2}~\rm{s}^{-1}~\rm{keV}^{-1}$}
  \tablenotetext{g}{null hypothesis probability}
\end{deluxetable}

\clearpage

\begin{deluxetable}{crrrr}
  \tablewidth{0pt}
  \tablecaption{Parameters for the Matched Filter Kernel\tablenotemark{a}
    \label{tbl:kernel}}
  \tablehead{
    \colhead{Instrument} & 
    \colhead{$c_0$} &
    \colhead{$c_1$} &
    \colhead{$c_2$} &
    \colhead{$c_3$}
  }
  \startdata
   SIS  & $4.43 \times 10^{-2}$ & $1.28 \times 10^{-2}$ & $2.52 \times 10^{-5}$ & \nodata \\
   GIS  & $4.13 \times 10^{-2}$ & $3.82 \times 10^{-2}$ & $-1.60 \times 10^{-3}$ & \nodata \\
   MECS & $5.20 \times 10^{-2}$ & $3.32 \times 10^{-2}$ & $-1.12 \times 10^{-3}$ & \nodata \\
   ACIS & $3.96 \times 10^{-2}$ & $5.28 \times 10^{-3}$ & $-1.31 \times 10^{-4}$ & \nodata \\
   MOS  & $2.50 \times 10^{-2}$ & $8.77 \times 10^{-3}$ & $-2.79 \times 10^{-4}$ & \nodata \\
   PN   & $3.42 \times 10^{-2}$ & $6.63 \times 10^{-3}$ & $-1.46 \times 10^{-4}$ & \nodata \\
   HEG  & $1.00 \times 10^{-2}$ & $-1.41 \times 10^{-4}$ & $1.46 \times 10^{-5}$ & $-1.963 \times 10^{-7}$ \\
   MEG  & $1.89 \times 10^{-2}$ & $-3.39 \times 10^{-4}$ & $2.98 \times 10^{-5}$ & $-4.110 \times 10^{-7}$ \\
   LEG  & $1.95 \times 10^{-2}$ & \nodata & \nodata & \nodata \\
  \enddata
  \tablenotetext{a}{The $c_i$ represent coefficients for the $i$ order term in
    the polynomial for $\sigma(E)$ in keV and $\sigma(\lambda)$ in \AA\ for
    non-dispersive and dispersive instruments, respectively.  See Equ.\
    \ref{eq:sigmaE} and \ref{eq:sigmaL} for details.}
\end{deluxetable}

\clearpage

\begin{deluxetable}{ccccrr}
  \tabletypesize{\scriptsize}
  \tablewidth{0pt}
  \tablecaption{Summary of Single- and Multi-trial Significances of Features
  with Single-Trial Significances $>$99.9\%\tablenotemark{a}\label{tbl:search_results}}
  \tablehead{
    \colhead{Burst} &
    \colhead{Instrument} &
    \colhead{Epoch/order} &
    \colhead{Energy Range (keV)\tablenotemark{b}} &
    \colhead{Single-trial} & 
    \colhead{Multi-trial}
  }
  \startdata
    991216  & MEG    & $|m| = 1$    & 1.317 -- 1.326 & $99.98$\% & $89.20$\% \\
    000214  & MECS   & \nodata      & 4.60 -- 4.80 & $99.92$\% & $99.03$\% \\
    001025  & MOS    & \nodata      & 2.60 -- 2.63 & $>99.99$\% & $99.23$\% \\
    011211  & PN     & $t < 5$~ksec & 0.88 -- 0.91 & $99.96$\% & $89.09$\% \\
    \nodata & PN     & $t > 5$~ksec & 0.51 -- 0.64 & $>99.99$\% & $>99.99$\% \\
    020321  & MOS    & \nodata  & 1.01 -- 1.03 & $99.95$\% & $94.85$\% \\
    \nodata & MOS    & \nodata  & 7.58 -- 7.66 & $>99.99$\% & $99.36$\% \\
    020322  & PN     & \nodata  & 6.15 -- 6.18 & $99.93$\% & $89.15$\% \\
    \nodata & MOS    & \nodata  & 2.09 -- 2.12 & $99.92$\% & $80.13$\% \\
    \nodata & MOS    & \nodata  & 4.97 -- 5.08 & $99.90$\% & $72.26$\% \\
    020405  & LETG   & $|m| = 1$ & 0.733 -- 0.740 & $99.97$\% & $98.23$\% \\
    \nodata & LETG   & $|m| = 1$ & 0.941 -- 0.953 & $99.96$\% & $96.98$\% \\
    020813  & HEG    & $|m| = 1$ & 1.303 -- 1.322 & $>99.99$\% & $99.40$\% \\
    021004  & ACIS-S & $|m| = 0$ & 1.60 -- 1.69 & $>99.99$\% & $99.92$\% \\
    030227  & PN     & total    & 1.51 -- 1.55 & $>99.99$\% & $98.92$\% \\
    \nodata & PN     & total    & 1.66 -- 1.69 & $>99.99$\% & $98.93$\% \\
    \nodata & PN     & total    & 1.80 -- 1.96 & $>99.99$\% & $98.99$\% \\
    \nodata & PN     & total    & 2.03 -- 2.11 & $>99.99$\% & $98.93$\% \\
    \nodata & PN     & total    & 3.95 -- 4.06 & $>99.99$\% & $99.96$\% \\
    \nodata & PN     & total    & 4.33 -- 4.39 & $>99.99$\% & $99.97$\% \\
    \nodata & MOS    & total    & 1.85 -- 1.88 & $>99.99$\% & $99.25$\% \\
    \nodata & MOS    & total    & 2.62 -- 2.66 & $>99.99$\% & $99.50$\% \\
    040106  & PN     & \nodata  & 0.54 -- 0.70 & $>99.99$\% & $>99.99$\% \\
    \nodata & PN     & \nodata  & 1.30 -- 1.37 & $>99.99$\% & $99.89$\% \\
    \nodata & PN     & \nodata  & 1.95 -- 1.96 & $>99.99$\% & $98.58$\% \\
    \nodata & PN     & \nodata  & 4.07 -- 4.11 & $99.94$\% & $93.49$\% \\
  \enddata

  \tablenotetext{a}{Only discrete features that are in excess of $99.9$\%
    single-trial significance ($3.29\sigma$) are listed with the exception of
    GRB030227, where we adopt a lower cut-off of $99.99$\% ($3.89\sigma$)
    owing to the relatively large number of features that lie beyond the
    $99.9$\% upper and lower limits.}

  \tablenotetext{b}{For features with single trial significances between
  99.9\% and 99.99\%, this corresponds to the range in energy where the data
  exceed the 99.9\% limits.  For features with single trial significances
  above 99.99\%, this corresponds to the range in energy where the data exceed
  the 99.99\% limits}

\end{deluxetable}


\cleardoublepage

\clearpage

\clearpage

\clearpage

\clearpage

\clearpage

\clearpage

\clearpage

\vspace{1cm}

\clearpage

\clearpage

\vspace{1cm}

\clearpage

\clearpage

\clearpage

\clearpage

\clearpage

\clearpage

\clearpage

\clearpage


\clearpage

\clearpage

\clearpage

\clearpage

\clearpage

\clearpage

\clearpage

\clearpage

\clearpage

\clearpage

\clearpage

\clearpage

\clearpage

\clearpage

\clearpage

\clearpage

\clearpage

\clearpage

\clearpage

\clearpage

\clearpage

\clearpage

\clearpage

\clearpage

\clearpage

\clearpage

\end{document}